\documentclass[useAMS,usenatbib]{mn2e}

\usepackage{graphicx}
\usepackage{lscape}
\usepackage{aalongtable}
\usepackage{ulem}

\title[s-Process in Low Metallicity Stars. III.]
{s-Process in Low Metallicity Stars. \\
III. Individual analysis of CEMP-$s$ and CEMP-$s/r$ with
AGB models.}
\author[S. Bisterzo, R. Gallino, O. Straniero, S. Cristallo, F. K\"appeler]
{S. Bisterzo$^{1}$\thanks{E-mail: bisterzo@ph.unito.it (AVR); sarabisterzo@gmail.com (ANO)}, 
R. Gallino$^{1,2}$,
O. Straniero$^{2}$,
S. Cristallo$^{3}$
and 
F. K\"appeler$^{4}$\\
$^{1}$Dipartimento di Fisica Generale, Universit\`{a} 
   di Torino, Via P. Giuria 1, 10125 Torino, Italy\\
$^{2}$INAF Osservatorio Astronomico di Collurania, via M. 
   Maggini, 64100 Teramo, Italy\\
$^{3}$Departamento de Fisica Teorica y del Cosmos, Universidad
de Granada, Campus de Fuentenueva, 18071 Granada, Spain\\
$^{4}$Karlsruhe Institute of Technology, Campus Nord, 
Institut f$\ddot{\rm u}$r Kernphysik, D-76021 Karlsruhe, Germany\\}

\begin{document}

\date{Accepted 1988 December 15. Received 1988 December 14; in original form 1988 October 11}

\pagerange{\pageref{firstpage}--\pageref{lastpage}} \pubyear{2002}

\maketitle

\label{firstpage}

\begin{abstract}

We provide an individual analysis of 94 carbon enhanced metal-poor stars 
showing an $s$-process enrichment (CEMP-$s$) collected from the literature. 
The $s$-process enhancement observed in these stars is ascribed to mass 
transfer by stellar winds in a binary system
from a more massive companion evolving faster toward the asymptotic 
giant branch (AGB) phase. The theoretical AGB nucleosynthesis models 
have been presented in Paper I.
Several CEMP-$s$ stars show an enhancement in both $s$ and $r$-process
elements (CEMP-$s/r$). In order to explain the peculiar abundances
observed in CEMP-$s/r$ stars, we assume that the molecular cloud from which CEMP-$s$ 
formed was previously enriched in $r$-elements by Supernovae pollution.
\\
A general discussion and the method adopted in order to interpret the observations 
 have been provided in Paper II.
We present in this paper a detailed study of spectroscopic observations of
individual stars.
We consider all elements from carbon to bismuth, with particular 
attention to the three $s$-process peaks, ls (Y, Zr), hs (La, Nd, Sm) and Pb,
and their ratios [hs/ls] and [Pb/hs]. The presence of an initial $r$-process 
contribution may be typically evaluated by the [La/Eu] ratio.
We found possible agreements between theoretical predictions and spectroscopic 
data. In general, the observed [Na/Fe] (and [Mg/Fe]) provide information on the AGB
initial mass, while [hs/ls] and [Pb/hs] are mainly indicators of the $s$-process efficiency.
A range of $^{13}$C-pocket strengths is required to interpret the
observations. 
However, major discrepancies between models and observations exist. 
We highlight star by star the agreements and the main problems encountered and, 
when possible, we suggest potential indications for further studies.
These discrepancies provide starting 
points of debate for unsolved problems in which spectroscopic and theoretical 
studies may intervene.

\end{abstract}



\section{Introduction}\label{intro}

 This paper is the third of a series of papers dedicated to the 
analysis of carbon rich metal-poor stars showing a $slow$ neutron capture process 
($s$-process) enhancement (CEMP-$s$).
The peculiar $s$-enrichment observed in these stars is currently explained by mass transfer through stellar winds 
from a more massive companion while suffering the thermally pulsing asymptotic 
giant branch (TP-AGB) phase, then evolved toward a (now invisible) white dwarf.
{This is because} CEMP-$s$ are main-sequence or giant stars, far from the AGB phase.
Then, the binary system scenario is the most plausible explanation
for the $s$-process enhancement detected in these old, low-metallicity 
stars ([Fe/H] $\la$ $-$2) of low initial mass 
($M$ $<$ 0.9 $M_\odot$). 
For a detailed discussion of $s$-process nucleosynthesis in AGB 
stars we refer to \citet{busso99}, \citet{herwig05}, \citet{straniero06}, \citet{kaeppeler11rmp}.
\\
A complete description of the AGB models adopted here has been presented by 
Bisterzo et al. (2010a, hereafter Paper I) and Bisterzo et al. (2011,
hereafter Paper II). Theoretical results are obtained 
with a post-process nucleosynthesis method \citep{gallino98}, based on full 
evolutionary FRANEC (Frascati Raphson-Newton Evolutionary Code, 
\citealt{chieffi89}) models, following the prescriptions by \citet{straniero03},
 as described in Paper I.
AGB models with initial masses $M$ = 1.3, 1.4, 1.5, 2 $M_\odot$, metallicities 
$-$3.6 $\leq$ [Fe/H] $\la$ $-$1.5, and a range of $^{13}$C-pockets are 
adopted. 
 The $^{13}$C-pocket is a thin $^{13}$C-rich layer, which forms when few 
protons diffuse after a thermal instability, once the bottom of the 
convective envelope penetrates into the top layers of the radiative He-intershell (the 
zone between the H- and He-shells). This recurrent phenomenon is called third 
dredge-up (TDU). At H-shell re-ignition, protons are captured by the abundant $^{12}$C
and the $^{13}$C-pocket is produced at the top of the He-intershell.
Subsequently, when the temperature increases up to 1 $\times$ 10$^8$ K, 
neutrons are released in radiative conditions via the $^{13}$C($\alpha$, 
n)$^{16}$O reaction. This is the major neutron source in low mass AGB 
stars. The physical environments involved in the formation of the $^{13}$C-pocket 
are still affected by large uncertainties. According to the description of our AGB
models provided in Paper I, we assumed a range of $^{13}$C-pocket strengths. 
%
This assumption is
based on the spectroscopic 
observations of different stellar populations (MS, S, C(N), Ba stars, CEMP-$s$ stars):
for a given metallicity, they show a spread that may be reproduced by AGB models 
if a range of $s$-process efficiencies is assumed 
\citep{busso95,busso01,abia01,abia02,SCG08,kaeppeler11rmp}. 
 Starting from the case ST defined by \citet{gallino98} in order to 
reproduce the solar main $s$-process component \citep{arlandini99}, we multiply 
or divide the $^{13}$C (and $^{14}$N) abundance in the pocket by different 
factors: from a minimum fraction of ST, 
below which the $s$-process contribution becomes negligible\footnote{Note 
that the minimum choice of the $^{13}$C-pocket depends
on the metallicity. Indeed, the $s$-process efficiency increases by decreasing 
the metallicity because of the lower number of iron seeds (see Paper I).}, up to a maximum
(ST$\times$2), because further proton ingestion gives rise to $^{14}$N instead of $^{13}$C. 
 Note that in our approximation, we did not account for a likely
decrease in mass of the $^{13}$C-pocket with the number of TDUs, as obtained by
recent FRANEC models (see \citealt{cristallo11}, and references therein). 
They introduced a physical algorithm for the treatment of the transition region 
between the convective envelope and the radiative H-exhausted core, based on a 
non-diffusive mixing scheme \citep{straniero06,cristallo09}.
This allows the formation of a tiny $^{13}$C-pocket,
which decreases in mass with the number of TDUs. 
The interpretation of the $s$-process spread observed in different stars
 with new FRANEC models \citep{cristallo11} is under investigation.
\\
 A second neutron source is
driven in the convective thermal pulse by the partial activation
of the $^{22}$Ne($\alpha$, n)$^{25}$Mg reaction.
\\
 When neutrons are released, both in the pocket and in the convective TP, 
a fraction is captured via the resonant reaction $^{14}$N(n, p)$^{14}$C. 
As discussed in Paper I, a complex chain of proton captures
occurs, affecting in particular the production of F (\citealt{lugaro08,abia10,gallino10}, and 
references therein).
\\
 Besides the formation of the $^{13}$C-pocket, additional uncertainties 
that affect AGB models and $s$-process nucleosynthesis
are the evaluation of the
mass loss and the efficiency of the TDU. 
Mass loss plays a key role during the AGB phase, and it affects 
several stellar properties, as the efficiency of the TDU (in which the surface is enriched 
with freshly synthesised $^{12}$C and $s$-process elements), the number of thermal pulses 
and, therefore, the duration of this evolutionary phase.
In Paper I, we discussed these topics and their effects on our theoretical
predictions. 

With the recent discovery of peculiar stars showing both $s$ and $r$-process
enhancements (CEMP-$s/r$) incompatible with a pure $s$-process
nucleosynthesis, a highly debated issue has begun. Indeed, the two processes
are synthesised in different astrophysical conditions. 
Several hypotheses have been proposed in order to justify the presence of the products of both 
processes in the same star (e.g., 
\citealt{cohen03,jonsell06,lugaro09pasa}, and references therein).
Our working scenario, already presented by \citet{SCG08} and \citet{bisterzo09pasa}, 
has been discussed in detail in Paper II. 
Considering the large spread observed for [Eu/Fe] in field halo stars,
we follow the hypothesis that the molecular cloud from which a given binary system formed 
was initially enriched in different proportions in $r$-process elements.
\\
The astrophysical site and physical conditions of the $r$-process are not
 understood and several theoretical models have been advanced 
in order to estimate the $r$ component.
As discussed in Paper II, a good approximation 
of the $r$-process contribution to isotopes from Ba to Bi is evaluated with the 
residual method, $N_{\rm r}$ = $N_\odot$ - $N_{\rm s}$, starting from the solar 
main and strong $s$-process contributions.
According to the spread observed for [Eu/Fe] in Galactic halo stars with [Fe/H]
$\la$ $-$2 (e.g., Paper II, Fig.~2), where Eu is a typical $r$-process element,
observations of CEMP-$s$ and CEMP-$s/r$ stars may be interpreted under the hypothesis of 
a range of initial $r$-enhancements ([r/Fe]$^{\rm ini}$ $\sim$ 0.0, 0.5, 1.0, 1.5, 
2.0, scaled to Eu).
The [La/Eu] ratio provides a good indicator of the initial $r$-process contribution 
in these stars. Indeed, 70\% of solar La is synthesised by
the $s$-process, while less than 5\% of solar Eu is produced by the $s$-process. 
On the basis of the initial $r$-enhancement adopted,
we consider as CEMP-$s/r$ those stars that need an [r/Fe]$^{\rm ini}$ from 1.0
to 2.0, while CEMP-$s$ are those stars that require an [r/Fe]$^{\rm ini}$ $<$ 1.
If not differently specified, CEMP-$s$ stars are interpreted by assuming an [r/Fe]$^{\rm ini}$ = 0.5,
in agreement with the average of [Eu/Fe] observed in field halo stars (Paper II, 
Section~3). 

 The goal of this paper is to provide a first step toward our understanding of these puzzling 
CEMP-$s$ and CEMP-$s/r$ stars, through a comparison between the AGB models presented in Paper
 I and the spectroscopic observations of each individual star. This analysis is of fundamental importance in order to
account for the uncertainties, number of lines, resolution spectra
obtained by different authors. This highlights possible agreements or discrepancies
between theory and observations and suggests starting points for future investigations. 
\\
 A general discussion of the sample considered here has been
provided in Paper II, in which we presented an extensive analysis 
of the sample of a 
hundred of CEMP-$s$ stars collected from the literature
(\citealt{aoki02a,aoki02c,aoki02d,aoki06,aoki07,aoki08}; 
\citealt{barbuy05};
\citealt{barklem05};
\citealt{behara10};
\citealt{cohen03,cohen06};
\citealt{goswami06};
\citealt{GA10};
\citealt{ivans05};
\citealt{ish10};
\citealt{israelian01};
\citealt{jonsell06}; 
\citealt{JB02,JB04};
\citealt{JP01};
\citealt{lucatello03,lucatello11astroph};
\citealt{lucatello04PhD};
\citealt{masseron06,masseron10};
\citealt{pereira09};
\citealt{PS01};
\citealt{roederer08,roederer10}; 
\citealt{schuler08};
\citealt{sneden03b};
\citealt{thompson08};
\citealt{tsangarides05};
\citealt{vaneck03};
\citealt{zhang09}).
\\
 At halo metallicities, major constraints on the AGB initial mass are provided by 
the observed [Na/Fe] (and [Mg/Fe]) and, in some cases, by [hs/ls] (see Paper II). As described in 
Paper I, the predicted [Na/Fe], [Mg/Fe] and [ls/Fe] mainly depend on the number of TDUs. Indeed, 
by decreasing the metallicity, an increasing amount of primary $^{22}$Ne in the advanced 
thermal pulses is produced starting from primary $^{12}$C (via $^{14}$N($\alpha$, 
$\gamma$)$^{18}$F($\beta$$^{+}$$\nu$)$^{18}$O and $^{18}$O($\alpha$, $\gamma$)$^{22}$Ne; 
see e.g., \citealt{mowlavi99,gallino06}). 
In our models, $^{23}$Na is mainly synthesised via neutron capture on the abundant 
$^{22}$Ne during the thermal pulses (Paper I). 
Starting from $^{23}$Na, Mg is also produced during the advanced TPs, via the reactions 
$^{23}$Na(n, $\gamma$)$^{24}$Mg, $^{22}$Ne($\alpha$, n)$^{25}$Mg and $^{22}$Ne($\alpha$, 
$\gamma$)$^{26}$Mg \citep{mowlavi99,gallino06,husti07}.
 \citet{goriely00} firstly suggested an additional important source of Na. It derives
from their choice of the proton profile in the pocket: they considered an exponentially 
decreasing proton profile, starting from the envelope abundance ({\it X(H)} $\sim$ 0.7) 
down to $\sim$ 10$^{-6}$. 
A similar range of H abundance has been introduced by new FRANEC (\citealt{cristallo09}, their Fig.~13).
In the outer layer of the pocket, when $^{12}$C get destroyed by the very high hydrogen 
fraction introduced in the pocket, and $^{22}$Ne becomes more abundant than $^{12}$C, 
proton capture on $^{22}$Ne feeds $^{23}$Na at a temperature close to 40 MK.
Instead, in our prescriptions, we adopted an H profile limited to the range {\it X(H)} $<$ 
1.8 $\times$ 10$^{-3}$, as described by 
\citet{gallino98} (their Fig.~1), which excludes the region in which a large amount 
of protons reaches the He-intershell. Consequently, we do not have in the pocket the 
 $^{14}$N-rich and $^{23}$Na-rich regions.
However, we have accounted for the contribution to $^{23}$Na by $^{22}$Ne that comes 
from the ashes of the H-burning shell. As anticipated above, during the TP almost all $^{14}$N 
left by H-burning is converted to primary $^{22}$Ne. 
This primary $^{22}$Ne, which is mixed with the envelope by subsequent TDU episodes, also affects 
the final $^{23}$Na production. Indeed, in the interpulse phase the H-burning shell advances in 
mass.
By adopting the reaction rates by NACRE, about 20\% of $^{22}$Ne is converted 
to $^{23}$Na via $^{22}$Ne(p, $\gamma$)$^{23}$Na during the H-shell.
This implies an increase of the [Na/Fe] on the surface by $\sim$ 0.1 dex.
\\
As well as $^{23}$Na, the large amount of primary $^{22}$Ne also produces
Sr, Y and Zr via the $^{22}$Ne($\alpha$, n)$^{25}$Mg neutron source. 
This results in an increase of the predicted [ls/Fe] with the AGB initial mass.
Note that, by decreasing the metallicity, the reduced abundance of the major seed $^{56}$Fe, 
together with large amount of primary $^{22}$Ne, highlights the effect of the
$^{22}$Ne(n, $\gamma$) reaction, which acts as efficient seed, besides as neutron poison
(see \citealt{gallino06}; Section~\ref{poisons}). 
\\
The two $s$-process indicators [hs/ls] and [Pb/hs] provide information 
about the choice of the $^{13}$C-pocket\footnote{As in Paper II, we define 
ls = $<$Y,Zr$>$, hs = $<$La,Nd,Sm$>$.}.
A dilution factor $dil$ (defined as the logarithm of the mass of the convective envelope 
of the observed star, $M^{\rm obs}_\star$, over the mass transferred from 
the AGB to the companion, $M^{\rm trans}_{\rm AGB}$) is adopted in order to 
simulate the mixing occurring in the envelope after the mass transfer from the AGB. 
The [hs/Fe] ratio gives the first assessment of the dilution of the C and 
$s$-rich material transferred from the AGB.
For subgiants/giants having suffered the first dredge-up (FDU, a large mixing 
involving about 80\% of the mass of the star), $dil$ $\ga$ 1 dex is needed.
 The method adopted to interpret the spectroscopic observations has been 
described in Paper II for three stars taken as example: the CEMP-$s$ giant 
HD 196944, the main-sequence CEMP-$s/r$ HE 0338--3945, and a CEMP-$s$ HE 
1135+0139 for which no lead is measured.
We extend this analysis here to 94 CEMP-$s$ and CEMP-$s/r$ stars
following the classification adopted in Paper II, in agreement with
the $s$- and $r$-process enhancements observed:

\begin{itemize}
	\item CEMP-$s$II stars show high $s$-process enhancement with [hs/Fe] $\ga$ 1.5 
	(Section~\ref{secCEMPsII});
	\item CEMP-$s$I stars show mild $s$-process enhancement with [hs/Fe] $<$ 1.5 
	(Section~\ref{secCEMPsI}); 
	\item CEMP-$s$II$/r$ are high $s$-process enhanced stars,
	also showing an $r$-process contribution not compatible with a pure
	AGB nucleosynthesis ([hs/Fe] $\ga$ 1.5; 0.0 $\la$ [La/Eu] $\la$ 0.4).
	Similarly to the $s$-process enhancement, we may distinguish two subgroups among 
	CEMP-$s$II$/r$ stars, according to the initial $r$-process enhancement (Section~\ref{secCEMPs/r}):
    \begin{itemize}
	      \item 	CEMP-$s$II$/r$II with 1.5 $\la$ [r/Fe]$^{\rm ini}$ $\leq$ 2.0 
	      \item 	CEMP-$s$II$/r$I with [r/Fe]$^{\rm ini}$ $\sim$ 1.0.
    \end{itemize}
	\end{itemize}
One would expect a further class of stars with both mild $s$- 
and $r$-process contributions, the CEMP-$s$I$/r$I. None of the stars of the sample
belongs to this group probably because these stars cannot be
distinguished from Galactic halo stars due to their low initial $r$-enhancement.
\\
Moreover, we consider in a separate category those stars without Eu detection, 
because a possible initial $r$-process enrichment can not be excluded: among them,
we separate CEMP-$s$II/$-$ (Section~\ref{secCEMPsIInoEu}) and
CEMP-$s$I/$-$ (Section~\ref{secCEMPsInoEu}) stars.
\\
 At the beginning of each Section we list the name of the stars analysed
in alphabetical order, with a distinction between main-sequence/turnoff subgiant 
stars before the FDU and subgiants/giants having suffered the FDU.
Stars with a large number of observations (collected in Table~2 and~10 of 
Paper II) are analysed in Sections~\ref{secCEMPs},~\ref{secCEMPs/r}, 
and~\ref{secCEMPsnoEu}.
Stars with a limited number of $s$-process observations (Table~3 and~11 of 
Paper II) are discussed in Appendix~A.
A summary of the results is given in Section~\ref{conclusions}.
 For convenience, we report in Tables~\ref{summary1} and~\ref{summary2} a list 
of the stars studied here, with their references, metallicity, classification
provided in Paper II, the number of the Figure associated to the theoretical 
interpretation.

\begin{table*}
\caption{Summary of the CEMP-$s$ and CEMP-$s/r$ stars analysed in this paper.
References are Aoki et al.
(2002a,c,d, 2006, 2007, 2008), A02a, A02c, A02d, A06, A07, A08;
Barbuy et al. (2005), BB05; Barklem et al. (2005), B05; Beers
et al. (2007), Beers07; Behara et al. (2010), B10; Cohen et al.
(2003, 2006), C03, C06; Drake \& Pereira (2008), DP08;
Goswami et al. (2006), G06; Goswami \&
Aoki (2010), GA10; Ivans et al. (2005), I05; Ishigaki et al. (2010),
I10; Israelian et al. (2001), I01; Jonsell et al. (2005, 2006), J05,J06; 
Johnson \& Bolte (2002, 2004), JB02, JB04; Johnson et al. (2007), J07; 
Junqueira \& Pereira (2001), JP01; Lai et al. (2007),
Lai07; Lai et al. (2004), Lai04; 
Lucatello (2004),
 L04; Lucatello et al. (2011), L11; Masseron et al.
(2006,2010), M06, M10; Pereira \& Drake (2009), P09; Preston \& Sneden
(2001), PS01; Roederer et al. (2008, 2010), R08, R10; Schuler
et al. (2007, 2008), Sch07, Sch08; 
Thompson et al. (2008), T08; Tsangarides (2005), T05; Van Eck
et al. (2003), VE03; Zhang et al. (2009), Z09.
Following Section~\ref{intro}, in column~4 we distinguish between CEMP-$s$I (sI),
CEMP-$s$II (sII), CEMP-$s$I/$r$I (sI/rI) and CEMP-$s$II/$r$II (sII/rII) stars. 
Labels sI/$-$ of sII/$-$ refer to CEMP-$s$I and CEMP-$s$II without Eu detection.
In column~5, label 'ms' means main-sequence, 'TO' turnoff, 'SG' subgiant and 'G' giant.
The number of the Figure and the page number associated to the 
theoretical interpretation are listed in column~6.}
\label{summary1}
\begin{center}
\resizebox{11cm}{!}{\begin{tabular}{lccccc}
\hline
Star            &  Ref.           & [Fe/H]      & Type     &  Phase       &  Page, Fig.   \\                                                            
(1)              &  (2)            & (3)         & (4)   & (5)        & (6)     \\ 
\hline                                                                                                                        
BD +04$^\circ$2466& P09,I10,Z09   &  -1.92,-2.10   & sI/$-$   &   G     & p.~\pageref{BD+pereira},  Fig.~\ref{BD+042466_pd09_ik10_z09_bab10d567m1p5z2m4rp0p5_alf0p5_dil0p9n5} \\ 
BS 16080--175     &  T05          &  -1.86         & sII      &   ms/TO & p.~\pageref{BS175}, Fig.~\ref{BS16080-175_tsangarides05_bab10d2d3m1p5z2m4rp0p7_diffdiln20} \\                          
BS 17436--058    &  T05           &  -1.90         & sI       &   G      & p.~\pageref{BS058}, Fig.~\ref{BS17436-058_tsangarides05_bab10d8d6d3m1p5z2m4rp0p7alf0p5_diffdiln5n9n20}  \\       
CS 22183--015    &  A07,C06       &  -2.75         & sII/rII    & SG     & p.~\pageref{HE0058}, Fig.~\ref{CS22183-015_cohen06_aoki07_JB02T05lai07}   \\   
CS 22880--074    &  A07,A02c,d    &  -1.93          & sI       & SG     & p.~\pageref{CS074}, Fig.~\ref{CS22880-074_aoki02+aoki07_bab10d346m1p5z2m4nralf0p5_dil0p85n5}    \\                     
CS 22881--036    &  PS01          &  -2.06          & sII      & ms/TO & p.~\pageref{CS036}, Fig.~\ref{CS22881-036_PS01_bab10d4d5d6m1p5z2m4rp0p5alf0p5_n4}   \\                        
CS 22887--048    &  T05           &  -1.70          & sII/rI    & ms/TO & p.~\pageref{CS048}, Fig.~\ref{CS22887-048_tsangarides05_bab10d1p5p1p25m1p5z5m4rp1alf0p5_diffdiln8n20}   \\                       
CS 22898--027    &  A07,A02c,d    &  -2.26         & sII/rII    & ms/TO &  p.~\pageref{CS027}, Fig.~\ref{CS22898-027_aoki02+aoki07+tsangarides05_bab10d8m1p5z1m4rp2alf0p5_n456}     \\                                  
CS 22942--019    &A02c,d,PS01,Sch08,M10,L11&  -2.64    & sI       & G  & p.~\pageref{CS019},  Fig.~\ref{CS22942-019_aoki02+aoki07+schuler08+masseron10_bab10d303540m2z5m5rp0p5alf0p5_dil0p7n26}   \\                              
CS 22948--27     &  BB05,A07,L11   &  -2.47         & sII/rII    &  G  & p.~\pageref{CS27}, Fig.~\ref{CS22948-27_barbuy05_aoki07_bab10}    \\                   
CS 22964--161A/B &  T08           &  -2.39          & sI       & ms/TO & p.~\pageref{CS161}, Fig.~\ref{CS22964-161_thompson08_bab10n3n5}   \\      
CS 29497--030    &  I05           &  -2.57          & sII/rII    & ms & p.~\pageref{CS030}, Fig.~\ref{CS29497-030_es}  \\  
CS 29497--34     &  BB05,A07,L11  &  -2.90         & sII/rII    & G &p.~\pageref{CS34}, Fig.~\ref{CS29497-34_barbuy05_aoki07_bab10d4d6d8m2z2m5_rp1p5_alf0p5_dil1p0n26}  \\ 
CS 29513--032    &  R10           &  -2.08          & sI       & SG & p.~\pageref{CS032}, Fig.~\ref{CS29513-032_R10_bab10d6d3m1p5z2m4rp0p3alf0p5_diffdiln5n20}       \\                                    
CS 29526--110    &A07,A02c,d,A08  &-2.38,-2.06      & sII/rII    & ms/TO &  p.~\pageref{CS110}, Fig.~\ref{CS29526-110_aoki02+aoki07+AOKI08_bab10d4m1p5z2m4rp1p5_n456}       \\                                          
CS 29528--028    &  A07           &  -2.86          & sII/$-$  & ms & p.~\pageref{CS028}, Fig.~\ref{CS29528-028_aoki07_bab10d5d7d10m1p5z2m5rp0p5_n26}                \\ 
CS 30301--015    & A07,A02c,d     &  -2.64         & sI       & G  & p.~\pageref{CS015}, Fig.~\ref{CS30301-015_aoki02+aoki07_bab10d4d6d8m1p5z5m5nralf0p5_dil1p8n20}       \\                          
CS 30322--023    &  M06,A07,M10,L11   &-3.50,-3.25      & sI       &G & p.~\pageref{CS023}, Fig.~\ref{CS30322-023_masseron06_aoki07_masseron10_bab10d1p3d2d3m1p5z5m6rm1srym1_dil2p49n20}\\                  
CS 31062--012    & I01,A07,A02c,d,A08 &  -2.55          & sII/rII   & ms  & p.~\pageref{CS012}, Fig.~\ref{mnras_CS31062-012_aoki02+aoki0708+Israelian01_bab10d18m1p5z5m5rp1p5_n345} \\                                     
CS 31062--050    &JB04,A07,A06,A02c,d &-2.42        & sII/rII    &  SG  & p.~\pageref{CS050}, Fig.~\ref{CS31062-050_JB04aoki0206BaOsIr07Na_bab10}      \\                               
HD 26            &  VE03,M10      &  -1.25,-1.02   & sII      &  G  & p.~\pageref{HD26}, Fig.~\ref{HD26_vaneck03_masseron10_bab10p1d1p3d1p6m1p5z1m3nr_dil1p0n20}         \\                                
HD 5223          &  G06,L11      &  -2.06         & sII/$-$  &  G  & p.~\pageref{HD5223}, Fig.~\ref{HD5223_goswami06_bab10d8d10d12m2z2m4rp0p5_dil1p2n26}      \\                                     
HD 187861        &  VE03,M10,L11  &  -2.30,-2.36   & sII/rI  &  G  & p.~\pageref{HD187861}, Fig.~\ref{HD187861_VE03_M10}    \\       
HD 189711        &  VE03          &  -1.80         & sI/$-$   & G  &  p.~\pageref{HD189711}, Fig.~\ref{HD189711_vaneck03_bab10d12d16d20m1p5z2m4rp0p5_dil0p9n20}    \\                        
HD 196944        &A07,A02c,d,VE03,M10&  -2.25      & sI       &  G  & p.~\pageref{HD196944},  Fig.~\ref{HD196944_aoki02+aoki06+vaneck03+masseron10_bab10d2d3d4m1p5z1m4_nr_diffdiln20}\\                   
HD 198269        &  VE03          &  -2.20         & sI/$-$   & G  &  p.~\pageref{HD198269}, Fig.~\ref{HD198269_vaneck03_bab10p1d1p3d2m1p5z1m4rp0p5_dil1p5n20}    \\                       
HD 201626        &  VE03          &  -2.10         & sII/$-$  &  G  & p.~\pageref{HD201626}, Fig.~\ref{HD201626_vaneck03_bab10d1p3d2d3m1p5z2m4rp0p5_dil1p3n20}        \\                          
HD 206983        & M10, JP01      &  -0.99,-1.43    & sI       & G  &  p.~\pageref{HD206983}, Fig.~\ref{HD206983_masseron10_bab10p1p3m1p5rp0p5_diffdiln5n20}  \\
HD 209621        &  GA10          &  -1.93         & sII/rI    & G  &  p.~\pageref{HD209621}, Fig.~\ref{HD209621_goswamiaoki10_bab10d8d10d12m2z2m4rp1_dil0p85n26}    \\                       
HD 224959        &  VE03,M10      &  -2.20         & sII/rII    &  G  & p.~\pageref{HD224959}, Fig.~\ref{HD224959_vaneck03_masseron10_bab10d1p3d1p6d2m1p5z1m4rp1p6_dil0p95n20}   \\                             
HE 0143--0441    &  C06           &  -2.31          & sII/rI    & ms/TO &  p.~\pageref{HE0143}, Fig.~\ref{HE0143-0441_cohen06_bab10d6m1p5d3m2z1m4rp1alf0p5_diffdiln5n26}  \\     
HE 0202--2204    &  B05           &  -1.98         & sI       &  G  & p.~\pageref{HE0202}, Fig.~\ref{HE0202-2204Bark05_bab10d6m1p5d4m2z2m4nr_diffdiln5n26}   \\ 
HE 0212-0557     &  C06           &  -2.27         & sII/$-$  & G  &  p.~\pageref{HE0212}, Fig.~\ref{HE0212-0557_cohen06_bab10d3d5d8m2z1m4rp0p5_dil0p8n26}   \\
HE 0231-4016     &  B05           &  -2.08         & sI/$-$   &  SG  & p.~\pageref{HE0231}, Fig.~\ref{HE0231-4016Bark05_bab10d8d3m1p5z2m4rp0p5_diffdiln3n20}    \\ 
HE 0336+0113     &  C06           &  -2.68          & sII      &  SG  & p.~\pageref{HE0336}, Fig.~\ref{HE0336+0113_cohen06_bab10d35m1p5d30m2z5m5rp0p5_diffdiln8n26}   \\          
HE 0338--3945    &  J06           &  -2.42          & sII/rII    & ms/TO & p.~\pageref{HE0338}, Fig.~\ref{HE0338-3945_Jonsell06_bab10d6710m1p5z1m4_heavyrp2_alf0p5_n5}     \\                                   
HE 0430--4404    &  B05           &  -2.07          & sI/$-$   & ms & p.~\pageref{HE0430-4404Bark05_bab10d6d2m1p5z2m4rp0p5_diffdiln3n20}, Fig.~\ref{HE0430-4404Bark05_bab10d6d2m1p5z2m4rp0p5_diffdiln3n20}  \\ 
HE 1031--0020    &  C06           &  -2.86         & sI/$-$   & G  &  p.~\pageref{HE1031}, Fig.~\ref{HE1031-0020_cohen06_bab10d3m1p5d3m2z2m5rp0p5alf0p5_diffdiln10n26} \\
HE 1105+0027     &  B05           &  -2.42          & sII/rII    & ms/TO  & p.~\pageref{HE1105}, Fig.~\ref{HE1105+0027Bark05_bab10d6m1p5d2m2z1m4rp1p8_diffdiln5n26}  \\ 
HE 1135+0139     &  B05           &  -2.33         & sI       &  G  & p.~\pageref{HE1135}, Fig.~\ref{HE1135+0139_Barklem05_bab10d16d4m1p5z1m4nr_diffdiln520}  \\   
HE 1152--0355    &  G06,L11       &  -1.27         & sI/$-$   &  G  & p.~\pageref{HE1152}, Fig.~\ref{HE1152-0355_goswami06_bab10p1p5p1d1p3m1p5z1m3nr_dil1p0n20}     \\ 
HE 1305+0007     &  G06,L11           &  -2.03        & sII/rII    & G  &  p.~\pageref{HE1305}, Fig.~\ref{HE1305+0007_goswami06_bab10d8d10d12m2z2m4rp2_dil0p35n26}    \\                                 
HE 1430--1123    &  B05           &  -2.71          & sII/$-$  & SG  &  p.~\pageref{HE1430}, Fig.~\ref{HE1430-1123Bark05_bab10d8m1p5d3m2z5m5rp0p5_diffdiln5n26}     \\
HE 1434-1442     &  C06           &  -2.39         & sI/$-$   &  SG  &  p.~\pageref{HE1434}, Fig.~\ref{HE1434-1442_cohen06_bab10d10m1p5z1m4d8m1p5z1m4rp0p5alf0p5_diffdiln5n8}  \\  
HE 1509--0806    &  C06           &  -2.91         & sII/$-$ &  G  & p.~\pageref{HE1509}, Fig.~\ref{HE1509-0806_cohen06_bab10d12m1p5d8m2z2m5rp0p5alf0p5_diffdiln8n26} \\     
HE 2148--1247    &  C03           &  -2.30          & sII/rII    & ms/TO  & p.~\pageref{HE2148}, Fig.~\ref{HE2148-1247_cohen03_bab10}               \\                          
HE 2150--0825    &  B05           &  -1.98          & sI/$-$   &  SG  & p.~\pageref{HE2150-0825Bark05_bab10d10d3m1p5z2m4rp0p5_diffdiln3n20}, Fig.~\ref{HE2150-0825Bark05_bab10d10d3m1p5z2m4rp0p5_diffdiln3n20}   \\ 
HE 2158--0348    &  C06           &  -2.70         & sII      &  G  & p.~\pageref{HE2158}, Fig.~\ref{HE2158-0348_cohen06_bab10d2d3d4m1p5z5m5rp0p5alf0p5_dil1p4n20}   \\                 
HE 2232--0603    &  C06           &  -1.85          & sI/$-$   &  SG  & p.~\pageref{HE2232}, Fig.~\ref{HE2232-0603_cohen06_bab10d8d8d6m1p5z2m4rp0p5_diffdiln3n5n20}  \\                                  
HKII 17435--00532&  R08           &  -2.23         & sI       &  G  &  p.~\pageref{HKII}, Fig.~\ref{HK_roederer08_bab10d4d8d12m1p5z1m4rp0p3alf0p5_dil1p8n20}       \\                   
LP 625--44       &  A02,A06       &  -2.70         & sII/rII    &  G  & p.~\pageref{LP44}, Fig.~\ref{LP625-44_aoki0206OsIr_bab10d3d5d7m1p5z5m5rp1p5_dil0p8n20}       \\                              
V Ari            &  VE03          &  -2.40         & sI/$-$   &  G  & p.~\pageref{VAri_vaneck03_bab10d16d20d24m1p5z1m4rp0p5_dil0p9n20}, Fig.~\ref{VAri_vaneck03_bab10d16d20d24m1p5z1m4rp0p5_dil0p9n20}    \\            
SDSS 0126+06     &  A08           &  -3.11         & sII/$-$  & ms  & p.~\pageref{SDSS+06}, Fig.~\ref{SDSS0126+06_aoki08_bab10d6d8d12m1p5z2m5rp0p5alf0p5_n8}       \\
SDSSJ 0912+0216  &  B10           &  -2.50         & sII/rI    & ms  & p.~\pageref{SDSSJ0912}, Fig.~\ref{SDSSJ0912+0216_behara10_astroph_bab9ltHTd12m1p5z5m5_rp1he25alf0p5_dil0p6n5}\\            
SDSSJ 1349--0229 &  B10           &  -3.00          & sII/rII    & ms  & p.~\pageref{SDSSJ1349}, Fig.~\ref{SDSSJ1349-0229_behara10_astroph_bab9ltHTd10m1p5z2m5_rp1p5he25alf0p5_n6}  \\           
\hline
\end{tabular}}
\end{center}
\end{table*}

\begin{table*}                  
\caption{The same as Table~\ref{summary1}, but 
for the CEMP-$s$ and CEMP-$s/r$ stars with a limited number of spectroscopic
observations. A discussion of these stars is provided in Appendix~A.
References are the same as Table~\ref{summary1}.
Additional references are Lucatello et al. (2003), L03 and Sneden et al. (2003b), S03.
In column~5, label 'ms' means main-sequence, 'TO' turnoff, 'SG' subgiant, 'G' giant
and 'HB' horizontal branch.
The page number and, when provided, the Figure associated to the 
theoretical interpretation are given in column~6.}
\label{summary2}       
\begin{center}   
\resizebox{11cm}{!}{\begin{tabular}{lccccc}                 
\hline    
Star                & Ref.   & [Fe/H] & Type & Phase & Page, Fig.   \\
(1)                 & (2)    & (3)    &  (4) & (5)  & (6) \\
\hline  
CS 22891--171       & M10    &  -2.25  & sII/rII  &  G  & p.~\pageref{CS171}, Fig.~\ref{mnras_CS22891-171_masseron10_bab10d30m2z5m5_rp1p8_dil0p3n26}    \\                 
CS 22956$-$28       &M10,S03 &-2.33,-2.08& sI/$-$    & ms/TO & p.~\pageref{CS28}, Fig.~\ref{CS22956-28_sneden03+masseron10_bab10d505560m1p5z1m4_rp0p5_n6}   \\                                                                                                                                                                                                                       
CS 22960$-$053      & A07    &  -3.14   & sI/$-$    &  G  &  p.~\pageref{CS053}  \\                                                                                                                                                                                                                   
CS 22967$-$07       & L04    &  -1.81    & sII(/$-$) & ms & p.~\pageref{CS07}  \\        
CS 29495$-$42       & L04    &  -1.88  & sI    & SG & p.~\pageref{CS42}   \\                                                                                                                                                                                                                                 
CS 29503$-$010      & A07    &  -1.06    & s(II)/$-$ & ms & p.~\pageref{CS010} \\                                                                                                                                                                                                          
CS 29509$-$027      & S03    &  -2.02   & sI/$-$    & ms &  p.~\pageref{CS027S03} \\                                                                                                                                                                                                              
CS 30315$-$91       & L04    &  -1.68   & sI/$-$    & SG & p.~\pageref{CS91}  \\                                                                                                                                                                                                                         
CS 30323$-$107      & L04    &  -1.75   & sII(/$-$) & ms & p.~\pageref{CS107}    \\                                                                                                                                                                                                                                 
CS 30338$-$089      & A07,L04&-2.45,-1.75& sII(/rII) &  G  & p.~\pageref{CS089}   \\                                                                                                                                                                                      
G 18$-$24           & I10    &  -1.62    & sI/$-$    & ms &  p.~\pageref{G18} \\ 
HE 0012$-$1441      & C06    &  -2.52    & sI/$-$    & SG & p.~\pageref{HE0012}  \\                                                                                                                                                                                                                 
HE 0024$-$2523      & L03    &  -2.70    & sII(/$-$) & ms & p.~\pageref{HE0024}, Fig.~\ref{HE0024-2523_lucatello03_bab10d6m1p5z5m5_nr_alf0p5_dil0p05n456}     \\                                                                                                                                                                                                                    
HE 0131$-$3953      & B05    &  -2.71    & sII/rII  & TO/SG & p.~\pageref{HE0131}, Fig.~\ref{HE0131-3953Bark05_bab10d7d8d10m1p5z5m5rp1p5_dil0p1n5}    \\                                                                                                                                                                                                                      
HE 0206$-$1916      & A07    &  -2.09   & sII/$-$   & SG &  p.~\pageref{HE0206}  \\                                                                                                                                                                                                                       
HE 0400$-$2030      & A07    &  -1.73   & sII/$-$   & SG & p.~\pageref{HE0400}  \\                                                                                                                                                                                                                                 
HE 0441$-$0652      & A07    &  -2.47   & sI/$-$    & G  & p.~\pageref{HE0441} \\                                                                                                                                                                                                                         
HE 0507$-$1653      & A07    &  -1.38   & sII/$-$   &  G  & p.~\pageref{HE0507} \\                                                                                                                                                                                                                         
HE 1001--0243       & M10    &  -2.88   & sI     &  G  & p.~\pageref{HE1001}, Fig.~\ref{HE1001-0243_masseron10_bab10d20d50m1p5z2m5_nr_diffdiln5}   \\   
HE 1005$-$1439      & A07    &  -3.17   & sI/$-$    & G  &  p.~\pageref{HE1005}  \\                                                                                                                                                                                                                        
HE 1157$-$0518      & A07    &  -2.34   & sII/$-$   &  G  & p.~\pageref{HE1157}  \\                                                                                                                                                                                                             
HE 1305+0132        & Sch08  & -1.92    & sI/$-$    & G/HB & p.~\pageref{HEsch}   \\  
HE 1319$-$1935      & A07    &  -1.74   & sII/$-$   &  G  & p.~\pageref{HE1319}   \\
HE 1410$-$0004      & C06    &  -3.02    & sI/$-$    &  SG  & p.~\pageref{HE1410}  \\                                                                                                                                                                                                                           
HE 1419--1324       & M10    &  -3.05   & sI     &  G  & p.~\pageref{HE1419}, Fig.~\ref{HE1419-1324_masseron10_bab10d9d1p3m1p5z2m5_rp0p5_diffdiln5n20}   \\  
HE 1429$-$0551      & A07    &  -2.47   & sII/$-$   &  G  &  p.~\pageref{HE1429}  \\                                                                                                                                                                                                     
HE 1443+0113        & C06    &  -2.07   & sI/$-$    &   G  &  p.~\pageref{HE1443}  \\                                                        
HE 1447+0102        & A07    &  -2.47   & sII/$-$   &   G  & p.~\pageref{HE1447}  \\                                                                                                                                                                                                           
HE 1523$-$1155      & A07    &  -2.15  & sII/$-$   &  G  &  p.~\pageref{HE1523} \\                                                                                                                                                                                                           
HE 1528$-$0409      & A07    &  -2.61   & sII/$-$   &  G  & p.~\pageref{HE1528} \\                                                                                                                                 
HE 2221$-$0453      & A07    &  -2.22   & sII/$-$   &  G  & p.~\pageref{HE2221} \\                                                                                                                                                                                                                  
HE 2227$-$4044      & B05    &  -2.32   & sI/$-$    &  SG  & p.~\pageref{HE2227} \\                                                                                                                                                                                             
HE 2228$-$0706      & A07    &  -2.41   & sII/$-$   &   G  & p.~\pageref{HE2228} \\                                                                                                                                                                                         
HE 2240$-$0412      & B05    &  -2.20    & sI/$-$    &  SG  &  p.~\pageref{HE2240} \\                                                                                                                                                                                                     
HE 2330$-$0555      & A07    &  -2.78   & sI/$-$   &   G  & p.~\pageref{HE2330} \\                                                                                                                                                          
SDSS 0817+26        & A08    &  -3.16    & (sI/$-$)    & ms &  p.~\pageref{SDSS+26} \\   
SDSS 0924+40	    & A08    &  -2.51    & sII/$-$   & ms &  p.~\pageref{SDSS+40}, Fig.~\ref{SDSS0924+40_aoki07B_bab10d6d3m1p5m2z5m5_rp0p5_diffdiln7n26} \\                                                                                                                                                                                                         
SDSS 1707+58	    & A08    &  -2.52    & sII/$-$   & ms &  p.~\pageref{SDSS+58}, Fig.~\ref{SDSS1707+58_aoki07B_bab10d8d12d16m2z5m5_rp0p5_alf0p5_n26}    \\                                                                                                                                                                                                       
SDSS 2047+00	    & A08    &  -2.05   & sII/$-$   & ms &  p.~\pageref{SDSS+00}, Fig.~\ref{SDSS2047+00_aoki07B_bab10d8d3m1p5z2m4_rp0p5_diffdiln3n20}    \\                                                                                                                                                                                                  
\hline                             
\end{tabular}}                     
\end{center}                               
\end{table*}


The initial chemical composition described in Paper I, Section~2.1 is adopted.
We assume initial negative Cr and Mn abundances ([Cr/Fe] = $-$0.2; [Mn/Fe] = $-$0.4),
in agreement with the average of unevolved halo stars \citep{cayrel04,francois04}.
Note that \citet{bergemann08} found that the observed Mn may increase up to
[Mn/Fe] $\sim$ $-$0.1 due to NLTE corrections.
Concerning Cu, we assume an initial solar-scaled value if
no observations are provided. 
Actually, because AGBs marginally produce Cu,
a better choice would be [Cu/Fe]$^{\rm ini}$ $\sim$ $-$0.7, 
which represents the average ratio observed in unevolved halo stars
 \citep{bisterzo04,romano07}.
We accounted for this effect when discussing the four CEMP-$s$ stars
for which Cu has been detected (HE 0338--3945 discussed in Paper II, Section~5.2,
CS 30322--023, CS 31062--050 and HD 206983).

The comparison between theoretical predictions and observations may help 
to establish the efficiency of non-convective mixing occurring in the envelope of the observed
star during their main-sequence phase (e.g., thermohaline, gravitational settlings, 
radiative levitation, see 
\citealt{richard02b,vauclair04,stancliffe07,stancliffe08,stancliffe09,thompson08,theado10}).
For stars on the red giant branch, having undergone the FDU,
all mixing processes occurred during the main-sequence phase are erased.



\section{CEMP-{\scriptsize s} stars} \label{secCEMPs}

In this Section we provide an individual analysis of CEMP-$s$II and CEMP-$s$I
stars (with Eu measured). 
As anticipated in Section~\ref{intro}, CEMP-$s$ are generally interpreted with an 
initial $r$-process enhancement [r/Fe]$^{\rm ini}$ = 0.5, on the basis of an average 
of the [Eu/Fe] observed in halo field stars (see Paper II, Fig.~2). 
Different [r/Fe]$^{\rm ini}$ from $-$1.0 to 0.7 are adopted for some peculiar stars,
as discussed below.

\subsection{CEMP-$s$II stars} \label{secCEMPsII}

There are six stars with an $s$-process enhancement higher than [hs/Fe] $\ga$ 1.5.
One star having not suffered the FDU, 
CS 22881--036 by \citet{PS01}, and a second star with uncertain 
occurrence of the FDU, HE 0336+0113  by \citet{cohen06},
(see discussion in Paper II, Section~2); 
two giants, 
HE 1509--0806 and 
HE 2158--0348 by \citet{cohen06}. 
Note that the low Eu upper limit detected for HE 1509--0806 
excludes a high initial $r$-process enhancement, and classifies
this star as a CEMP-$s$II.
In addition, we describe in this Section the CEMP-$s$II subgiant
BS 16080--175, for which spectroscopic data have been analysed by 
\citet{tsangarides05} (PhD Thesis).
HD 26 \citep{vaneck03,masseron10}, which 
shows a higher metallicity ([Fe/H] = $-$1), is discussed separately 
in Section~\ref{CH}.

\begin{figure}
\includegraphics[angle=-90,width=8.5cm]{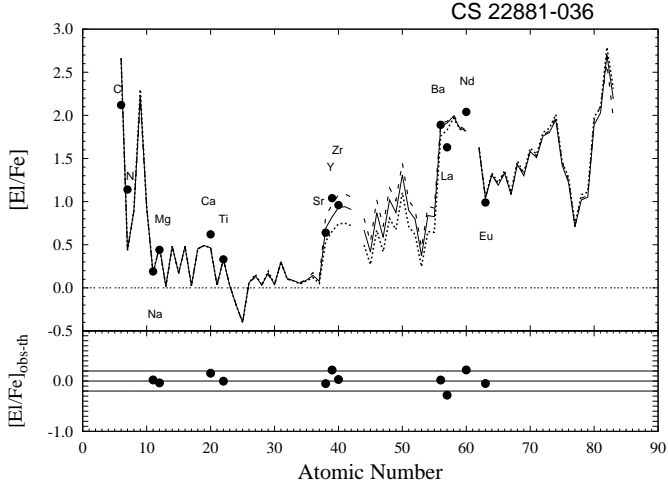}
\vspace{2mm}
\caption{Spectroscopic [El/Fe] observations of the main-sequence/turnoff star CS 22881--036
 ([Fe/H] = $-$2.06; {\textit T}$_{\rm eff}$ = 6200 K; log $g$ = 4.0)
 compared with AGB models of $M^{\rm AGB}_{\rm ini}$ = 
1.3 $M_{\odot}$, ST/6 (dotted line), ST/8 (solid line) and ST/9 (dashed line),
no dilution. 
Observations are from \citet{PS01}. This star shows [hs/ls] = 0.76. 
No error bars are provided by the authors for single species (see text). 
Here and in the following Figures, the lower panel displays the difference 
between observations and the AGB model represented with solid line
([El/Fe]$_{\rm obs-th}$). C and N are not included because they may be affected
by deep-mixing (see text).
The range between the two lines corresponds to a typical model uncertainty of
0.2 dex. 
An initial $r$-process enrichment of [r/Fe]$^{\rm ini}$ = 0.5 is adopted. 
Here and in the following Figures, both theoretical and spectroscopic
abundances have been normalised to the solar photospheric values
by \citet{lodders09}. This explains possible discrepancies with respect to
the observations listed in Paper II (Tables~2 and~3), in which we adopted
the normalisations provided by the different authors.}
\label{CS22881-036_PS01_bab10d4d5d6m1p5z2m4rp0p5alf0p5_n4}
\end{figure}

\subsubsection{CS 22881--036 (Fig.~\ref{CS22881-036_PS01_bab10d4d5d6m1p5z2m4rp0p5alf0p5_n4})}
\label{CS036}

This turnoff star with [Fe/H] = $-$2.06, {\textit T}$_{\rm eff}$ = 6200 K
and log $g$ = 4.0, was studied by \citet{PS01}. 
After the first radial
velocity study by \citet{PS00}, \citet{preston09pasa} find variations 
over a period of 16 years, with $P$ = 378 days. 
\\
Following the method explained in Paper II (Section~5), the choice of the 
$^{13}$C-pocket is made according to the observed $s$-process indicator 
[hs/ls] = 0.76. \citet{PS01} suggest typical uncertainties of 0.2 -- 0.3 
dex for species with many transitions. No lead is detected in this star.
In Fig.~\ref{CS22881-036_PS01_bab10d4d5d6m1p5z2m4rp0p5alf0p5_n4}, 
we show possible solutions with AGB models of $M^{\rm AGB}_{\rm ini}$
= 1.3 $M_{\odot}$, cases ST/6, ST/8 and ST/9, and no dilution.
Here and in the following Figures
the name of the star, its metallicity, the literature 
of the spectroscopic data, and the characteristics of the AGB models 
adopted (i.e. initial mass, $^{13}$C-pocket, dilution factor and 
initial $r$-process enhancement) are given in the caption.
In the lower panel the differences between observations and AGB predictions 
[El/Fe]$_{\rm obs-th}$ are represented. 
The two lines placed at $\pm$ 0.2 dex are
to indicate possible uncertainties of the model or of the initial abundances
of light elements.
\\
As discussed in Paper II, Section~5, the [C/Fe], [N/Fe] and $^{12}$C/$^{13}$C
ratios measured in CEMP-$s$ stars are largely overestimated by AGB models.
Indeed, AGB models predict a large amount of $^{12}$C in the envelope already 
after the first thermal pulses with TDU (Paper I).
The occurrence of extra-mixing like the cool bottom processing (CBP) 
have been hypothesised in order to interpret observations
in AGB stars and pre-solar grains of AGB origin 
\citep{nollett03,dominguez04a,dominguez04b,wasserburg06,zinner06,cristallo07,busso10}.
Several physical processes may concur in this mixing (e.g., 
rotation, magnetic fields, thermohaline mixing), and their efficiency
is difficult to estimate. 
An additional deep mixing may reduce the $^{12}$C/$^{13}$C ratio and increase
 $^{14}$N in low-mass low-metallicity AGB stars ([Fe/H] $\la$ $-$2.5): during 
the first fully developed TP, an ingestion of protons from the envelope down to the convective 
He-intershell occurs \citep{hollowell90,iwamoto04,campbell08,cristallo09pasa,campbell10}. 
This leads to efficient proton captures on CNO isotopes,
with a large production of $^{13}$C and, at lower level, of $^{14}$N.
This episode may also modify the $s$-process pattern. 
There is a maximum mass (which increases with decreasing the metallicity) below
which AGB models undergo proton ingestion. The assumption of an initial enhanced
distribution (e.g., of $\alpha$-elements compatible with observations of
metal-poor stars) would further decrease this mass limit \citep{straniero10}.
Further studies on this topic are desirable. 
For these reasons, C and N are not included in the lower panel of 
Fig.~\ref{CS22881-036_PS01_bab10d4d5d6m1p5z2m4rp0p5alf0p5_n4}
and in the following Figures. We refer to 
Paper II for a general discussion about C and N. 
We are planning to reconsider this topic in a forthcoming study.
However, we underline that high uncertainties affect the spectroscopic 
determination of C and N, because of NLTE or 3D model corrections 
\citep{asplund05,collet07,GAS07,asplund09,caffau09,frebelnorris11}.
\\
The AGB models displayed in 
Fig.~\ref{CS22881-036_PS01_bab10d4d5d6m1p5z2m4rp0p5alf0p5_n4}
 predict [Pb/Fe]$_{\rm th}$ $\sim$ 2.7.
The uncertainty estimated for the [hs/Fe] ratio could decrease
the lead prediction down to [Pb/Fe]$_{\rm th}$ $\sim$ 1.4 (case ST/18). 
Pb detection is highly desirable in this star.
In agreement with the average of field halo stars (for which a [Eu/Fe]
ratio about 0.5 is observed), we adopt an initial $r$-process enrichment
[r/Fe]$^{\rm ini}$ = 0.5. Note that
the solution without initial $r$-enhancement ([r/Fe]$^{\rm ini}$ = 0)
would decrease the predicted [Eu/Fe] by about 0.1 dex, still 
in agreement with the observation. 
The $s$-elements can be matched by AGB models with higher initial 
mass ($M^{\rm AGB}_{\rm ini}$ $\sim$ 1.5 -- 2 $M_{\odot}$), but the
estimated [Na/Fe] is too high.
The most plausible interpretations are found with $M^{\rm AGB}_{\rm ini}$ =
1.3 $M_{\odot}$ and $dil$ = 0.0 -- 0.3 dex.
The range of dilution accounts for the spread observed in the hs elements:
indeed, with a slightly lower $^{13}$C-pocket efficiency and $dil$ = 0.3 dex, we may
still interpret the two $s$-peaks.
This indicates that mixing during the main-sequence phase were not efficient
in this CEMP-$s$.
In order to estimate plausible variations of the dilution resulting from the uncertainties 
of the $^{13}$C-pocket, we test the effect on the abundances
of a variation of the mass involved in the pocket. 
 We recall that in our models, the mass 
of the pocket is 5 $\times$ 10$^{-4}$ $M_{\odot}$, that is about 1/20 of the typical mass involved in 
a thermal pulse. A mass of the pocket of the order of 1/10 of the mass involved in 
the thermal pulse can be considered as an extreme case. This would increase the [El/Fe] 
of the $s$-process elements of about 0.3 dex. Therefore, we found that the uncertainty of the mass 
of the $^{13}$C-pocket only marginally affects the dilution (up to $\sim$ 0.2 dex, given 
that the previous example is an extreme case).
AGB models with low initial mass undergo a limited number of thermal pulses.
In these models, the [El/Fe] abundances predicted in the envelope after two subsequent 
TDUs differ by $\sim$ 0.2 dex, corresponding to an increase of the initial mass of about +0.025 
$M_{\odot}$.
This is not the case of AGB models with initial mass $M$ $>$ 1.4 $M_{\odot}$, for which negligible 
differences are observed in the $s$-process distribution after the 12$^{\rm th}$ TDU
(see Paper I, Fig.~4, top panel).
Summing up, for lower AGB initial masses, the dilution may increase up to $\sim$ 0.4 dex, 
owing to a plausible increase of the mass of the pocket, or to an additional TDU. 
This value may be considered in agreement with a moderate mixing.

\begin{figure}
\includegraphics[angle=-90,width=8.5cm]{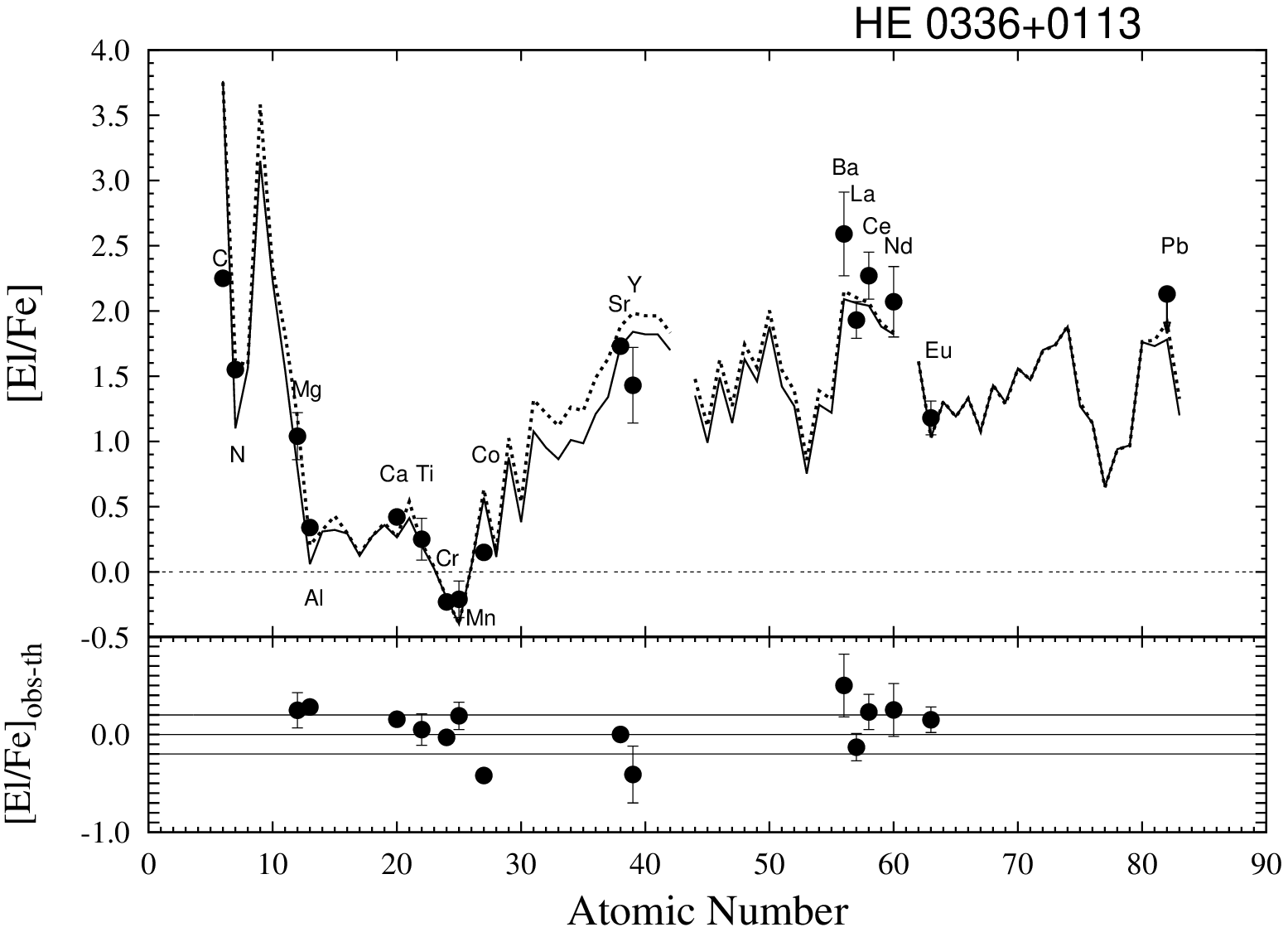}
\vspace{2mm}
\caption{Spectroscopic [El/Fe] abundances of the subgiant HE 0336+0113 
([Fe/H] = $-$2.68; {\textit T}$_{\rm eff}$ = 5700 K; log $g$ = 3.5, uncertain FDU,
see text)
 compared with AGB models of $M^{\rm AGB}_{\rm ini}$ = 1.4 $M_{\odot}$, 
ST/55, no dilution (solid line), or $M^{\rm AGB}_{\rm ini}$ = 2 $M_{\odot}$, 
ST/45 and $dil$ = 0.3 dex (dotted line).
Observations are from \citet{cohen06}.
This star shows [hs/ls] $\sim$ 0.2 and [Pb/hs] $\leq$ 0.2.
No Zr and Sm are detected and the observed [hs/ls] ratio is very
uncertain.
The differences [El/Fe]$_{\rm obs-th}$ refer to a model for $M^{\rm AGB}_{\rm ini}$ = 1.4 
$M_{\odot}$ (solid line).
An initial $r$-process enrichment of [r/Fe]$^{\rm ini}$ = 0.5 is adopted. }
\label{HE0336+0113_cohen06_bab10d35m1p5d30m2z5m5rp0p5_diffdiln8n26}
\end{figure}

\subsubsection{HE 0336+0113 (Fig.~\ref{HE0336+0113_cohen06_bab10d35m1p5d30m2z5m5rp0p5_diffdiln8n26})}
\label{HE0336}

HE 0336+0113 was analysed by \citet{cohen06}, who found [Fe/H] = $-$2.7,
{\textit T}$_{\rm eff}$ = 5700 ($\pm$ $\sim$ 150) K and log $g$ = 3.5 ($\pm$ $\sim$ 0.4). 
As anticipated in Paper II (Section~2), the onset of the FDU for subgiants 
may involve marginal mixing of the accreted AGB wind 
material with the original envelope of the observed star, unless
efficient thermohaline mixing is at play. 
The high Mg observed ([Mg/Fe] $\sim$ 1 dex) can be interpreted only by AGB models 
with initial masses higher than 1.3 $M_{\odot}$. 
A high [Na/Fe] is predicted.
Observations of Na lines are highly desirable.
Among ls elements, Zr is not detected, and among hs elements no Sm is 
measured, making the [hs/ls] ratio very uncertain.
The low upper limit detected for Pb ([Pb/hs] $\leq$ 0.2) indicates low
$^{13}$C-pocket efficiencies. In case of very low $^{13}$C-pocket choices,
the predicted [hs/Fe] does not exceed $\sim$ 2 dex, in agreement with
a low dilution.
In Fig.~\ref{HE0336+0113_cohen06_bab10d35m1p5d30m2z5m5rp0p5_diffdiln8n26},
we show solutions with $M^{\rm AGB}_{\rm ini}$ = 1.4 and 2 $M_{\odot}$,
cases close to ST/50, and $dil$ = 0.0 -- 0.3 dex. The low dilution 
suggests that the envelope did not reach yet its maximum penetration during the FDU episode. 
For these stars we can not properly quantify the efficiency of mixing during the 
main-sequence phase (see Section~\ref{CS036}).
We adopt an initial $r$-process enrichment [r/Fe]$^{\rm ini}$ = 0.5 in 
agreement with the average of field halo stars to obtain
[Eu/Fe]$^{\rm s+r}_{\rm th}$ = 1.03. 
 A similar value within 0.1 dex is predicted by a model without
initial $r$-process enhancement.
C and N are very uncertain for this star and no systematic 
errors are provided by \citet{cohen06}. 
Solutions with $M^{\rm AGB}_{\rm ini}$ = 1.5 $M_{\odot}$ give even worse
interpretations for C, N and Mg.

\begin{figure}
\includegraphics[angle=-90,width=8.5cm]{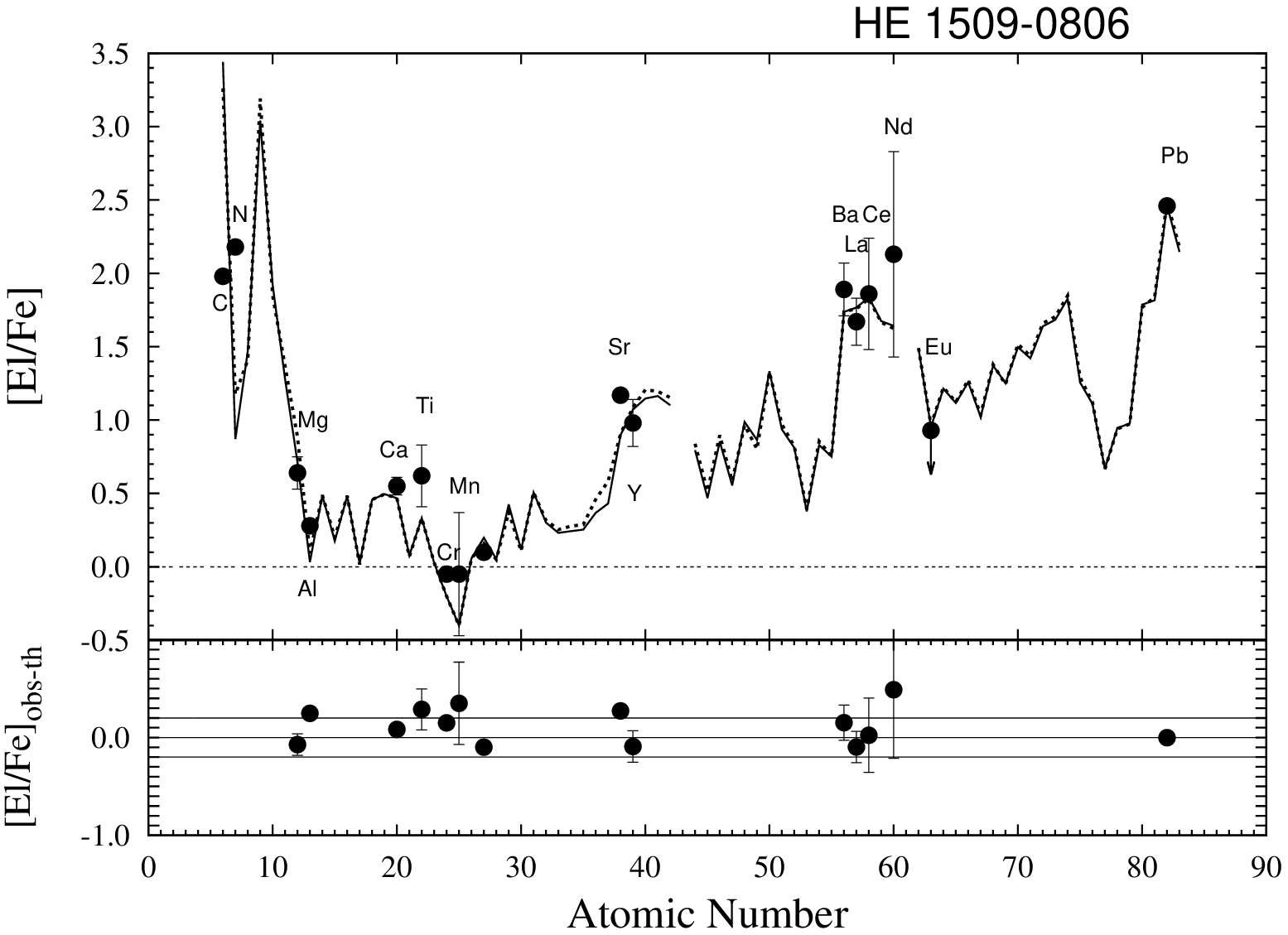}
\vspace{2mm}
\caption{Spectroscopic [El/Fe] abundances of the giant HE 1509--0806
([Fe/H] = $-$2.91; {\textit T}$_{\rm eff}$ = 5185 K; log $g$ = 2.5)
compared with AGB models of $M^{\rm AGB}_{\rm ini}$ = 1.4 $M_{\odot}$ (pulse 8$^{\rm th}$), 
ST/18, $dil$ = 0.7 dex (solid line), or $M^{\rm AGB}_{\rm ini}$ = 2 $M_{\odot}$, 
ST/12 and $dil$ = 1.2 dex (dotted line).
Observations are from \citet{cohen06}.
\citet{cohen06} observed [hs/ls] = 0.70 and [Pb/hs] =0.83.
The differences [El/Fe]$_{\rm obs-th}$ refer to a model for $M^{\rm AGB}_{\rm ini}$ = 1.4 
$M_{\odot}$ (solid line).
An initial $r$-process enrichment of [r/Fe]$^{\rm ini}$ = 0.5 is adopted.}
\label{HE1509-0806_cohen06_bab10d12m1p5d8m2z2m5rp0p5alf0p5_diffdiln8n26}
\end{figure}

\subsubsection{HE 1509--0806 (Fig.~\ref{HE1509-0806_cohen06_bab10d12m1p5d8m2z2m5rp0p5alf0p5_diffdiln8n26})}
\label{HE1509}

This is one of the coolest giants studied by \citet{cohen06}
([Fe/H] = $-$2.91; {\textit T}$_{\rm eff}$ = 5185 K; log $g$ = 2.5). 
HE 1509--0806 already suffered the FDU and theoretical interpretations with negligible dilutions 
would not account for this large mixing involving 
about 80\% of the convective envelope of the observed star (see Paper II). 
AGB models with low initial mass undergo a limited number of thermal pulses with TDU
(e.g., n3, n4, n5 for $M^{\rm AGB}_{\rm ini}$ $\sim$ 1.2 -- 1.3 $M_{\odot}$ models) 
and, consequently, the [El/Fe] distribution predicted in the envelope is low.
Therefore, low dilutions must be applied, in contrast with a giant after the FDU.
By increasing the AGB initial mass, 
plausible interpretations are found, as shown in
Fig.~\ref{HE1509-0806_cohen06_bab10d12m1p5d8m2z2m5rp0p5alf0p5_diffdiln8n26},
with $M^{\rm AGB}_{\rm ini}$ = 1.4 and 2 $M_{\odot}$ models, cases ST/18 and ST/12,
$dil$ = 0.7 -- 1.2 dex, respectively. Both models may equally interpret the
spectroscopic abundances ([hs/ls] = 0.70 and [Pb/hs] =0.83). 
Nd is highly uncertain and no Sm lines have been detected.
[Mg/Fe] is slightly overestimated by a $M^{\rm AGB}_{\rm ini}$ = 1.5 $M_{\odot}$ 
model. 
The low Eu upper limit permits to exclude high initial $r$-process enhancements:
0 $\leq$ [r/Fe]$^{\rm ini}$ $\leq$ 0.5 dex may equally
interpret the observations.

\begin{figure}
\includegraphics[angle=-90,width=8.5cm]{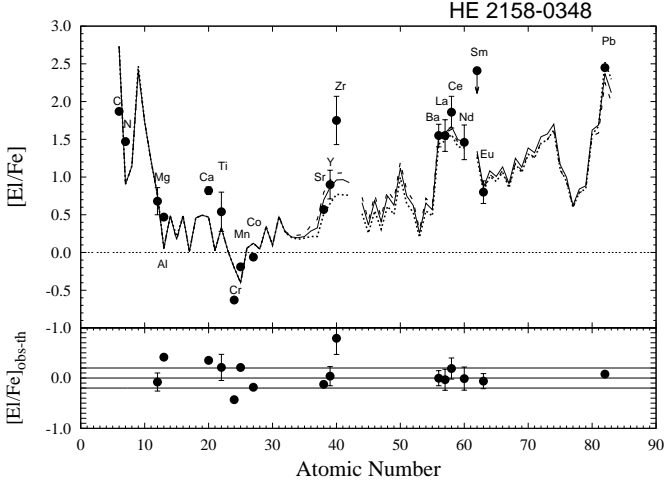}
\vspace{2mm}
\caption{Spectroscopic [El/Fe] abundances of the giant HE 2158--0348 
([Fe/H] = $-$2.70; {\textit T}$_{\rm eff}$ = 5215 K; log $g$ = 2.5)
compared with AGB models of $M^{\rm AGB}_{\rm ini}$ = 1.5 $M_{\odot}$, 
cases ST/3 (dotted line), ST/5 (solid line), ST/6 (dashed line), and $dil$ = 1.4 dex.  
Observations are from \citet{cohen06}.
This star shows [hs/ls] = 0.25 and [Pb/hs] = 1.13. [Zr/Fe] (two lines)
is about 0.8 dex higher than [Y/Fe] (three lines), which agrees with
AGB models (see text).
An initial $r$-process enrichment of [r/Fe]$^{\rm ini}$ = 0.5 is adopted.}
\label{HE2158-0348_cohen06_bab10d2d3d4m1p5z5m5rp0p5alf0p5_dil1p4n20}
\end{figure}

\subsubsection{HE 2158--0348 (Fig.~\ref{HE2158-0348_cohen06_bab10d2d3d4m1p5z5m5rp0p5alf0p5_dil1p4n20})}
\label{HE2158}

Similarly to HE 1509--0806, 
AGB models of low initial mass are excluded for this CEMP-$s$II 
giant ([Fe/H] = $-$2.70; {\textit T}$_{\rm eff}$ = 5215 K and log $g$ = 2.5; 
\citealt{cohen06}), because a low dilution does not agree 
with a star having suffered the FDU.
Significant differences among the ls elements are observed,
[Zr/Y] $\sim$ 0.8 dex (two lines are detected for Zr and three lines for Y). 
As displayed in  
Fig.~\ref{HE2158-0348_cohen06_bab10d2d3d4m1p5z5m5rp0p5alf0p5_dil1p4n20} 
for AGB models with $M^{\rm AGB}_{\rm ini}$ = 1.5 $M_{\odot}$ (ST/3,
ST/5 and ST/6) and $dil$ = 1.4 dex, we predict [Zr/Y]$_{\rm th}$ $\sim$ 0
and [hs/ls]$_{\rm th}$ $\sim$ 0.56.
A similar result is provided by a $M^{\rm AGB}_{\rm ini}$ = 2 
$M_{\odot}$ model (case ST/9).
[Zr/Fe] is 0.8 dex higher than the AGB prediction. 
The observed [Zr/Fe] may be interpreted by decreasing the choice of the 
$^{13}$C-pocket, but [Pb/Fe]$_{\rm th}$ would be about 1 dex lower than observed.

\begin{figure}
\includegraphics[angle=-90,width=8.5cm]{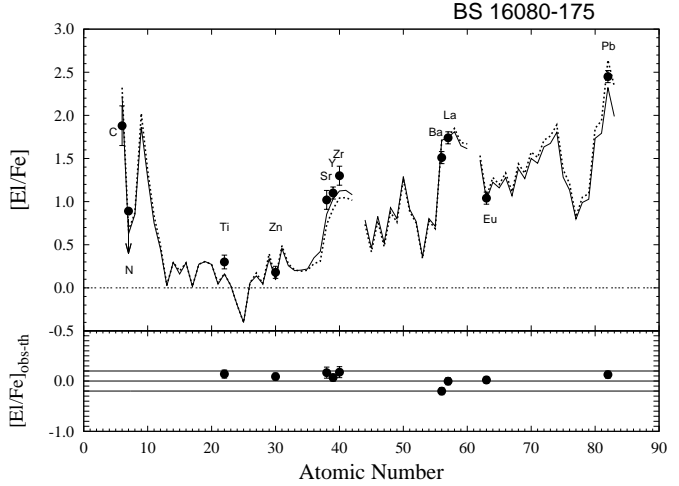}
\vspace{2mm}
\caption{Spectroscopic [El/Fe] abundances of the main-sequence/turnoff star BS 16080--175 
([Fe/H] = $-$1.86; {\textit T}$_{\rm eff}$ = 6240 K; log $g$ = 3.7)
compared with AGB models of $M^{\rm AGB}_{\rm ini}$ = 1.5 $M_{\odot}$, 
 cases ST/3 and ST/5 with $dil$ = 1.2 -- 1.3 dex, dashed and solid lines, respectively.
Observations are from \citet{tsangarides05}, ([hs/ls] = 0.42 and [Pb/hs] = 0.98).
A similar solution is obtained by $M^{\rm AGB}_{\rm ini}$ 
= 2 $M_{\odot}$. An initial $r$-process enrichment [r/Fe]$^{\rm ini}$ = 
0.7 is adopted.}
\label{BS16080-175_tsangarides05_bab10d2d3m1p5z2m4rp0p7_diffdiln20}
\end{figure}

\subsubsection{BS 16080--175 (Fig.\ref{BS16080-175_tsangarides05_bab10d2d3m1p5z2m4rp0p7_diffdiln20})}
\label{BS175}

The main-sequence/turnoff star BS 16080--175 was studied by \citet{tsangarides05},
with [Fe/H] = $-$1.86, {\textit T}$_{\rm eff}$ = 6240 K and log $g$ = 3.7.  
The $s$-process indicators show [hs/ls] = 0.4 and [Pb/hs] = 1.0.
Note that only Ba and La are detected among the hs elements.
In Fig.~\ref{BS16080-175_tsangarides05_bab10d2d3m1p5z2m4rp0p7_diffdiln20},
we show possible theoretical interpretations with $M^{\rm AGB}_{\rm ini}$ = 1.5 
$M_{\odot}$ models, two $s$-process efficiencies (ST/3 and ST/5) and a 
large dilution ($dil$ $\sim$ 1.2 dex).
Similar solutions can be obtained with $M^{\rm AGB}_{\rm ini}$ = 1.35 
and 2 $M_{\odot}$ models ($dil$ = 0.6 and 1.2 dex; cases ST/9 and ST/6, respectively).
An $r$-process enhancement [r/Fe]$^{\rm ini}$ = 0.7 is adopted in order to
interpret the observed [La/Eu] = 0.7. 
AGB models with $M^{\rm AGB}_{\rm ini}$ = 1.3 $M_{\odot}$ do not reproduce
both the [hs/ls] and [Pb/hs] observed: a case ST/6 and $dil$ = 0.4 dex predicts a low
first $s$-process peak ([ls/Fe]$_{\rm th}$ = 0.6), while a case ST/15 
predicts [Pb/hs]$_{\rm th}$ $\sim$ 0, about 1 dex lower than observed.
 This excludes interpretations with $M^{\rm AGB}_{\rm ini}$ = 1.3 $M_{\odot}$
models and low dilutions, suggesting
the occurrence of mixing during the main-sequence phase.
 Measurement of Na and Mg are highly desirable.

\subsection{CEMP-$s$I} 
\label{secCEMPsI}

Twelve stars show mild $s$-process enhancement.
The turnoff star CS 22964--161 by \citet{thompson08};  
two stars for which the occurrence of the FDU is 
uncertain, CS 22880--074 by \citet{aoki02c,aoki02d,aoki07}
and CS 29513--032 by \citet{roederer10}; 
seven giants,
CS 22942--019 by \citet{aoki02c,aoki02d,PS01},
CS 30301--015 by \citet{aoki02c,aoki02d,aoki07},
CS 30322--023 by \citet{masseron06,aoki07},
HD 196944 by \citet{aoki02c,aoki02d,aoki07,masseron10} 
(already discussed in Paper II, Section~5),
HE 0202--2204 and 
HE 1135+0139 by \citet{barklem05} (already discussed in Paper II, Section~5),
HK II 17435--00532 by \citet{roederer08}.
In addition, we described in this Section a further CEMP-$s$I giant,
BS 17436--058, for which spectroscopic data have been analysed by 
\citet{tsangarides05} (PhD Thesis).
Similarly to HD 26, the giant HD 206983 studied by \citet{JP01} 
and \citet{masseron10} has metallicity [Fe/H] $\sim$ $-$1
and will be discussed in Section~\ref{CH}.

\begin{figure}
\includegraphics[angle=-90,width=8.5cm]{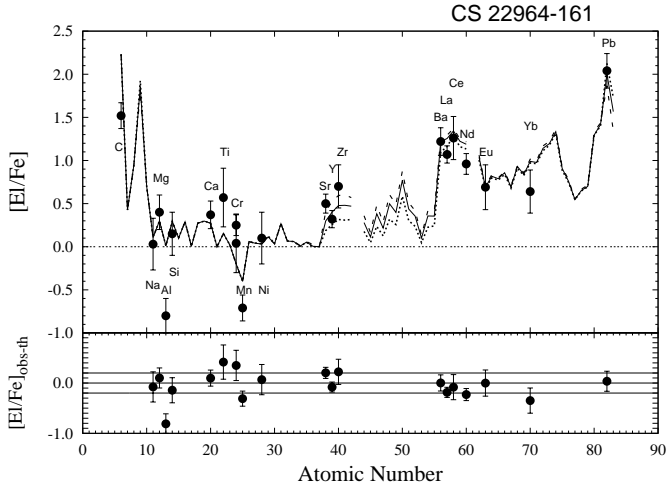}
\vspace{2mm}
\caption{Spectroscopic [El/Fe] abundances of main-sequence/turnoff star CS 22964--161
([Fe/H] = $-$2.39; {\textit T}$_{\rm eff}$ = 6050 K; log $g$ = 3.7)
compared with AGB models of $M^{\rm AGB}_{\rm ini}$ = 1.3 $M_{\odot}$, cases
ST/9 (dotted line), ST/12 (solid line), ST/15 (dashed line), and $dil$ = 0.9 dex. 
Observations are from \citet{thompson08}. This star shows [hs/ls] = 0.65 and [Pb/hs] = 1.15.
A solution with $dil$ $\sim$ 0.4 dex may be obtained with a $M^{\rm AGB}_{\rm ini}$ 
$\sim$ 1.2 $M_{\odot}$ model at the 3$^{\rm rd}$ pulse (case ST/15).
An initial $r$-process enrichment of [r/Fe]$^{\rm ini}$ = 0.5 is adopted.}
\label{CS22964-161_thompson08_bab10n3n5}
\end{figure}

\subsubsection{CS 22964--161 (Fig.~\ref{CS22964-161_thompson08_bab10n3n5})}
\label{CS161}

This main-sequence/turnoff star is a double-lined spectroscopic binary
([Fe/H] = $-$2.39; {\textit T}$_{\rm eff}$ = 6050 and 5950 K, log $g$ = 3.7 
and 4.1, for primary and secondary, respectively, \citealt{thompson08}),
with enhanced lithium (log $\epsilon$ (Li) = 2.0 $\pm$ 0.2).
The abundances of the secondary star are more uncertain, but an $s$-process 
enhancement is observed in both stars, with [Pb/Fe] $\sim$ 2 dex.
Theoretical interpretations with AGB models have been widely discussed by 
\citet{thompson08}.
Satisfactory solutions for the three $s$-process peaks are found with 
1.3 $\leq$ $M/M_{\odot}$ $\leq$ 2. 
The close to solar [Na/Fe] agrees with AGB models of initial
mass $M$ = 1.3 $M_{\odot}$, cases ST/9, ST/12, ST/15, and $dil$ = 0.9 dex,
as displayed in Fig.~\ref{CS22964-161_thompson08_bab10n3n5}.
A large dilution is applied in order to interpret the mild 
$s$-process enhancement observed ([ls/Fe] $\sim$ 0.5 and [hs/Fe] $\sim$ 1). 
However, theoretical interpretations with lower dilution ($dil$ $\sim$ 0.4 dex)
may be obtained by AGB models with $M^{\rm AGB}_{\rm ini}$ $\sim$ 
1.2 $M_{\odot}$ at the 3$^{\rm rd}$ pulse (case ST/15).
\citet{thompson08} conclude that only moderate thermohaline mixing could occur
in this star.
Indeed, gravitational settling,
which involved the star in the first
3 -- 4 Gyr before the mass accretion of the AGB, offsets the thermohaline 
efficiency, producing a mean molecular weight ($\mu$) barrier below the 
convective zone, which confined the thermohaline convection (see also 
\citealt{richard02a}). This is sustained by the high Li observed (log $\epsilon$ 
(Li) = +2.09 $\pm$ 0.20), which otherwise would be depleted, because 
of the higher temperature reached in the inner layers of the star.
However, these computations do not include radiative levitation, hence 
they do not represent the conclusive step of investigations of mixing on the secondary star.

\begin{figure}
\includegraphics[angle=-90,width=8.5cm]{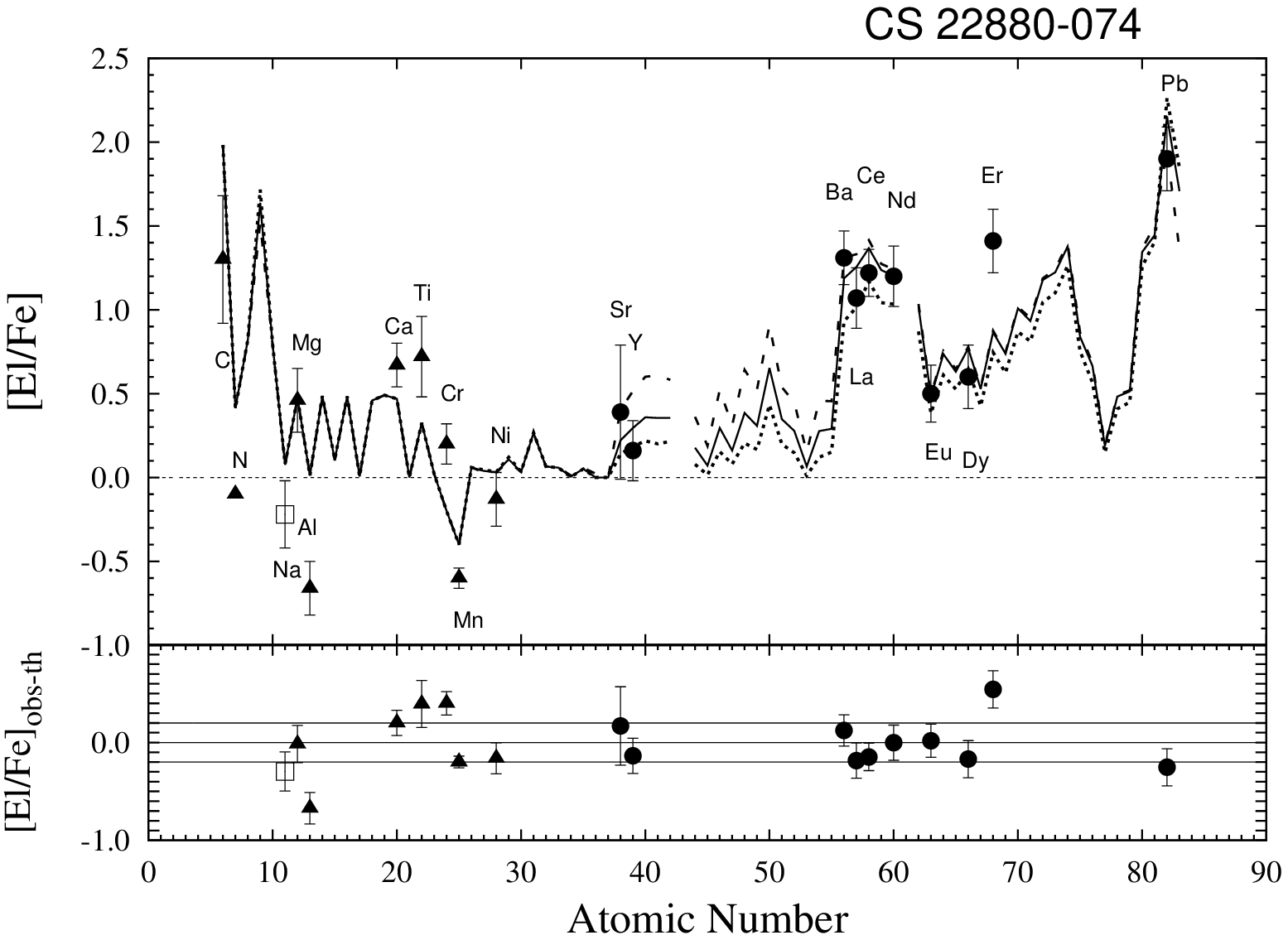}
\vspace{2mm}
\caption{Spectroscopic [El/Fe] abundances of the subgiant CS 22880--074
([Fe/H] = $-$1.93; {\textit T}$_{\rm eff}$ = 5850 K; log $g$ = 3.8, uncertain FDU, see text)
compared with AGB models of $M^{\rm AGB}_{\rm ini}$ = 1.3 $M_{\odot}$,
cases ST/5 (dotted line), ST/6 (solid line), ST/9 (dashed line), and $dil$ = 0.9 dex. 
Observations are from \citet{aoki02c}, (filled triangles), \citet{aoki02d}, (filled circles), 
\citet{aoki07}, (empty square).
\citet{aoki02d} detected [hs/ls] = 0.84 and [Pb/hs] = 0.80. 
\citet{aoki02c} highlight the unreliability of N in this star.
An initial [r/Fe]$^{\rm ini}$ = 0.0 is adopted.
The observed [Er/Fe] is about 0.5 dex higher than AGB models (see text).}
\label{CS22880-074_aoki02+aoki07_bab10d346m1p5z2m4nralf0p5_dil0p85n5}
\end{figure}

\subsubsection{CS 22880--074 (Fig.~\ref{CS22880-074_aoki02+aoki07_bab10d346m1p5z2m4nralf0p5_dil0p85n5})}
\label{CS074}

This subgiant ([Fe/H] = $-$1.93, {\textit T}$_{\rm eff}$ = 5850 K and log $g$ = 3.8)
was analysed by 
\citet{aoki02c,aoki02d}. 
CS 22880--074 lies in the region of the HR diagram in which the occurrence 
of the FDU is uncertain. 
We remind that in very metal-poor stars Na may be affected by strong 
uncertainties due to non-local thermodynamic equilibrium (NLTE) corrections 
(\citealt{andr07,aoki07,barbuy05} and references therein). 
For this star, the NLTE correction for the Na I D lines decreases the
[Na/Fe] abundance by 0.7 dex \citep{aoki07}.
We show in Fig.~\ref{CS22880-074_aoki02+aoki07_bab10d346m1p5z2m4nralf0p5_dil0p85n5} 
 theoretical interpretations using $M^{\rm AGB}_{\rm ini}$ 
= 1.3 $M_{\odot}$ models (cases ST/5, ST/6 and ST/9, in agreement with the
observed [hs/ls] = 0.84 and [Pb/hs] = 0.80). 
Due to the mild $s$-enhancement, a large dilution is needed (0.9 dex). 
With an AGB model of lower initial mass, $M^{\rm AGB}_{\rm ini}$ = 1.2 
$M_{\odot}$ at the 3$^{\rm rd}$ pulse, a similar result is obtained with 
$dil$ = 0.4 dex. 
Dilutions of the order of 0.4 -- 0.9 dex suggest that internal 
mixing occurred in this star (e.g., FDU or thermohaline, gravitational settling; 
 see Section~\ref{intro}).
Lower dilutions may only be obtained by AGB models 
suffering a lower number of TDUs (simulated with a $M^{\rm AGB}_{\rm ini}$ 
$\sim$ 1.2 $M_{\odot}$ model at the 2$^{\rm nd}$ pulse).
These solutions slightly overestimated the observed subsolar [Na/Fe]. 
Models with $M^{\rm AGB}_{\rm ini}$ = 1.5 
-- 2 $M_{\odot}$ ($dil$ $\sim$ 2 dex) predict higher Na and Y ([Na/Fe]$_{\rm th}$ 
$\sim$ 0.3; [Y/Fe]$_{\rm th}$ $\sim$ 0.4). 
AGB models disagree with the negative [N/Fe] observed: \citet{aoki02c} underline 
that N is unreliable in this star.  
Only one line is detected for the three $r$-process elements Eu, Dy and Er. 
The observed [Eu,Dy/Fe] are about 0.9 dex lower
than [Er/Fe] and suggest no initial $r$-process enhancement.
Note that the [Er/Fe] measured is even higher than [hs/Fe]. 
Further investigations on the $r$-elements would be desirable.
\\
No radial velocity variations were registered over
a time of 16 years (\citealt{PS01,aoki02c,preston09pasa}).

\begin{figure}
\includegraphics[angle=-90,width=8.5cm]{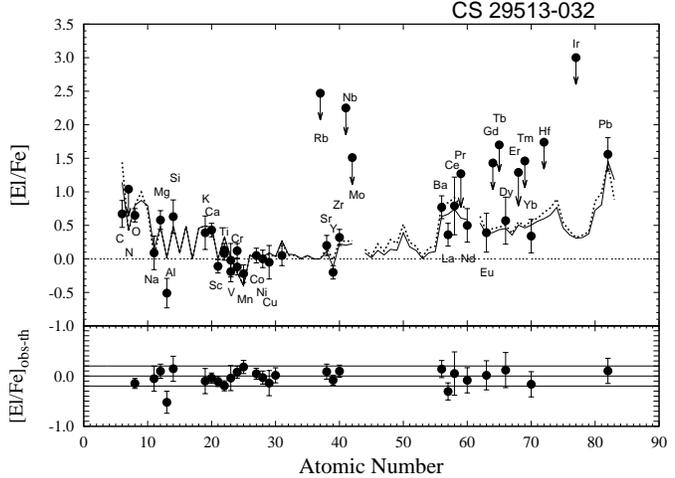}
\vspace{2mm}
\caption{Spectroscopic [El/Fe] abundances of the subgiant CS 29513--032 
([Fe/H ] = $-$2.08; {\textit T}$_{\rm eff}$ = 5810 K; log $g$ = 3.3, uncertain FDU)
compared
with $M^{\rm AGB}_{\rm ini}$ = 1.3 or 1.5 $M_{\odot}$ models, cases ST/9 or 
ST/3 and $dil$ = 1.4 or 2.4 dex, (dotted or solid lines, respectively).
Observations are from \citet{roederer10}, who detected [hs/ls] = 0.45 and 
[Pb/hs] = 1.29.
In order to interpret the negative [Y/Fe] value, we assume [Y/Fe]$^{\rm ini}$ 
= $-$0.5 (see text). Solar-scaled initial abundances for Sr and Zr are adopted.
The observed [La/Fe] is slightly overestimated by both models. 
The differences [El/Fe]$_{\rm obs-th}$ represented in the lower panel refer
to the $M^{\rm AGB}_{\rm ini}$ = 1.5 $M_{\odot}$ model (solid line).
An initial $r$-process enrichment of [r/Fe]$^{\rm ini}$ = 0.3 is adopted.} 
\label{CS29513-032_R10_bab10d6d3m1p5z2m4rp0p3alf0p5_diffdiln5n20}
\end{figure}

\subsubsection{CS 29513--032 (Fig.~\ref{CS29513-032_R10_bab10d6d3m1p5z2m4rp0p3alf0p5_diffdiln5n20})}
\label{CS032}

This subgiant with [Fe/H ] = $-$2.08, {\textit T}$_{\rm eff}$ = 5810 K and 
log $g$ = 3.3, has been recently studied by 
\citet{roederer10}. The occurrence of the FDU is uncertain in this star.
It is a member of a stellar stream identified by
\citet{helmi99}, probably
originating from the disruption 
of a former Milky Way satellite galaxy. 
\citet{roederer10} studied twelve of these stars, but only CS 29513--032
is $s$-process enhanced.
The authors firstly have hypothesised a contribution from an AGB companion.
Despite the presence of a definite, albeit moderate, $s$-process signature, 
CS 29513--032 shows [C/Fe] = 0.63 ($\pm$ 0.2 dex), lower than usually observed 
in CEMP-$s$ stars.
Actually, some authors consider CEMP those stars with [C/Fe] $\geq$ 1, according to the 
definition of \citet{beers05}; we include among CEMP all objects with [C/Fe] $\ga$ 0.5.
For this mild $s$-rich star a very large dilution is necessary in order to interpret
the observations ([ls/Fe] $\sim$ 0, [hs/Fe] $\sim$ 0.5, [Pb/Fe] $\sim$ 1.6)
even for AGB models with low initial mass: already at the 5$^{\rm th}$ TDU (with 
$M^{\rm AGB}_{\rm ini}$ = 1.3 $M_{\odot}$) a dilution of 1.4 dex is needed.
Lower dilutions may be obtained at the 2$^{\rm nd}$ TDU 
($M^{\rm AGB}_{\rm ini}$ $\sim$ 1.2 $M_{\odot}$; $dil$ = 0.3 dex).
This may imply that efficient mixing have taken place in this subgiant
(e.g., FDU or thermohaline, gravitational settling; see Section~\ref{intro}).
The low observed [Na/Fe] accounts for NLTE corrections. Owing to the large
dilution applied, a low [Na/Fe] is also predicted by models with 
$M^{\rm AGB}_{\rm ini}$ = 1.5 and 2 $M_{\odot}$. For this star, Na and can not 
provide constraints on the AGB initial mass and
all AGB models in the range 1.3 $\la$ $M/M_\odot$ $\leq$ 2 
may equally fit the observations. 
In Fig.~\ref{CS29513-032_R10_bab10d6d3m1p5z2m4rp0p3alf0p5_diffdiln5n20}, two solutions
are shown, $M^{\rm AGB}_{\rm ini}$ = 1.3 and 1.5 $M_{\odot}$ models, cases ST/9
and ST/3, $dil$ = 1.4 and 2.4 dex, respectively. We adopt a negative initial 
[Y/Fe]$^{\rm ini}$ = $-$0.5, which is compatible with the spread of [Y/Fe]
observed in field halo stars (e.g., \citealt{francois07}),
in order to interpret the subsolar [Y/Fe] observed.
Under this assumption, a model of $M^{\rm AGB}_{\rm ini}$ = 1.5 $M_{\odot}$ that requires
a large dilution, provides [Y/Fe]$_{\rm th}$ = $-$0.12.
[La/Fe] (4 lines detected) is overestimated by both models.  
A similar solution is obtained for a $M^{\rm AGB}_{\rm ini}$ = 2 $M_{\odot}$ 
model (case ST/5 and $dil$ = 2.5 dex). 
\\
No radial velocity variations have been detected in a span of 
three months \citep{roederer10} and further investigations are desirable. 

\begin{figure}
\includegraphics[angle=-90,width=8.5cm]{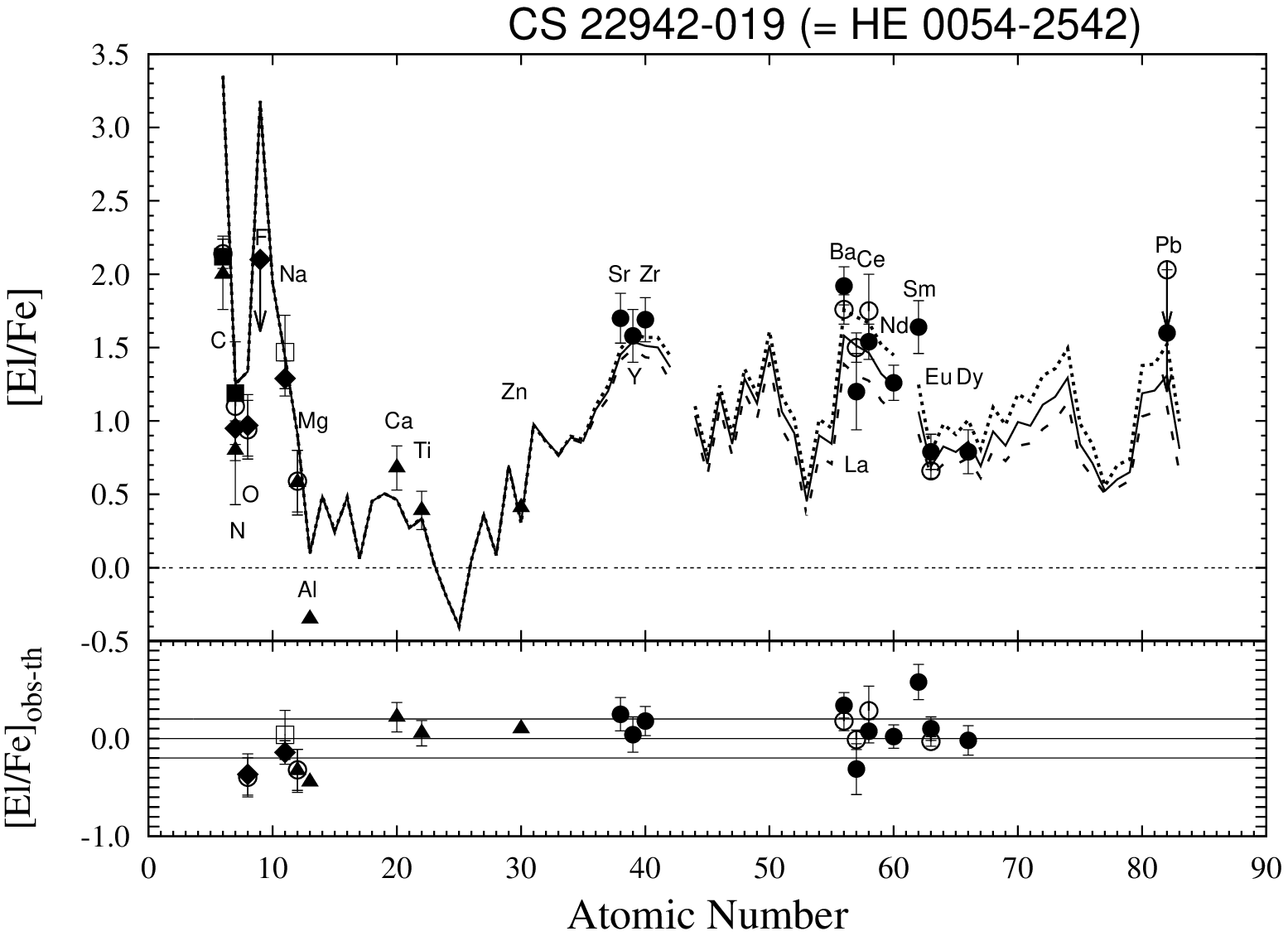}
\vspace{2mm}
\caption{Spectroscopic [El/Fe] abundances of the giant CS 22942--019 
([Fe/H ] = $-$2.43; {\textit T}$_{\rm eff}$ = 5100 K; log $g$ = 2.5) 
compared with AGB 
models of $M^{\rm AGB}_{\rm ini}$ = 2 $M_{\odot}$, cases ST/45 (dotted line), ST/50 (solid line), 
ST/60 (dashed line), and $dil$ = 0.7 dex.
Observations are from \citet{aoki02c} (filled triangles), \citet{aoki02d} (filled circles), 
\citet{PS01} (empty square), \citet{schuler08} (filled squares), \citet{masseron10} (empty
circles), \citet{lucatello11astroph} (filled diamonds). 
This star shows [hs/ls] = $-$0.36, while an upper limit is
observed for Pb. 
The spread observed among the hs elements is discussed in the text.
An initial $r$-process enrichment of [r/Fe]$^{\rm ini}$ = 0.5 is adopted.} 
\label{CS22942-019_aoki02+aoki07+schuler08+masseron10_bab10d303540m2z5m5rp0p5alf0p5_dil0p7n26}
\end{figure}

\subsubsection{CS 22942--019 ($\equiv$ HE 0054-2542),
(Fig.~\ref{CS22942-019_aoki02+aoki07+schuler08+masseron10_bab10d303540m2z5m5rp0p5alf0p5_dil0p7n26})} 
\label{CS019}

The giant CS 22942--019 is a long period binary ($P$ = 2800 d; 3616 d; \citealt{PS01,lucatello09pasa})
with [Fe/H ] = $-$2.43, {\textit T}$_{\rm eff}$ = 5100 K and 
log $g$ = 2.5.
Spectroscopic observations of several elements have been investigated
by different authors: \citet{aoki02c,aoki02d,schuler08,masseron10,lucatello11astroph}.
A high [Na/Fe] was detected by \citet{PS01}, [Na/Fe] = 1.44.
However, \citet{PS01} do not consider NLTE corrections that could reduce 
the observed [Na/Fe].
The low upper limit measured for lead ([Pb/Fe] $\leq$ 1.6; [Pb/hs] $\la$ 
0.3) is interpreted by an AGB model of initial mass $M$ = 
2 $M_{\odot}$, as shown in 
Fig.~\ref{CS22942-019_aoki02+aoki07+schuler08+masseron10_bab10d303540m2z5m5rp0p5alf0p5_dil0p7n26}. 
The low [Pb/Fe] upper limit is interpreted by a case ST/50. 
Together with HE 0336+0113 (Section~\ref{HE0336}), CS 22942--019 is one of
the stars with the lowest $^{13}$C-pocket choice, because of the decreasing
behaviour of [ls/Fe] $>$ [hs/Fe] $>$ [Pb/Fe] (see also Section~\ref{conclusions}).
The observed [hs/Fe] shows a spread of about 0.6 dex.
AGB models with low $^{13}$C-pockets predict a slightly decrease, by increasing the atomic
 number, of the abundances from Ba to Sm.
The observed Ba and Sm are overestimated by models. We consider Ce and Nd (6 and 7 lines
detected, respectively) more reliable than La and Sm (2 and 1 lines, 
respectively)\footnote{The number of lines are taken from \citet{aoki02c,aoki02d}. 
The observational data by \citet{masseron10} reported in 
Fig.~\ref{CS22942-019_aoki02+aoki07+schuler08+masseron10_bab10d303540m2z5m5rp0p5alf0p5_dil0p7n26}
 will be discussed by the authors in Masseron et al., in preparation}.
A dilution of 0.7 dex is applied. AGB models with lower initial mass 
($M^{\rm AGB}_{\rm ini}$ $\leq$ 1.5 $M_{\odot}$) would need $dil$ $\leq$ 0.5 dex, 
which disagrees with the large mixing occurring in a giant. 
\\
An upper limit for fluorine is detected by \citet{lucatello11astroph} ([F/Fe] $<$ 2.1).
AGB models predict higher [F/Fe] abundance
(almost at the same order than [C/Fe], see Paper I).
Further spectroscopic investigations are desirable in order to constrain theoretical
predictions.

\begin{figure}
\includegraphics[angle=-90,width=8.5cm]{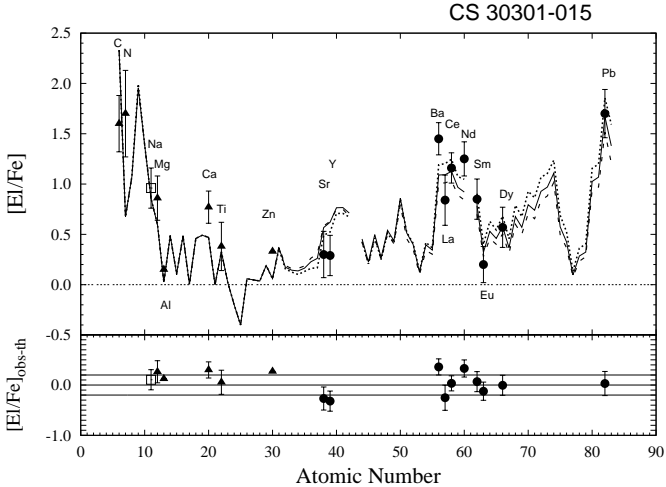}
\vspace{2mm}
\caption{Spectroscopic [El/Fe] abundances of the giant CS 30301--015 
([Fe/H] = $-$2.64; {\textit T}$_{\rm eff}$ = 4750 K; log $g$ = 0.8)
compared with AGB models of $M^{\rm AGB}_{\rm ini}$ 
= 1.5 $M_{\odot}$, cases ST/6 (dotted line), ST/9 (solid line), ST/12 (dashed line), 
and $dil$ = 1.8 dex.
Observations are from \citet{aoki02c} (filled triangles), \citet{aoki02d} 
(filled circles), \citet{aoki07} (empty squared).
\citet{aoki02d} detected [hs/ls] = 0.45 and [Pb/hs] = 0.72.
An initial [r/Fe]$^{\rm ini}$ = 0 is adopted. }
\label{CS30301-015_aoki02+aoki07_bab10d4d6d8m1p5z5m5nralf0p5_dil1p8n20}
\end{figure}

\subsubsection{CS 30301--015 (Fig.~\ref{CS30301-015_aoki02+aoki07_bab10d4d6d8m1p5z5m5nralf0p5_dil1p8n20})}
\label{CS015}

The spectra of cool giants are highly contaminated by  
molecular bands. This is the case of CS 30301--015, showing
[Fe/H] = $-$2.64, {\textit T}$_{\rm eff}$ = 4750 K and log $g$ = 0.8   \citep{aoki02c,aoki02d,aoki07}. The CN molecular 
bands are too strong for a reliable N abundance determination. 
The high [Na/Fe] and [Mg/Fe] observed by \citet{aoki07} ([Na/Fe] = 1.09 and 
[Mg/Fe] = 0.80) exclude interpretations with $M^{\rm AGB}_{\rm ini}$ $\leq$ 1.3 
$M_{\odot}$ models. By increasing the number of TDUs, solutions with 
$M^{\rm AGB}_{\rm ini}$ = 1.5 $M_{\odot}$ (ST/6, ST/9 and ST/12) and $dil$ = 
1.8 dex are shown in
Fig.~\ref{CS30301-015_aoki02+aoki07_bab10d4d6d8m1p5z5m5nralf0p5_dil1p8n20}.
 The spread observed among the hs elements does not agree with 
AGB predictions. \citet{aoki02c} detected 13 lines for Ce, 4 for Ba,
and 3 for La, Nd and Sm. 
Solutions with $M^{\rm AGB}_{\rm ini}$ = 2 $M_{\odot}$ are excluded 
because they would underestimate the observed [Na/Fe] and [Mg/Fe] by about 0.4 dex.
As highlighted by \citet{aoki02c}, N is very uncertain in this star.

\begin{figure}
\includegraphics[angle=-90,width=8.5cm]{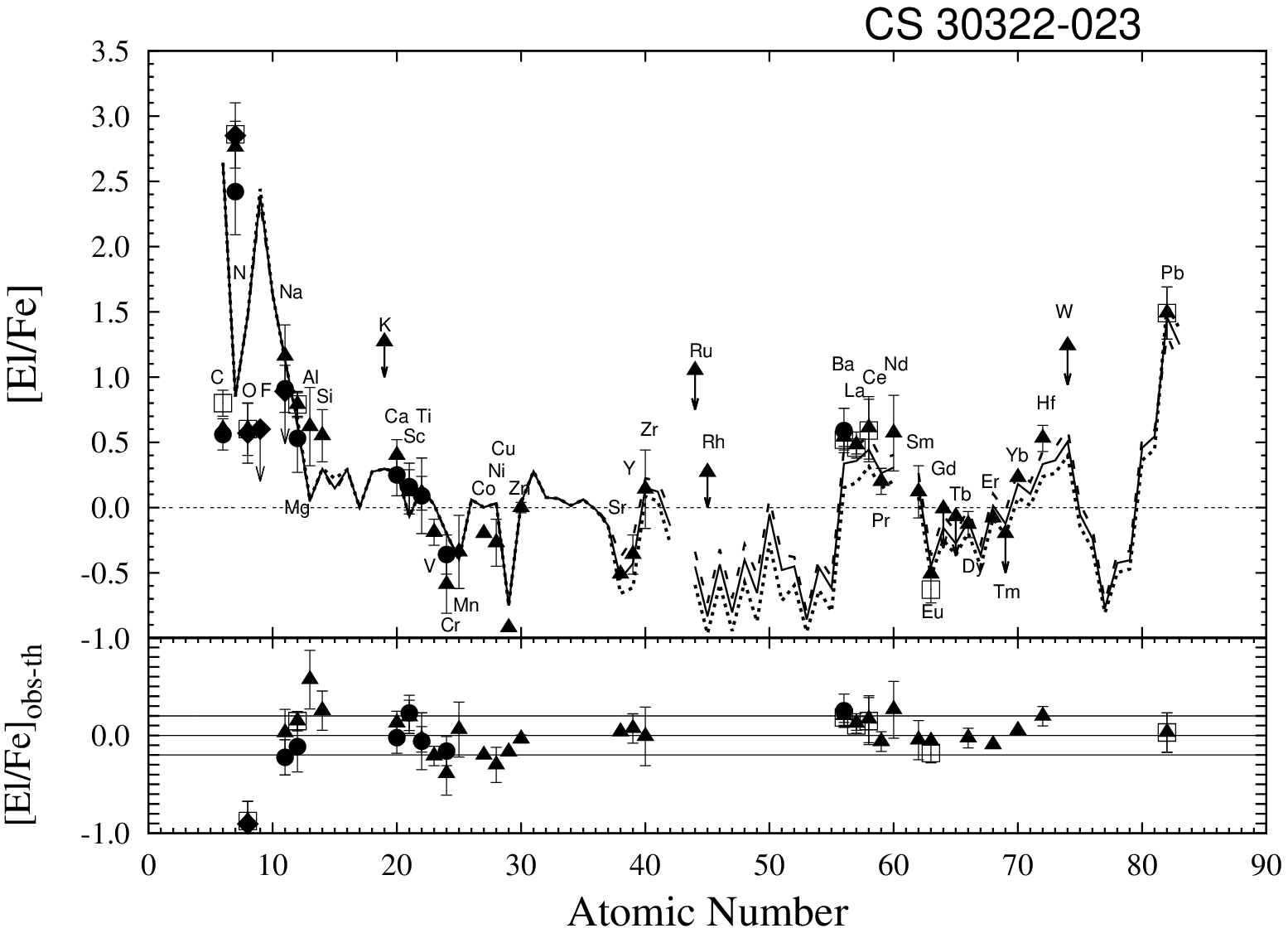}
\vspace{2mm}
\caption{Spectroscopic [El/Fe] abundances of the giant CS 30322--023 
([Fe/H] $\sim$ $-$3.4, {\textit T}$_{\rm eff}$ = 4100 K, log $g$ = $-$0.3 \citealt{masseron06,masseron10}; 
[Fe/H] = $-$3.25, {\textit T}$_{\rm eff}$ = 4300 K, log $g$ = 1.0 \citealt{aoki07})
 compared with AGB models of 
$M^{\rm AGB}_{\rm ini}$ = 1.5 $M_{\odot}$, cases ST/2 (dotted line), ST/3 (solid line), 
ST/5 (dashed line), and $dil$ = 2.5 dex. 
Observations are from \citet{masseron06} (filled triangles), \citet{masseron10} 
(empty squares), \citet{aoki07} (filled circles), \citet{lucatello11astroph} (filled
diamonds). This star shows [hs/ls] = 0.66 and [Pb/hs] = 0.96.
We assumed [Sr,Y/Fe]$^{\rm ini}$ = $-$1 in order to interpret the observed negative values. 
An [r/Fe]$^{\rm ini}$ = $-$1 is adopted.}
\label{CS30322-023_masseron06_aoki07_masseron10_bab10d1p3d2d3m1p5z5m6rm1srym1_dil2p49n20}
\end{figure}

\subsubsection{CS 30322--023 (Fig.~\ref{CS30322-023_masseron06_aoki07_masseron10_bab10d1p3d2d3m1p5z5m6rm1srym1_dil2p49n20})}
\label{CS023}

The giant CS 30322--023 is the most metal-poor star of the sample 
([Fe/H] = $-$3.5, $-$3.39 \citealt{masseron06,masseron10}; 
[Fe/H] = $-$3.25 \citealt{aoki07}).
Due to its low surface gravity ({\textit T}$_{\rm eff}$ = 4100 K; log $g$ = 
$-$0.3)\footnote{The uncertainties of the atmospheric parameters 
($\Delta${\textit T}$_{\rm eff}$ = 250 K and $\Delta$log $g$ = 
1.0 dex) is due to the use of LTE instead of NLTE atmospheric models. In fact, 
the gravities of metal-poor giants derived from LTE iron-line analysis are 
most probably underestimated by 0.5 up to 1 dex (\citealt{thevenin99}; \citealt{israelian01,israelian04}; 
\citealt{korn03}).}, the hypothesis of an 
intrinsic AGB star at the beginning of its TP-AGB phase was advanced by \citet{masseron06}.
 \citet{aoki07} measured
{\textit T}$_{\rm eff}$ = 4300 $\pm$ 100 K and log $g$ = 1.0 $\pm$ 0.3 dex, 
supporting instead the binary scenario. 
\\
CS 30322--023 shows a low carbon enhancement with respect to other CEMP-$s$
stars ([C/Fe] $\sim$ 0.6), together with a very strong nitrogen overabundance 
([N/Fe] $\sim$ 2.5 -- 2.8). While negative [Sr/Fe] and [Y/Fe] are detected, 
the observed [Zr/Fe] is slightly higher than solar, as other 
stars classified CEMP-$no/s$ by  \citet{sivarani06}, CS 29528--041 and CS 31080--095,
and possibly SDSS J1036+1212 \citep{behara10}, (see Paper II, Section~1).
\citet{cui07} hypothesised an AGB donor with initial mass in the range 
2 $\leq$ $M/M_{\odot}$ $\leq$ 4, due to the high [N/C] ratio\footnote{The 
condition $M^{\rm AGB}_{\rm ini}$ $>$ 2 $M_{\odot}$ comes from the 
hypothesis of the Hot Bottom Burning process in order to explain [N/Fe] $>$ [C/Fe], 
while the condition $M^{\rm AGB}_{\rm ini}$ $<$ 4 $M_{\odot}$ was adopted by 
the authors to justify the absence of an $r$-process overabundance, for which 
the AGB companion cannot evolve as Type 1.5 Supernova.}.
IMS stars with [Fe/H] $\la$ $-$2.3, undergo an extremely deep TDU,
in which the envelope reaches almost the bottom of the He intershell
\citep{sugimoto71,iben73,karakas03,ventura05}, possibly 
modifying the structure and the evolution of the star.
Nucleosynthesis models including hot TDU are under study.
\\
In Fig.~\ref{CS30322-023_masseron06_aoki07_masseron10_bab10d1p3d2d3m1p5z5m6rm1srym1_dil2p49n20}, 
we present possible interpretations with AGB models of 
$M^{\rm AGB}_{\rm ini}$ = 1.5 $M_{\odot}$ (cases ST/2, ST/3, ST/5) and a
very large dilution ($dil$ = 2.5 dex). Similar solutions are obtained with 
$M^{\rm AGB}_{\rm ini}$ = 1.4 -- 2 $M_{\odot}$ models and $dil$ = 1.7 to 3 dex.
The observed [Na/Fe] = 1.04 is underestimated by AGB models of lower initial
mass ($M^{\rm AGB}_{\rm ini}$ = 1.3 $M_{\odot}$).\footnote{Note that the observed 
Na accounts for NLTE and 3D corrections \citep{aoki07}.}
A low upper limit is detected for [F/Fe] by \citet{lucatello11astroph} ([F/Fe] 
$<$ 0.6).
AGB models predict larger fluorine abundance (see Paper I; Section~\ref{CS019}).

\noindent As introduced in Section~\ref{intro},  
we assume an initial [Cu/Fe]$^{\rm ini}$ = $-$0.7 according 
to the average of metal-poor Galactic stars. 
At halo metallicity, a spread is also observed
for [Sr/Fe] and [Y/Fe] in unevolved Galactic stars \citep{francois07}, 
and we adopt initial [Sr,Y/Fe]$^{\rm ini}$ = $-$1. 
[Eu/Fe] observations in unevolved stars show a large spread for [Fe/H] 
$\leq$ $-$2.0, from [Eu/Fe] $\sim$ $-$1 to +2 dex (Paper II, Fig.~2). 
The Eu detected in CS 30322--023 is negative ([Eu/Fe] = $-$0.6), requiring 
a negative initial $r$-process composition of the molecular cloud 
([r/Fe]$^{\rm ini}$ = $-$1).
A $dil$ = 2.5 dex means that the mass transferred from the AGB is 300
 times lower than the mass of the convective envelope of the secondary 
star; then, the $s$-process contribution is very low, and negative values 
for [Sr,Y/Fe], as well as for $r$-process elements as [Eu,Gd,Tb,Dy/Fe] are 
obtained. Otherwise, for $dil$ $\sim$ 1 dex the negative initial [El/Fe] 
abundances are overcome by the $s$-process contribution. 
\\
Because of the peculiarity of this star, caution in the interpretation
of the spectroscopic data is suggested. In fact, due to the very low metallicity, 
the AGB nucleosynthesis may differ from the canonical scenario due to the
occurrence of a proton ingestion episode (see Section~\ref{CS036}).

\begin{figure}
\includegraphics[angle=-90,width=8.5cm]{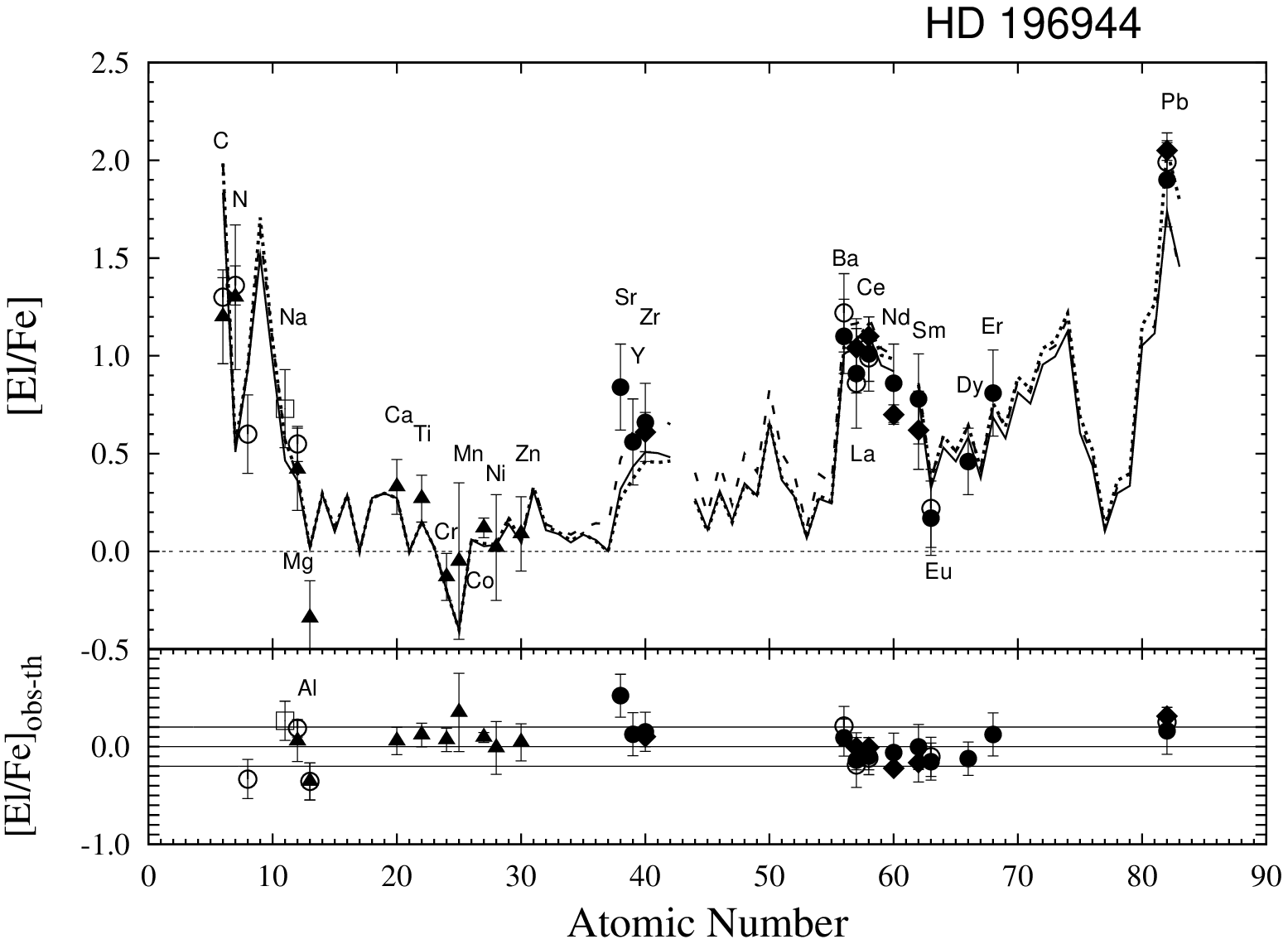}
\vspace{2mm}
\caption{Spectroscopic [El/Fe] abundances of the giant HD 196944
([Fe/H] = $-$2.25; {\textit T}$_{\rm eff}$ = 5250 K; log $g$ = 1.8)
compared with AGB models of $M^{\rm AGB}_{\rm ini}$ = 1.5 $M_{\odot}$, 
cases ST/3 (dotted line), ST/5 (solid line) and ST/6 (dashed line), $dil$ = 2.0 dex 
[hs/ls]$_{\rm obs}$ =  0.3; [Pb/hs]$_{\rm obs}$ = 1.0).
Observations are from \citet{aoki02c} (filled circles),
\citet{aoki02d} (filled triangles),
\citet{aoki07} (empty square),
\citet{vaneck03} (filled diamonds),
\citet{masseron10} (empty circles). 
No initial $r$-process enhancement is adopted.
This star was discussed in Paper II as representative of the CEMP-$s$ stars (see Fig.~12), 
and it is reported here for completeness.}
\label{HD196944_aoki02+aoki06+vaneck03+masseron10_bab10d2d3d4m1p5z1m4_nr_diffdiln20}
\end{figure}

\subsubsection{HD 196944 (Fig.~\ref{HD196944_aoki02+aoki06+vaneck03+masseron10_bab10d2d3d4m1p5z1m4_nr_diffdiln20})}
\label{HD196944} 

This CEMP-$s$I giant (\citealt{aoki02c,aoki02d,aoki07} and \citealt{vaneck03};
[Fe/H] = $-$2.25, {\textit T}$_{\rm eff}$ = 5250 K and log $g$ = 1.8)
has been analysed in Paper II, Section~5.
An AGB model of initial mass $M$ = 1.5 $M_{\odot}$, case ST/5, $dil$ $\sim$ 2.0
dex and no initial $r$-process enhancement provides a plausible theoretical
interpretation for this star 
(Fig.~\ref{HD196944_aoki02+aoki06+vaneck03+masseron10_bab10d2d3d4m1p5z1m4_nr_diffdiln20}). 
The observed [Na/Fe] constrains the AGB initial mass, while the
$s$-process elements ([hs/ls] = 0.3; [Pb/hs] = 1.0) may be equally interpreted 
by $M^{\rm AGB}_{\rm ini}$ = 1.3 and 2 $M_{\odot}$ models.

\begin{figure}
\includegraphics[angle=-90,width=8.5cm]{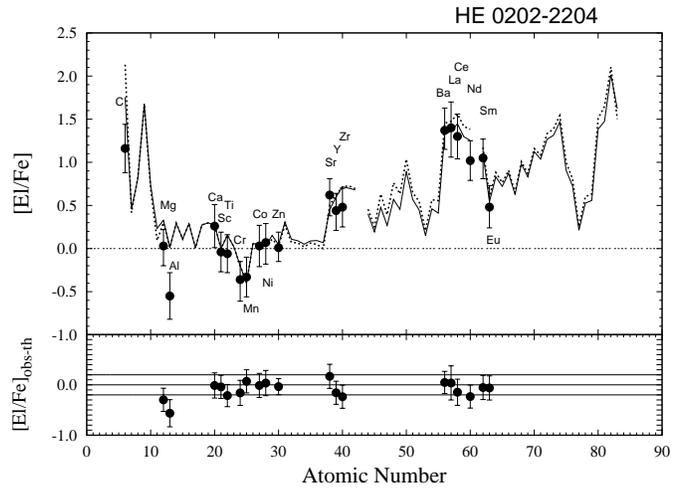}
\vspace{2mm}
\caption{Spectroscopic [El/Fe] abundances of the giant HE 0202--2204 
([Fe/H] = $-$ 1.98; {\textit T}$_{\rm eff}$ = 5280 K; log $g$ = 1.65)
compared with AGB models of $M^{\rm AGB}_{\rm ini}$ = 1.3 or 2 $M_{\odot}$, 
cases ST/9 or ST/6, $dil$ = 0.7 or 1.7 dex, (dotted or solid lines, respectively; 
[hs/ls]$_{\rm obs}$ = 0.67). Observations are from \citet{barklem05}. 
The [La/Eu]$_{\rm obs}$ is better interpreted by a pure $s$-process distribution
([r/Fe]$^{\rm ini}$ = 0.0, as shown here); however, the initial $r$-process enhancement
 [r/Fe]$^{\rm ini}$ = 0.5 generally adopted in this paper for CEMP-s stars
 agrees within the errors with the observations.
These models predict [Pb/Fe]$_{\rm th}$ $\sim$ 2.  
The differences [El/Fe]$_{\rm obs-th}$
displayed in the lower panel represent the $M^{\rm AGB}_{\rm ini}$ = 2 $M_{\odot}$ model.
No initial $r$-process enhancement is adopted.}
\label{HE0202-2204Bark05_bab10d6m1p5d4m2z2m4nr_diffdiln5n26}
\end{figure}

\subsubsection{HE 0202--2204 (Fig.~\ref{HE0202-2204Bark05_bab10d6m1p5d4m2z2m4nr_diffdiln5n26})}
\label{HE0202}

The giant HE 0202--2204 ([Fe/H] = $-$ 1.98, {\textit T}$_{\rm eff}$ = 5280 K 
and log $g$ = 1.65) was studied by \citet{barklem05}. 
In Fig.~\ref{HE0202-2204Bark05_bab10d6m1p5d4m2z2m4nr_diffdiln5n26}, we show
two AGB models in agreement with the observed [hs/ls] = 0.67:
 $M^{\rm AGB}_{\rm ini}$ = 1.3 and 2 $M_{\odot}$, cases ST/9 and ST/6
(solid and dashed lines, respectively). 
Both cases are in agreement with a giant having suffered the FDU because of the high
dilutions applied ($dil$ = 0.7 -- 1.7 dex). These models predict [Pb/Fe]$_{\rm th}$ 
$\sim$ 2. With higher $s$-process efficiencies (case ST) and $M^{\rm AGB}_{\rm ini}$
= 1.5 $M_{\odot}$ ($dil$ = 0.9 dex), the estimated Pb is very high ([Pb/Fe]$_{\rm th}$ 
$\sim$ 3.2): this solution is discarded because [Mg/Fe] would be largely overestimated 
 ([Mg/Fe]$_{\rm th}$ = 0.7).
No initial $r$-process enhancement is adopted in 
Fig.~\ref{HE0202-2204Bark05_bab10d6m1p5d4m2z2m4nr_diffdiln5n26}, but an 
[r/Fe]$^{\rm ini}$ = 0.5, generally assumed for CEMP-$s$ stars,
 still agrees within the [La/Eu] uncertainty.

\begin{figure}
\includegraphics[angle=-90,width=8.5cm]{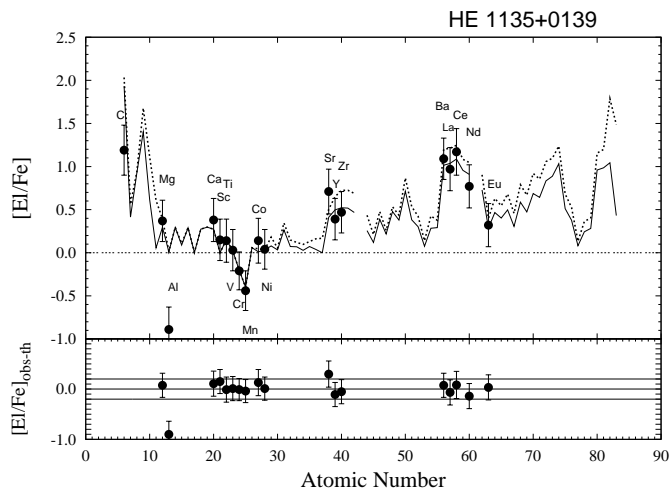}
\vspace{2mm}
\caption{Spectroscopic [El/Fe] abundances of the giant HE 1135+0139 
([Fe/H] = $-$2.33; {\textit T}$_{\rm eff}$ = 5487 K; log $g$ = 1.8)
compared with AGB models of $M^{\rm AGB}_{\rm ini}$ = 
1.3 $M_{\odot}$, case ST/24 and $dil$ = 1.2 dex (solid line) and $M^{\rm AGB}_{\rm ini}$ = 
1.5 $M_{\odot}$, case ST/6 and $dil$ = 1.8 dex (dotted line).
Observations are from \citet{barklem05}, who found [hs/ls] = 0.48. 
We predict [Pb/Fe]$_{\rm th}$ $\sim$ 1.0 -- 1.8. 
No initial $r$-process enrichment is assumed. 
The differences [El/Fe]$_{\rm obs-th}$ displayed in the 
lower panel refer to the AGB model represented with solid line.
This star was discussed in Paper II as representative of those CEMP-$s$ stars
without Pb detection (see Fig.~14), 
and it is reported here for completeness.}
\label{HE1135+0139_Barklem05_bab10d16d4m1p5z1m4nr_diffdiln520}
\end{figure}

\subsubsection{HE 1135+0139 (Fig.~\ref{HE1135+0139_Barklem05_bab10d16d4m1p5z1m4nr_diffdiln520})} 
\label{HE1135}

This giant, with [Fe/H] = $-$2.33, {\textit T}$_{\rm eff}$ = 5487 K 
and log $g$ = 1.8 \citep{barklem05}, has been analysed in Paper II, Section~5. 
Two possible theoretical interpretations were shown: $M^{\rm AGB}_{\rm ini}$ = 
1.3 $M_{\odot}$, case ST/24 and $dil$ = 1.2 dex and $M^{\rm AGB}_{\rm ini}$ = 
1.5 $M_{\odot}$, case ST/6 and $dil$ = 1.8 dex
(Fig.~\ref{HE1135+0139_Barklem05_bab10d16d4m1p5z1m4nr_diffdiln520}).
Similar solutions are obtained with $M^{\rm AGB}_{\rm ini}$ = 2 $M_{\odot}$. 
The high dilutions applied agree with a giant having suffered the FDU.
We predict [Pb/Fe]$_{\rm th}$ $\sim$ 1.0 -- 1.8.

\begin{figure}
\includegraphics[angle=-90,width=8.5cm]{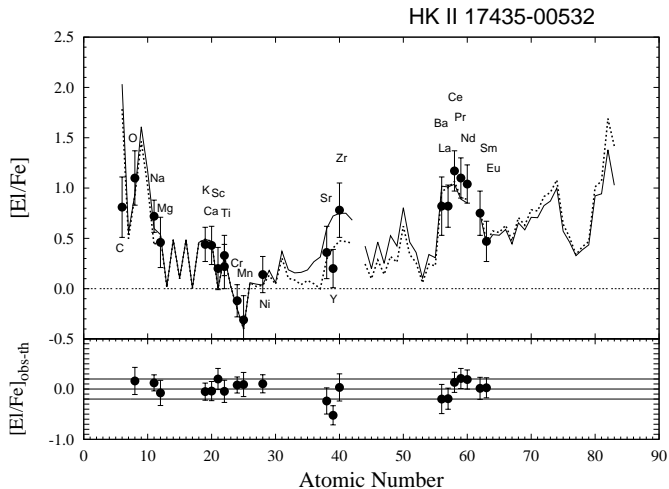}
\vspace{2mm}
\caption{Spectroscopic [El/Fe] abundances of the giant HK II 17435--00532 
([Fe/H] = $-$2.23; {\textit T}$_{\rm eff}$ = 5200 K; log $g$ = 2.15)
compared with $M^{\rm AGB}_{\rm ini}$ = 1.5 $M_\odot$ models, cases ST/5 and ST/12, 
$dil$ = 2.1 and 1.8 dex (dashed and solid lines, respectively). 
Observations are from \citet{roederer08}, who detected [hs/ls] = 0.55.
We predict [Pb/Fe]$_{\rm th}$ $\sim$ 1.4 -- 1.7.
The differences [El/Fe]$_{\rm obs-th}$ displayed in the 
lower panel refer to case ST/12, represented with solid line.
An [r/Fe]$^{\rm ini}$ = 0.3 is adopted.}
\label{HK_roederer08_bab10d4d8d12m1p5z1m4rp0p3alf0p5_dil1p8n20}
\end{figure}

\subsubsection{HK II 17435--00532 (Fig.~\ref{HK_roederer08_bab10d4d8d12m1p5z1m4rp0p3alf0p5_dil1p8n20})}
\label{HKII}

A complete analysis of this mild $s$-process giant was provided by
 \citet{roederer08} ([Fe/H] = $-$2.23, {\textit T}$_{\rm eff}$ = 5200 K 
and log $g$ = 2.15). 
They found an unexpected high amount of lithium.
So far HK II 17435--00532 does not seem to be member of a binary system.
Further radial velocity measurements are required because the observation was 
done over a time span of about 180 days, which does not permit to
discover very long periods. 
\\
Theoretical interpretations with AGB models have been widely discussed
by \citet{roederer08}.
In Fig.~\ref{HK_roederer08_bab10d4d8d12m1p5z1m4rp0p3alf0p5_dil1p8n20},
we show solutions with $M^{\rm AGB}_{\rm ini}$ = 1.5 $M_\odot$ models.
The solid line represents the case ST/12 and $dil$ = 1.8 dex examined by 
\citet{roederer08}, in agreement with the observed [hs/ls] = 0.55. 
No lead is detected. [Y/Fe]$_{\rm obs}$ is 0.5 dex lower than the 
AGB prediction. 
We present here an additional interpretation with a M$^{\rm AGB}_{\rm ini}$ 
= 1.5 $M_\odot$ model, case ST/5 and $dil$ = 2.1 dex (dashed line).
This case agrees better with the observed [Y/Fe], 
but predicts a low [Na/Fe] (about 0.3 dex). Even by decreasing [Na/Fe]
of about 0.1 dex owing to NLTE corrections \citep{roederer08}, [Na/Fe]
would be slightly higher than the value predicted by the 
dashed line. 
We estimate [Pb/Fe]$_{\rm th}$ $\sim$ 1.4 -- 1.7. 
AGB solutions with lower initial mass (e.g., $M^{\rm AGB}_{\rm ini}$ 
= 1.3 $M_\odot$) are discarded, because [Na/Fe] would be underestimated by such models
([Na/Fe]$_{\rm th}$ $\sim$ 0). 
This star requires a low $r$-process enrichment ([r/Fe]$^{\rm ini}$ = 0.3).

\begin{figure}
\includegraphics[angle=-90,width=8.5cm]{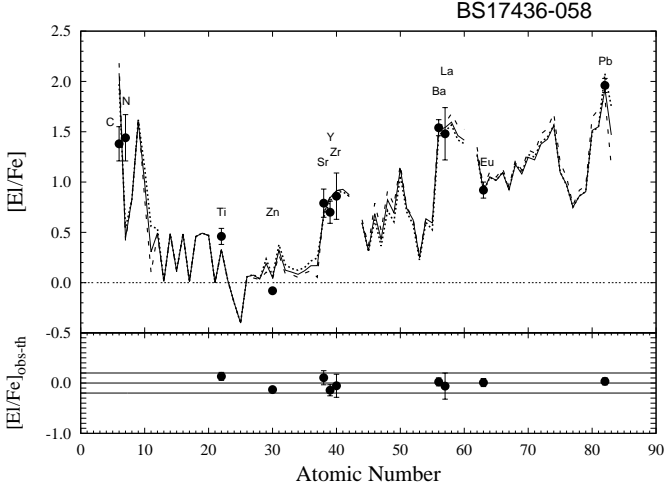}
\vspace{2mm}
\caption{Spectroscopic [El/Fe] abundances of the giant BS 17436--058 
([Fe/H] = $-$1.9; {\textit T}$_{\rm eff}$ = 5390 K; log $g$ = 2.2)
compared with three AGB stellar models: $M^{\rm AGB}_{\rm ini}$ = 1.3 $M_{\odot}$, 
ST/12, $dil$ = 0.7 dex (dashed line), $M^{\rm AGB}_{\rm ini}$ = 1.4 $M_{\odot}$, 
ST/9, $dil$ = 1.2 dex (solid line), $M^{\rm AGB}_{\rm ini}$ = 1.5 $M_{\odot}$, 
ST/5, with $dil$ = 1.6 dex (dotted line).
Observations are from \citet{tsangarides05}, who detected [hs/ls] = 0.60 and 
[Pb/hs] = 0.76. 
The differences [El/Fe]$_{\rm obs-th}$ displayed in lower panel refer to 
a model for $M^{\rm AGB}_{\rm ini}$ = 1.4 $M_{\odot}$ (solid line).
An [r/Fe]$^{\rm ini}$ = 0.7 is adopted (see text).}
\label{BS17436-058_tsangarides05_bab10d8d6d3m1p5z2m4rp0p7alf0p5_diffdiln5n9n20}
\end{figure}

\subsubsection{BS 17436--058 (Fig.~\ref{BS17436-058_tsangarides05_bab10d8d6d3m1p5z2m4rp0p7alf0p5_diffdiln5n9n20})}
\label{BS058}

BS 17436--058 is a giant with [Fe/H] = $-$1.9, {\textit T}$_{\rm eff}$ = 5390 K 
and log $g$ = 2.2 \citep{tsangarides05}.
AGB models with an initial mass from 1.3 to 2 $M_{\odot}$ ($dil$ = 0.7 -- 1.6 dex) may
equally interpret the observations.  
Measurement of Na and Mg are highly desirable.
In Fig.~\ref{BS17436-058_tsangarides05_bab10d8d6d3m1p5z2m4rp0p7alf0p5_diffdiln5n9n20},
we show three possible solutions:  $M^{\rm AGB}_{\rm ini}$ = 1.3 $M_{\odot}$, 
ST/12, $dil$ = 0.7 dex, $M^{\rm AGB}_{\rm ini}$ = 1.4 $M_{\odot}$, 
ST/9, $dil$ = 1.2 dex, $M^{\rm AGB}_{\rm ini}$ = 1.5 $M_{\odot}$, 
ST/5, $dil$ = 1.6 dex, all in agreement with the two $s$-process indicators
[hs/ls] = 0.60 and [Pb/hs] = 0.76.
The observed [La/Eu] is interpreted with an $r$-process enhancement 
[r/Fe]$^{\rm ini}$ = 0.7. Note that an [r/Fe]$^{\rm ini}$ = 0.5, generally 
adopted for CEMP-$s$ stars, still agrees within the uncertainties.
No velocity variations are detected; this does not necessarily disprove 
the presence of a companion, since this star could have an orbital axis 
inclination which does not permit observations of velocity variations
\citep{tsangarides05}.


\section{CEMP-{\scriptsize s/r} stars} \label{secCEMPs/r}

This Section is dedicated to CEMP-$s$II$/r$ stars,
divided in two smaller groups following their $r$-process
enhancement:
CEMP-$s$II$/r$II with [r/Fe]$^{\rm ini}$ included between $\sim$ 1.5
and 2 (Section~\ref{secCEMPs/rIIwrII} and~\ref{secCEMPs/rIIwr1p5})
and CEMP-$s$II$/r$I with [r/Fe]$^{\rm ini}$ = 1.0 (Section~\ref{secCEMPs/rIIwrI}).

\subsection{CEMP-$s$II$/r$II with [r/Fe]$^{\rm ini}$ $\sim$ 2} 
\label{secCEMPs/rIIwrII}

We discuss in this Section six stars showing very high $s$- 
and $r$-process enhancements ([hs/Fe] $\sim$ [Eu/Fe] $\sim$ 2) 
interpreted with an initial $r$-enrichment of the molecular cloud 
[r/Fe]$^{\rm ini}$ $\sim$ 2. 
Five are main-sequence/turnoff stars,
CS 22898--027 by \citet{aoki02c,aoki02d,aoki07} with [La/Eu] = 0.25, 
CS 29497--030 by \citet{ivans05} with [La/Eu] = 0.23, 
HE 0338--3945 by \citet{jonsell06} with [La/Eu] = 0.34
(discussed in Paper II, Section~5), 
HE 1105+0027 by \citet{barklem05} with [La/Eu] = 0.29, 
HE 2148--1247 by \citet{cohen03} with [La/Eu] = 0.40;
one is a giant HE 1305+0007 by \citet{goswami06} with [La/Eu] = 0.59 
(and very high [La/Fe] = 2.56).

\begin{figure}
\includegraphics[angle=-90,width=8.5cm]{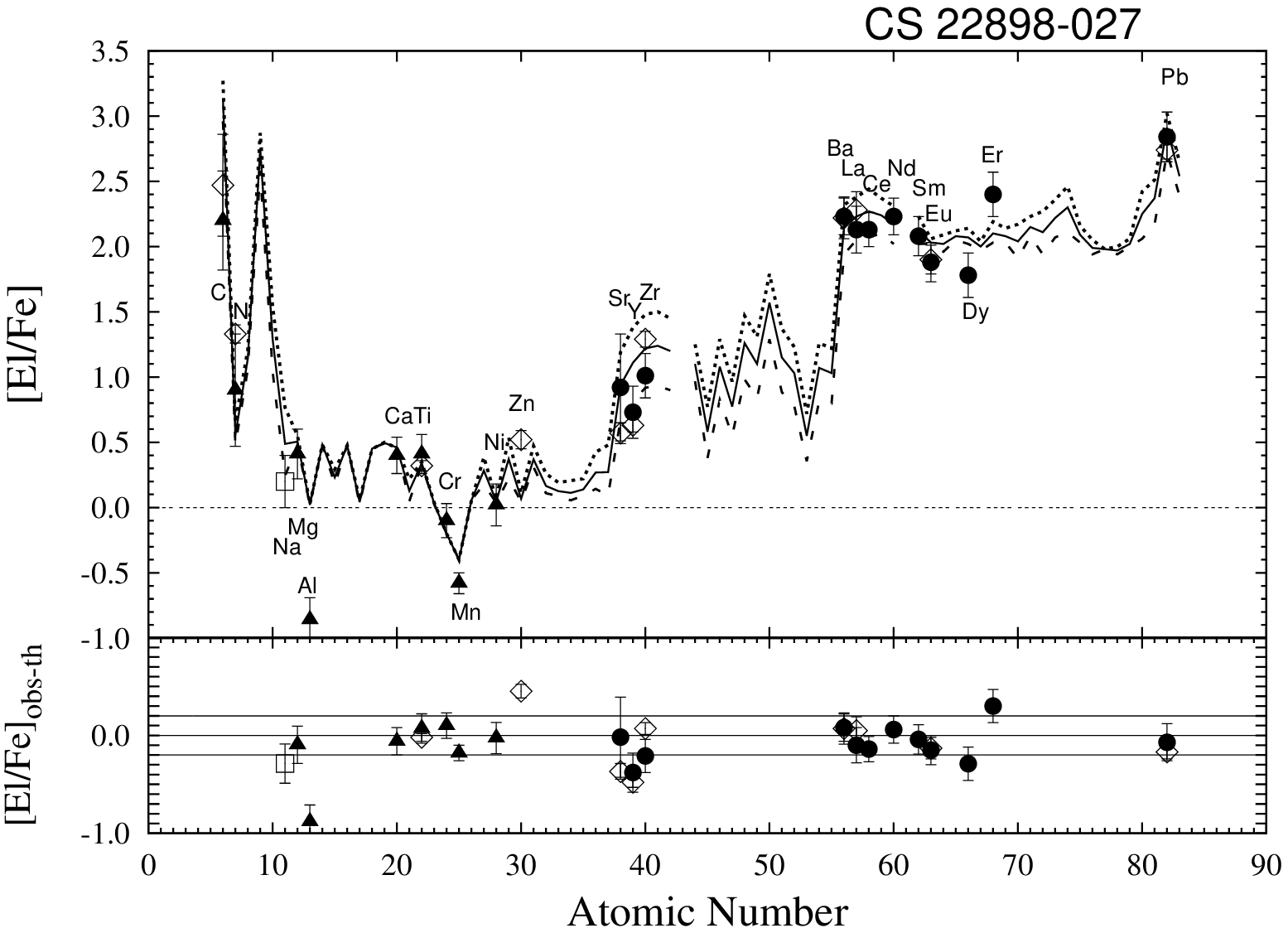}
\vspace{2mm}
\caption{Spectroscopic [El/Fe] abundances of the turnoff/subgiant CS 22898--027 
 ([Fe/H] = $-$2.26; {\textit T}$_{\rm eff}$ = 6250 K; log $g$ = 3.7, before the FDU)
compared with AGB models
of $M^{\rm AGB}_{\rm ini}$ $\sim$ 1.3 $M_{\odot}$, case ST/12 and no dilution.
Observations are from \citet{aoki02c} (filled triangles), \citet{aoki02d} (filled
circles), \citet{aoki07} (empty square), \citet{tsangarides05} (empty diamonds).
\citet{aoki02c} detected [hs/ls] = 1.30 and [Pb/hs] = 0.67. 
Three thermal pulses are represented, pulses 4 (dashed line), 5 (solid line) and 
6 (dotted line), corresponding to an increase of the AGB initial mass of 
$\Delta M$ $\sim$ 0.025 $M_{\odot}$  (see text).
An initial $r$-process enrichment [r/Fe]$^{\rm ini}$ = 2.0 is assumed 
([La/Eu]$_{\rm s+r}$ = 0.2; [La/Eu]$_{\rm s}$ = 0.9).}
\label{CS22898-027_aoki02+aoki07+tsangarides05_bab10d8m1p5z1m4rp2alf0p5_n456}
\end{figure}

\subsubsection{CS 22898--027 (Fig.~\ref{CS22898-027_aoki02+aoki07+tsangarides05_bab10d8m1p5z1m4rp2alf0p5_n456})}
\label{CS027}

This turnoff star ([Fe/H] = $-$2.26; {\textit T}$_{\rm eff}$ = 6250 K 
and log $g$ = 3.7, before the occurrence of the FDU) has been analysed by 
\citet{PS01}, \citet{aoki02c,aoki02d}, and \citet{aoki07}.
No radial velocity variations are found for this star
\citep{PS01,aoki02c,tsangarides05,preston09pasa}. 
\\
A theoretical interpretation of the spectroscopic abundances
was presented by \citet*{SCG08}.
In Fig.~\ref{CS22898-027_aoki02+aoki07+tsangarides05_bab10d8m1p5z1m4rp2alf0p5_n456},
we provide similar solutions with updated models.
The observed [Na/Fe] is low while the second $s$-peak is high ([hs/Fe] $\la$ 2),
 as in several main-sequence stars. This agrees  
with a $M^{\rm AGB}_{\rm ini}$ $\sim$ 1.3 $M_{\odot}$ model,
ST/12 and no dilution.
 Three thermal pulses with TDU are shown.
They represent a plausible decrease (n4) or increase (n6) in mass of about 0.025 $M_{\odot}$ for 
$M^{\rm AGB}_{\rm ini}$ $\sim$ 1.3 $M_{\odot}$ models (see Section~\ref{CS036}). 
An additional TDU corresponds to an increase of the abundances of $\sim$ 
0.2 dex. 
The model with 4 TDUs seems to better interpret the low [Na/Fe] and [Y/Fe] observed.
 However, only one line has been detected for Na, while 2 lines for Y (as for Sr and Zr),
 and both elements agree with theoretical predictions within an uncertainty of 0.2 dex.
An initial $r$-enrichment [r/Fe]$^{\rm ini}$ = 2.0 is adopted in order to 
interpret the average among Eu, Dy and Er (3, 2 and 4 lines, respectively). 
By increasing the number of TDUs ($M^{\rm AGB}_{\rm ini}$ = 1.5 or 2 
$M_{\odot}$; case ST/4.5; $dil$ $\sim$ 1 dex), the observed [Na/Fe] and [ls/Fe] would be 
overestimated by AGB models. 
 This excludes the possibility that the star underwent efficient mixing 
during its main-sequence phase.

\begin{figure}
\includegraphics[angle=-90,width=8.5cm]{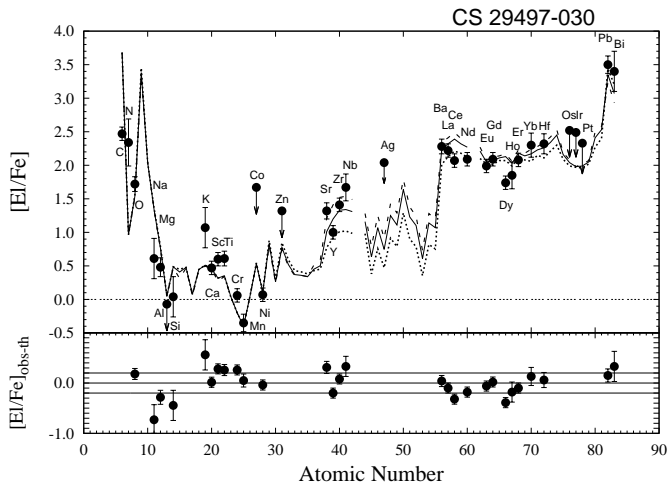}
\vspace{2mm}
\caption{Spectroscopic [El/Fe] abundances of the main-sequence star CS 29497--030 
([Fe/H] = $-$2.57; {\textit T}$_{\rm eff}$ = 7000 K; log $g$ = 4.1)
compared with AGB models of $M^{\rm AGB}_{\rm ini}$ = 1.35 $M_{\odot}$
(pulse 7, see Sections~\ref{CS036} and~\ref{CS027}), 
cases ST/6 (dotted line), ST/9 (solid line), ST/12 (dashed line), 
and no dilution. Observations are from \citet{ivans05}, who found [hs/ls] = 1.02 
and [Pb/hs] = 1.46.
Note that the values represented for O, Na, Al, and K do not account for  NLTE corrections.
An initial $r$-process enrichment [r/Fe]$^{\rm ini}$ = 2 is assumed
([La/Eu]$_{\rm s+r}$ = 0.3; [La/Eu]$_{\rm s}$ = 0.9).}
\label{CS29497-030_es}
\end{figure}

\subsubsection{CS 29497$-$030 (Fig.~\ref{CS29497-030_es})}
\label{CS030}

\citet{PS00} investigated a sample of 62 blue metal$-$poor 
(BMP) stars, including the main-sequence star CS 29497--030, 
arguing that half are binaries. 
This is the case of CS 29497--030, a BMP star subject to 
different studies (\citealt{sneden03b}; \citealt{sivarani04}; 
\citealt{ivans05}). \citet{sneden03b} confirm the binary scenario. 
\\
For this analysis, we consider only the most recent observations provided
by \citet{ivans05} ([Fe/H] = $-$2.57; {\textit T}$_{\rm eff}$ = 7000 K 
and log $g$ = 4.1). 
Theoretical AGB interpretations have been presented by \citet{ivans05},
\citet{bisterzo08FSIII} and \citet{kaeppeler11rmp}. 
In Fig.~\ref{CS29497-030_es}, solutions with $M^{\rm AGB}_{\rm ini}$
$\sim$ 1.35 $M_{\odot}$ models (at the seventh TDU, n = 7), cases ST/6, ST/9 
and ST/12, and no dilution interpret the observed [hs/ls] = 1.02 and [Pb/hs] = 1.46. 
CS 29497--030 has the highest lead observed so far, 
[Pb/Fe] = 3.65.
For the first time in metal-poor stars Bi is detected, with a high
overabundance, in agreement with AGB predictions at these low 
metallicities: stars with a huge amount of lead are also expected to 
exhibit a high $s$-process abundance of bismuth. In fact, despite the solar 
bismuth is mainly produced by the $r$-process ($\sim$ 80\%), at 
[Fe/H] $\sim$ $-$2.6 and for a given $^{13}$C-pocket, the number 
of neutrons per iron seed is $\sim$ 400 times higher than solar, 
directly feeding the third $s$-process peak (Paper I). 
Also Nb was detected in this star supporting the binary  
scenario ([Zr/Nb] $\sim$ 0; \citealt{ivans05}, see also Paper I). 
[Na/Fe] is overestimated by AGB models in Fig.~\ref{CS29497-030_es}. 
We recall that Na may have a large uncertainty (of 0.6 dex or more) due 
to poorly understood NLTE effects on Na line formation and for 3D 
atmospheric models.
By increasing the AGB initial mass, and therefore the number 
of TDUs, a dilution factor must be applied in order to reproduce the 
observed values, but both Na and Mg would be highly overestimated
by theoretical models \citep{bisterzo08FSIII}. 
Interpretations with negligible dilutions are compatible with 
moderate mixing during the main-sequence phase.
An initial $r$-enrichment of 2 dex is assumed. 
Note that only an upper limit has been detected for Ag and at present
we do not adopt initial $r$-process contributions for isotopes lower 
than Ba (see Paper II, Section~3).
The low [Y/Fe] observed does not agrees with AGB predictions.
The hypothesis of an initial subsolar [Y/Fe] does not change sensibly the 
final [Y/Fe] prediction, because of the high $s$-process contribution
to Y together with no dilution \citep{ivans05}.

\begin{figure}
\includegraphics[angle=-90,width=8.5cm]{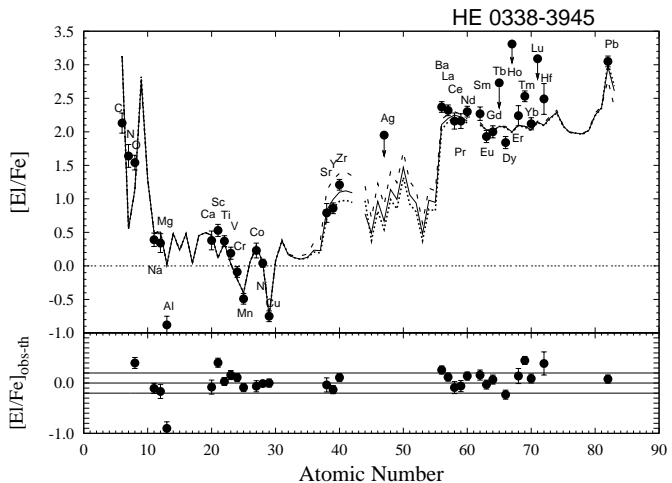}
\vspace{2mm}
\caption{Spectroscopic [El/Fe] abundances of the main-sequence star HE 0338--3945 
([Fe/H] = $-$2.42; {\textit T}$_{\rm eff}$ = 6160 K; log $g$ = 4.1)
compared with AGB models of $M^{\rm AGB}_{\rm ini}$ = 1.3 $M_{\odot}$ 
cases ST/9 (dotted line), ST/11 (solid line), ST/15 (dashed line), and no dilution. 
Observations are from \citet{jonsell06}, who found [hs/ls] = 1.24 and [Pb/hs] = 0.81. 
An initial $r$-process enrichment [r/Fe]$^{\rm ini}$ = 2.0 is assumed
(([La/Eu]$_{\rm s+r}$ = 0.12; [La/Eu]$_{\rm s}$ = 0.87). 
We assumed an initial [Cu/Fe] = $-$0.7, in agreement with unevolved halo stars 
at this metallicity (see Section~\ref{intro}).
This star was discussed in Paper II as representative of the CEMP-$s/r$ stars (see Fig.~15), 
and is reported here for completeness.}
\label{HE0338-3945_Jonsell06_bab10d6710m1p5z1m4_heavyrp2_alf0p5_n5}
\end{figure}

\subsubsection{HE 0338--3945 (Fig.~\ref{HE0338-3945_Jonsell06_bab10d6710m1p5z1m4_heavyrp2_alf0p5_n5})} 
\label{HE0338}

This main-sequence star, with [Fe/H] = $-$2.42,
{\textit T}$_{\rm eff}$ = 6160 K and log $g$ = 4.1,
has been widely discussed in Paper II, Section~5.
An AGB model of initial mass $M$ = 1.3 $M_{\odot}$, case ST/11, 
no dilution and [r/Fe]$^{\rm ini}$ = 2.0 provides a plausible theoretical
interpretation for the observed [hs/ls] = 1.24, [Pb/hs] = 0.81 and [La/Eu]
= 0.34 (Fig.~\ref{HE0338-3945_Jonsell06_bab10d6710m1p5z1m4_heavyrp2_alf0p5_n5}). 
Solutions with no dilution suggest that only
negligible or moderate mixing occurred in this star.
The main constraints about the AGB initial mass are provided by the low 
[Na/Fe] and [ls/Fe] observed. 
Among the heavy elements, Ba, Dy and Hf lie within 0.2 dex of AGB model uncertainty, 
while the observed [Tm/Fe] is 0.4 dex larger than AGB predictions. The four Tm lines 
analysed in this star have oscillator strengths from \citet{kurucz95}, which, according to
\citet{sneden96}, are rescaled laboratory data from \citet{corliss62}. 
Tm should be reconsidered with the high-quality $gf$-values published by \citet{wickliffe97}.
A negative [Cu/Fe] is observed, in agreement with unevolved halo stars
(see Section~\ref{intro}).

\begin{figure}
\includegraphics[angle=-90,width=8.5cm]{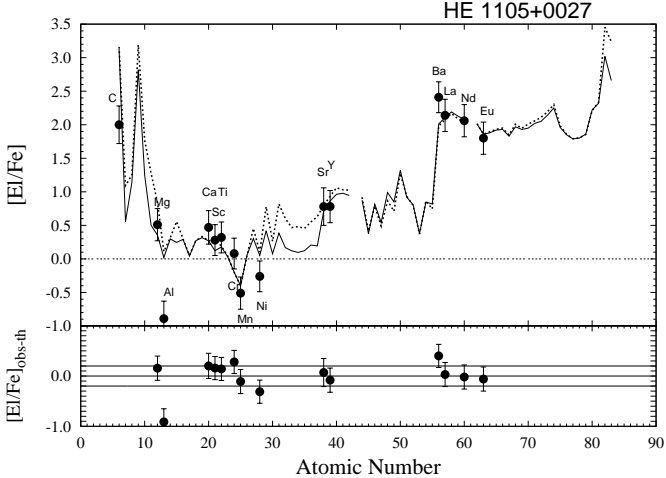}
\vspace{2mm}
\caption{Spectroscopic [El/Fe] abundances of the main-sequence/turnoff star HE 1105+0027 
([Fe/H] = $-$2.42; {\textit T}$_{\rm eff}$ = 6132 K; log $g$ = 3.45)
compared with AGB models of $M^{\rm AGB}_{\rm ini}$ = 1.3 $M_{\odot}$ 
(ST/9 and no dilution; solid line) or $M^{\rm AGB}_{\rm ini}$ = 2 $M_{\odot}$ 
(ST/3 and $dil$ = 0.6 dex; dotted line). 
Observations are from \citet{barklem05}, who found [hs/ls] = 1.15.
For lead we predict [Pb/Fe]$_{\rm th}$ $\sim$ 3. 
The differences [El/Fe]$_{\rm obs-th}$ displayed in the lower panel refer to 
the $M^{\rm AGB}_{\rm ini}$ = 1.3 $M_{\odot}$ model, represented by the solid line. 
An initial $r$-process enrichment [r/Fe]$^{\rm ini}$ = 1.8
is assumed. }
\label{HE1105+0027Bark05_bab10d6m1p5d2m2z1m4rp1p8_diffdiln5n26}
\end{figure}

\subsubsection{HE 1105+0027 (Fig.~\ref{HE1105+0027Bark05_bab10d6m1p5d2m2z1m4rp1p8_diffdiln5n26})}
\label{HE1105}

HE 1105+0027 lies close to the turnoff 
([Fe/H] = $-$2.42; {\textit T}$_{\rm eff}$ = 6132 K and log $g$ = 3.45; \citealt{barklem05}).
Two possible theoretical interpretations for HE 1105+0027 are shown in
Fig.~\ref{HE1105+0027Bark05_bab10d6m1p5d2m2z1m4rp1p8_diffdiln5n26}, 
with $M^{\rm AGB}_{\rm ini}$ = 1.3 $M_{\odot}$, case ST/9 and no dilution, 
and $M^{\rm AGB}_{\rm ini}$ = 2 $M_{\odot}$, case ST/3 and $dil$ = 0.6 dex.
A Na measurement is highly desirable. [Mg/Fe] (0.5 $\pm$ 0.24) agrees 
within the errors with both AGB models, even if the solution with a lower
number of TDUs better interprets the observations.
[Ba/Fe]$_{\rm obs}$ (with 2 lines detected) is underestimated by models. 
La (6 lines) and Nd (9 lines) are considered more reliable ([hs/ls] = 1.15). 
We predict [Pb/Fe]$_{\rm th}$ $\sim$ 3. 
An $r$-process enrichment of [r/Fe]$^{\rm ini}$ = 1.8 dex is adopted.

\begin{figure}
\includegraphics[angle=-90,width=8.5cm]{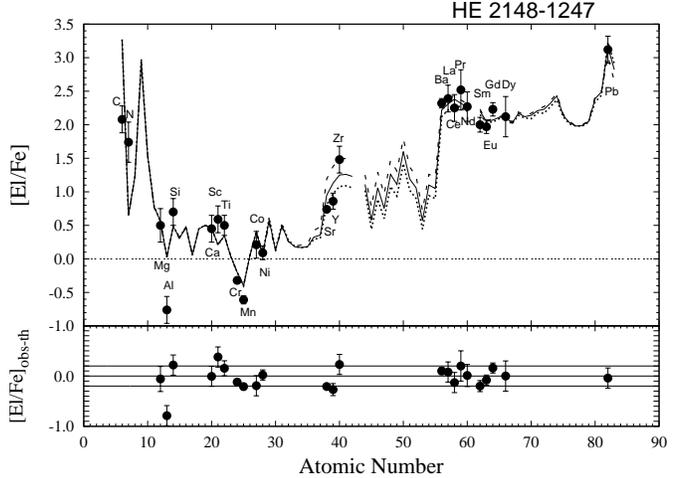}
\vspace{2mm}
\caption{Spectroscopic [El/Fe] abundances of the main-sequence star HE 2148--1247
([Fe/H] = $-$2.3; {\textit T}$_{\rm eff}$ = 6380 K; log $g$ = 3.9)
compared with AGB models of $M^{\rm AGB}_{\rm ini}$ $\sim$ 1.35 $M_{\odot}$ 
(pulse 6, see Sections~\ref{CS036} and~\ref{CS027}), 
cases ST/8 (dotted line), ST/9 (solid line), ST/12 (dashed line), 
and no dilution.
Observations are from \citet{cohen03}, who detected [hs/ls] = 1.21 and [Pb/hs] 
= 0.84. 
An initial $r$-process enrichment of [r/Fe]$^{\rm ini}$ = 2.0 is
adopted ([La/Eu]$_{\rm s+r}$ = 0.26; [La/Eu]$_{\rm s}$ = 0.86).} 
\label{HE2148-1247_cohen03_bab10}
\end{figure}

\subsubsection{HE 2148--1247 (Fig.~\ref{HE2148-1247_cohen03_bab10})}
\label{HE2148}

The main-sequence star HE 2148--1247 ([Fe/H] = $-$2.3, {\textit T}$_{\rm eff}$ = 6380
 K and log $g$ = 3.9) was the first showing high
enhancements in both $s$ and $r$-process elements \citep{cohen03},
with [Eu/Fe] = 2, [La/Eu] = 0.4, [hs/Eu] = 0.26.
\citet{cohen03} classify this star as a small-amplitude long-period 
binary.

In Fig.~\ref{HE2148-1247_cohen03_bab10}, we show theoretical interpretations
with $M^{\rm AGB}_{\rm ini}$ $\sim$ 1.35 $M_{\odot}$ models (cases ST/8, ST/9, and ST/12)
and no dilution. 
Note the difference between the observed [Y/Fe] and [Zr/Fe] ($\sim$ 0.6 dex;
5 and 3 lines detected, respectively), which disagrees with AGB predictions. 
Similar solutions may be obtained by $M^{\rm AGB}_{\rm ini}$ = 2 $M_{\odot}$ 
models with $dil$ = 0.7 dex and case ST/6. 
At present, the theoretical interpretations shown in Fig.~\ref{HE2148-1247_cohen03_bab10} 
seem to better agree with the observed [Mg/Fe] and [ls/Fe], sustaining the hypothesis of 
negligible or moderate mixing. However, we may not exclude solutions with 
$M^{\rm AGB}_{\rm ini}$ = 2 $M_{\odot}$, which predicts [Mg/Fe]$_{\rm th}$ = 0.8
and [ls/Fe]$_{\rm th}$ = 1.4. 
AGB models with $M^{\rm AGB}_{\rm ini}$ = 1.5 $M_{\odot}$ ($dil$ = 0.5 dex; 
case ST/3), are in accord with [hs/ls]$_{\rm obs}$ = 1.21 and [Pb/hs]$_{\rm obs}$ 
= 0.84, but [Mg/Fe]$_{\rm th}$ = 1.16 would be about 0.5 dex higher than observed.
A Na detection would help to assess the AGB initial mass. 
An initial $r$-process enrichment of [r/Fe]$^{\rm ini}$ = 2.0 is
adopted.

\begin{figure}
\includegraphics[angle=-90,width=8.5cm]{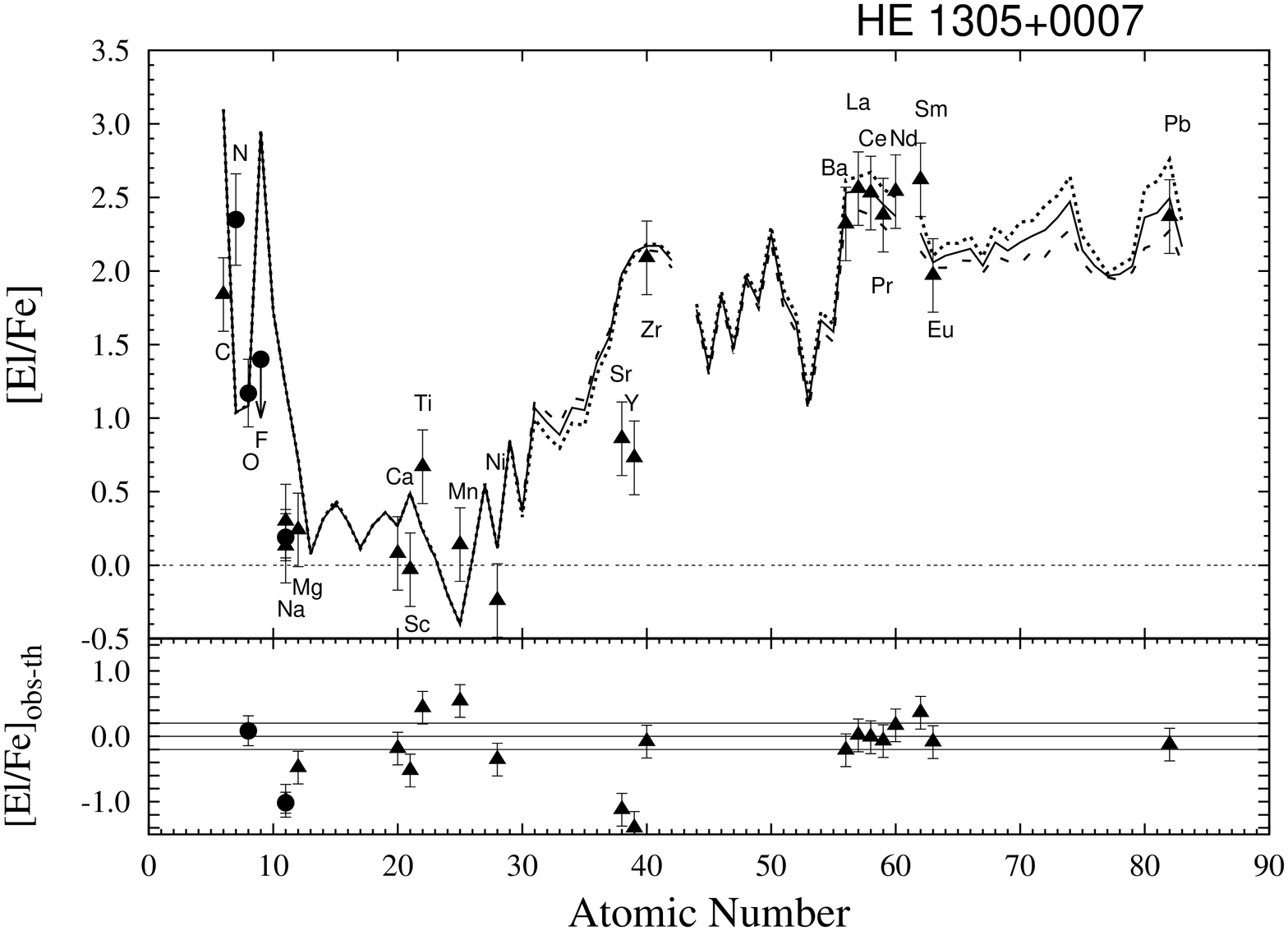}
\vspace{2mm}
\caption{Spectroscopic [El/Fe] abundances of the giant HE 1305+0007
([Fe/H] = $-$2.03; {\textit T}$_{\rm eff}$ = 4750 K; log $g$ = 2.0)
compared with AGB models of $M^{\rm AGB}_{\rm ini}$ = 2 $M_{\odot}$, cases ST/12 (dotted line), 
ST/15 (solid line), ST/18 (dashed line), and $dil$ = 0.4 dex.
Observations are from \citet{goswami06} (filled triangles)
and \citet{lucatello11astroph} (filled circles).
Typical error bars of $\pm$ 0.3 dex are shown for data by \citet{goswami06}, 
in agreement with the maximum fitting errors estimated by the authors.
This star shows [hs/ls] = 1.17 and [Pb/hs] = $-$0.21.
Note that AGB models can not predict [Zr/Y] $\sim$ 1.3: we consider 
[ls/Fe]$_{\rm th}$ $\sim$ [Zr/Fe]$_{\rm th}$ $\sim$ 2, overestimating 
the observed [Sr/Fe] and [Y/Fe] by about 1 dex. 
Both the [Na/Fe] abundance calculated from the resonance doublet Na I D lines
are shown.
[N/Fe], [O/Fe], [F/Fe] by \citet{lucatello11astroph} 
have been evaluated by adopting lower atmospheric parameters
and metallicity than \citet{goswami06} ($\Delta$[Fe/H] = 0.5, $\Delta$log
$g$ = 1).
An [r/Fe]$^{\rm ini}$ = 2.0 is adopted.}
\label{HE1305+0007_goswami06_bab10d8d10d12m2z2m4rp2_dil0p35n26}
\end{figure}

\subsubsection{HE 1305+0007 (Fig.~\ref{HE1305+0007_goswami06_bab10d8d10d12m2z2m4rp2_dil0p35n26})}
\label{HE1305}

The giant HE 1305+0007 shows [Fe/H] = $-$2.03, {\textit T}$_{\rm eff}$ = 4750
K and log $g$ = 2.0 \citep{goswami06}. 
The spectrum is dominated by molecular absorption lines of CH, CN, C$_2$.
Note the large spread covered by the ls elements, 
with [Zr/Fe] about 1.4 dex higher than [Y/Fe]\footnote{A similar behaviour 
has been also observed in the other two cold giants 
studied by \citet{goswami06}: HD 5223 for ls-peak with [Zr/Y] $\sim$ 1 
(Section~\ref{secCEMPsIInoEu}) and HE 1152--0355 for hs-peak
with [La/Nd] $\sim$ 1.3 (Section~\ref{secCEMPsInoEu}).}. 
Instead, AGB models predict [Zr/Y] and [La/Nd] $\sim$ 0.
The second $s$-peak is very enhanced ([hs/Fe] $\sim$ 2.5), similar to 
[Zr/Fe] and [Pb/Fe] ([Pb/hs] $\sim$ 0).
By considering [ls/Fe] $\sim$ [Zr/Fe],
possible theoretical interpretations are displayed in 
Fig.~\ref{HE1305+0007_goswami06_bab10d8d10d12m2z2m4rp2_dil0p35n26}, with
$M^{\rm AGB}_{\rm ini}$ = 2 $M_{\odot}$ models and low $s$-process efficiencies
(cases ST/12, ST/18 and ST/24).
The dilution adopted ($dil$ = 0.4 dex) is low for a giant having
suffered the FDU; moreover, this solution overestimates the observed 
[Na/Fe] by about 1 dex.
The sodium abundance is calculated from the resonance doublet (Na I D lines 
at 5890 and 5896 $\dot{\rm A}$: \citealt{goswami06}) found [Na/Fe] = +0.26, based on
the Na I D$_{2}$ line, and an overabundance [Na/Fe] = +0.43, based on
the Na I D$_{1}$ line. We show both values in the Figure.
Also the observed [Mg/Fe] is $\sim$ 0.5 dex lower than predicted.
However, we must exclude AGB models with lower initial mass that need even lower
dilutions or do not reach the high [hs/Fe] observed.
By excluding the observed Zr from the ls elements, the [hs/ls] ratio would
increase. AGB models with [hs/ls]$_{\rm th}$ $\geq$ 0.5 would predict 
[Pb/Fe]$_{\rm th}$ $\ga$ 3 at [Fe/H] = $-$2 (see Paper I, Figs.~15 and~16, middle panels). 
Note that the Pb line at 4057 ${\rm \AA}$ is strongly affected by 
molecular absorption lines. The interpretation for this giant remains a problem, 
 for Na, Mg, the ls elements and for the low dilution applied.
\citet{beers07} detected at low resolution C and N in this giant:
they found [C/Fe] = 2.4	$\pm$ 0.35 (0.6 dex higher than \citealt{goswami06}) 
and [N/Fe] =	1.90 $\pm$ 0.46.
\citet{beers07} adopt [Fe/H] = $-$2.5 and log $g$ = 1.0, about 0.5 and
1 dex lower than \citet{goswami06}.
Recently \citet{lucatello11astroph} derive N, O, $^{12}$C/$^{13}$C and an
upper limit for F ([F/Fe] $<$ 1.4) based on high-resolution spectra 
($R$ = 50\,000), by adopting the atmospheric parameters by \citet{beers07}. 
AGB models predict too large [C/Fe] and [F/Fe] (see Paper I; 
Section~\ref{CS019}).
Further investigations are desirable.

\subsection{CEMP-$s$II$/r$II with [r/Fe]$^{\rm ini}$ $\sim$ 1.5} 
\label{secCEMPs/rIIwr1p5}

This class includes ten stars showing very high $s$-process enhancement 
([hs/Fe] $\sim$ 2) and an initial $r$-enrichment [r/Fe]$^{\rm ini}$ $\sim$ 1.5.
\\
Three are main-sequence stars,
CS 29526--110, CS 31062--012 by \citet{aoki02c,aoki02d,aoki07},
and SDSS J1349--0229 by \citet{behara10}.
One star is a subgiant having not suffered the FDU,
CS 31062--050 by \citet{JB04} and \citet{aoki07}.
Five stars are giants,
CS 22948--27 and CS 29497--34 by \citet{barbuy05} and \citet{aoki07},
HD 187861 and HD 224959 by \citet{vaneck03} and \citet{masseron10},
LP 625--44 by \citet{aoki02a,aoki02d,aoki06}.
For the last star CS 22183--015, discrepant atmospheric parameters 
have been estimated by different authors \citep{JB02,cohen06},
and the occurrence of the FDU remains uncertain.

\begin{figure}
\includegraphics[angle=-90,width=8.5cm]{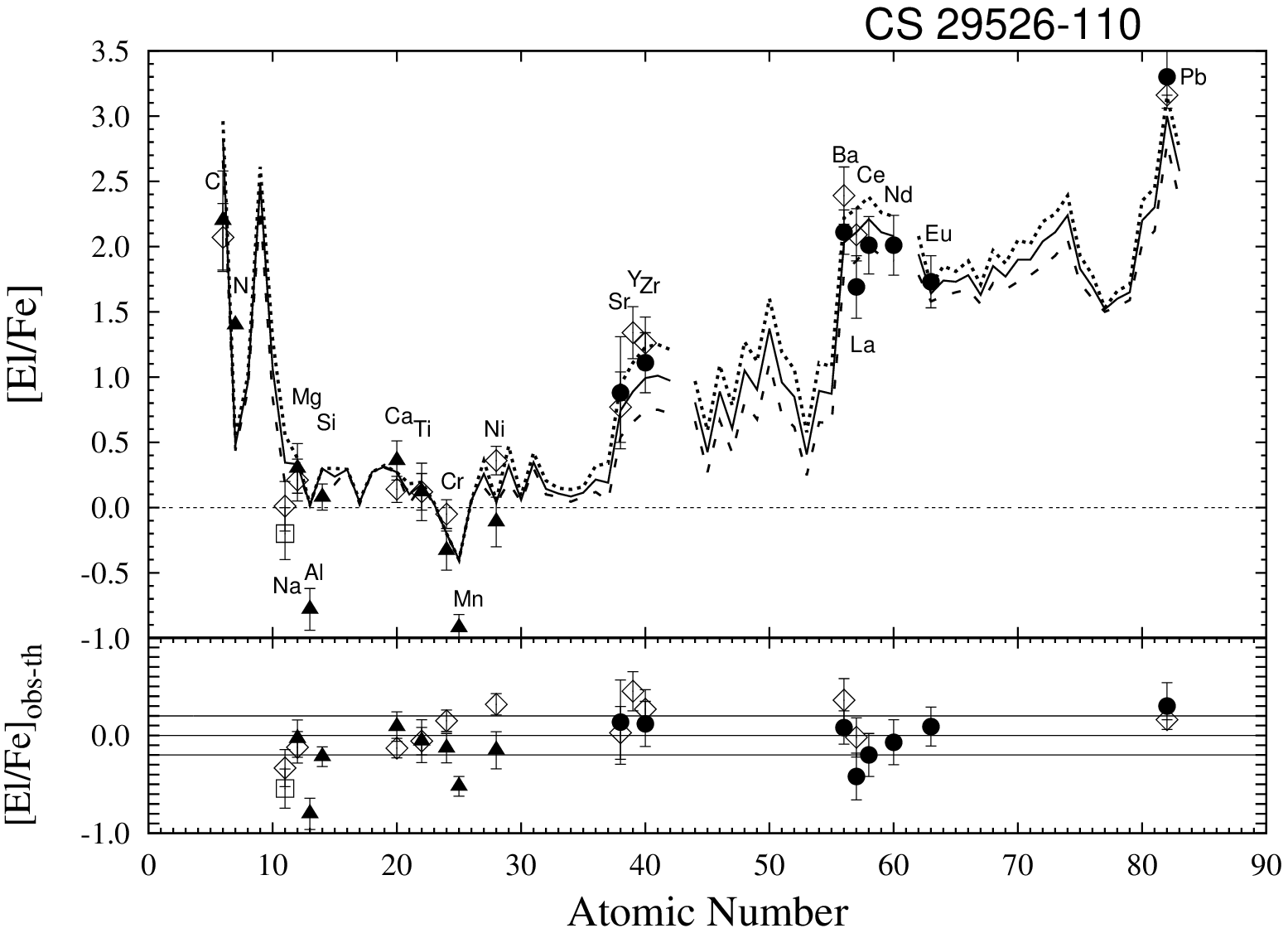}
\vspace{2mm}
\caption{Spectroscopic [El/Fe] abundances of the main-sequence star CS 29526--110 
([Fe/H] = $-$2.06; {\textit T}$_{\rm eff}$ = 6800 K; log $g$ = 4.1, \citealt{aoki08})
compared with AGB models 
of $M^{\rm AGB}_{\rm ini}$ $\sim$ 1.3 $M_{\odot}$, case ST/6, and no dilution. 
Three thermal pulses are displayed, pulse 4 (dashed line), 5 (solid line) and 6 (dotted line),
(Sections~\ref{CS036} and~\ref{CS027}).
 Observations are from \citet{aoki02c} (filled triangles), \citet{aoki02d} (filled circles), 
\citet{aoki07} (empty square), \citet{aoki08} (empty diamonds). 
This star shows [hs/ls] = 0.88 and [Pb/hs] = 1.42.
An initial $r$-process enrichment of [r/Fe]$^{\rm ini}$ = 1.5 is adopted.}
\label{CS29526-110_aoki02+aoki07+AOKI08_bab10d4m1p5z2m4rp1p5_n456}
\end{figure}

\subsubsection{CS 29526--110 (Fig.~\ref{CS29526-110_aoki02+aoki07+AOKI08_bab10d4m1p5z2m4rp1p5_n456})}
\label{CS110}

The main-sequence star CS 29526--110 was subject to different studies 
\citep{aoki02c,aoki02d,aoki07,aoki08}. 
It is a single-lined binary \citep{aoki02d,tsangarides05}, although its
period remains unknown. Different effective temperatures are estimated from 
V$-$K and B$-$V ({\textit T}$_{\rm eff}$($B-V$) = 6500 K; 
{\textit T}$_{\rm eff}$($V-K$) = 6800 K). We report the most recent values
by \citet{aoki08} ({\textit T}$_{\rm eff}$ = 6800 $\pm$ 150 K;
log $g$ = 4.1 $\pm$ 0.3), with [Fe/H] = $-$2.06, about 0.3 dex higher than 
previous studies. Nitrogen is difficult to detect because the spectra are
contaminated by CN molecular bands, and the value provided by \citet{aoki02d} 
is very uncertain. 
The solution which interprets the observed $s$ indicators [hs/ls] = 0.88 
and [Pb/hs] = 1.42 is
shown in Fig.~\ref{CS29526-110_aoki02+aoki07+AOKI08_bab10d4m1p5z2m4rp1p5_n456}
and corresponds to $M^{\rm AGB}_{\rm ini}$ $\sim$ 1.3 $M_{\odot}$, 
case ST/6 and no dilution. Three thermal pulses are represented 
(pulses 4, 5 and 6).
The most recent [Ba/Fe] measurement ($\sim$ 0.3 dex higher than that measured by
\citealt{aoki02c}) is based on two new 
red lines which are suitable for abundance determination, as well as the 
two very strong resonance lines previously considered \citep{aoki08}.
The solution shown in 
Fig.~\ref{CS29526-110_aoki02+aoki07+AOKI08_bab10d4m1p5z2m4rp1p5_n456} 
corresponding to the 6$^{\rm th}$ TDU seems to better interpret the recent 
$s$-process measurements by \citet{aoki08}, but predicts [Na/Fe] about 0.5
dex higher than observed (the [Na/Fe] estimated by \citealt{aoki07} 
includes NLTE corrections).
By adopting $M^{\rm AGB}_{\rm ini}$ $\geq$ 1.4 $M_{\odot}$ (and larger dilution), 
both the observed [Na/Fe] and [Mg/Fe] would be overestimated by AGB models. 
The lack of interpretations with dilution agrees 
with the occurrence of negligible or moderate mixing in this star. 
The [La/Eu] ratio indicates an initial $r$-process enrichment
[r/Fe]$^{\rm ini}$ = 1.5.

\begin{figure}
\includegraphics[angle=-90,width=8.5cm]{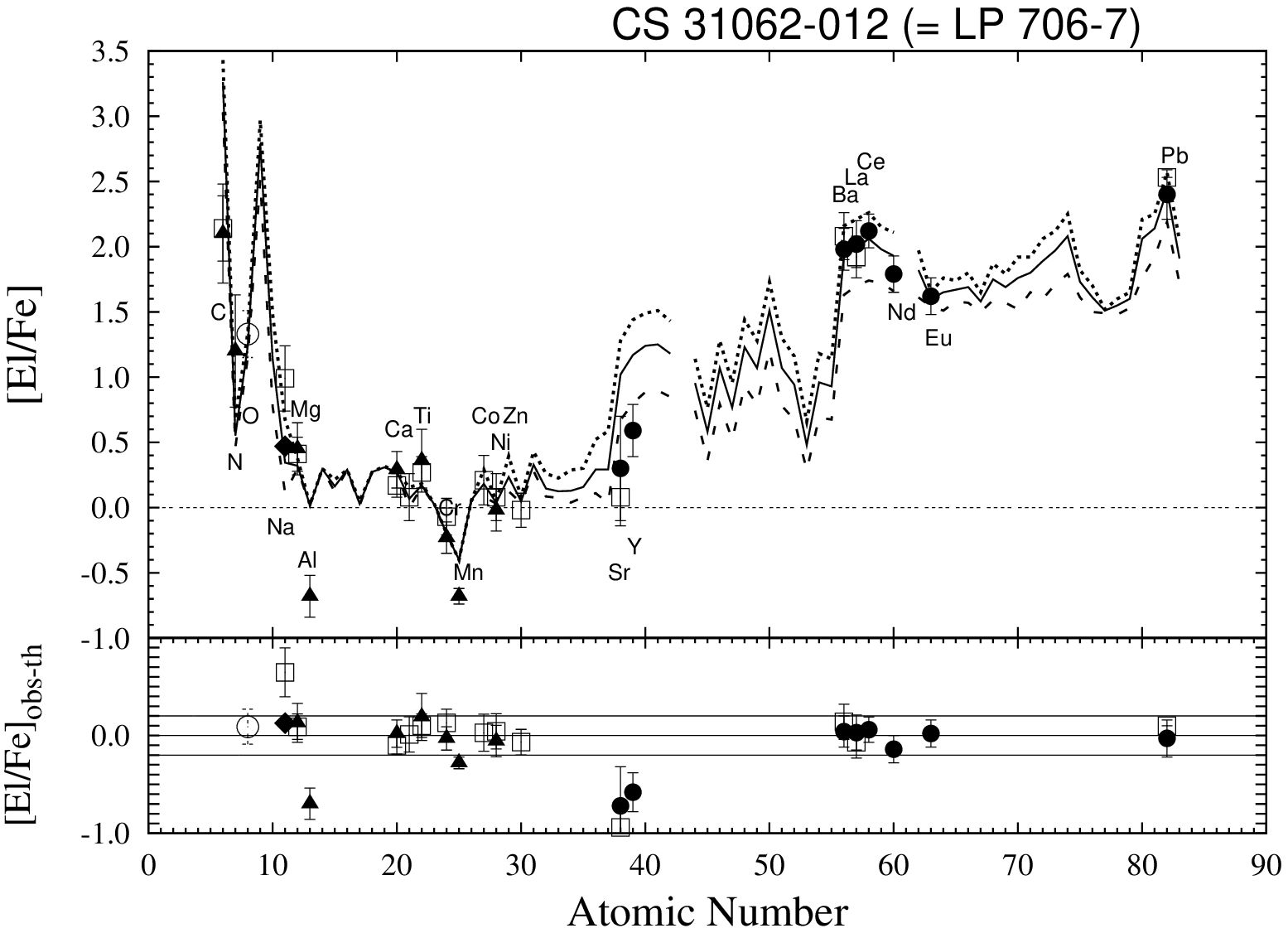}
\vspace{2mm}
\caption{Spectroscopic [El/Fe] abundances of the main-sequence star CS 31062--012 
 ([Fe/H] = $-$2.55; {\textit T}$_{\rm eff}$ = 6250 K; log $g$ = 4.5)
compared with AGB models of $M^{\rm AGB}_{\rm ini}$ $\sim$ 1.3 $M_{\odot}$, 
case ST/30, and no dilution. 
Three thermal pulses with TDU are displayed: pulse 3 (dashed line), 4 
(solid line) and 5 (dotted line),
(see Sections~\ref{CS036} and~\ref{CS027}).
Observations are from \citet{israelian01} (empty circle), \citet{aoki02c} 
(filled triangles), \citet{aoki02d} (filled circles),
\citet{aoki07} (filled diamond), \citet{aoki08} (empty squares).
This star shows [hs/Y]$_{\rm obs}$ $\sim$ 1.5 and [Pb/hs] = 0.48.
See text for discussion about Sr and Y.
An initial $r$-process enrichment of [r/Fe]$^{\rm ini}$ = 1.5 is adopted.}
\label{mnras_CS31062-012_aoki02+aoki0708+Israelian01_bab10d18m1p5z5m5rp1p5_n345}
\end{figure}

\subsubsection{CS 31062--012 ($\equiv$ LP 706$-$7), (Fig.~\ref{mnras_CS31062-012_aoki02+aoki0708+Israelian01_bab10d18m1p5z5m5rp1p5_n345})}
\label{CS012}

The main-sequence star CS 31062--012 ([Fe/H] = $-$2.55; {\textit T}$_{\rm eff}$ = 6250 K;
log $g$ = 4.5) has been analysed 
by \citet{norris97}, \citet{aoki01}, \citet{israelian01}, 
\citet{aoki02c,aoki02d,aoki07,aoki08}.
CS 31062--012 does not show significant radial velocity variations
\citep{norris97,aoki02c}, even with the extended period of 6000 days of 
observation \citep{aoki08}. 
Despite that, the high [hs/Fe] ($\sim$ 2) and the detection of [Pb/Fe] 
= 2.4, implies a significant contribution from an AGB companion.
Spectroscopic data are interpreted with AGB models of $M^{\rm AGB}_{\rm ini}$ =
1.3 $M_{\odot}$, case ST/30 and no dilution 
(Fig.~\ref{mnras_CS31062-012_aoki02+aoki0708+Israelian01_bab10d18m1p5z5m5rp1p5_n345}). 
Because of the high [hs/ls] observed ($\sim$ 1.5 dex),
the first $s$-peak is about 0.5 dex lower than theoretical predictions.
No improvement may be obtained under the hypothesis of an initial 
[Sr,Y/Fe]$^{\rm ini}$ = $-$1, compatible with the spread observed 
in field stars (e.g., \citealt{francois07}),
 because the $s$-process contribution prevails if no dilution is applied.
Moreover, with an initial $r$-process enhancement of [r/Fe]$^{\rm ini}$ = 1.5,
the [hs/ls] ratio does not increase appreciably (Paper II).
However, only 2 lines for Sr and 1 line for Y have been detected.
The Na measured in 2007 is lower with respect to the value
detected in 2008, due to NLTE corrections ($\Delta$[Na/Fe]$_{\rm LTE-NLTE}$ 
0.7 dex, see \citealt{aoki07}, Table~13). 
Interpretations with 
$M^{\rm AGB}_{\rm ini}$ $\geq$ 1.4 $M_{\odot}$ models, which predict even higher 
[Na/Fe] and [ls/Fe], are excluded. 
This discards solutions with large dilutions, in agreement with the occurrence
of negligible or moderate mixing during the main-sequence phase.

\begin{figure}
\includegraphics[angle=-90,width=8.5cm]{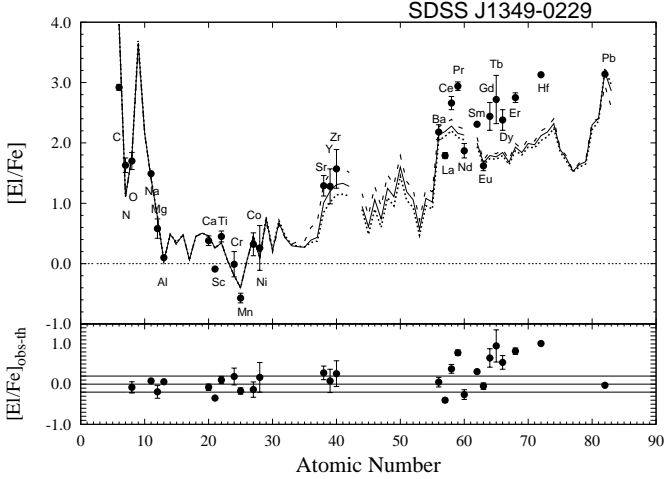}
\vspace{2mm}
\caption{Spectroscopic [El/Fe] abundances of the main-sequence star SDSS J1349-0229
([Fe/H] = $-$3.0; {\textit T}$_{\rm eff}$ = 6200 K; log $g$ = 4.0)
compared with AGB model of $M^{\rm AGB}_{\rm ini}$ = 1.35 
$M_{\odot}$ (pulse 6$^{\rm th}$; 
see Sections~\ref{CS036} and~\ref{CS027}), 
cases ST/12 (dotted line), ST/15 (solid line), ST/24 (dashed line), 
and no dilution.
Observations are from \citet{behara10}, who found [hs/ls] = 0.57 and [Pb/hs] = 1.09.
See text for discussion about the spread shown by elements from Ba to Hf.
[C/Fe] and [N/Fe] derived by CH and NH bands are displayed.
3D model atmospheres calculations 
reduce the observed [C/Fe] and [N/Fe] by 0.73 and 0.93 dex,
respectively (see text).
An initial $r$-process enrichment of [r/Fe]$^{\rm ini}$ = 1.5 is adopted.}
\label{SDSSJ1349-0229_behara10_astroph_bab9ltHTd10m1p5z2m5_rp1p5he25alf0p5_n6}
\end{figure}

\subsubsection{SDSS J1349--0229 (Fig.~\ref{SDSSJ1349-0229_behara10_astroph_bab9ltHTd10m1p5z2m5_rp1p5he25alf0p5_n6})}
\label{SDSSJ1349}

SDSS J1349--0229 is a main-sequence/turnoff star recently studied by 
\citet{behara10} ([Fe/H] = $-$3.0; {\textit T}$_{\rm eff}$ = 6200 K;
log $g$ = 4.0). 
It shows evidence for highly enhanced neutron capture elements, from both $s$- 
and $r$-process contributions.
In Fig.~\ref{SDSSJ1349-0229_behara10_astroph_bab9ltHTd10m1p5z2m5_rp1p5he25alf0p5_n6},
we display a solution with $M^{\rm AGB}_{\rm ini}$ $\sim$ 1.35 $M_{\odot}$ model 
(case ST/15 and no dilution).
A large spread of the order of 1 dex is detected among the hs elements
(e.g, [Pr/La] = 1.13). 
An average among Ba, La, Ce, Pr and Nd is chosen 
in Fig.~\ref{SDSSJ1349-0229_behara10_astroph_bab9ltHTd10m1p5z2m5_rp1p5he25alf0p5_n6}
as representative of the hs peak. 
However, we underline the large discrepancy affecting the elements of the second $s$-peak 
and the $s$-process indicators [hs/ls] = 0.57 and [Pb/hs] = 1.09.
We weight the initial $r$-enhancement on the observed [Eu/Fe], because 
the other $r$-elements (Gd, Tb, Dy and Er) are 
generally affected by larger uncertainties.
An initial $r$-process enrichment of [r/Fe]$^{\rm ini}$ = 1.5 is adopted.
However, we underline the enhancement observed in the $r$-elements as Gd, Tb, Dy and Er
with respect to Eu ([e.g., [Er/Eu] $\sim$ 1).
Note that about 60\% of solar Hf is produced by the $s$-process (\citealt{arlandini99},
Paper II), but the observed [Hf/Fe] is very uncertain and no error bars are
provided by the authors.
This star shows high Na and Mg ([Na/Fe] = 1.5 and [Mg/Fe] = 0.6).
At [Fe/H] = $-$3, a theoretical [Na/Fe] $\sim$ 1.5 is predicted by AGB models
already at the 6$^{\rm th}$ TDU,
as shown in Fig.~\ref{SDSSJ1349-0229_behara10_astroph_bab9ltHTd10m1p5z2m5_rp1p5he25alf0p5_n6}
(see Paper I).
The observed [Na/Fe] accounts for the 
 NLTE corrections from \citet{gratton99},
using the Na D resonance lines at 588.995 and 589.592 nm. 
\citet{behara10} performed 3D model atmospheres calculations to determine
the abundances of C and N, which decrease the values in 
Fig.~\ref{SDSSJ1349-0229_behara10_astroph_bab9ltHTd10m1p5z2m5_rp1p5he25alf0p5_n6}
by 0.73 and 0.93 dex, respectively: the final values estimated
from the CH lines are [C/Fe]$_{\rm 3D}$ = 2.09 and [N/Fe]$_{\rm 3D}$  = 0.67. 
This low [N/Fe] is not interpreted by the AGB model shown in  
Fig.~\ref{SDSSJ1349-0229_behara10_astroph_bab9ltHTd10m1p5z2m5_rp1p5he25alf0p5_n6},
which predicts [N/Fe]$_{\rm th}$ = 1.11.
AGB models with $M^{\rm AGB}_{\rm ini}$ = 1.3 or 1.5 $M_{\odot}$ (cases ST/15 and ST/3) 
would predict an [Na/Fe] lower or higher than observed, respectively.
However, both N and Na are affected by large uncertainties in CEMP-$s$ stars.
The present study suggests that no efficient mixing had occurred
in this star.\\
Further investigations are strongly desirable for this star. Indeed, serious
problems are found in the theoretical intepretation with AGB models of both hs and 
$r$-process elements.

\begin{figure}
\includegraphics[angle=-90,width=8.5cm]{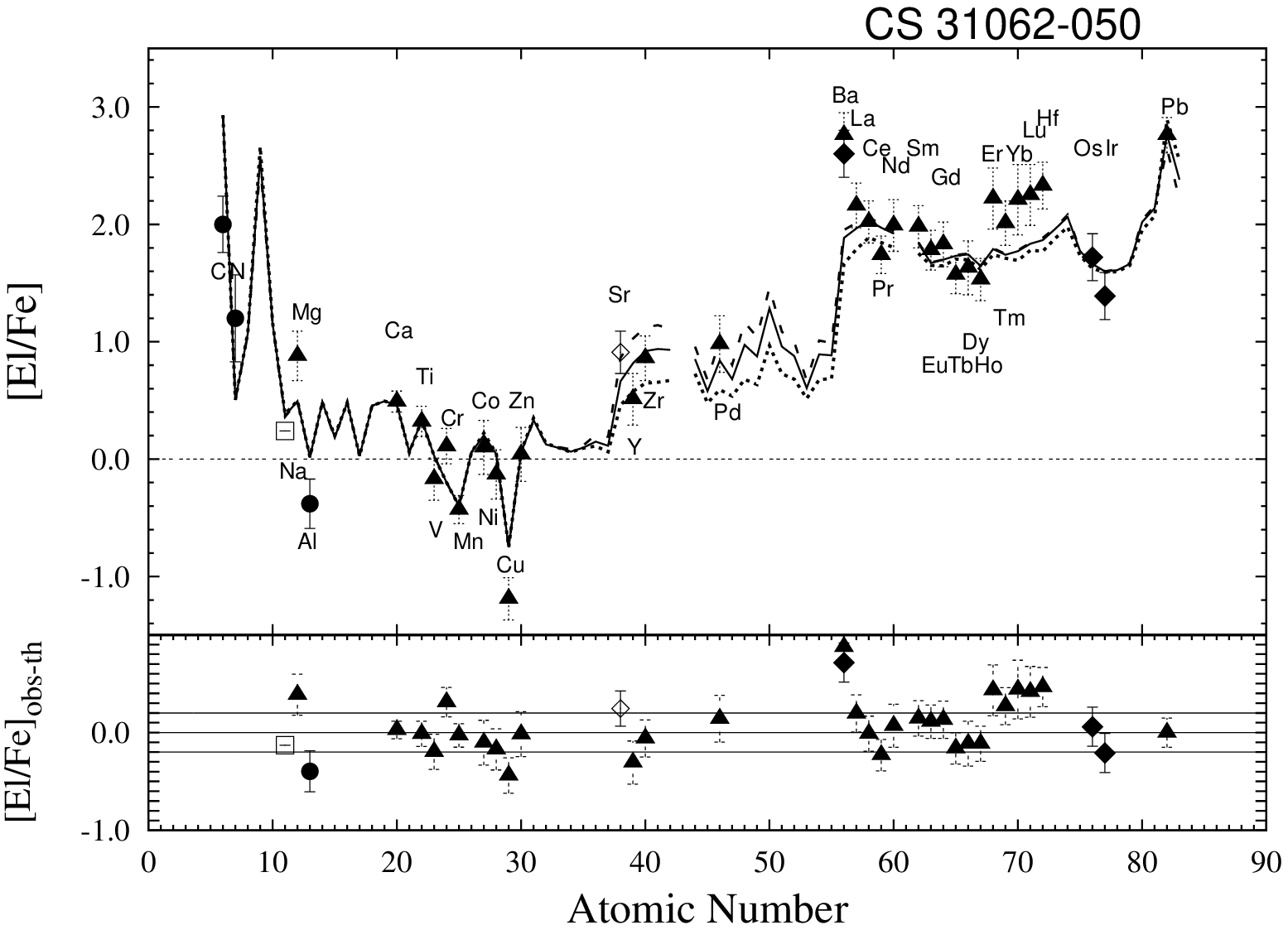}
\vspace{2mm}
\caption{Spectroscopic [El/Fe] abundances of the subgiant CS 31062--050
([Fe/H] = $-$2.42; {\textit T}$_{\rm eff}$ = 5600 K; log $g$ = 3.0, uncertain FDU)
compared with AGB models of $M^{\rm AGB}_{\rm ini}$ = 1.3 $M_{\odot}$, cases ST/8 (dotted line), 
ST/12 (solid line), ST/15 (dashed line), and $dil$ = 0.2 dex.
Observations are from \citet{aoki02c} (filled circles), \citet{aoki02d} (empty diamond),
\citet{aoki06} (filled diamonds), \citet{aoki07} (empty square), \citet{JB04} (filled triangles).
This star shows [hs/ls] = 1.40 and [Pb/hs] = 0.79.
The low [Na/Fe] observed agrees with a star having not suffered the FDU episode.
An initial $r$-process enhancement [r/Fe]$^{\rm ini}$ = 1.6 is adopted.
In order to interpret the low [Pd/Fe] observed,
we adopt an initial $r$-process enhancement [light-r/Fe]$^{\rm ini}$ = 0.5 for the elements
from Mo to Cs (see text).}
\label{CS31062-050_JB04aoki0206BaOsIr07Na_bab10}
\end{figure}

\subsubsection{CS 31062--050 (Fig.~\ref{CS31062-050_JB04aoki0206BaOsIr07Na_bab10})}
\label{CS050}

The subgiant CS 31062--050 was examined by \citet{aoki02c,aoki02d,aoki06,aoki07} and 
\citet{JB04}, ([Fe/H] = $-$2.42; {\textit T}$_{\rm eff}$ = 5600 K; log $g$ = 3.0). 
 The occurrence of the FDU is uncertain for this star.
Many elements have been observed, among them Os and Ir lines were
detected by \citet{aoki06} for the first time in CEMP-$s$ stars. 
Moreover, \citet{aoki03} and \citet{tsangarides05}
found radial velocity variations, confirming the binary scenario. 

\noindent Ba is higher than the other hs elements, even if the result of \citet{JB04}
is reduced by 0.2 dex according to \citet{aoki06}, who used two weaker lines, which are 
less sensitive to hyperfine splitting.
CS 31062--050 has been discussed in detail in the review by \citet{kaeppeler11rmp}.
The low [Na/Fe] (which accounts for NLTE corrections) agrees with a 
$M^{\rm AGB}_{\rm ini}$ = 1.3 $M_{\odot}$ model with $dil$ = 0.2 dex
(Fig.~\ref{CS31062-050_JB04aoki0206BaOsIr07Na_bab10}), according to a star 
before the FDU and in agreement with moderate mixing during the main-sequence
phase. 
This solution predicts a [Mg/Fe] about 0.4 dex lower than observed.
AGB models with $M^{\rm AGB}_{\rm ini}$ = 1.5 and 2 $M_{\odot}$ (case ST/3 and $dil$ = 
1.1 dex) provide similar solutions for the $s$-process distribution
([hs/ls] = 1.40; [Pb/hs] = 0.79), but the predicted [Na/Fe] is about 1 dex 
higher than observed. 
Despite only one line was detected for Na, which is affected by a high 
uncertainty because of the severe contamination from interstellar 
Na absorption, \citet{aoki07} excluded [Na/Fe] observations higher than
0.8 dex.
Os and Ir, whose $r$-process fractions in the solar system 
material are 88\% and 98\%, respectively (see Paper II), 
are an important confirmation of the $r$-process enhancement.
The initial $r$-process enrichment [r/Fe]$^{\rm ini}$ = 1.6 accounts for 
the observed low [Ir/Fe] with correspondingly lower estimates for 
[Er,Tm,Yb,Lu/Fe].
The observed [Hf/Fe] is higher than our theoretical prediction.
We recall that Hf is mainly produced by the
$s$-process (about 60\% of solar Hf, Paper II).
Therefore, larger initial r-process enhancements would not affect the [Hf/Fe] prediction.
This is the only star among CEMP-$s$ and CEMP-$s/r$ with a measurement among 
the light-$r$-elements from Mo to Cs: [Pd/Fe] = 0.98 \citep{JB04}.
About 50\% of solar Pd is produced by the $s$-process (see Paper II, Table~5), 
while 50\% of solar Pd is ascribed to the $r$-process. 
An [r/Fe]$^{\rm ini}$ = 1.6 would provide [Pd/Fe]$_{\rm th}$ = 1.4, 
about 0.4 dex higher than observed.
Lower initial light-$r$-enhancements are assumed in order to interpret
Pd, [light-r/Fe]$^{\rm ini}$ = 0.5 -- 1.0, corresponding to 
[Pd/Fe]$_{\rm th}$ = 0.8 -- 1.0, respectively.
In Fig.~\ref{CS31062-050_JB04aoki0206BaOsIr07Na_bab10},
a [light-r/Fe]$^{\rm ini}$ = 0.5 is shown for elements from Mo to Cs.
The exact site of nucleosynthesis of the $r$-process remains still 
unknown and a possible explanation of this difference 
comes from the hypothesis of a multiplicity
of the $r$-process contributions (see Paper II, Section~3;
\citealt{travaglio04}, 
\citealt{QW08},  
\citealt{SCG08}). 
\\
Cu and Al are not produced in AGB stars, as confirmed by the observations. 
In particular, the negative [Cu/Fe] value is 
consistent with the observations of unevolved halo stars 
in the same range of metallicity (see Section~\ref{intro}).

\begin{figure}
\includegraphics[angle=-90,width=8.5cm]{Fig27top.ps}
\vspace{3mm}
\includegraphics[angle=-90,width=8.5cm]{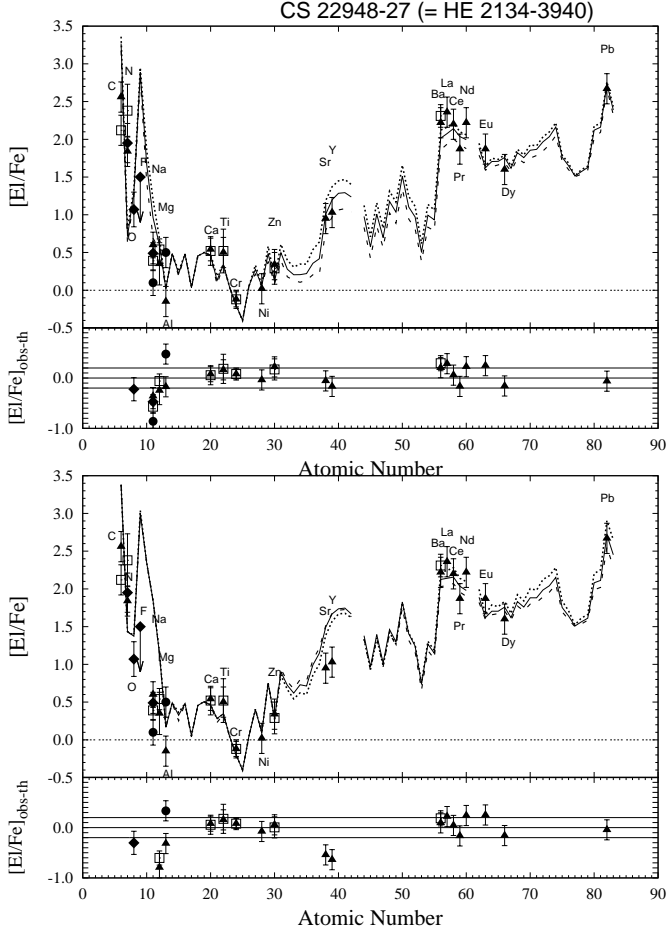}
\vspace{2mm}
\caption{Spectroscopic [El/Fe] abundances of the giant CS 22948--27 
([Fe/H] = $-$2.47; {\textit T}$_{\rm eff}$ = 4800 K; log $g$ = 1.8, \citealt{barbuy05}; 
[Fe/H] = $-$2.21; {\textit T}$_{\rm eff}$ = 5000 K; log $g$ = 1.9; \citealt{aoki07})
compared with AGB models of two initial masses: $M^{\rm AGB}_{\rm ini}$ $\sim$ 
1.35 $M_{\odot}$, ST/15,
$dil$ = 0.4 dex and three thermal pulses, 6, 7, 8, see Sections~\ref{CS036} 
and~\ref{CS027}) (dashed, solid and dotted
lines, respectively; \textit{top panel});
$M^{\rm AGB}_{\rm ini}$ = 1.5 $M_{\odot}$, cases ST/6 (dotted line), ST/9
(solid line), ST/12 (dashed line), and $dil$ = 0.8 dex (\textit{bottom panel}). 
Observations are from \citet{barbuy05} (filled triangles; filled circles), \citet{aoki07} 
(empty squares), \citet{lucatello11astroph} (filled diamonds). 
Na measured by \citet{aoki07} takes account of NLTE 
corrections, while values with and without NLTE are shown for the Na measured by
\citet{barbuy05} (filled circles and triangles, respectively).
This star has [hs/ls] = 0.93; [Pb/hs] = 0.56.
The differences [El/Fe]$_{\rm obs-th}$ displayed in the lower panels refer to the AGB
models represented by solid lines.
An initial $r$-process enrichment of [r/Fe]$^{\rm ini}$ = 1.5 is adopted. }
\label{CS22948-27_barbuy05_aoki07_bab10}
\end{figure}

\subsubsection{CS 22948--27 ($\equiv$ HE 2134--3940), (Fig.~\ref{CS22948-27_barbuy05_aoki07_bab10})}
\label{CS27}
 
\citet{barbuy97} and \citet{hill00},
analysed the spectrum of this cool giant, which is heavily   
contaminated by CH, CN, and C$_{2}$ molecular bands.  
Recently, \citet{barbuy05} reviewed this star using high-resolution 
spectra that permit the detection of lead.
Na and Al lines are sensitive to NLTE effects, which 
decreases the abundance of Na by 0.5 dex and increases the abundance of Al
by 0.65 dex \citep{barbuy05}. 
\citet{aoki07} confirmed the results by \citet{barbuy05}, providing 
updated values for Na and Mg and adopting a slightly higher effective temperature
({\textit T}$_{\rm eff}$ = 5000 K 
instead of {\textit T}$_{\rm eff}$ = 4800 K; log $g$ = 1.8). 
Strong molecular absorption lines remain the main characteristics
of this very cool star.
Despite that, the [El/Fe] ratios provided by the different authors agree within the 
quoted uncertainties; the only exceptions are the metallicity
([Fe/H] = $-$2.47 by \citealt{barbuy05}; [Fe/H] = $-$2.21 by \citealt{aoki07})
and C and N, explained by the different effective temperatures adopted.
\\
Theoretical interpretations of the spectroscopic abundances
([hs/ls] = 0.93; [Pb/hs] = 0.56) are shown in 
Fig.~\ref{CS22948-27_barbuy05_aoki07_bab10}. 
A large dilution would be required by a giant having suffered the FDU.
Instead, the solutions that better fit to the low observed
[Na/Fe] and [Sr,Y/Fe] ratios contrast with this hypothesis: 
we displayed AGB models of $M^{\rm AGB}_{\rm ini}$ $\sim$ 1.35 $M_{\odot}$
(pulse 7$^{\rm th}$), case ST/15 and $dil$ = 0.4 dex (\textit{top panel}). 
AGB models of $M^{\rm AGB}_{\rm ini}$ = 1.5 $M_{\odot}$, cases ST/6, ST/9, ST/12
and $dil$ = 0.8 dex (see \textit{bottom panel}), would predict [Na/Fe] and [ls/Fe]
$\sim$ 0.5 dex higher than observed. Similar results
are obtained by $M^{\rm AGB}_{\rm ini}$ = 2 $M_{\odot}$ models.
AGB models predict large [F/Fe] abundances, incompatible with the upper limit 
detected by \citet{lucatello11astroph} (see Paper I; Section~\ref{CS019}).
Further measurements are desirable.
\\
\citet{PS01}\footnote{The first measurement by \citet{hill00} could not find any
clear radial velocity variation due to the low resolution spectra.} first discovered 
a radial velocity variation for CS 22948--27 with $P$ = 505 d.
\citet{barbuy05} determine a period of 426 days. 
\citet{aoki07} confirm this last value, but further investigations are desirable.

\begin{figure}
\includegraphics[angle=-90,width=8.5cm]{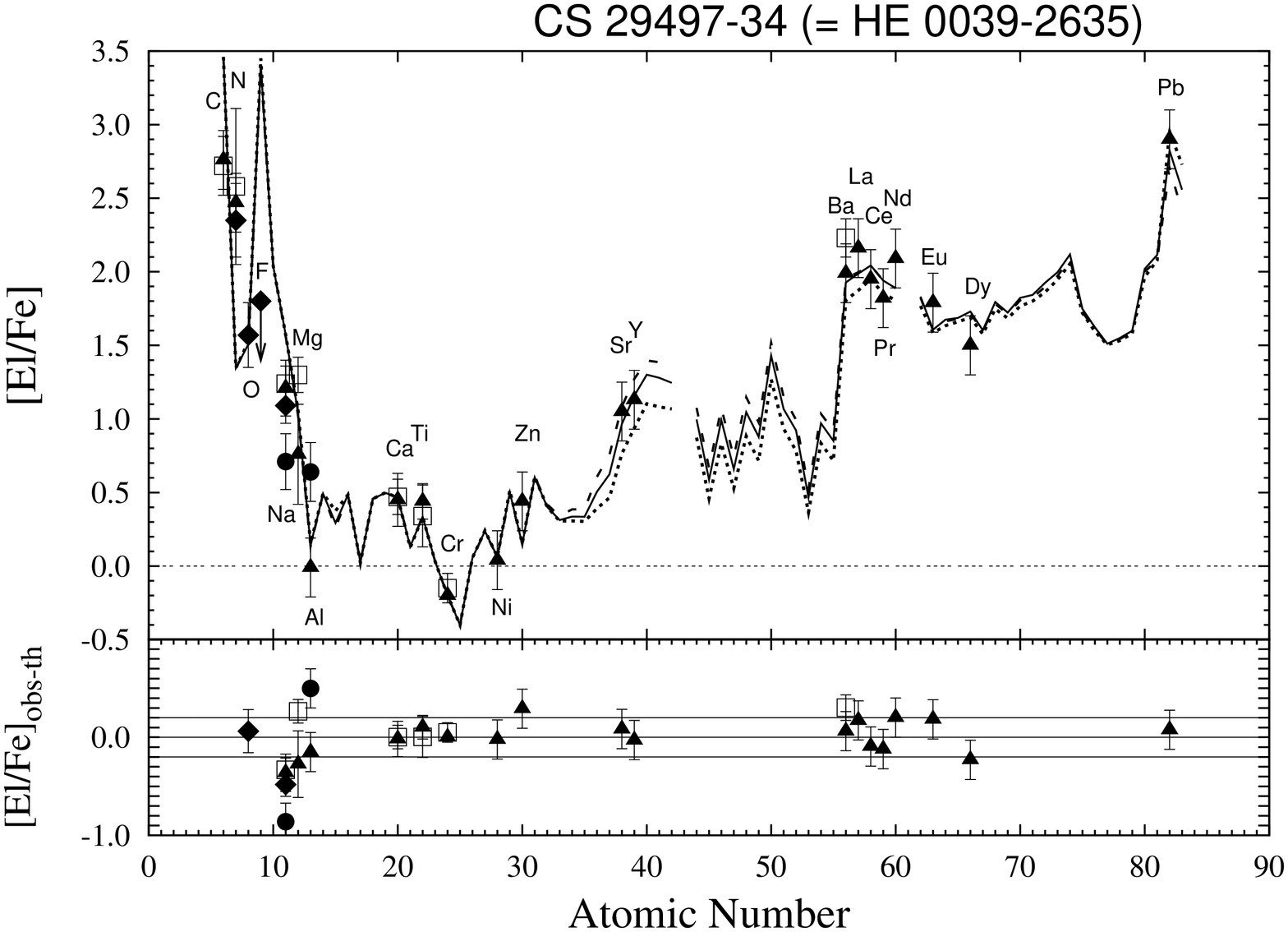}
\vspace{2mm}
\caption{Spectroscopic [El/Fe] abundances of the giant CS 29497--34
([Fe/H] = $-$2.90; {\textit T}$_{\rm eff}$ = 4800 K; log $g$ = 1.8) 
compared with AGB models of $M^{\rm AGB}_{\rm ini}$
= 2 $M_{\odot}$, cases ST/6 (dotted line), ST/9 (solid line), ST/12
(dashed line), and $dil$ = 1.0 dex.
Observations are from \citet{barbuy05} (filled triangles; filled circles),
\citet{aoki07} (empty squares), \citet{lucatello11astroph} (filled diamonds).
This star shows [hs/ls]$_{\rm obs}$ = 0.81 and [Pb/hs]$_{\rm obs}$ = 0.93.
An initial $r$-process enrichment of [r/Fe]$^{\rm ini}$ = 1.5 is adopted. }
\label{CS29497-34_barbuy05_aoki07_bab10d4d6d8m2z2m5_rp1p5_alf0p5_dil1p0n26}
\end{figure}

\subsubsection{CS 29497$-$34 ($\equiv$ HE 0039$-$2635),
(Fig.~\ref{CS29497-34_barbuy05_aoki07_bab10d4d6d8m2z2m5_rp1p5_alf0p5_dil1p0n26})}
\label{CS34}

CS 29497--34 ([Fe/H] = $-$2.90) is a long period binary, with $P$ = 4130 days 
\citep{PS01}.
This very cool giant ({\textit T}$_{\rm eff}$ = 4800 K and 
log $g$ = 1.8), affected by strong C$_{\rm 2}$ and CH molecular bands,
 was analysed by \citet{barbuy97}, \citet{hill00},
\citet{lucatello04PhD} and \citet{barbuy05}. 
\citet{aoki07} adopted a slightly higher effective temperature
({\textit T}$_{\rm eff}$ = 4900 K), confirming the 
results by \citet{barbuy05}, and provided updated values for Na and Mg.
\\
In 
Fig.~\ref{CS29497-34_barbuy05_aoki07_bab10d4d6d8m2z2m5_rp1p5_alf0p5_dil1p0n26}
we show theoretical interpretations with $M^{\rm AGB}_{\rm ini}$ = 
2 $M_{\odot}$ (ST/6, ST/9 and ST/12) and $dil$ = 1 dex, in agreement 
with the observed [hs/ls] = 0.81 and [Pb/hs] = 0.93.
The abundances found by the two studies are in 
agreement within the uncertainties with the exception of
Na and Mg, for which \citet{aoki07} found 0.4 and 0.5 dex higher values
than \citet{barbuy05}.
\citet{aoki07} exclude that these discrepancies are due to the small 
differences in the atmospheric parameters adopted, 
and highlight the uncertainty of the Na abundance, for which only
two very strong lines are available. However, an excess in [Na/Fe]
is confirmed by both authors.
The enhanced Na observed ([Na/Fe] = 1.37, \citealt{aoki07}) is about 0.2 dex 
lower than that predicted by the AGB model shown in
Fig.~\ref{CS29497-34_barbuy05_aoki07_bab10d4d6d8m2z2m5_rp1p5_alf0p5_dil1p0n26}.
AGB models of $M^{\rm AGB}_{\rm ini}$ = 1.5 $M_{\odot}$ (case ST/5 and $dil$ = 1.0 dex)
overestimate the observed [Na/Fe] by about 0.6 dex.
Both Na and Al lines are sensitive to the NLTE effects, 
 which decrease and increase the observations by 0.4 -- 0.5 dex and 0.65 dex, 
respectively \citep{barbuy05,aoki07}. 
Solutions with AGB models of low initial mass and negligible dilution are excluded 
for this giant having suffered the FDU.
As for CS 22948--27 (see also Paper I; Section~\ref{CS019}), 
AGB models predict a [F/Fe] lower than the value detected by \citet{lucatello11astroph}.
Further measurement are desirable.

\begin{figure}
\includegraphics[angle=-90,width=8cm]{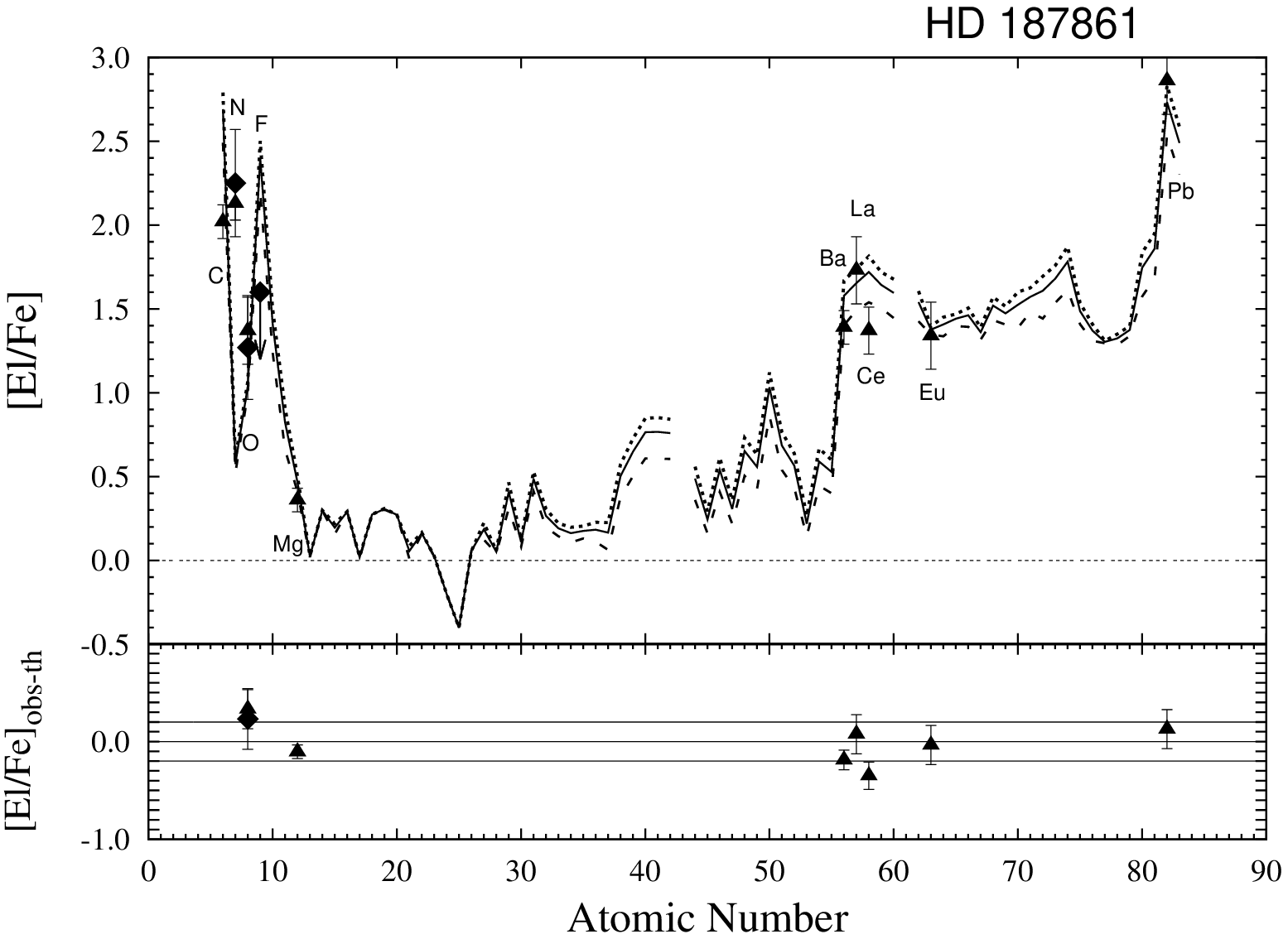}
\vspace{2mm}
\caption{Spectroscopic [El/Fe] abundances of the giant HD 187861 
([Fe/H] = $-$2.36; {\textit T}$_{\rm eff}$ = 4600 K; log $g$ = 1.7)
compared with AGB models of $M^{\rm AGB}_{\rm ini}$ = 1.4 $M_{\odot}$, ST/5, 
and $dil$ = 0.8 (dotted line), 0.9 (solid line) and 1.1 dex (dashed line). 
Observations are from \citet{masseron10} (filled triangles) 
and \citep{lucatello11astroph} (filled diamonds).
This giant shows [Pb/hs] = 1.28, while no ls elements
have been detected by \citep{masseron10} (see text).
An initial $r$-process enrichment of [r/Fe]$^{\rm ini}$ = 1.3 is adopted
(see text). }
\label{HD187861_VE03_M10}
\end{figure}

\subsubsection{HD 187861 (Fig.~\ref{HD187861_VE03_M10})}
\label{HD187861}

This giant was firstly studied by \citet{vanture92c}.
Subsequently, high-resolution spectra were analysed by
\citet{vaneck03} and \citet{masseron10}.
The abundances by \citet{vaneck03} are uncertain because only few 
lines veiled by molecular bands are available.
\citet{masseron10} estimate [Fe/H] = $-$2.36, {\textit T}$_{\rm eff}$ = 4600 K
and log $g$ = 1.7.
Solutions with $M^{\rm AGB}_{\rm ini}$ = 1.4 $M_{\odot}$ models (case ST/5 and 
$dil$ $\sim$ 1 dex) are shown in Fig.~\ref{HD187861_VE03_M10}, compared with 
spectroscopic observations by \citet{masseron10} ([Pb/hs] = 1.28). 
Recently \citep{lucatello11astroph}, detected N, O and an upper limit
for F (see Section~\ref{CS019}).
[ls/Fe]$_{\rm th}$ $\sim$ 0.7 dex is predicted by this model. 
\citet{vaneck03} detected [Zr/Fe] = 1.3 by adopting an effective temperature
700 K higher than \citet{masseron10}, which may explain a difference of about 0.4
dex in [El/Fe].
An [r/Fe]$^{\rm ini}$ = 1.3 dex is adopted in order to interpret
the observed [La/Eu]. The average of [Ba,La,Ce/Eu] suggests even larger 
initial $r$-process enhancement.
Theoretical interpretations with $M^{\rm AGB}_{\rm ini}$ $<$ 1.4 $M_{\odot}$ 
and negligible dilution would not agree with a giant having suffered the FDU,
while $M^{\rm AGB}_{\rm ini}$ = 1.5 and 2 $M_{\odot}$ would overestimated the 
observed [Mg/Fe].
The observations by \citet{masseron10} shown in Fig.~\ref{HD187861_VE03_M10} will be 
discussed by the authors in Masseron et al., in preparation.

\begin{figure}
\includegraphics[angle=-90,width=8cm]{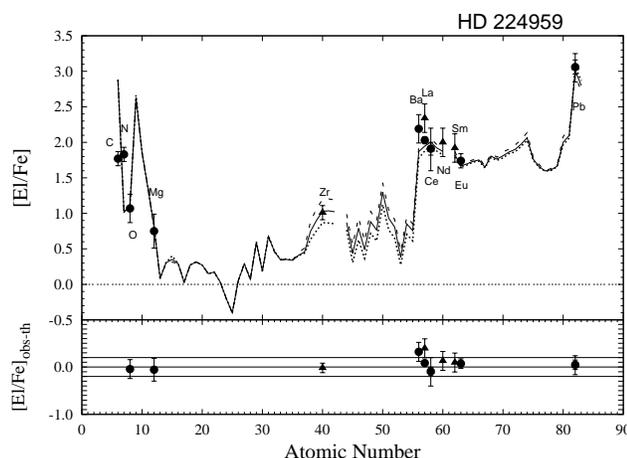}
\vspace{2mm}
\caption{Spectroscopic [El/Fe] abundances of the giant HD 224959 
([Fe/H] = $-$2.06; {\textit T}$_{\rm eff}$ = 4900 K; log $g$ = 2.0)
compared with AGB models of $M^{\rm AGB}_{\rm ini}$ = 1.5 $M_{\odot}$, 
cases ST/2 (dotted line), ST/2.5 (solid line), ST/3 (dashed line), and $dil$ $\sim$ 1.0 dex.
Observations are from \citet{vaneck03} (filled triangles) and 
\citet{masseron10} (filled circles).
This star shows [hs/ls] = 1.12 and [Pb/hs] = 1.03.
An [r/Fe]$^{\rm ini}$ = 1.6 dex is adopted.}
\label{HD224959_vaneck03_masseron10_bab10d1p3d1p6d2m1p5z1m4rp1p6_dil0p95n20}
\end{figure}

\subsubsection{HD 224959
(Fig.~\ref{HD224959_vaneck03_masseron10_bab10d1p3d1p6d2m1p5z1m4rp1p6_dil0p95n20})}
\label{HD224959}

\citet{vaneck03} analysed this giant ([Fe/H] = $-$2.2; {\textit T}$_{\rm eff}$ = 5200 K; 
log $g$ = 1.9), providing spectroscopic observations
for Zr, La, Ce, Nd, Sm and Pb ([hs/ls] = 1.12 and [Pb/hs] = 1.03).
The recent analysis by \citet{masseron10}
([Fe/H] = $-$2.06, {\textit T}$_{\rm eff}$ = 4900 K and log $g$ = 2.0)
 confirms the previous results 
and classifies this star as a CEMP-$s/r$, with [La/Eu] $\sim$ 0.3.
An initial $r$-process enhancement of 1.6 dex is adopted.
In Fig.~\ref{HD224959_vaneck03_masseron10_bab10d1p3d1p6d2m1p5z1m4rp1p6_dil0p95n20},
we show theoretical interpretations with AGB models of initial mass 1.5 $M_{\odot}$ 
and $dil$ = 1 dex (case $\sim$ ST/3), compared with both spectroscopic studies
\citep{vaneck03,masseron10}. 
Analogous interpretations are obtained by AGB models of initial mass 2 
$M_{\odot}$ (ST/3 and $dil$ = 1 dex). Solutions with initial masses $M$ $<$ 
1.5 $M_{\odot}$ and lower dilutions are discarded, because they would
be in contrast with a giant after the FDU.

\begin{figure}
\includegraphics[angle=-90,width=8.5cm]{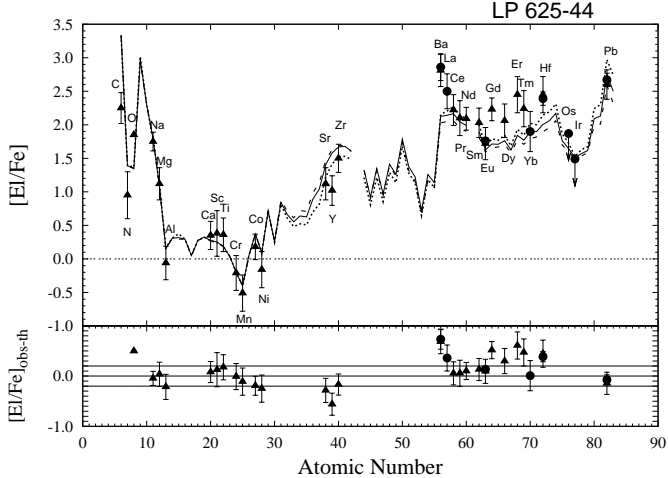}
\vspace{2mm}
\caption{Spectroscopic [El/Fe] abundances of the giant LP 625--44 
([Fe/H] = $-$2.7; {\textit T}$_{\rm eff}$ = 5500 K; log $g$ = 2.5)
compared with AGB models of initial mass 1.5 $M_{\odot}$, cases ST/5 (dotted line), 
ST/8 (solid line), ST/11 (dashed line), and $dil$ = 0.8 dex.
Observations are from \citet{aoki02a} (filled triangles) and \citet{aoki06} (filled circles).
This star shows [hs/ls]$_{\rm obs}$ = 0.93 and [Pb/hs]$_{\rm obs}$ = 0.46. 
Note that the [hs/ls]$_{\rm th}$ is lower than observed,
because [Y/Fe] and [La/Fe] are lower and higher than theoretical predictions 
(see text).
An initial $r$-process enrichment of [r/Fe]$^{\rm ini}$ = 1.5 is adopted.}
\label{LP625-44_aoki0206OsIr_bab10d3d5d7m1p5z5m5rp1p5_dil0p8n20}
\end{figure}

\subsubsection{LP 625--44 (Fig.~\ref{LP625-44_aoki0206OsIr_bab10d3d5d7m1p5z5m5rp1p5_dil0p8n20})}
\label{LP44}

\citet{aoki02a} carried out a detailed analysis of 
this giant ([Fe/H] = $-$2.7; {\textit T}$_{\rm eff}$ = 5500 K; log $g$ = 2.5)
 with the High Dispersion
Spectrograph (HDS) of the Subaru Telescope, subsequently
improved by the detection of upper limits 
for two $r$-process elements, Os and Ir \citep{aoki06}. 
The binarity was confirmed by radial 
velocity monitoring \citep{norris97,aoki00}, strongly 
supporting the mass transfer scenario.
The period has not been estimated yet,
and further measurements of the radial velocity are 
required to infer orbital parameters for this 
star.
\\
Very enhanced Na and Mg are detected: [Na/Fe] = 1.75 and [Mg/Fe] = 1.12. 
The Na abundance was based on five weak lines as well as the strong D lines. 
The observed [Na/Fe] and [Mg/Fe] are underestimated by AGB models with initial mass
 $M^{\rm AGB}_{\rm ini}$ $\leq$ 1.35 $M_{\odot}$ and case ST/30
(see \citealt{bisterzo09pasa}, Fig.~5, bottom panel). 
Moreover, these models require negligible dilution, in contrast with
a giant after the FDU.
Fig.~\ref{LP625-44_aoki0206OsIr_bab10d3d5d7m1p5z5m5rp1p5_dil0p8n20}
shows theoretical interpretations with $M^{\rm AGB}_{\rm ini}$ = 1.5 $M_{\odot}$ (cases 
ST/5, ST/8, ST/11) and $dil$ = 0.8 dex, in agreement with the observed [Na/Fe] 
and [Mg/Fe].
The observed [Y/Fe] is about 0.4 dex lower than AGB predictions, 
while [Ba/Fe] is 0.6 dex higher. 
Note that the error bar shown for La accounts for uncertainties due to 
fitting of synthetic spectra and atmospheric parameters.
The [Pb/hs] ratio is low in this star ($\sim$ 0.46 dex), and low $^{13}$C-pockets
are needed in order to interpret the $s$-process distribution.
A similar solution is obtained with $M^{\rm AGB}_{\rm ini}$ = 2 $M_{\odot}$, but 
the observed [Na/Fe] is slightly higher than AGB prediction. 
To match the $r$-process abundances, we adopt an [r/Fe]$^{\rm ini}$ = 1.5 dex. 
This choice was assessed on the observed [Eu/Fe] \citep{aoki02a,aoki06},
as well as on the recent Yb detection and on the upper limit of Ir
\citep{aoki06}.
The observed $r$-process elements Gd, Er and Tm (7, 6 and 3 lines) are 
underestimated with the initial $r$-enhancement assumed in
Fig.~\ref{LP625-44_aoki0206OsIr_bab10d3d5d7m1p5z5m5rp1p5_dil0p8n20}.
We underline that four among the neutron capture elements (Y, Ba, Gd, Er) 
are not interpreted by AGB models.
The interpretation of this star remains problematic and further
investigations both on spectroscopic and on theoretical point of view
are desirable.

\begin{figure}
\includegraphics[angle=-90,width=8.5cm]{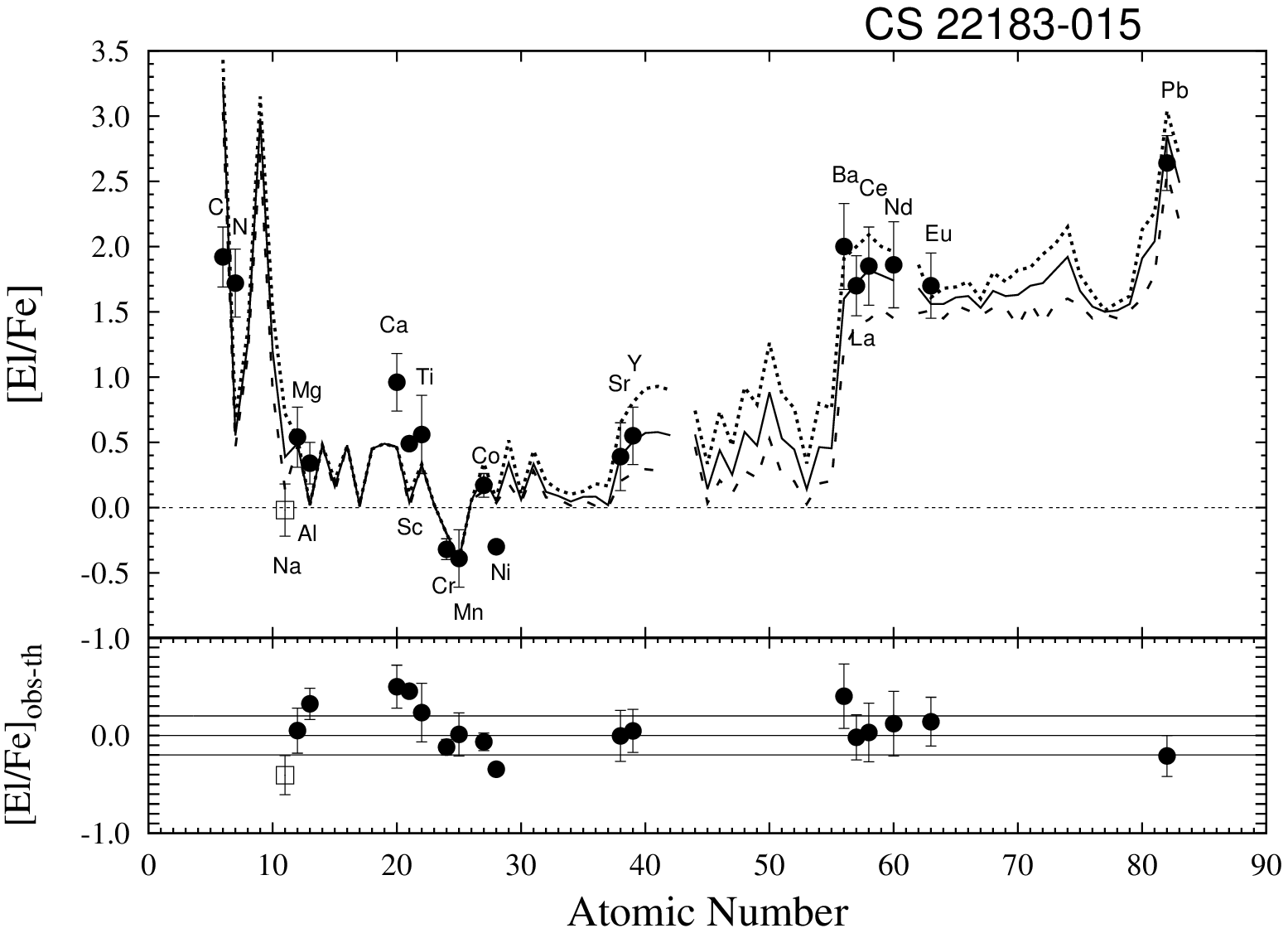}
\vspace{2mm}
\caption{CS 22183--015 has uncertain atmospheric parameters
(\citealt{cohen06,aoki07} and \citealt{JB02,tsangarides05}; see text).
We show here spectroscopic [El/Fe] abundances by 
\citet{cohen06} (filled circles) and \citet{aoki07} (empty square), 
([Fe/H] = $-$2.75; {\textit T}$_{\rm eff}$ = 5620 K and log $g$ = 3.4, uncertain FDU; 
[hs/ls] = 1.21; [Pb/hs] = 1.03), 
compared with AGB models of $M^{\rm AGB}_{\rm ini}$ $\sim$  
1.3 $M_{\odot}$, case ST/12, no dilution. 
Three thermal pulses are shown: pulses 3 (dashed line), 4 (solid line), 5 (dotted line),
 (see Sections~\ref{CS036} and~\ref{CS027}).
 An initial $r$-process enrichment of [r/Fe]$^{\rm ini}$ = 1.5 is adopted.}
\label{CS22183-015_cohen06_aoki07_JB02T05lai07}
\end{figure}

\subsubsection{CS 22183--015 ($\equiv$ HE 0058--0244); 
(Fig.~\ref{CS22183-015_cohen06_aoki07_JB02T05lai07})}
\label{HE0058}

CS 22183--015 has been studied by 
\citet{JB02}, \citet{lucatello04PhD}, \citet{tsangarides05},  
\citet{cohen06} and \citet{aoki07}.
\citet{tsangarides05} performed radial velocity measurements, 
but could not confirm the binarity of this star due to the low velocity
amplitude. 
Discrepancies between metallicity and atmospheric 
parameters are measured by different authors: \citet{JB02}
report [Fe/H] = $-$3.12, {\textit T}$_{\rm eff}$ = 5200 K
and log $g$ = 2.5, while \citet{cohen06} give [Fe/H] = $-$2.75, {\textit 
T}$_{\rm eff}$ = 5620 K and log $g$ = 3.4. More recently,
\cite{lai04,lai07} analysed CS 22183--015 with $R$ $\sim$ 7\,000, and found 
 [Fe/H] = $-$3.17, {\textit T}$_{\rm eff}$ = 5178 K, log $g$ = 2.69,
similarly to \citet{JB02}.
\citet{aoki07} adopted the atmospheric parameters by \citet{cohen06}
and determined the Na abundance accounting for 
 NLTE corrections 
($\Delta$[NLTE] = $-$0.5 dex). 
The strong differences in the atmospheric parameters indicate that the evolutionary 
phase of this star is uncertain: following 
\citet{JB02} and \citet{lai07} it is a giant, while for \citet{cohen06} it 
lies still on the early subgiant phase, where the FDU may not have 
occurred yet. 
\\
In 
Fig.~\ref{CS22183-015_cohen06_aoki07_JB02T05lai07}, 
we present possible solutions by adopting the spectroscopic observations
by \citet{cohen06} and \citet{aoki07}. 
The low [Na/Fe] observed together with a high [hs/Fe] are interpreted by
AGB models of $M^{\rm AGB}_{\rm ini}$ $\sim$ 1.3 $M_{\odot}$ 
(case ST/12, three TDUs, 3, 4, 5) and $dil$ = 0 dex,
in agreement with a subgiant having not suffered the FDU episode. 
An initial $r$-process enhancement of [r/Fe]$^{\rm ini}$ = 1.5 is adopted. 
Theoretical interpretations with AGB models of higher initial mass and 
dilution ($M$ = 1.4 to 2 $M_{\odot}$; $dil$ = 0.6 -- 1.1 dex; cases ST/12 -- ST/2) 
would predict a too high [Na/Fe] and [Sr,Y/Fe].

\subsection{CEMP-$s$II$/r$I with [r/Fe]$^{\rm ini}$ $\sim$ 1.0} 
\label{secCEMPs/rIIwrI}

This Section includes four stars with high $s$-process enhancement
([hs/Fe] $\sim$ 2) together with an initial $r$-enhancement 
[r/Fe]$^{\rm ini}$ $\sim$ 1.0.
Three stars lie on the main-sequence/turnoff phase, 
HE 0143--0441 by \citet{cohen06}, SDSS J0912+0216 by \citet{behara10},
and CS 22887--048, studied by \citet{tsangarides05} (PhD Thesis).
One star is a giant, HD 209621 by \citet{GA10}.

\begin{figure}
\includegraphics[angle=-90,width=8.5cm]{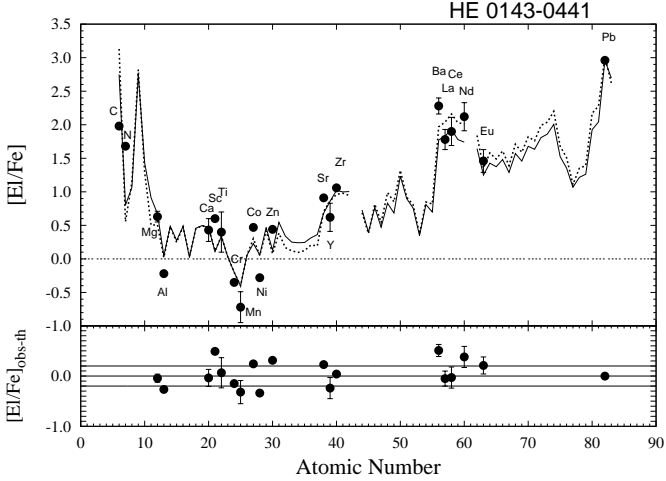}
\vspace{2mm}
\caption{Spectroscopic [El/Fe] abundances of the main-sequence/turnoff star
HE 0143--0441 ([Fe/H] = $-$2.31; {\textit T}$_{\rm eff}$ = 6240 K; log $g$ = 3.7)
compared with two AGB models: $M^{\rm AGB}_{\rm ini}$ = 1.3 $M_{\odot}$, case 
ST/9 and $dil$ = 0.0 dex (dotted line); $M^{\rm AGB}_{\rm ini}$ = 2 $M_{\odot}$, 
case ST/5 and $dil$ = 1.0 dex (solid line).
Observations are from \citet{cohen06}.
The observed $s$-process indicators are [hs/ls] = 1.12 and [Pb/hs] = 1.25.
A spread is observed among the hs elements (see text).
The differences [El/Fe]$_{\rm obs-th}$ displayed in the lower panel refer to 
the $M^{\rm AGB}_{\rm ini}$ = 2 $M_{\odot}$ model (solid line).
We adopt an [r/Fe]$^{\rm ini}$ = 1.0. }
\label{HE0143-0441_cohen06_bab10d6m1p5d3m2z1m4rp1alf0p5_diffdiln5n26}
\end{figure}

\subsubsection{HE 0143--0441 
(Fig.~\ref{HE0143-0441_cohen06_bab10d6m1p5d3m2z1m4rp1alf0p5_diffdiln5n26})}
\label{HE0143}

This is a main-sequence/turnoff star ([Fe/H] = $-$2.31; {\textit T}$_{\rm eff}$
= 6240 
K; log $g$ = 3.7), 
analysed by \citet{cohen04,cohen06}.
Only the most recent data by \citet{cohen06} are considered here.
We present possible theoretical interpretations with AGB models of different 
initial masses. Two solutions are shown in
Fig.~\ref{HE0143-0441_cohen06_bab10d6m1p5d3m2z1m4rp1alf0p5_diffdiln5n26}: 
the dotted line corresponds to an AGB model of $M^{\rm AGB}_{\rm ini}$ 
= 1.3 $M_{\odot}$ (case ST/9 and no dilution), the 
solid line represents an AGB model of $M^{\rm AGB}_{\rm ini}$ 
= 2 $M_{\odot}$ (case ST/5 and $dil$ = 1 dex).
Both models interpret the observed $s$-process indicators
[hs/ls] = 1.12 and [Pb/hs] = 1.25.
A spread is observed among the hs elements ([Nd/La] $\sim$ 0.4 dex;
4 lines are detected for La, 3 lines for Nd as for Ba and Ce),
while [Y/Fe] is about 0.2 lower than AGB predictions.
Similar solutions may be obtained with
AGB models of initial mass $M$ = 1.5 $M_{\odot}$ and $dil$ = 0.6 dex.
A Na measurement would help to discriminate the AGB initial 
 mass ([Na/Fe]$_{\rm th}$ = 0.5 is predicted by 
 $M^{\rm AGB}_{\rm ini}$ = 1.3 $M_{\odot}$, pulse 5 and negligible dilution; 
[Na/Fe]$_{\rm th}$ = 1.1 with $M^{\rm AGB}_{\rm ini}$ = 1.5 $M_{\odot}$, $dil$ $\sim$ 1
dex). This would also provide information about the efficiency of the mixing during the
main-sequence phase. 
An initial $r$-process enrichment [r/Fe]$^{\rm ini}$ = 1 is adopted.

\begin{figure}
\includegraphics[angle=-90,width=8.5cm]{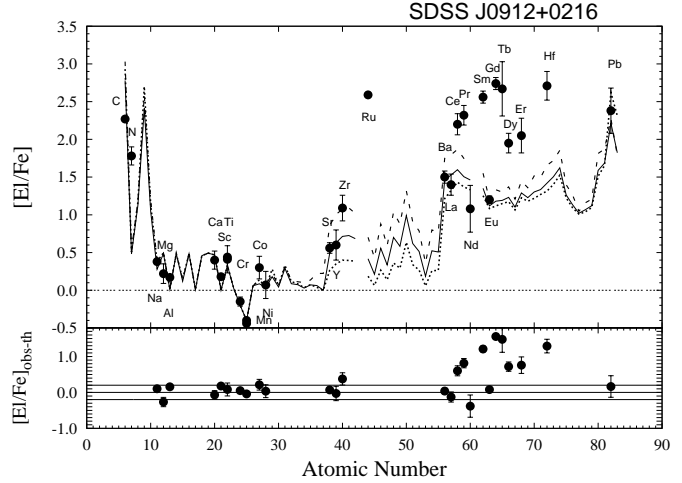}
\vspace{2mm}
\caption{Spectroscopic [El/Fe] abundances of the main-sequence star SDSS J0912+0216
 ([Fe/H] = $-$2.5; {\textit T}$_{\rm eff}$ = 6500 K; log $g$ = 4.5)
compared with AGB models of $M^{\rm AGB}_{\rm ini}$ = 1.3 $M_{\odot}$, 
cases ST/9 (dotted line), ST/18 (solid line), ST/24 (dashed line), 
$dil$ $\sim$ 0.6 dex. 
Observations are from \citet{behara10}.
For a discussion of 
 the differences between observations and predictions for Zr, Ru and 
elements from Ba to Hf see text.
On average [hs/ls]$_{\rm obs}$ = 0.84 and [Pb/hs]$_{\rm obs}$ = 0.64.
[C/Fe] and [N/Fe] from CH bands are displayed. 
3D model atmospheres calculations reduce these values to 1.67 and 1.07
dex, respectively \citep{behara10}.
An initial $r$-process enrichment of [r/Fe]$^{\rm ini}$ = 1.0 is adopted,
in agreement with the observed [Eu/Fe] (see text). }
\label{SDSSJ0912+0216_behara10_astroph_bab9ltHTd12m1p5z5m5_rp1he25alf0p5_dil0p6n5}
\end{figure}

\subsubsection{SDSS J0912+0216
(Fig.~\ref{SDSSJ0912+0216_behara10_astroph_bab9ltHTd12m1p5z5m5_rp1he25alf0p5_dil0p6n5})}
\label{SDSSJ0912}

This main-sequence star was studied by \citet{behara10}
with $R$ = 30\,000
([Fe/H] = $-$2.5; {\textit T}$_{\rm eff}$ = 6500 K; log $g$ = 4.5 dex).
The neutron capture elements are highly enhanced.
Large differences are detected among the hs peak elements
(e.g., [Ce/La] $\sim$ 1) and among the $r$-elements (e.g., [Gd/Eu] $\sim$ 1.5),
similarly to the star SDSS J1349--0229 (Section~\ref{SDSSJ1349}).
We assess the initial $r$-enhancement by the observed [Eu/Fe],
while La is chosen as the most representative among the hs peak. 
In Fig.~\ref{SDSSJ0912+0216_behara10_astroph_bab9ltHTd12m1p5z5m5_rp1he25alf0p5_dil0p6n5},
SDSS J0912+0216 is interpreted with a $M^{\rm AGB}_{\rm ini}$ = 1.3 
$M_{\odot}$ model (case ST/18 and $dil$ = 0.6 dex). 
This star lies on the main-sequence, and the dilution provided by
this model suggests that mixing has been efficient during this phase.
However, the large spread affecting the hs elements leave this conclusion
very uncertain. Solutions
with negligible dilutions may be obtained by decreasing the
number of TDUs (e.g., $M^{\rm AGB}_{\rm ini}$ = 1.2 $M_{\odot}$, pulse number 3),
but the predicted [Na/Fe] would be 0.3 dex lower than that observed.
The detected [La/Eu] ratio 
needs an initial $r$-process enhancement of [r/Fe]$^{\rm ini}$ = 1 dex.
{However, we are not able to interpret the discrepancy observed among
Eu and the other $r$-elements.}
Similarly, Ru is highly enhanced, at the same level of Gd and Tb.
The low [Na/Fe] would exclude models with $M^{\rm AGB}_{\rm ini}$ $\geq$ 1.5 
$M_{\odot}$.
Note that by increasing the AGB initial mass and with a proper choice of 
the $^{13}$C-pocket and dilution factor, we may find theoretical solutions for the 
high [Ce,Pr/Fe] observed, but several observed elements ([Na, Mg, Sr, Y, Ba, La,
Nd, Eu/Fe]) would be overestimated by models.
\citet{behara10} provide 3D atmospheric model corrections for 
C and N, which decrease the observations shown in 
Fig.~\ref{SDSSJ0912+0216_behara10_astroph_bab9ltHTd12m1p5z5m5_rp1he25alf0p5_dil0p6n5}
by 0.5 and 0.67 dex, respectively.\\
Further investigations are strongly desirable for this star, especially in 
the light of the large discrepancies outlined both among hs and $r$-elements that
 can not be explained by AGB models.

\begin{figure}
\includegraphics[angle=-90,width=8.5cm]{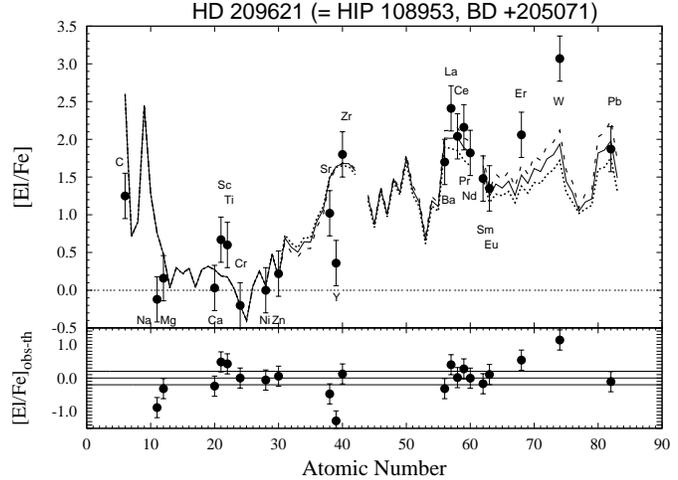}
\vspace{2mm}
\caption{Spectroscopic [El/Fe] abundances of the giant HD 209621
([Fe/H] = $-$1.93; {\textit T}$_{\rm eff}$ = 4500 K; log $g$ = 2.0)
compared with AGB models of $M^{\rm AGB}_{\rm ini}$ =
2 $M_{\odot}$, cases ST/12 (dashed line), ST/15 (solid line), ST/18 
(dotted line). We adopt $dil$ = 0.9 dex.
Observations are from \citet{GA10}.
A spread in the ls elements is observed, while [Pb/hs] $\sim$ 0.
For discussions about differences between observed and predicted
Na, Y, Er and W see text.
An initial $r$-process enrichment of [r/Fe]$^{\rm ini}$ = 1.0 is adopted. }
\label{HD209621_goswamiaoki10_bab10d8d10d12m2z2m4rp1_dil0p85n26}
\end{figure}

\subsubsection{HD 209621 ($\equiv$ HIP 108953, BD +205071), 
(Fig.~\ref{HD209621_goswamiaoki10_bab10d8d10d12m2z2m4rp1_dil0p85n26})}
\label{HD209621}

This giant with [Fe/H] $\sim$ $-$1.93 ({\textit T}$_{\rm eff}$ = 4500 
K and log $g$ = 2.0) 
has been recently observed by \citet{GA10}. 
HD 209621 was one of the CH stars analysed by \citet{vanture92c}  
with lower resolution spectra ($R$ = 20\,000 instead of 50\,000 by
\citealt{GA10}).
The metallicity of the star was estimated as [Fe/H] = $-$0.9,
significantly higher than that derived by \citet{GA10}
([Fe/H] = $-$1.93).
HD 209621 was found to be a long-period binary by \citet{MC90}, with $P$
$\sim$ 400 d. This supports the hypothesis of mass transfer of $s$-rich
 material from an AGB companion.
The spectra of this star are dominated by molecular
absorption lines of CH, CN and C$_2$.
\citet{GA10} carefully considered 
only the unblended lines for the determination
of the spectroscopic abundances. They found a large spread 
among the observed ls element (e.g.,
[Zr/Y] = 1.4). This disagrees with theoretical AGB models, which predict
[Y/Fe] $\sim$ [Zr/Fe]. However, we recall that only one reliable line has been 
adopted for the analysis of Sr, Y e Zr in this cool giant.
In Fig.~\ref{HD209621_goswamiaoki10_bab10d8d10d12m2z2m4rp1_dil0p85n26}
we select to fit Zr among the ls elements, 
according with a low $^{13}$C-pocket (ST/15), which interpret both
[hs/Zr] and [Pb/hs].
$M^{\rm AGB}_{\rm ini}$ = 2 $M_{\odot}$ models with a large dilution 
in agreement with a giant ($dil$ = 0.9 dex) are shown.
The low [Na/Fe] is calculated with the resonance doublet Na I D lines 
at 5890 and 5896 ${\rm \AA}$ with a LTE analysis.
Lower AGB initial mass and lower dilutions are discarded because
in contrast with a giant after the FDU (e.g., a model of $M^{\rm AGB}_{\rm ini}$ = 1.5
$M_{\odot}$ and ST/15 would imply a $dil$ = 0.6 dex).
An [r/Fe]$^{\rm ini}$ = 1 is adopted to interpret the [hs/Eu] ratio.
The observed [Er/Fe] is lower than theoretical predictions, 
but only one line is detected for this element.
Note that W, similarly to Hf, is mostly an $s$-process element ($\sim$ 60\% 
of solar W is produced by the main-$s$ process). Therefore, our theoretical
prediction would not be largely affected by higher initial $r$-process 
enhancements. However, \citet{GA10} explicitly mention a potential overestimation of 
the [W/Fe] abundance owing to a possible blending.
\\
\citet{GA10} provided theoretical interpretations with a parametric model,
based on the solar system $s$ and $r$-process isotopic abundances (scaled 
to the metallicity of the star) of each isotope
provided by \citet{arlandini99}. This method does not account for the 
dominant contribution to Pb and Bi by the strong component at low metallicities. 
Indeed, in order to estimate the solar $r$-process percentage of Pb and Bi,
we adopt a Galactic Chemical Evolution model, which accounts for the $s$-process
contribution of all AGB masses and all metallicities (\citealt{travaglio04}, Paper II).

\begin{figure}
\includegraphics[angle=-90,width=8.5cm]{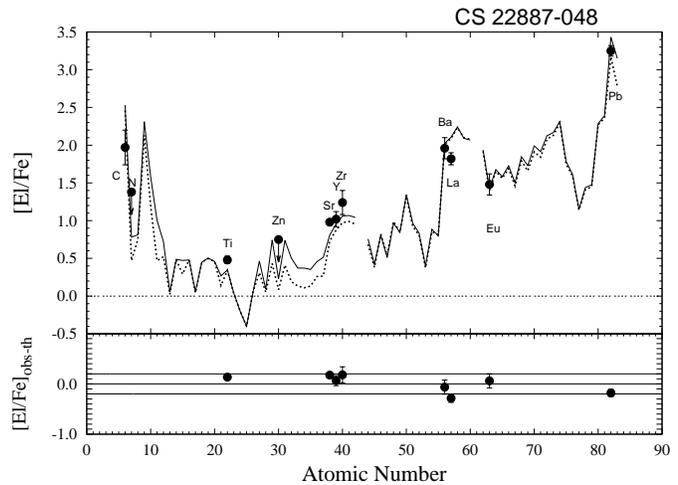}
\vspace{2mm}
\caption{Spectroscopic [El/Fe] abundances of the main-sequence/turnoff star CS 22887--048
([Fe/H] = $-$1.7; {\textit T}$_{\rm eff}$ = 6500 K; log $g$ = 3.35)
compared with different AGB models: $M^{\rm AGB}_{\rm ini}$ 
= 1.4 $M_{\odot}$, case ST/2 and no dilution (dotted line), 
or $M^{\rm AGB}_{\rm ini}$ = 1.5 $M_{\odot}$, case ST$\times$1.2 
and $dil$ = 0.3 dex (solid line).
Observations are from \citet{tsangarides05}, who found
 [hs/ls] = 0.80 and [Pb/hs] = 1.49.
The differences [El/Fe]$_{\rm obs-th}$ displayed in the lower panel
refer to the $M^{\rm AGB}_{\rm ini}$ = 1.5 $M_{\odot}$ model.
An initial $r$-process enrichment of [r/Fe]$^{\rm ini}$ = 1.0 is adopted.} 
\label{CS22887-048_tsangarides05_bab10d1p5p1p25m1p5z5m4rp1alf0p5_diffdiln8n20}
\end{figure}

\subsubsection{CS 22887--048 
(Fig.~\ref{CS22887-048_tsangarides05_bab10d1p5p1p25m1p5z5m4rp1alf0p5_diffdiln8n20})}
\label{CS048} 

CS 22887--048, with [Fe/H] = $-$1.7, was studied by 
\citet{tsangarides05}, PhD thesis.
 This star shows an effective temperature typical of 
main-sequence/turnoff stars ({\textit T}$_{\rm eff}$ = 
6500 K), while its surface gravity is rather low (log $g$ 
= 3.35). 
\citet{johnson07} included this star in a sample of metal-poor N-rich candidates.
They estimated a metallicity 1 dex lower than \citet{tsangarides05}, ([Fe/H] = $-$2.79).
 Although the period remains unknown,
\citet{tsangarides05} confirms the binarity of this star,  
supporting the mass transfer scenario.
\\
In Fig.~\ref{CS22887-048_tsangarides05_bab10d1p5p1p25m1p5z5m4rp1alf0p5_diffdiln8n20}, 
possible interpretations with AGB models 
are shown: $M^{\rm AGB}_{\rm ini}$  = 1.4 $M_{\odot}$, case ST/2, and no dilution, 
and $M^{\rm AGB}_{\rm ini}$ = 1.5 $M_{\odot}$, case ST$\times$1.2 and $dil$ = 0.3 dex.
Note that this star needs the most efficient $^{13}$C-pocket strength of
all CEMP-$s$ (and CEMP-$s/r$) sample.
 Both solutions sustain that moderate mixing had occurred in this star.
An initial $r$-process enhancement [r/Fe]$^{\rm ini}$
 = 1.0 is adopted in order to interpret a [Ba/Eu] ratio of 0.5 dex.
[La/Fe] is $\sim$ 0.3 dex lower than AGB models: a smaller 
[hs/Fe]$_{\rm th}$ value would be obtained by a lower $^{13}$C-pocket efficiency
(e.g. the case ST/3 for $M^{\rm AGB}_{\rm ini}$ = 1.5 
 $M_{\odot}$), with a significant decrease in lead ([Pb/Fe]$_{\rm th}$ = 2.7). 
Note that [La/Eu] = 0.24 would require a higher initial $r$-process 
enrichment. Further AGB model constraints may be obtained by 
detecting Na and additional hs elements.

%


\section{CEMP-{\scriptsize s} stars with no E{\scriptsize u} measurement} \label{secCEMPsnoEu}

In this Section, we analyse CEMP-$s$II and CEMP-$s$I stars with no Eu detection.
These stars are interpreted by an initial
$r$-process enrichment [r/Fe]$^{\rm ini}$ = 0.5, chosen as representative
of the average of [Eu/Fe] observed in halo field stars (see Section~\ref{intro}).

\subsection{CEMP-$s$II with no Eu} 
\label{secCEMPsIInoEu}

Six stars show a high $s$-process enhancement 
and no Eu measurement:
three stars are main-sequence/turnoff,
CS 29528--028 by \citet{aoki07},
HE 1430--1123 by \citet{barklem05},
and SDSS 0126+06 by \citet{aoki08}.
Other three stars are giants,
HD 201626 by \citet{vaneck03},
HD 5223 by \citet{goswami06},
and HE 0212--0557 by \citet{cohen06}.

\begin{figure}
\includegraphics[angle=-90,width=8.5cm]{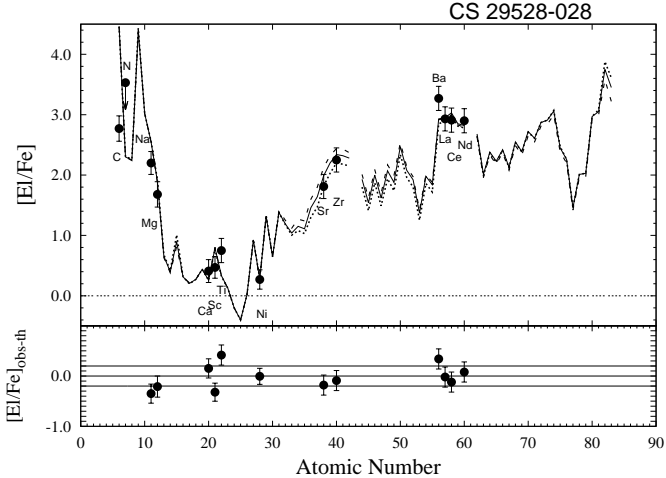}
\vspace{2mm}
\caption{Spectroscopic [El/Fe] abundances of the main-sequence star CS 29528--028
([Fe/H] = $-$2.86; {\textit T}$_{\rm eff}$ = 6800 K; log $g$ = 4.0)
compared with AGB models of $M^{\rm AGB}_{\rm ini}$ = 
2 $M_{\odot}$, cases ST/8 (dotted line), ST/12 (solid line), ST/15 
(dashed line), and no dilution.
Observations are from \citet{aoki07}, who detected [hs/ls] = 0.62.
This model predicts [Pb/Fe]$_{\rm th}$ $\sim$ 3.8.
An [r/Fe]$^{\rm ini}$ = 0.5 is adopted. }
\label{CS29528-028_aoki07_bab10d5d7d10m1p5z2m5rp0p5_n26}
\end{figure}

\subsubsection{CS 29528--028
(Fig.~\ref{CS29528-028_aoki07_bab10d5d7d10m1p5z2m5rp0p5_n26})}
\label{CS028}

CS 29528--028 is the hottest main-sequence star among the sample studied by \citet{aoki07},
([Fe/H] = $-$2.86; {\textit T}$_{\rm eff}$ = 6800 K; log $g$ = 4.0). 
The authors detected several $s$-elements, Sr, Zr, Ba, La, Ce and Nd,  
showing a great $s$-enhancement, about 1 dex higher than the average 
of CEMP-$s$ stars: [ls/Fe] $\sim$ 2 and [hs/Fe] $\sim$ 3.
Moreover, this star exhibits high Na and Mg ([Na/Fe] = 2.33 and [Mg/Fe] = 1.69). 
No solutions are attained using AGB 
models of initial mass $M^{\rm AGB}_{\rm ini}$ $\la$ 1.4 $M_{\odot}$,
because the observed [Na,Mg,ls,hs/Fe] would be too high.
In Fig.~\ref{CS29528-028_aoki07_bab10d5d7d10m1p5z2m5rp0p5_n26},
interpretations with $M^{\rm AGB}_{\rm ini}$ = 2 $M_{\odot}$ are shown, 
(cases ST/8, ST/12, ST/15 and no dilution).
This model predicts [Pb/Fe]$_{\rm th}$ $\sim$ 3.8.
AGB models of $M^{\rm AGB}_{\rm ini}$ = 1.5 $M_{\odot}$
are excluded, because they overestimate the observed Na and Mg
([Na/Fe]$_{\rm th}$ = 3.00; [Mg/Fe]$_{\rm th}$ = 2.25,
case ST/5 and no dilution). 
The lack of dilution suggests that no efficient mixing had occurred.
This star shows the highest [ls/Fe] and [hs/Fe] known.
Observations of Eu and Pb would be very useful. 
A similar behaviour has been observed in SDSS 1707+58 by \citet{aoki08}, 
([Sr/Fe] = 2.3 and [Ba/Fe] = 3.4), as described in Appendix~A.
Further investigations on these two stars are strongly suggested.

\begin{figure}
\includegraphics[angle=-90,width=8.5cm]{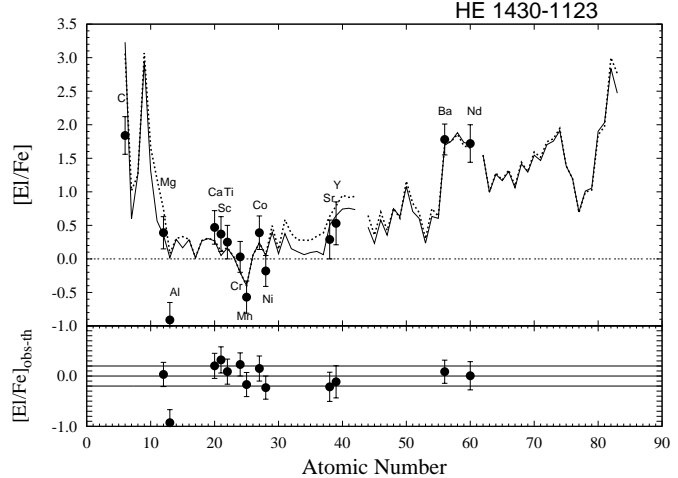}
\vspace{2mm}
\caption{Spectroscopic [El/Fe] abundances of the turnoff/subgiant HE 1430--1123 
([Fe/H] = $-$2.71; {\textit T}$_{\rm eff}$ = 5915 K; log $g$ = 3.75, before the FDU)
compared with AGB models of initial mass
1.3 $M_{\odot}$ (ST/12 and $dil$ = 0.2 dex; solid line) and 2 $M_{\odot}$ 
(ST/5 and $dil$ = 1.0 dex; dotted line). 
Observations are from \citet{barklem05}, who found [hs/ls] = 0.94.
An [r/Fe]$^{\rm ini}$ = 0.5 is adopted. }
\label{HE1430-1123Bark05_bab10d8m1p5d3m2z5m5rp0p5_diffdiln5n26}
\end{figure}

\subsubsection{HE 1430--1123
(Fig.~\ref{HE1430-1123Bark05_bab10d8m1p5d3m2z5m5rp0p5_diffdiln5n26})}
\label{HE1430}

This turnoff/subgiant shows [Fe/H] = $-$2.71, {\textit T}$_{\rm eff}$ = 5915 K
and log $g$ = 3.75 \citep{barklem05}.
In Fig.~\ref{HE1430-1123Bark05_bab10d8m1p5d3m2z5m5rp0p5_diffdiln5n26},
theoretical interpretations are shown using AGB models of 
$M^{\rm AGB}_{\rm ini}$ = 2 $M_{\odot}$, case ST/5 and 
 $dil$ = 1.0 dex (dotted line) and
$M^{\rm AGB}_{\rm ini}$ = 1.3 $M_{\odot}$, case ST/12 and a
negligible dilution ($dil$ = 0.2 dex, solid line). 
The model with a lower number of TDUs better
interprets the observed [Mg/Fe], even if the value predicted by 
$M^{\rm AGB}_{\rm ini}$ = 2 $M_{\odot}$ still agrees within the errors.
Solutions with $M^{\rm AGB}_{\rm ini}$ = 1.5 $M_{\odot}$ predict
a higher [Mg/Fe] than observed. 
A Na measurement would help to provide constraints on the AGB initial mass
and on the efficiency of possible mixing.

\begin{figure}
\includegraphics[angle=-90,width=8.5cm]{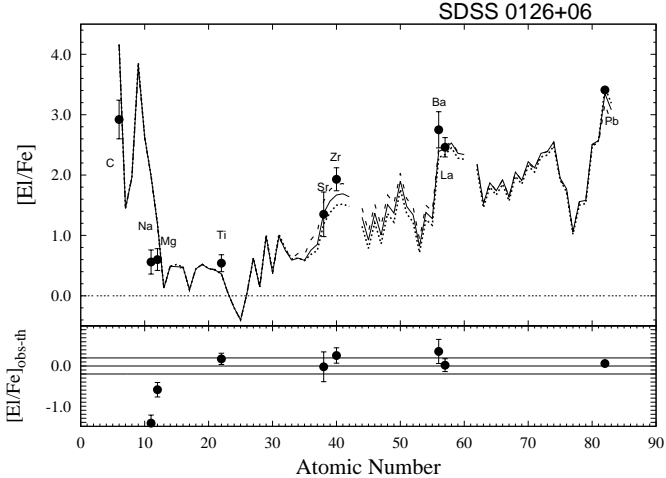}
\vspace{2mm}
\caption{Spectroscopic [El/Fe] abundances of the main-sequence star SDSS 0126+06
([Fe/H] = $-$3.11; {\textit T}$_{\rm eff}$ = 6600 K; log $g$ = 4.1)
compared with AGB models of
initial mass $M^{\rm AGB}_{\rm ini}$ = 1.4 $M_{\odot}$, cases
ST/9 (dotted line), ST/12 (solid line), ST/18 (dashed line),
 and no dilution.
Observations are from \citet{aoki08}.
This star shows a very high $s$-process enhancement, [La/Fe] = 2.46
([hs/ls] = 0.58 and [Pb/hs] = 1.08),
impossible to interpret with AGB models of lower initial mass
($M^{\rm AGB}_{\rm ini}$ = 1.2 -- 1.3 $M_{\odot}$, which undergo a limited
number of TDUs).
This leads to an overestimation of the observed [Na/Fe].
An [r/Fe]$^{\rm ini}$ = 0.5 is adopted. }
\label{SDSS0126+06_aoki08_bab10d6d8d12m1p5z2m5rp0p5alf0p5_n8}
\end{figure}

\subsubsection{SDSS 0126+06
(Fig.~\ref{SDSS0126+06_aoki08_bab10d6d8d12m1p5z2m5rp0p5alf0p5_n8})}
\label{SDSS+06}

A high $s$-process enhancement is shown by the main-sequence star SDSS 0126+06
([Fe/H] = $-$3.11, {\textit T}$_{\rm eff}$ = 6600 K, log $g$ = 4.1):
[Zr/Fe] $\sim$ 1.9, [La/Fe] $\sim$ 2.4 and [Pb/Fe] $\sim$ 3.4 
\citep{aoki08}.
A possible theoretical interpretation is displayed in 
Fig.~\ref{SDSS0126+06_aoki08_bab10d6d8d12m1p5z2m5rp0p5alf0p5_n8}
by models of $M^{\rm AGB}_{\rm ini}$ = 1.4 $M_{\odot}$, no dilution and 
three $^{13}$C-pockets (cases ST/9, ST/12 and ST/18).
Note that the observed [Na/Fe] and [Mg/Fe] are largely overestimated
by models.
Similar solutions are found by $M^{\rm AGB}_{\rm ini}$ = 
1.5 and 2 $M_{\odot}$ models (with $dil$ $\sim$ 0.4 dex, suggesting
a moderate mixing), but the observed [Na/Fe] and [Mg/Fe] would be 1 dex (or more)
higher than AGB predictions.
Models of $M^{\rm AGB}_{\rm ini}$ = 1.3 $M_{\odot}$, which undergo
5 TDUs, would provide lower [Na/Fe] and [Mg/Fe], but they can not reach 
the high [La/Fe] observed.

\begin{figure}
\includegraphics[angle=-90,width=8cm]{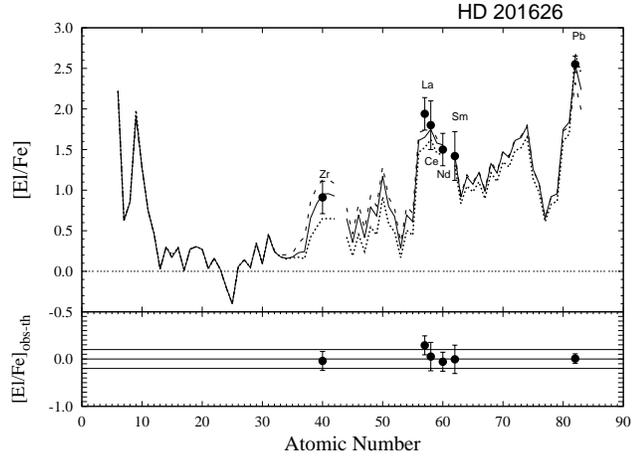}
\vspace{2mm}
\caption{Spectroscopic [El/Fe] abundances of the giant HD 201626 
([Fe/H] = $-$2.1; {\textit T}$_{\rm eff}$ = 5190 K; log $g$ = 2.25)
compared with AGB models of $M^{\rm AGB}_{\rm ini}$ = 1.5 $M_{\odot}$, 
cases ST/2 (dotted line), ST/3 (solid line), ST/5 (dashed line), and $dil$ = 1.3 dex.
Observations are from \citet{vaneck03}, who detected
[hs/ls] = 0.73 and [Pb/hs] = 1.00.
An [r/Fe]$^{\rm ini}$ = 0.5 is adopted. }
\label{HD201626_vaneck03_bab10d1p3d2d3m1p5z2m4rp0p5_dil1p3n20}
\end{figure}

\subsubsection{HD 201626
(Fig.~\ref{HD201626_vaneck03_bab10d1p3d2d3m1p5z2m4rp0p5_dil1p3n20})}
\label{HD201626}

For this giant ([Fe/H] = $-$2.1; {\textit T}$_{\rm eff}$ = 5190 K; log $g$ = 2.25),
Zr, La, Ce, Nd, Sm and Pb have been detected by \citet{vaneck03}
([hs/ls] = 0.73 and [Pb/hs] = 1.00).
The only constraint for AGB models is provided by the occurrence of 
the FDU, which imply a dilution of the order of 1 dex or more.
Possible theoretical interpretations are shown in 
Fig.~\ref{HD201626_vaneck03_bab10d1p3d2d3m1p5z2m4rp0p5_dil1p3n20}, 
for $M^{\rm AGB}_{\rm ini}$ = 1.5 $M_{\odot}$, cases ST/2, ST/3,
ST/5, and $dil$ = 1.3 dex.
Analogous solutions are obtained for $M^{\rm AGB}_{\rm ini}$ = 2
 $M_{\odot}$ (ST/3 and $dil$ = 1.3 dex).
Models with $M^{\rm AGB}_{\rm ini}$ $<$ 1.4 $M_{\odot}$ are excluded 
because dilutions lower than $\sim$ 0.5 dex would be in contrast with
a giant after the FDU. 
Observations of Na or Mg would provide information on the AGB
initial mass.

\begin{figure}
\includegraphics[angle=-90,width=8.5cm]{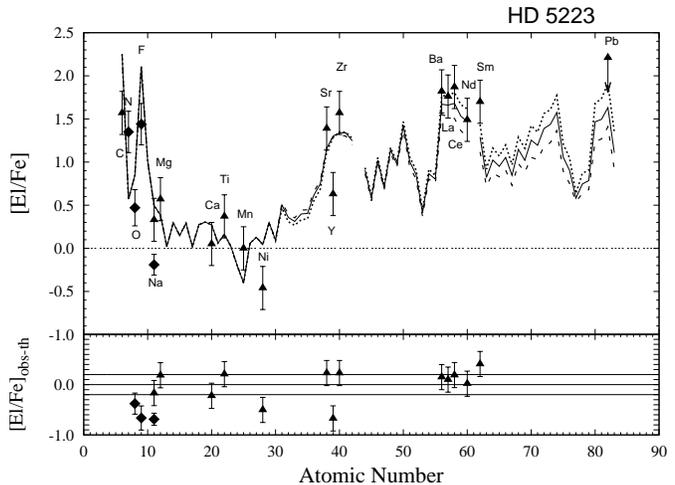}
\vspace{2mm}
\caption{Spectroscopic [El/Fe] abundances of the giant HD 5223 
([Fe/H] = $-$2.06; {\textit T}$_{\rm eff}$ = 4500 K; log $g$ = 1.0) 
compared with AGB models of $M^{\rm AGB}_{\rm ini}$ = 2 $M_{\odot}$, 
cases ST/12 (dotted line), ST/15 (solid line), ST/18 (dashed line), 
and $dil$ = 1.2 dex.
Observations are from \citet{goswami06} (filled circles) 
and \citet{lucatello11astroph} (filled diamonds).  
A spread in the ls elements is observed.
See text for the discussion about Y and Na.
An [r/Fe]$^{\rm ini}$ = 0.5 is adopted.  }
\label{HD5223_goswami06_bab10d8d10d12m2z2m4rp0p5_dil1p2n26}
\end{figure}

\subsubsection{HD 5223 
(Fig.~\ref{HD5223_goswami06_bab10d8d10d12m2z2m4rp0p5_dil1p2n26})}
\label{HD5223}

HD 5223 is a cold giant ([Fe/H] = $-$2.06; {\textit T}$_{\rm eff}$ = 4500 K;
log $g$ = 1.0), analysed by \citet{goswami06}. 
The high hs-peak ([hs/Fe] = 1.8) confirms the evidence of an
AGB contribution, also supported by a binary scenario \citep{MC90}.
Large uncertainties affect the ls elements ([Zr/Y] $\sim$ 0.8 dex),
in disagreement with AGB theoretical predictions\footnote{Two other
stars studied by \citet{goswami06} show a
similar spectroscopic behaviour: HE 1305+0007,
discussed in Section~\ref{HE1305} with [Zr/Y] $\sim$ 1.3,
and HE 1152--0355 for hs-peak
with [La/Nd] $\sim$ 1.3 (see Section~\ref{secCEMPsInoEu}).}.
Only an upper limit for Pb is detected ([Pb/hs] $\leq$ 0.55).
Note that NLTE effects were not considered to determine the abundances.
The single Pb line at 4057 {\rm \AA} is strongly affected by molecular
absorption bands.
Possible interpretations with AGB models of $M^{\rm AGB}_{\rm ini}$ = 
2 $M_{\odot}$, cases ST/12, ST/15, ST/18, and $dil$ = 1.2 dex are displayed 
in Fig.~\ref{HD5223_goswami06_bab10d8d10d12m2z2m4rp0p5_dil1p2n26}.                                   
Higher $^{13}$C-pocket efficiencies may explain the observed [Y/Fe],
but the upper limit of Pb would be overestimated by these models ([Pb/Fe]$_{\rm th}$ $\sim$ 2.9, 
case ST/3, $dil$ $\sim$ 1 dex).
We exclude solutions with $M^{\rm AGB}_{\rm ini}$ = 1.5 
$M_{\odot}$ models, since the observed [Na/Fe] would be overestimated. 
On the other side, AGBs with lower initial mass ($M^{\rm AGB}_{\rm ini}$
 = 1.3 $M_{\odot}$; case ST/24) have negligible dilutions, in contrast with 
a giant having suffered the FDU. The [Na/Fe]
observed by \citet{lucatello11astroph} is about 0.5 dex lower than
the AGB predictions.
\citet{lucatello11astroph} detected [F/Fe] in this star\footnote{Fluorine
has been measured in two other CEMP-$s$ stars: HE 1152--0355 by
\citet{lucatello11astroph} ([F/Fe] = 0.64; Section~\ref{HE1152}) and HE
1305+0132 by \citet{schuler08} (see Section~\ref{HEsch}).}.
The AGB models shown in Fig.~\ref{HD5223_goswami06_bab10d8d10d12m2z2m4rp0p5_dil1p2n26}
overestimated the observed [F/Fe] by about 0.5 dex.

\begin{figure}
\includegraphics[angle=-90,width=8.5cm]{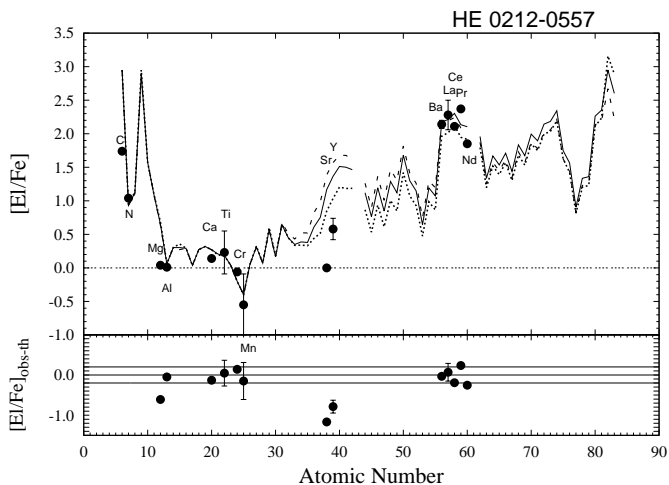}
\vspace{2mm}
\caption{Spectroscopic [El/Fe] abundances of the giant HE 0212--0557
 ([Fe/H] = $-$2.27; {\textit T}$_{\rm eff}$ = 5075 K; log $g$ = 2.15)
compared with AGB models of $M^{\rm AGB}_{\rm ini}$ 
= 2 $M_{\odot}$, cases ST/5 (dotted line), ST/8 (solid line), ST/12 (dashed line),
 and $dil$ = 0.8 dex.
Observations are from \citet{cohen06}.
The first $s$-peak is very uncertain, because Zr is not measured
and only one line is available for Sr. Three lines are detected for Y.
An [r/Fe]$^{\rm ini}$ = 0.5 is adopted. }
\label{HE0212-0557_cohen06_bab10d3d5d8m2z1m4rp0p5_dil0p8n26}
\end{figure}

\subsubsection{HE 0212--0557 (Fig.~\ref{HE0212-0557_cohen06_bab10d3d5d8m2z1m4rp0p5_dil0p8n26})}
\label{HE0212}

This giant HE 0212--0557 ([Fe/H] = $-$2.27; {\textit T}$_{\rm eff}$ = 5075 K;
log $g$ = 2.15) has been analysed by \citet{cohen06}.
It is a cool star, affected by molecular absorption from CH and CN bands 
\citep{cohen06}.
The Na I D line is too strong for a reliable abundance determination. 
Concerning the ls elements, Zr is not detected, one line is available 
for Sr, and three lines for Y.
\\
Possible solutions with an AGB initial mass of $M^{\rm AGB}_{\rm ini}$ 
= 2 $M_{\odot}$ are shown in 
Fig.~\ref{HE0212-0557_cohen06_bab10d3d5d8m2z1m4rp0p5_dil0p8n26},
(cases ST/5, ST/8, ST/12 and $dil$ = 0.8 dex).  
These models predict [Pb/Fe]$_{\rm th}$ $\sim$ 3 dex. 
The main problem is the observed [Y/Fe] ([hs/Y]$_{\rm obs}$ = 1.5), 
about 1 dex lower than the AGB predictions\footnote{Note that in Paper II we 
calculated [hs/ls] = 1 (see Table~2), by considering [ls/Fe] as an 
average between [Y/Fe]$_{\rm obs}$ and [Zr/Fe]$_{\rm th}$.}.
A better solution for [Y/Fe] would be obtained with a case ST/3, but
the low dilution ($\sim$ 0.5 dex) barely agrees with a giant after the FDU.
Although the low [Mg/Fe] observed would better interpret the value predicted by 
$M^{\rm AGB}_{\rm ini}$ = 1.3 $M_{\odot}$ models, solutions with no dilution are 
in contrast with a giant having suffered the FDU.

\subsection{CEMP-$s$I with no Eu} 
\label{secCEMPsInoEu}

Eleven stars show mild $s$-process enhancement ([hs/Fe] $<$ 1.5)
for which Eu measurements are not available.
Four stars lie before the occurrence of the FDU:
HE 0231--4016, HE 0430--4404 and HE 2150--0825 by \citet{barklem05},
HE 2232--0603 by \citet{cohen06}.
Six stars are giants:
BD +04$^{\circ}$2466 by (\citealt{pereira09,zhang09,ish10}),
HD 189711, HD 198269 and V Ari by \citet{vaneck03},
HE 1031--0020, HE 1434--1442 by \citet{cohen06}.
Similarly to HD 26 and HD 206983, the giant HE 1152--0355, with
[Fe/H] = $-$1.3 \citep{goswami06}, will
be discussed in Section~\ref{CH}.

\begin{figure}
\includegraphics[angle=-90,width=8.5cm]{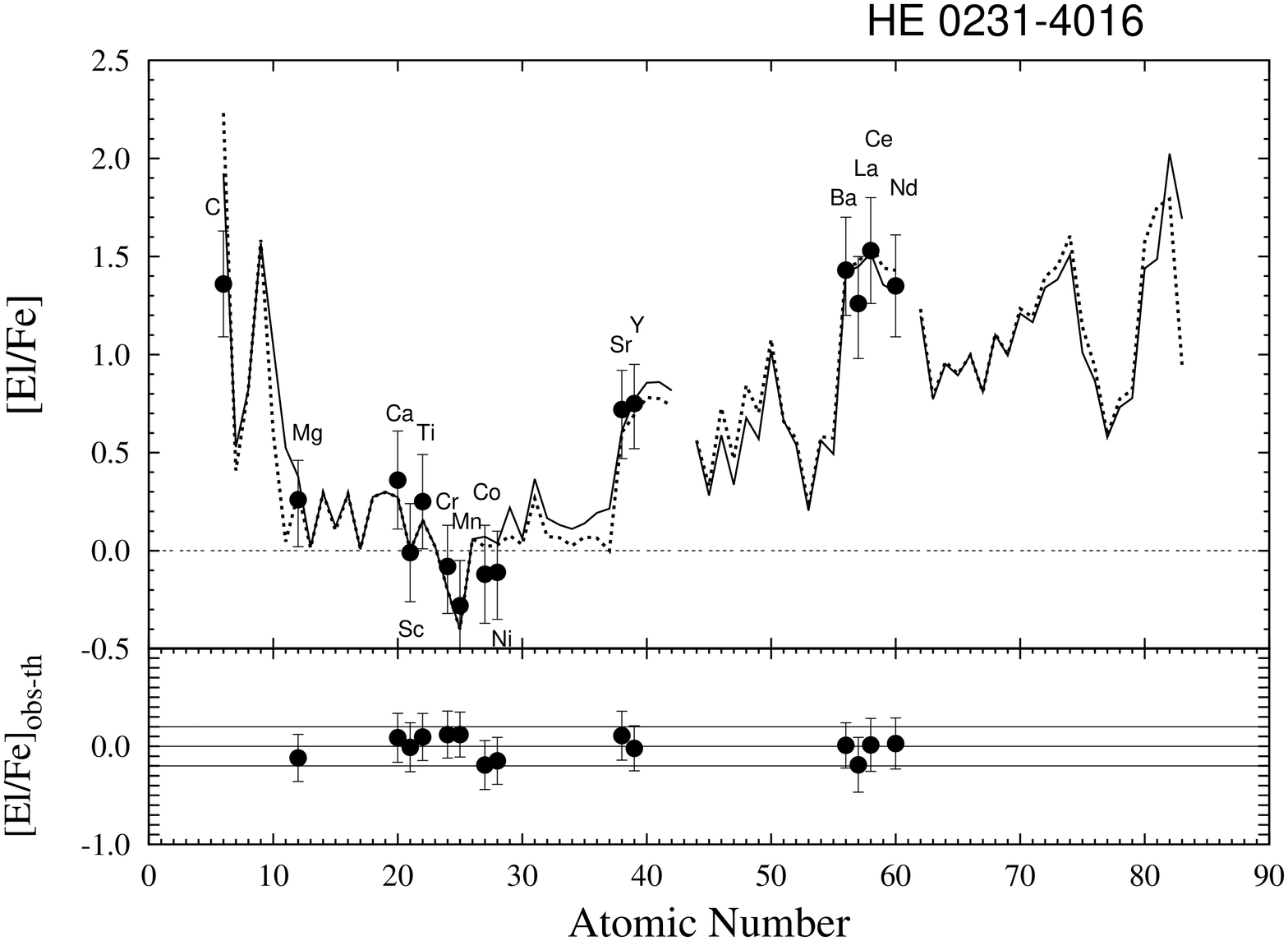}
\vspace{2mm}
\caption{Spectroscopic [El/Fe] abundances of the subgiant HE 0231--4016
([Fe/H] = $-$2.08; {\textit T}$_{\rm eff}$ = 5972 K; log $g$ = 3.59, before 
the occurrence of the FDU) 
compared with two AGB models: $M^{\rm AGB}_{\rm ini}$ = 1.2 $M_{\odot}$, case ST/12, 
$dil$ = 0.2 dex (dotted line), or $M^{\rm AGB}_{\rm ini}$ = 1.5 $M_{\odot}$,
case ST/5, $dil$ = 1.6 dex (solid line).
Observations are from \citet{barklem05}, who found [hs/ls] = 0.5.
We predict [Pb/Fe]$_{\rm th}$ $\sim$ 1.9. 
The differences [El/Fe]$_{\rm obs-th}$ displayed in the lower panel refer
to the $M^{\rm AGB}_{\rm ini}$ = 1.5 $M_{\odot}$ model (solid line).
Similar solutions are obtained for two other CEMP-$s$, HE 2150--0825 and HE 0430--4404,
as described in Appendix~A.
An [r/Fe]$^{\rm ini}$ = 0.5 is adopted. }
\label{HE0231-4016Bark05_bab10d8d3m1p5z2m4rp0p5_diffdiln3n20}
\end{figure}

\subsubsection{HE 0231--4016 
(Fig.~\ref{HE0231-4016Bark05_bab10d8d3m1p5z2m4rp0p5_diffdiln3n20}), 
HE 0430--4404 
(Fig.~\ref{HE0430-4404Bark05_bab10d6d2m1p5z2m4rp0p5_diffdiln3n20}) and 
HE 2150--0825
(Fig.~\ref{HE2150-0825Bark05_bab10d10d3m1p5z2m4rp0p5_diffdiln3n20})}
\label{HE0231}

These three stars, analysed by \citet{barklem05}, are discussed together because
they show similar abundance patterns ([hs/ls] $\sim$ 0.5 -- 0.6).
\\
HE 0231--4016 ([Fe/H] = $-$2.08; {\textit T}$_{\rm eff}$ = 5972 K; log $g$ = 3.59) 
and HE 2150--0825 ([Fe/H] = $-$1.98; {\textit T}$_{\rm eff}$ = 5960 K; log $g$ = 3.67)
lie on the subgiant phase or close to the turnoff,
while HE 0430--4404 is a main-sequence star ([Fe/H] = $-$2.07;
{\textit T}$_{\rm eff}$ = 6214 K; log $g$ = 4.27). 
No lead has been detected.
In Fig.~\ref{HE0231-4016Bark05_bab10d8d3m1p5z2m4rp0p5_diffdiln3n20}, we 
show theoretical interpretations for HE 0231--4016, using AGB 
models of $M^{\rm AGB}_{\rm ini}$ = 1.2 $M_{\odot}$ (case ST/12 and 
$dil$ = 0.2 dex; dotted line) and $M^{\rm AGB}_{\rm ini}$ = 1.5 $M_{\odot}$ 
(case ST/5 and $dil$ = 1.6 dex; solid line).
All AGB initial masses in the range
 1.2 $M_{\odot}$ $\la$ $M^{\rm AGB}_{\rm ini}$ $\leq$ 2 $M_{\odot}$
can equally interpret the spectroscopic observations with a proper 
$^{13}$C-pocket and dilution.
Similar $^{13}$C-pocket efficiencies and dilutions 
are adopted for HE 0430--4404 and HE 2150--0825 (see 
Fig.~\ref{HE0430-4404Bark05_bab10d6d2m1p5z2m4rp0p5_diffdiln3n20} 
and \ref{HE2150-0825Bark05_bab10d10d3m1p5z2m4rp0p5_diffdiln3n20}, 
Appendix~A).
For these three stars we predict [Pb/Fe]$_{\rm th}$ $\sim$ 1.6 -- 2.2. 
A Na investigation may help to discriminate the AGB initial mass
and the efficiency of mixing during the main-sequence phase.

\begin{figure}
\includegraphics[angle=-90,width=8.5cm]{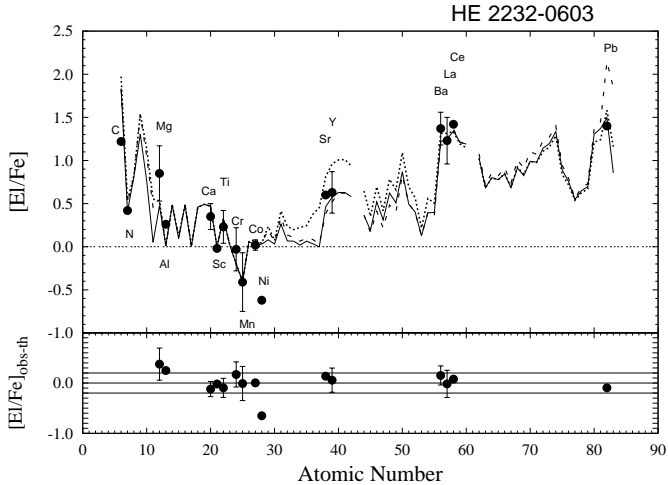}
\vspace{2mm}
\caption{Spectroscopic [El/Fe] abundances of the subgiant HE 2232--0603 
 ([Fe/H] = $-$1.85; {\textit T}$_{\rm eff}$ = 5750 K; log $g$ = 3.5, uncertain FDU)
compared with an AGB model of $M^{\rm AGB}_{\rm ini}$ = 1.3 $M_{\odot}$, 
case ST/12 and $dil$ = 1.0 dex shown by solid line. 
Observations are from \citet{cohen06}.
This star shows [hs/ls] = 0.53 and [Pb/hs] = 0.25.  
Solutions with $M^{\rm AGB}_{\rm ini}$ = 1.5 $M_{\odot}$ are also 
shown: cases ST/5 and ST/3, $dil$ $\sim$ 1.7 dex, represented with 
dotted and dashed lines, respectively (see text). 
Large uncertainties affect the observed [C/Fe] and [N/Fe],
for which no error bars are provided by the authors.
An [r/Fe]$^{\rm ini}$ = 0.5 is adopted. }
\label{HE2232-0603_cohen06_bab10d8d8d6m1p5z2m4rp0p5_diffdiln3n5n20}
\end{figure}

\subsubsection{HE 2232--0603
(Fig.~\ref{HE2232-0603_cohen06_bab10d8d8d6m1p5z2m4rp0p5_diffdiln3n5n20})}
\label{HE2232}%

The mild $s$-process subgiant HE 2232--0603 ([Fe/H] = $-$1.85; 
{\textit T}$_{\rm eff}$ = 5750 K; log $g$ = 3.5, with uncertain FDU)
has been analysed by \citet{cohen06}.
It shows [hs/Fe] $\sim$ 1.15 and [Pb/Fe] $\sim$ 1.4.
An AGB model of $M^{\rm AGB}_{\rm ini}$ = 1.3 $M_{\odot}$ (case ST/12 
and $dil$ = 1.0 dex) interprets the neutron capture elements, as displayed in 
Fig.~\ref{HE2232-0603_cohen06_bab10d8d8d6m1p5z2m4rp0p5_diffdiln3n5n20} by
the solid line. Similar results may be obtained by $M^{\rm AGB}_{\rm ini}$ = 1.2
and 1.4 $M_{\odot}$ models ($dil$ = 0.4 and 1.3 dex, respectively).
By increasing the number of thermal pulses we can not find solutions 
for all three $s$-peaks: e.g., for $M^{\rm AGB}_{\rm ini}$ = 
1.5 $M_{\odot}$, the observed [Sr/Fe] and [Y/Fe] would be lower than the AGB  
models (case ST/5; dotted lines) or the predicted [Pb/Fe] would be $\sim$ 
0.7 dex higher than observed (case ST/3; dashed line), as shown in 
Fig.~\ref{HE2232-0603_cohen06_bab10d8d8d6m1p5z2m4rp0p5_diffdiln3n5n20}. 
However, \citet{cohen06} remarked the uncertainty of the observed Pb, with 
only one line detected.
The high observed [Mg/Fe] ratio agrees within the error bars with the AGB predictions
 ([Mg/Fe] = 0.85 $\pm$ 0.32, 5 lines).

\begin{figure}
\includegraphics[angle=-90,width=8.5cm]{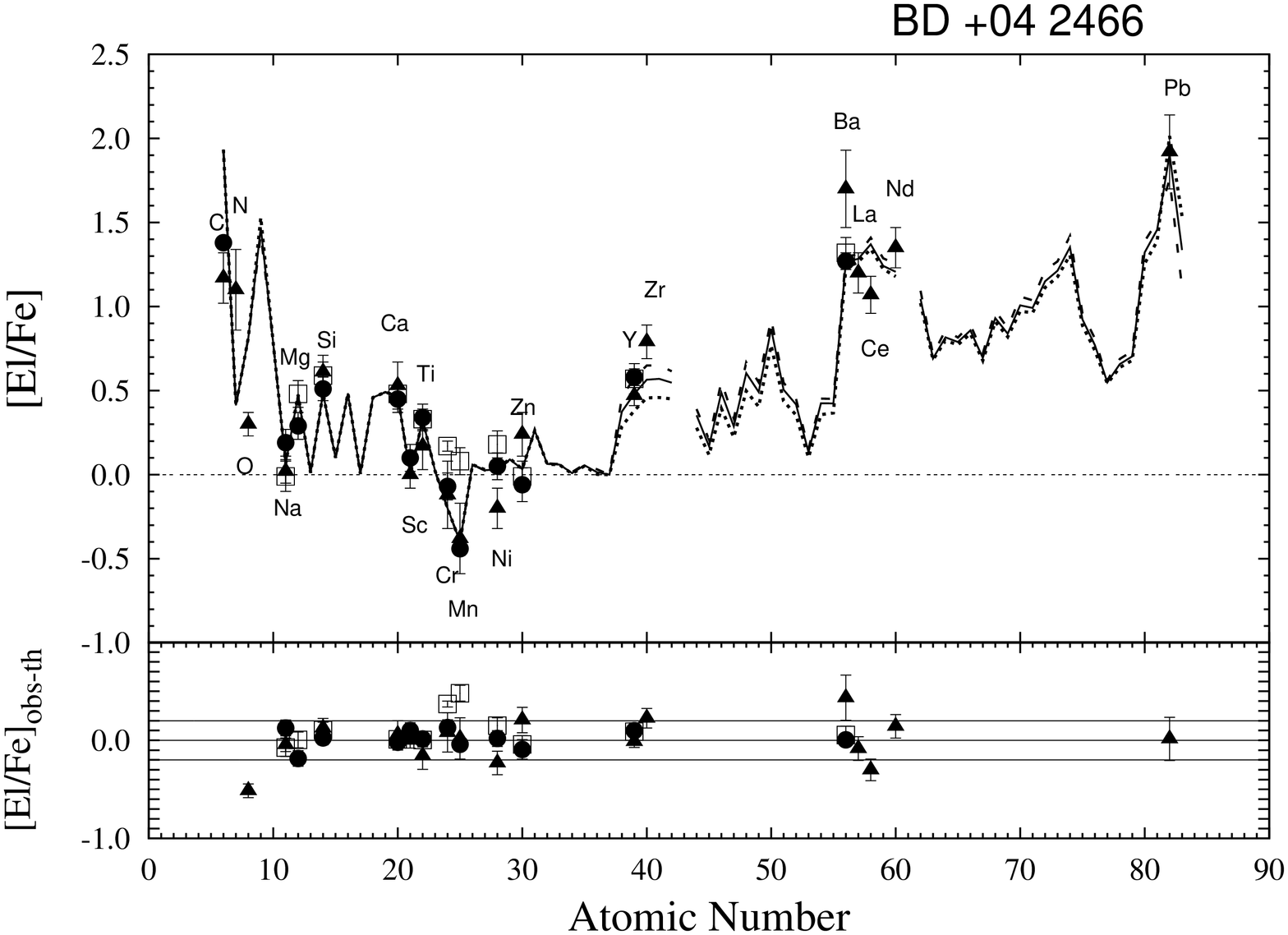}
\vspace{2mm}
\caption{Spectroscopic [El/Fe] abundances of the giant BD+04$^\circ$2466
([Fe/H] $\sim$ $-$2; {\textit T}$_{\rm eff}$ = 5100 K; log $g$ = 1.8)
 compared with AGB models of $M^{\rm AGB}_{\rm ini}$ = 1.3 $M_{\odot}$, 
cases ST/6 (dotted line), ST/9 (solid line), ST/12 (dashed line), and $dil$ = 0.9 dex.
Observations are from \citet{pereira09} (filled triangles), \citet{zhang09} (filled
circles), \citet{ish10} (empty squares).
The two $s$-process indicators are [hs/ls] = 0.6 and [Pb/hs] = 0.7.
The observed [O/Fe] is about 0.5 dex lower than predicted by the models (see text).
An [r/Fe]$^{\rm ini}$ = 0.5 is adopted. }
\label{BD+042466_pd09_ik10_z09_bab10d567m1p5z2m4rp0p5_alf0p5_dil0p9n5}
\end{figure}

\subsubsection{BD+04$^\circ$2466 ($\equiv$ HIP 55852), 
(Fig.~\ref{BD+042466_pd09_ik10_z09_bab10d567m1p5z2m4rp0p5_alf0p5_dil0p9n5})}
\label{BD+pereira}

This giant has been recently analysed by different authors:
\citet{pereira09}, \citet{zhang09}, \citet{ish10}, ([Fe/H] $\sim$ $-$2;
{\textit T}$_{\rm eff}$ = 5100 K; log $g$ = 1.8).
It was firstly identified as a metal-poor star by \citet{bond80}, 
and afterwards it was classified as CH star or "metal-deficient barium 
star" by \citet{LB91}. \citet{jorissen05} confirm the binary scenario,
finding radial velocity variations with a period $P$ = 4593 days.
\\
Large dilutions are needed in order to interpret the mild $s$-process 
observed ([ls/Fe] = 0.6, [hs/Fe] = 1.2, [Pb/Fe] = 1.9), even using AGB models with 
low initial mass.
In Fig.~\ref{BD+042466_pd09_ik10_z09_bab10d567m1p5z2m4rp0p5_alf0p5_dil0p9n5},
we show solutions with $M^{\rm AGB}_{\rm ini}$ = 1.3 $M_{\odot}$, 
cases ST/6, ST/9, ST/12, and $dil$ = 0.9 dex.
Among the hs elements, the observed [Ce/Fe], with 5 detected lines,
is about 0.3 dex lower than the AGB predictions. Nd is the most
reliable with 12 detected lines. 
The [Na/Fe] $\sim$ 0 observed by \citet{ish10}, would be slightly overestimated 
by AGB models with higher initial mass, although the $s$-process elements are equally 
well interpreted: for instance, $M^{\rm AGB}_{\rm ini}$ = 1.5 $M_{\odot}$
predicts [Na/Fe]$_{\rm th}$ = 0.3 (case ST/3; $dil$ = 1.8 dex). 
The observed [C/Fe] is $\sim$ 0.6 dex lower than the AGB prediction.
The low $^{12}$C/$^{13}$C ratio detected (15$^{+5}_{-3}$)
confirms that efficient mixing has taken place.
The predicted [O/Fe] is $\sim$ 0.5 dex higher than observed.
Note that the uncertainty shown for [O/Fe]
 in Fig.~\ref{BD+042466_pd09_ik10_z09_bab10d567m1p5z2m4rp0p5_alf0p5_dil0p9n5}, 
may be underestimated, because C, N, and O measured by \citet{pereira09} 
are interdependent.

\begin{figure}
\includegraphics[angle=-90,width=8cm]{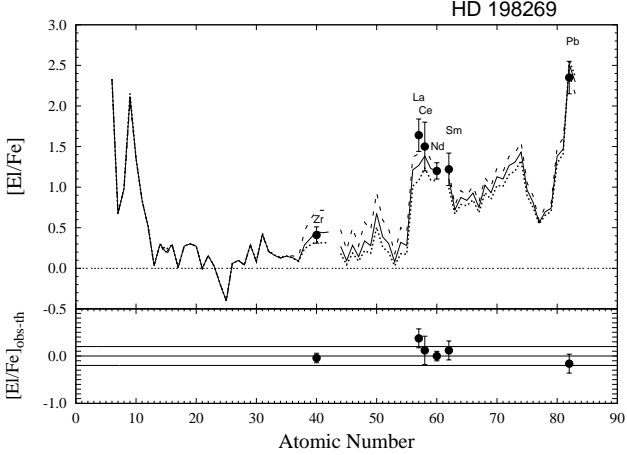}
\vspace{2mm}
\caption{Spectroscopic [El/Fe] abundances of the giant HD 198269 
 ([Fe/H] $\sim$ $-$2.2; {\textit T}$_{\rm eff}$ = 4800 K; log $g$ = 1.3)
compared with AGB models of $M^{\rm AGB}_{\rm ini}$ = 1.5 $M_{\odot}$, 
cases ST/1.5 (dotted line), ST/2 (solid line), ST/3 (dashed line), 
and $dil$ = 1.5 dex.
Observations are from \citet{vaneck03}, who found [hs/ls] = 0.94 and [Pb/hs] = 1.07.
An [r/Fe]$^{\rm ini}$ = 0.5 is adopted. }
\label{HD198269_vaneck03_bab10p1d1p3d2m1p5z1m4rp0p5_dil1p5n20}
\end{figure}

\subsubsection{HD 198269
(Fig.~\ref{HD198269_vaneck03_bab10p1d1p3d2m1p5z1m4rp0p5_dil1p5n20})}
\label{HD198269}

The mild $s$-process enhanced giant HD 198269, with [Fe/H] $\sim$ $-$2.2,
{\textit T}$_{\rm eff}$ = 4800 K and log $g$ = 1.3, has been studied by \citet{vaneck03}.
Only Zr is measured among the ls elements, while the hs peak is better determined
with La, Ce, Nd and Sm ([hs/ls] = 0.94). 
One line is detected for Pb ([Pb/hs] = 1.07).
As said by the authors, the derived abundances may be uncertain owing to 
the presence of molecular lines in the spectra.
The most reliable is Ce, for which more lines are available.
Possible theoretical interpretations are displayed in 
Fig.~\ref{HD198269_vaneck03_bab10p1d1p3d2m1p5z1m4rp0p5_dil1p5n20},
for AGB models with initial mass $M$ = 1.5 $M_{\odot}$,
cases ST/1.5, ST/2, ST/3 and $dil$ = 1.5 dex.
Analogous solutions are obtained for $M^{\rm AGB}_{\rm ini}$ = 2
 $M_{\odot}$ (ST/2 and $dil$ = 1.5 dex).
Models with $M^{\rm AGB}_{\rm ini}$ $<$ 1.4 $M_{\odot}$ are excluded 
because dilutions lower than $\sim$ 0.5 dex would be in contrast with
a giant after the FDU. No light elements are available for this giant.
 Observations of Na or Mg would provide key information on the AGB
initial mass.

\subsubsection{HD 189711
(Fig.~\ref{HD189711_vaneck03_bab10d12d16d20m1p5z2m4rp0p5_dil0p9n20});
V Ari
(Fig.~\ref{VAri_vaneck03_bab10d16d20d24m1p5z1m4rp0p5_dil0p9n20})}
\label{HD189711}

We analyse these two mild $s$-process enhanced stars together,
because they show similar $s$-process distributions
([hs/ls] $\sim$ 0.2; [Pb/hs] $\sim$ $-$0.3).
\\
As said before, the abundances by \citet{vaneck03} are uncertain because 
only few lines are available in a very crowded region of the 
spectrum, veiled by molecular lines. 
The difficulties increase for the coolest stars, as the giants HD 189711
([Fe/H] $\sim$ $-$1.8; {\textit T}$_{\rm eff}$ = 3500 K; log $g$ = 0.5) 
and V Ari ([Fe/H] $\sim$ $-$2.4; {\textit T}$_{\rm eff}$ = 3580 K; 
log $g$ = $-$0.2), for which the abundance determination represent a
real challenge for the spectroscopists, and the errors are of
the order of $\pm$ 0.4 dex.
Caution is suggested for these stars.
An attempt to interpret the [El/Fe] observations is given in 
Figs.~\ref{HD189711_vaneck03_bab10d12d16d20m1p5z2m4rp0p5_dil0p9n20} 
and~\ref{VAri_vaneck03_bab10d16d20d24m1p5z1m4rp0p5_dil0p9n20} (Appendix A).
Very low $s$-process efficiencies are adopted (ST/24 and ST/30),
in order to obtain [Pb/hs] $\sim$ $-$0.42 and $-$0.22, respectively.
For HD 189711, \citet{kipper96} measured [Eu/Fe] = 1.45, which may suggest
a possible initial $r$-process enhancement. 
However, the metallicities detected by \citet{vaneck03} and \citet{kipper96} 
are discrepant ([Fe/H] = $-$1.8 and $-$1.15, respectively), and 
further Eu investigations are desirable in this star.

\begin{figure}
\includegraphics[angle=-90,width=8.5cm]{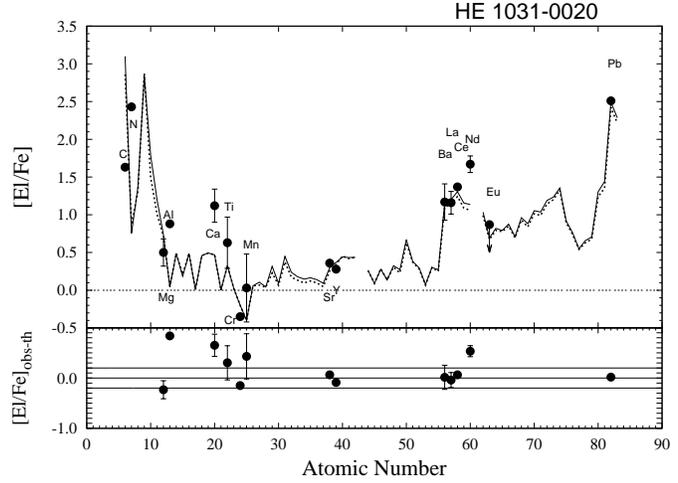}
\vspace{2mm}
\caption{Spectroscopic [El/Fe] abundances of the giant HE 1031--0020
([Fe/H] = $-$2.86; {\textit T}$_{\rm eff}$ = 5080 K; log $g$ = 2.2)
 compared with two AGB models: 
$M^{\rm AGB}_{\rm ini}$ = 1.4 $M_{\odot}$, case ST/5 and $dil$ = 1.2 dex 
(solid line), or $M^{\rm AGB}_{\rm ini}$ = 2 $M_{\odot}$, case ST/5 and
$dil$ = 1.6 dex (dotted line). 
Observations are from \citet{cohen06}, who detected [hs/ls] = 0.94
and [Pb/hs] = 1.37. For discussion about Nd see text.
The differences [El/Fe]$_{\rm obs-th}$ displayed in the lower panel refer to the
$M^{\rm AGB}_{\rm ini}$ = 1.4 $M_{\odot}$ model (solid line).
An [r/Fe]$^{\rm ini}$ = 0.5 is adopted. }
\label{HE1031-0020_cohen06_bab10d3m1p5d3m2z2m5rp0p5alf0p5_diffdiln10n26}
\end{figure}

\subsubsection{HE 1031--0020
(Fig.~\ref{HE1031-0020_cohen06_bab10d3m1p5d3m2z2m5rp0p5alf0p5_diffdiln10n26})}
\label{HE1031}%

The giant HE 1031--0020 is one of the coolest stars analysed by \citet{cohen06}, 
([Fe/H] = $-$2.86; {\textit T}$_{\rm eff}$ = 5080 K; log $g$ = 2.2). 
It is affected by molecular absorption 
from CH and CN bands, and the metallicity is probably underestimated.
The Na I D line is too strong and no reliable abundance can be obtained. 
A large uncertainty is reported by \citet{cohen06} for Ti ($\pm$0.4 dex) 
with 15 detected lines.

The detection of Mg is, in general, accurate, with an uncertainty of about 
$\pm$ 0.2 dex.
Caution is suggested by \citet{cohen06} for Sr and Ba detections.
This giant shows mild enhancement in ls and hs: [ls/Fe] = 0.35 and 
[hs/Fe] = 1.29.
Instead, a very high [Pb/Fe] is observed ([Pb/Fe] = 2.66; [Pb/hs] = 1.37).
However, Pb is uncertain in this star, because the single line detected 
is strongly blended by CH features. 
Possible solutions with two AGB initial masses are shown in 
Fig.~\ref{HE1031-0020_cohen06_bab10d3m1p5d3m2z2m5rp0p5alf0p5_diffdiln10n26}:
for $M^{\rm AGB}_{\rm ini}$ = 1.4 $M_{\odot}$, $dil$ =
 1.2 dex is required (solid line), while for
$M^{\rm AGB}_{\rm ini}$ = 2 $M_{\odot}$ a $dil$ = 1.6 dex is adopted 
(dotted line). A case ST/5 is needed for both models.
AGBs with lower initial mass may equally interpret the spectroscopic 
data: $M^{\rm AGB}_{\rm ini}$ = 1.3 $M_{\odot}$, case 
ST/15 and $dil$ = 0.8 dex, in agreement with a giant after the FDU.
Among the hs elements, a difference of about 0.5 dex is observed 
between La and Nd (four lines detected for both elements).
This discrepancy contrasts with AGB predictions. In 
Fig.~\ref{HE1031-0020_cohen06_bab10d3m1p5d3m2z2m5rp0p5alf0p5_diffdiln10n26} we consider
 La as more reliable among the hs elements. With a proper choice of 
the $^{13}$C-pocket and dilution factor, we may find theoretical solutions for 
the observed [Nd/Fe], but in this case the observed [Mg/Fe] and [ls/Fe] would be overestimated.
AGB models of $M^{\rm AGB}_{\rm ini}$ = 1.5 $M_{\odot}$ would predict a
larger [Mg/Fe] than observed.
No error bars are provided by the authors for C and N.

\begin{figure}
\includegraphics[angle=-90,width=8.5cm]{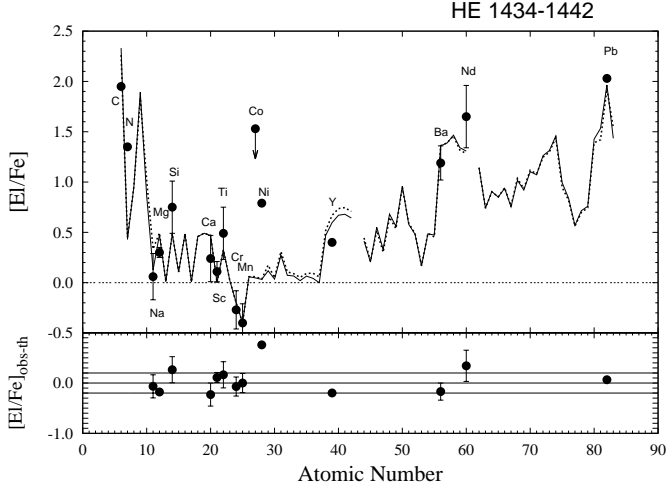}
\vspace{2mm}
\caption{Spectroscopic [El/Fe] abundances of the subgiant HE 1434--1442 
([Fe/H] = $-$2.39; {\textit T}$_{\rm eff}$ = 5420 K; log $g$ = 3.15, having suffered
the FDU) compared with two AGB models:
 $M^{\rm AGB}_{\rm ini}$ = 1.3 $M_{\odot}$, case ST/15, 
$dil$ = 0.8 dex (solid line), or $M^{\rm AGB}_{\rm ini}$ = 1.4 $M_{\odot}$,
case ST/12, $dil$ = 1.2 dex (dotted line).
Observations are from \citet{cohen06}.
 This solution is indicative owing to the 
limited number of $s$-elements observed ([hs/ls] $\sim$ 0.9; [Pb/hs] $\sim$
0.8, see text).
The differences [El/Fe]$_{\rm obs-th}$ displayed in the lower panel refer to the
$M^{\rm AGB}_{\rm ini}$ = 1.3 $M_{\odot}$ model.
An [r/Fe]$^{\rm ini}$ = 0.5 is adopted. }
\label{HE1434-1442_cohen06_bab10d10m1p5z1m4d8m1p5z1m4rp0p5alf0p5_diffdiln5n8}
\end{figure}

\subsubsection{HE 1434--1442
(Fig.~\ref{HE1434-1442_cohen06_bab10d10m1p5z1m4d8m1p5z1m4rp0p5alf0p5_diffdiln5n8})}
\label{HE1434}%

The giant HE 1434--1442 has [Fe/H] = $-$2.39, {\textit T}$_{\rm eff}$ = 5420 K
and log $g$ = 3.15 \citep{cohen06}.
As other stars studied by \citet{cohen06}, it is affected by molecular 
absorption from CH and CN bands.
A limited number of $s$ elements is observed:
Ba (three lines), Y and Nd (two lines), and Pb (one line).
Fig.~\ref{HE1434-1442_cohen06_bab10d10m1p5z1m4d8m1p5z1m4rp0p5alf0p5_diffdiln5n8} 
shows possible theoretical interpretations with AGB models of 
low initial mass ($M^{\rm AGB}_{\rm ini}$ = 1.3 and
1.4 $M_{\odot}$, solid and dotted line, respectively), in agreement
with [hs/ls] = 0.89 and [Pb/hs] = 0.77. 
These solutions interpret the observed [Na/Fe] $\sim$ 0.
This star lies on the subgiant phase after the FDU, according to the 
large dilution adopted ($dil$ = 0.8 -- 1.2 dex).
AGB models of $M^{\rm AGB}_{\rm ini}$ = 1.5 $M_{\odot}$ (case $\sim$ ST/6)
are excluded because
they predict a high Na abundance ([Na/Fe]$_{\rm th}$ = 0.6). 
A lower Na value is obtained by $M^{\rm AGB}_{\rm ini}$ = 2 $M_{\odot}$,
 case ST/9 and $dil$ = 1.6 dex
([Na/Fe]$_{\rm th}$ = 0.4).
While the neutron capture elements may be interpreted with 
all AGB initial masses in the range
 1.3 $M_{\odot}$ $\leq$ $M^{\rm AGB}_{\rm ini}$ $\leq$ 2 $M_{\odot}$, 
Na agrees with $M^{\rm AGB}_{\rm ini}$ $\leq$ 1.4 $M_{\odot}$ models.
 However, Na is explicitly mentioned by the authors as very 
uncertain for this cool star.
C and N are not very reliable, because no error bars are provided by 
\citet{cohen06}.

\section{Stars at [F{\scriptsize e}/H] $\sim$ $-$1.2}\label{CH}

In this Section, we consider three stars with [Fe/H] $\ga$ $-$1.4: 
HD 26 ([Fe/H] = $-$1.25,$-$1.02; \citealt{vaneck03}, \citealt{masseron10}), 
HD 206983 ([Fe/H] = $-$0.99, $-$1.43; \citealt{masseron10}, \citealt{JP01}), 
HE 1152--0355 ([Fe/H] = $-$1.27; \citealt{goswami06});
for two other stars, CS 29503--010 ([Fe/H] = $-$1.06; \citealt{aoki07})
and HE 0507--1653 ([Fe/H] = $-$1.38, $-$1.42; \citealt{aoki07,schuler08}), 
only Ba is detected among the $s$-process elements.
All these stars show enhanced [C/Fe] and $s$-process elements. 
Their metallicity is close to disc stars 
and their atmospheric parameters are far from the TP-AGB phase.
These stars may be considered as a link between CEMP-$s$ 
and $s$-rich giants or dwarfs of disc metallicity (e.g., barium stars\footnote{Barium stars 
are giants (and dwarfs) showing Ba and Sr overabundances by the presence of singly ionised barium, 
Ba II, at $\lambda$ = 4554 ${\rm \AA}$ and Sr II ($\lambda$ = 4077 ${\rm \AA}$, $\lambda$ 
= 4215 ${\rm \AA}$), as well as CH G band and CN bands also enhanced \citep{allen06,smij07}. 
Values of [Pb/Fe] $>$ 1 dex have been observed for the first time in barium stars
by \citet{allen06}, explained with efficient $^{13}$C-pockets in the AGB companion.
Theoretical interpretations of barium stars have been discussed by 
\citet{husti08FSIII,husti09pasa}.}, CH stars, MS/S stars with no Tc,
symbiotic stars), because the carbon and $s$-enrichment on their surface is commonly 
ascribed to a binary scenario with mass transfer by stellar winds. 
Some of these stars are not CH stars,  
and, starting from \citet{LB91}, they have been classified 
as "metal-poor barium stars".

\begin{figure}
\includegraphics[angle=-90,width=8cm]{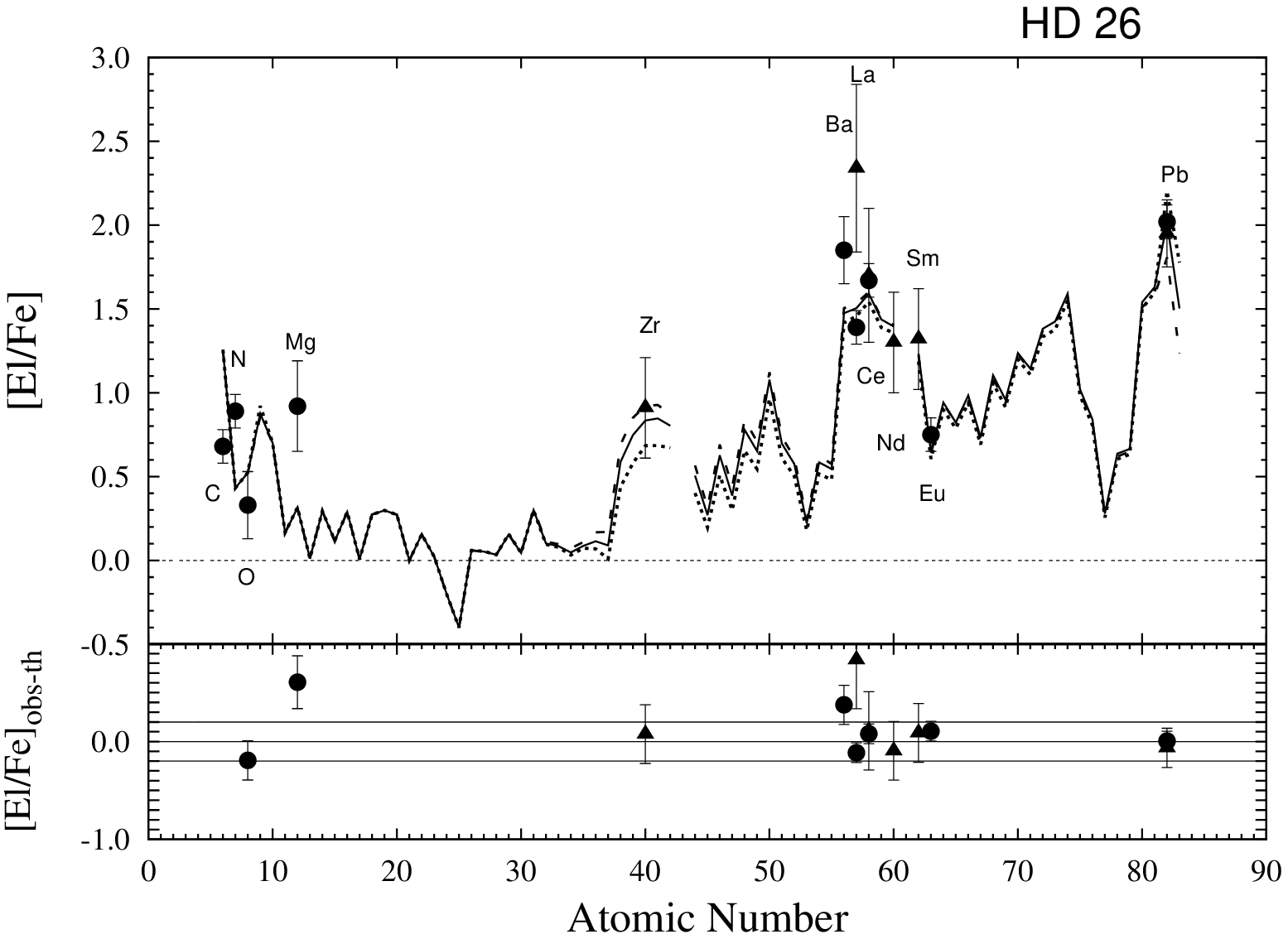}
\vspace{2mm}
\caption{Spectroscopic [El/Fe] abundances of the giant HD 26 
 ([Fe/H] = $-$1.25, {\textit T}$_{\rm eff}$ = 5170 K, log $g$ = 2.2, \citealt{vaneck03};
[Fe/H] = $-$1.02, {\textit T}$_{\rm eff}$ = 4900 K, log $g$ = 1.5, \citet{masseron10})
compared with AGB models of initial mass $M$ = 1.5 $M_{\odot}$, 
cases ST/1.5 (dotted line), ST/2 (solid line), ST/2.5 (dashed line), 
and $dil$ = 1 dex. Observations are from
\citet{vaneck03} (filled triangles) and \citet{masseron10} (filled circles).
This giant  shows [hs/ls] = 0.5 and [Pb/hs] = 0.7.
An [r/Fe]$^{\rm ini}$ = 0.0 is adopted. }
\label{HD26_vaneck03_masseron10_bab10p1d1p3d1p6m1p5z1m3nr_dil1p0n20}
\end{figure}

\subsection{HD 26 
(Fig.~\ref{HD26_vaneck03_masseron10_bab10p1d1p3d1p6m1p5z1m3nr_dil1p0n20})}
\label{HD26}

\citet{vanture92b,vanture92c} studied for the first time HD 26,
deriving spectroscopic abundances for C, N, O, and heavier elements. 
They classified this giant as a CH star.
Later on, \citet{vaneck03} and \citet{masseron10} reported spectroscopic
observations obtained with high-resolution spectra. 
We discuss here only these most recent values.
A model constraint is given by the occurrence of the FDU
({\textit T}$_{\rm eff}$ = 5170 K and log $g$ = 2.2 by \citealt{vaneck03};
{\textit T}$_{\rm eff}$ = 4900 K and log $g$ = 1.5 by \citealt{masseron10}),
which requires a dilution of the order of 1 dex.
The abundances by \citet{vaneck03} are uncertain because only few 
lines veiled by molecular bands are available; they refer to 
Ce as the most reliable element among the hs-peak.
Besides five neutron capture elements (Zr, La, Ce, Nd, and Sm), 
they detected the Pb I line at 4057.812 {\rm \AA}, clearly resolved thanks 
to the high-resolution $R$ = $\lambda/\Delta \lambda$ = 135\,000.  
\citet{masseron10} provided new observations for C, N, O, Mg,
and Eu, as well as updated results for Ba, La, Ce, and Pb
([hs/ls] = 0.5; [Pb/hs] = 0.7).
In Fig.~\ref{HD26_vaneck03_masseron10_bab10p1d1p3d1p6m1p5z1m3nr_dil1p0n20} 
we show theoretical interpretations with AGB models 
of $M^{\rm AGB}_{\rm ini}$ = 1.5 $M_{\odot}$, cases ST/1.5, ST/2,
ST/2.5, and $dil$ = 1 dex.
Similar solutions are obtained with higher AGB initial masses 
(e.g., $M^{\rm AGB}_{\rm ini}$ = 2 $M_{\odot}$).
No initial $r$-process enhancement is adopted for this star.
The observed [Mg/Fe] is higher than the 
 AGB predictions.
The observations by \citet{masseron10} shown in 
Fig.~\ref{HD26_vaneck03_masseron10_bab10p1d1p3d1p6m1p5z1m3nr_dil1p0n20} will be 
discussed by the authors in 
Masseron et al., in preparation.

\begin{figure}
\includegraphics[angle=-90,width=8cm]{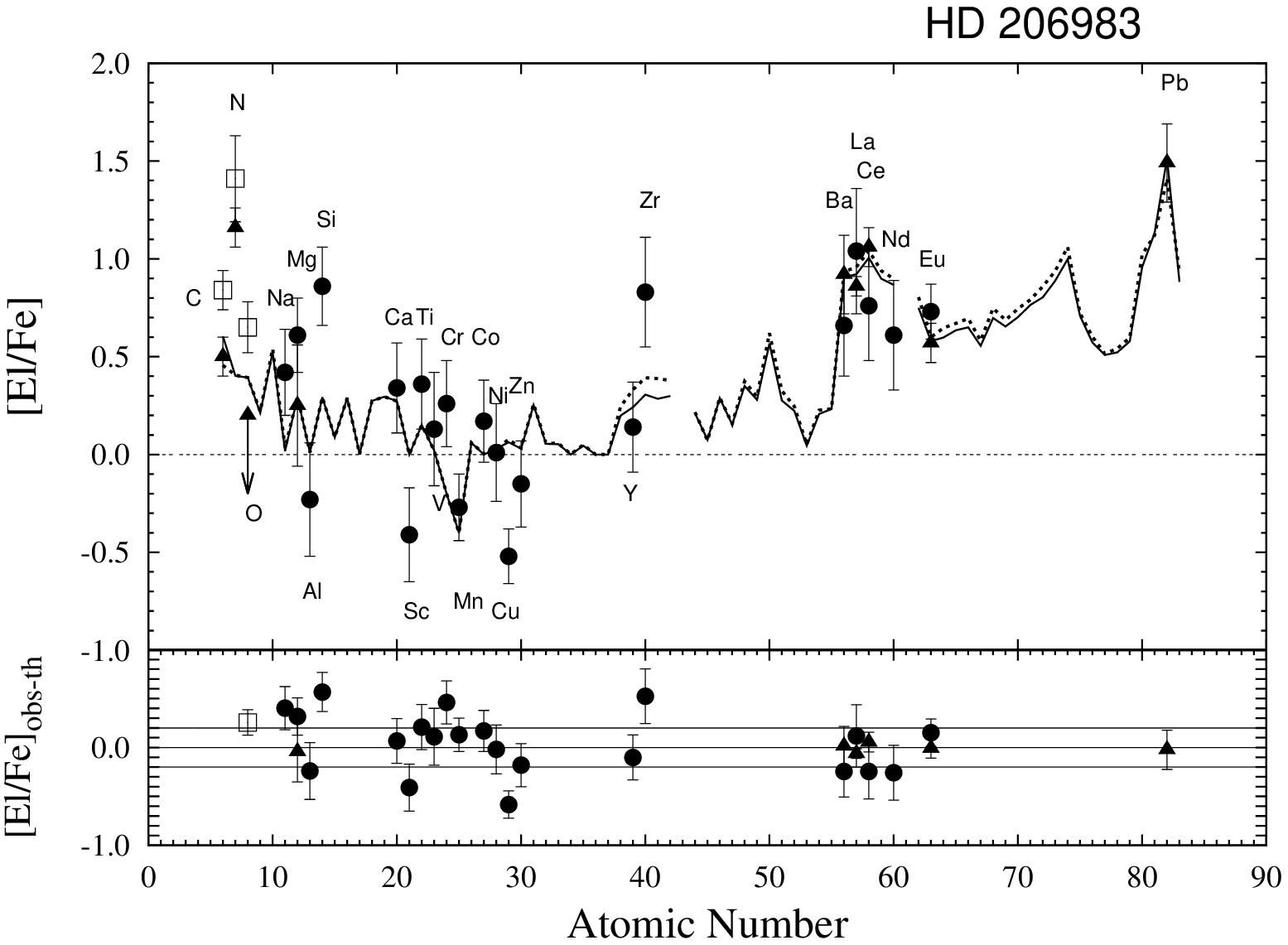}
\vspace{2mm}
\caption{Spectroscopic [El/Fe] abundances of the giant HD 206983 
([Fe/H] = $-$0.99; {\textit T}$_{\rm eff}$ $\sim$ 4200 K; log $g$ = 0.6, \citealt{masseron10})
compared with AGB models of $M^{\rm AGB}_{\rm ini}$ = 1.3 or 1.5 $M_{\odot}$, 
case ST, $dil$ = 0.7 or 1.6 dex (solid or dotted lines, respectively).
Observations are from \citet{masseron10} (filled triangles), \citet{drakepereira08} 
(empty squares), \citet{JP01} (filled circles).
This giant shows [hs/ls] = 0.38 and [Pb/hs] = 0.67 \citep{masseron10}.
See text for discussion about Cu.
An [r/Fe]$^{\rm ini}$ = 0.5 is adopted. }
\label{HD206983_masseron10_bab10p1p3m1p5rp0p5_diffdiln5n20}
\end{figure}

\subsection{HD 206983 
(Fig.~\ref{HD206983_masseron10_bab10p1p3m1p5rp0p5_diffdiln5n20})}
\label{HD206983}

This giant ({\textit T}$_{\rm eff}$ $\sim$ 4200 K, log $g$ = 0.6)
has been analysed by \citet{masseron10}, \citet{drakepereira08} 
(who studied C, N, O), and \citet{JP01}.
\\
In Fig.~\ref{HD206983_masseron10_bab10p1p3m1p5rp0p5_diffdiln5n20}, we show
theoretical interpretations using AGB models of initial masses 
$M$ = 1.3 and 1.5 $M_{\odot}$, case ST, $dil$ = 0.7 and 1.6 dex, respectively.
We considered Y (seven detected lines) more reliable than Zr (four lines available),
 \citep{JP01}.
Discrepant [C/Fe] and [O/Fe] ratios are found by \citet{drakepereira08} 
and \citet{masseron10}: note that \citet{drakepereira08} adopted 
the stellar parameters obtained by \citet{JP01}, which provided
a metallicity 0.4 dex lower than [Fe/H] = $-$0.99 by \citet{masseron10}. 
In Fig.~\ref{HD206983_masseron10_bab10p1p3m1p5rp0p5_diffdiln5n20}, we 
display solutions with AGB models of [Fe/H]$_{\rm th}$ = $-$1,
which agree with [C/Fe] detected by \citet{masseron10}.
Instead, the observed [N/Fe] is about 0.7 dex higher than AGB predictions.
Note that, by decreasing metallicity, a larger primary amount of
[C/Fe] is predicted. 
C and N are also affected by uncertainty in 3D atmosphere models 
that may decrease the observations \citep{asplund09}.
[Na/Fe]$_{\rm obs}$ (two lines) is 0.4 dex higher than AGB models, but no NLTE
corrections are considered by \citet{JP01}.
Note that \citet{JP01} detected [Cu/Fe] = $-$0.5, by assuming 
[Fe/H] = $-$1.4. At this metallicity, spectroscopic observations
of unevolved stars show on average similar values.
Indeed, as discussed in Section~\ref{intro}, for halo stars ([Fe/H] $\la$ $-$1.7)
[Cu/Fe] shows a negative constant ratio close to $-$0.7 dex.
By increasing the metallicity, [Cu/Fe] increases starting
from [Fe/H] $\sim$ $-$1.5 and reaching solar values at [Fe/H] $\sim$ $-$0.5.

\begin{figure}
\includegraphics[angle=-90,width=8.5cm]{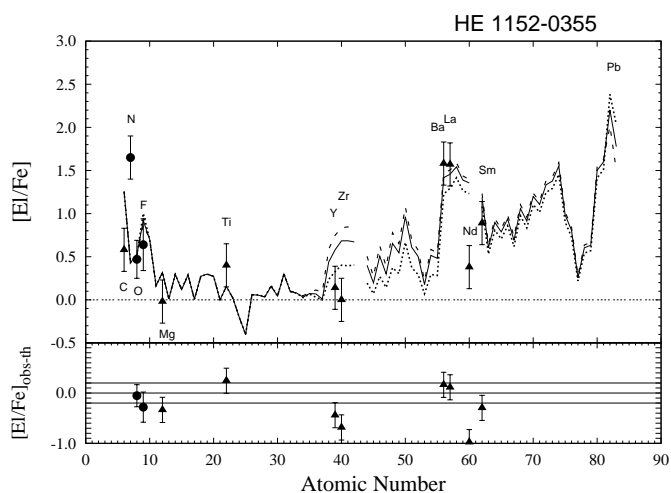}
\vspace{2mm}
\caption{Spectroscopic [El/Fe] abundances of the giant HE 1152--0355 
([Fe/H] = $-$1.27; {\textit T}$_{\rm eff}$ = 4000 K; log $g$ = 1.0)
compared with a AGB stellar models of 1.5
$M_{\odot}$, cases ST (dotted line), ST/1.5 (solid line), ST/2
(dashed line), and $dil$ = 1.0 dex.
Observations are from \citet{goswami06} (filled triangles) and 
\citet{lucatello11astroph} (filled circles).  
The observed [Y/Fe], [Zr/Fe] and [Nd/Fe] ratios are overestimated by 
AGB models (see text). 
An [r/Fe]$^{\rm ini}$ = 0.0 is adopted.}
\label{HE1152-0355_goswami06_bab10p1p5p1d1p3m1p5z1m3nr_dil1p0n20}
\end{figure}

\subsection{HE 1152--0355 
(Fig.~\ref{HE1152-0355_goswami06_bab10p1p5p1d1p3m1p5z1m3nr_dil1p0n20})}
\label{HE1152}

The giant HE 1152--0355 is a very cool star ([Fe/H] = $-$1.27; 
$T_{\rm eff}$ = 4000 K; log $g$ = 1.0) 
affected by strong molecular contaminations, probably inducing the spread
observed among the hs elements ([La/Nd] $\sim$ 1 dex).  
Moreover, \citet{goswami06} could not estimate Na for the highly blended
lines present in the spectra. 
A possible interpretation is shown in 
Fig.~\ref{HE1152-0355_goswami06_bab10p1p5p1d1p3m1p5z1m3nr_dil1p0n20},
for AGB models of $M^{\rm AGB}_{\rm ini}$ = 1.5 $M_{\odot}$ (case ST,
ST/1.5, ST/2 and $dil$ = 1.0 dex). Note the low [Nd/Fe] with respect to 
[Ba,La/Fe]: we have considered La as the most reliable among the hs elements. 
These models predict [Pb/Fe]$_{\rm th}$ $\sim$ 2. 
By considering lower [hs/Fe]$_{\rm th}$, we may interpret the observed
[ls/Fe] under the hypothesis of higher dilutions.
However, the large uncertainties of the observations do not permit
to provide accurate theoretical discussions.
\citet{lucatello11astroph} determined [F/Fe] abundance for this star,
in agreement with theoretical AGB predictions.

\subsection{CS 29503--010}
\label{CS010}

This main-sequence star ({\textit T}$_{\rm eff}$ $\sim$ 6500 K and log $g$ = 4.5;
[Fe/H] = $-$1.06) has only Ba detected among the $s$-elements, with [Ba/Fe] = 1.5.
AGB models with initial mass in the range 1.3 $\leq$ $M/M_\odot$ 
$\leq$ 2 may equally interpret the observations, by adopting different
$^{13}$C-pockets and dilutions. 
However, at present, this star seems not to belong to binary system
\citep{tsangarides05}, and further spectroscopic investigations
are desirable.

\subsection{HE 0507--1653}
\label{HE0507}

Only barium among the $s$-process elements is measured for this giant
with {\textit T}$_{\rm eff}$ $\sim$ 5000 K and log $g$ = 2.4
([Fe/H] = $-$1.38, $-$1.42; \citealt{aoki07,schuler08}). 
\citet{schuler08} detected C and N, in agreement with
previous results by \citet{aoki07}.  
Due to the limited number of spectroscopic observations,
a range higher than 1 dex may be predicted for the ls peak and 
Pb.

\section{Effect of $^{22}$N{\scriptsize e}, $^{12}$C and $^{16}$O at low metallicity} \label{poisons}

The aim of this Section is to provide a more detailed discussion about the impact of the
$^{22}$Ne(n, $\gamma$)$^{23}$Ne 
reaction, both as neutron poison and as neutron seed, and about the effect of the 
$^{12}$C(n, $\gamma$)$^{13}$C and $^{16}$O(n, $\gamma$)$^{17}$O reactions as neutron poisons. 
Additional tests with respect to Paper I (see Section~4 and Appendix~C) are presented, by changing the 
AGB initial mass and the $^{13}$C-pocket. 

The $s$-process distribution observed in several stars can be interpreted by AGB models 
with different initial masses and a proper choice of the $^{13}$C-pocket.
In particular, at [Fe/H] $\sim$ $-$2.5, comparable [hs/ls] and [Pb/hs] are obtained with $M$ = 1.3 
$M_{\odot}$ model and $\sim$ST/12 or $M$ = 1.5 $M_{\odot}$ and case $\sim$ST/3.
This is mainly due to the large amount of primary $^{22}$Ne at low metallicities,
which acts both as neutron poison and as neutron seed. 
In addition to $^{22}$Ne, neutron poisons by primary $^{12}$C 
and $^{16}$O also affect the $s$-process abundances.
This result highlights the importance of a study focused on the production of the light elements, 
the reactions involved and their uncertainties, as well as their effects on the $s$-process path.

\subsection{$^{22}$Ne}

\begin{figure}
\includegraphics[angle=-90,width=8.5cm]{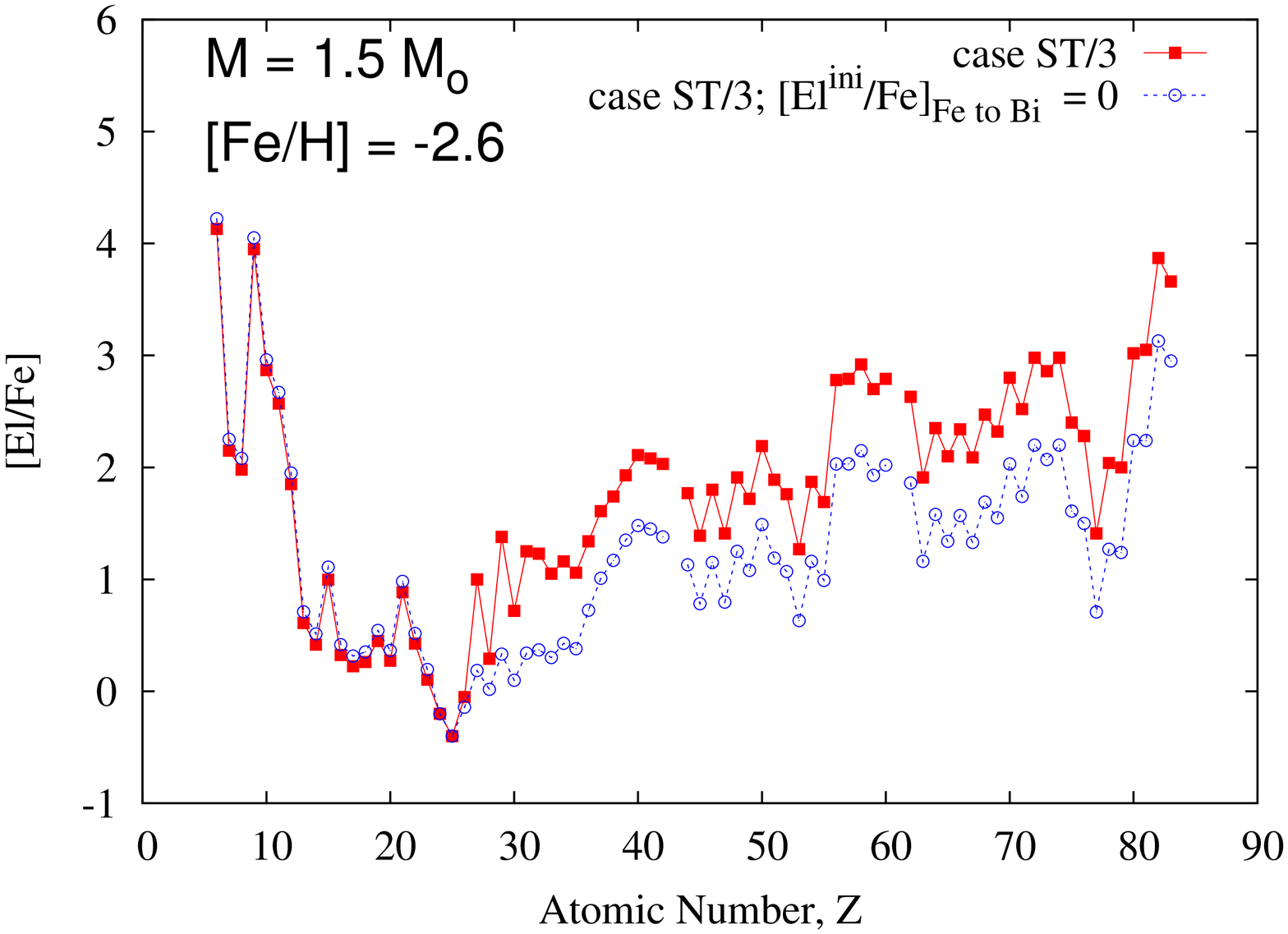}
\includegraphics[angle=-90,width=8.5cm]{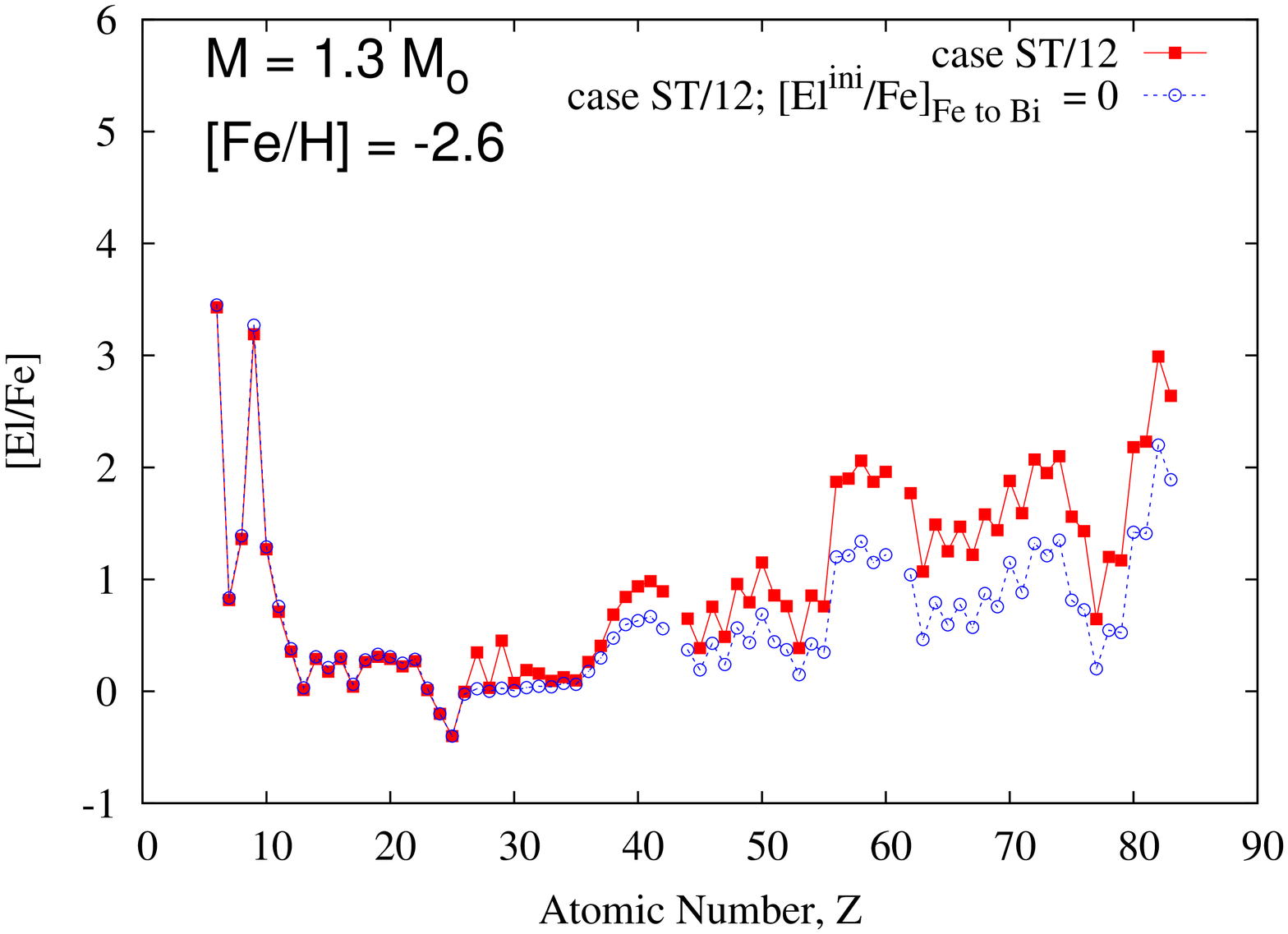}
\vspace{2mm}
\caption{Left panel: we compare an AGB model of $M$ = 1.5 $M_{\odot}$ at [Fe/H] = $-$2.6 (red solid line)
and case ST/3, with a test case in which we set to zero the initial 
abundances of all isotopes from $^{56}$Fe to $^{209}$Bi (blue dotted line).
{\it (See the electronic paper for a colour version of the figures of this Section.}
Right panel: the same as left panel but for an AGB model of $M$ = 1.3 $M_{\odot}$ and case ST/12.}
\label{52}
\end{figure}

\begin{figure}
\includegraphics[angle=-90,width=8.5cm]{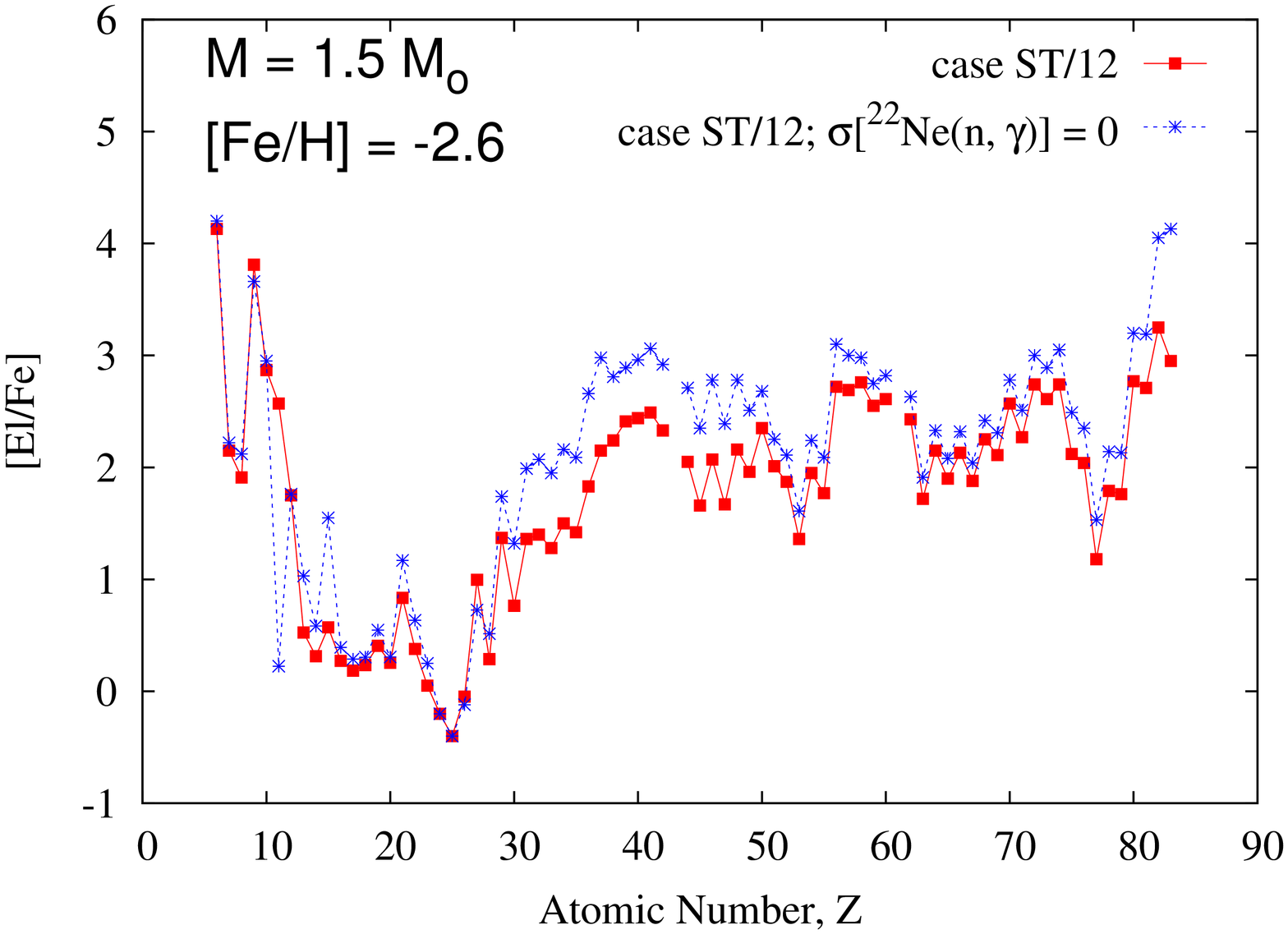}
\includegraphics[angle=-90,width=8.5cm]{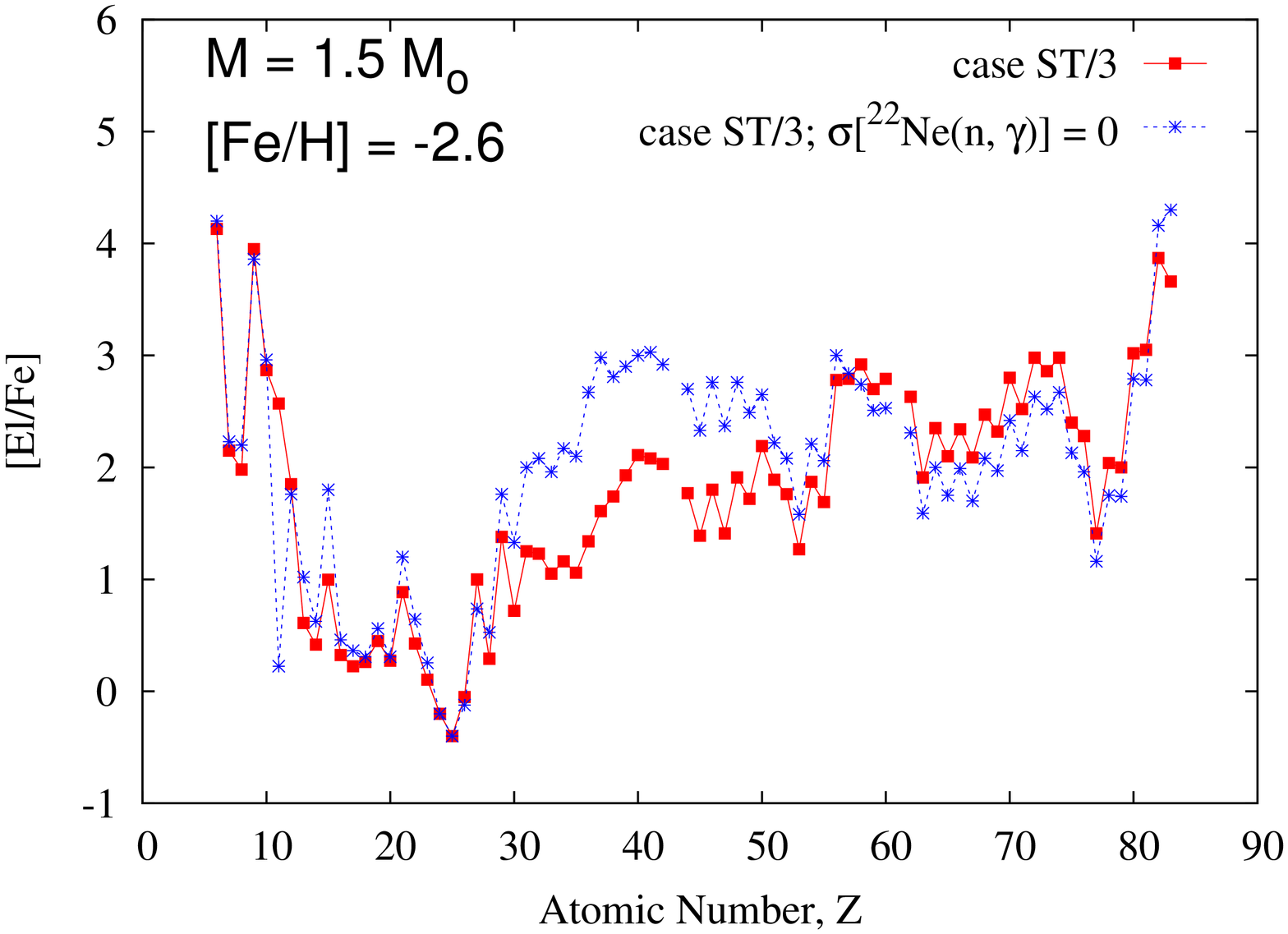}
\vspace{2mm}
\caption{AGB models of $M$ = 1.5 $M_{\odot}$ (20$^{\rm th}$ TDU) at 
[Fe/H] = $-$2.6 (red solid lines), compared with a test case in which the  $^{22}$Ne(n, $\gamma$)$^{23}$Ne 
channel is set to zero (blue dotted lines). Two $^{13}$C-pockets are considered: ST/12 (left panel) and ST/3
(right panel).}
\label{53}
\end{figure}

At low metallicities, $^{22}$Ne has two effects: it acts both as neutron seed and as neutron poison. 
The first effect leads to a production of $^{56}$Fe by neutron captures on $^{22}$Ne; then, $^{56}$Fe 
becomes seed for the nucleosynthesis of the $s$-elements.
Therefore, both iron seeds and the number of neutrons released change.
\\
Concerning the effect of $^{22}$Ne as neutron seed, we report in Fig.~\ref{52}, left panel, a comparison of 
the envelope abundances for an AGB model of $M$ = 1.5 $M_{\odot}$ at [Fe/H] = $-$2.6 and a case ST/3 
(red solid line), with a test case in which we set to zero the initial abundances of all isotopes from 
$^{56}$Fe to $^{209}$Bi (blue dotted line).
Owing to the abundant primary $^{22}$Ne, the $^{22}$Ne(n, $\gamma$)$^{23}$Ne reaction drives a neutron 
chain that extends up to $^{56}$Fe and beyond, producing a large amount of $s$-elements.
A similar effect is obtained for an AGB model of initial mass $M$ = 1.3 $M_{\odot}$ and case ST/12
(Fig.~\ref{52}, right panel). 
\\
 Concerning the effect of neutron poison, we made additional tests by setting to zero 
the $^{22}$Ne(n, $\gamma$)$^{23}$Ne reaction.
\\
We report in Fig.~\ref{53}, the results of AGB models of $M$ = 1.5 $M_{\odot}$ (20 TDUs) at 
[Fe/H] = $-$2.6 (red solid lines), compared with a test case in which the  $^{22}$Ne(n, $\gamma$)$^{23}$Ne 
channel is set to zero (blue dotted lines). Two $^{13}$C-pockets are considered: ST/12 (left panel) and ST/3
(right panel). 
By excluding the $^{22}$Ne(n, $\gamma$)$^{23}$Ne channel, the production of the three $s$-process peaks 
in general increases.
When a case ST/3 is adopted, major effects are observed for the ls elements, while the hs peak is almost 
unchanged; for case ST/12, the effect on Pb is large. 
\\
 For AGB models with $M$ = 1.4 $M_{\odot}$ (10 TDUs) we obtain results similar to 
$M$ = 1.5 $M_{\odot}$ 
models.
\\
For AGB models of initial mass $M$ = 1.3 $M_{\odot}$, the effect of the $^{22}$Ne(n, $\gamma$)$^{23}$Ne 
reaction is marginal owing to the limited number of thermal pulses (5 TDUs).
\\
 For AGB models with $M$ $\sim$ 1.35 $M_{\odot}$ at the 6$^{\rm th}$ and 7$^{\rm th}$ TDU, 
we obtain results similar to $M$ = 1.3 $M_{\odot}$ models.

\subsection{$^{12}$C and $^{16}$O}

\begin{figure}
\includegraphics[angle=-90,width=8.5cm]{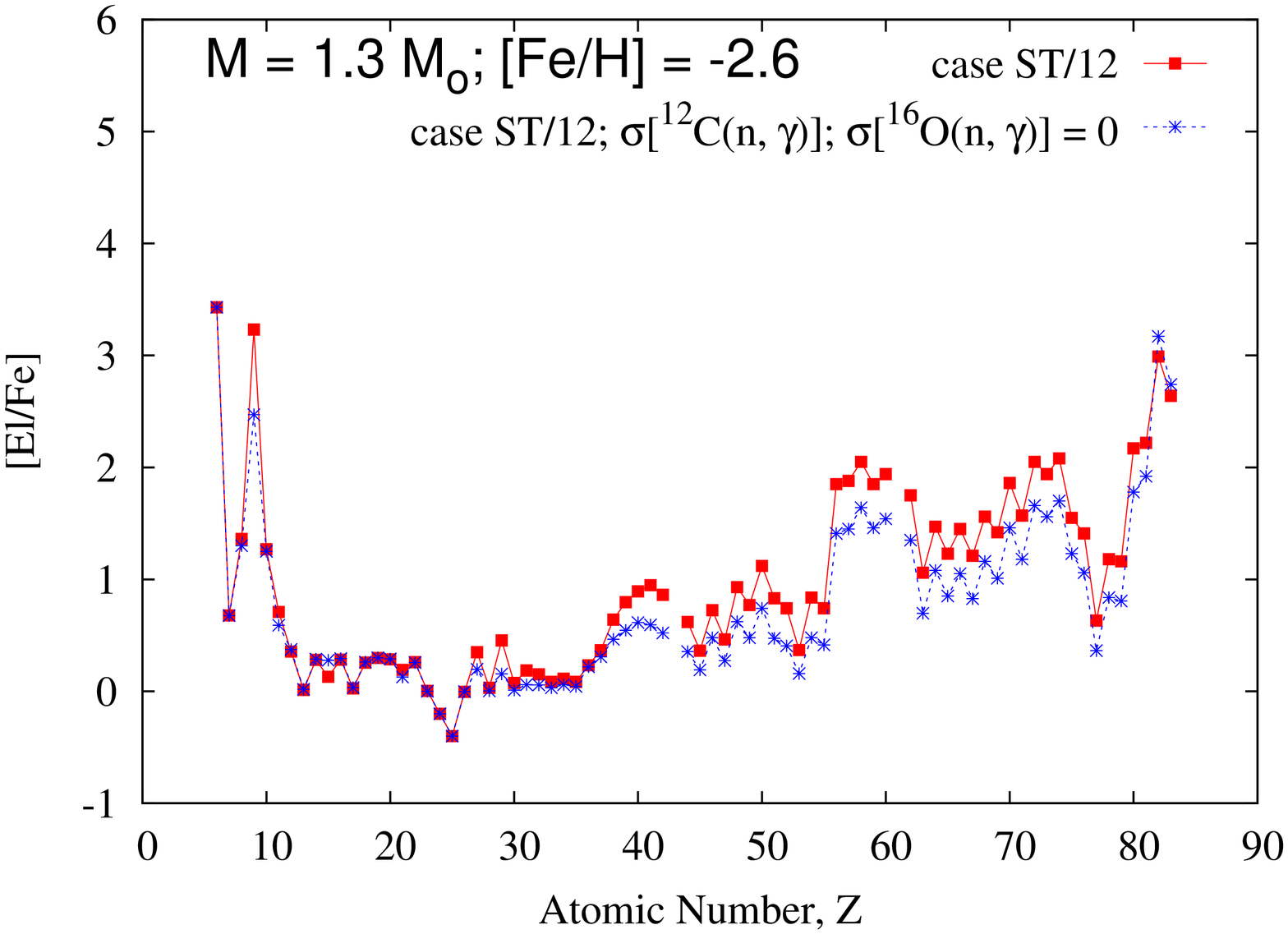}
\includegraphics[angle=-90,width=8.5cm]{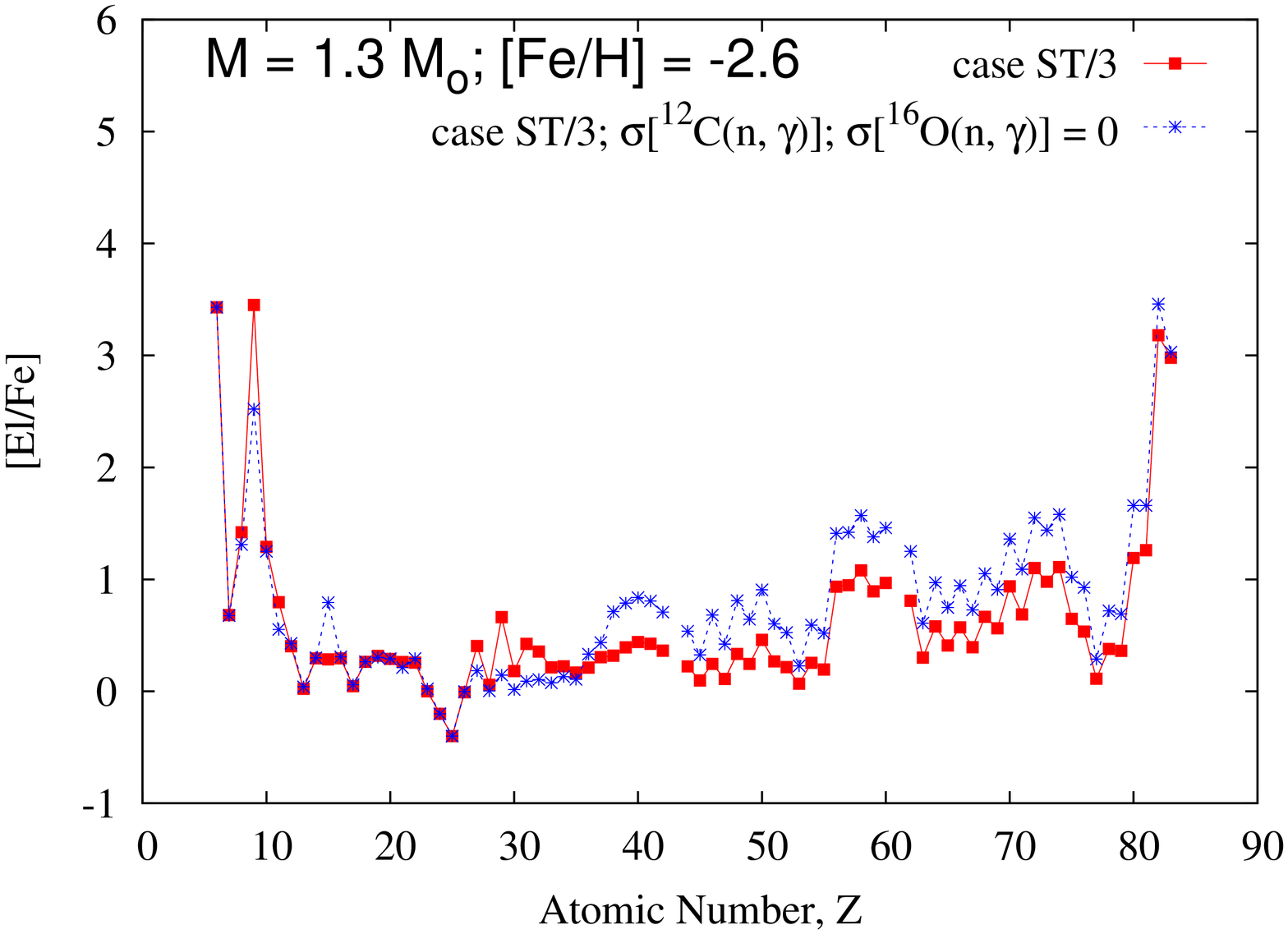}
\vspace{2mm}
\caption{AGB model of $M$ = 1.3 $M_{\odot}$ at [Fe/H] = $-$2.6 (red solid lines), 
compared with a test case in which both the $^{12}$C(n, $\gamma$)$^{13}$C and $^{16}$O(n, 
$\gamma$)$^{17}$O 
channels have been set to zero (blue dotted lines). 
Two $^{13}$C-pockets are considered: ST/12 (left panel) and ST/3 (right panel).}
\label{54}
\end{figure}

In Fig.~\ref{54}, we compare an AGB model of $M$ = 1.3 $M_{\odot}$ at [Fe/H] = $-$2.6 (red solid lines), 
with a test case in which both $^{12}$C(n, $\gamma$)$^{13}$C and $^{16}$O(n, $\gamma$)$^{17}$O channels 
have been set to zero (blue dotted lines). Two $^{13}$C-pockets are considered: ST/12 (left panel) and ST/3 
(right panel).
The variation of the ls and hs elements is evident because of 
the low [ls/Fe] and [hs/Fe] abundances. 
\\
For AGB models with $M$ $\sim$ 1.35 $M_{\odot}$ at the 6$^{\rm th}$ and 7$^{\rm th}$ TDU, 
we obtain results similar to $M$ = 1.3 $M_{\odot}$ models. 
\\
Instead, by setting to zero both the $^{12}$C(n, $\gamma$)$^{13}$C and $^{16}$O(n, 
$\gamma$)$^{17}$O channels in AGB models of $M$ = 1.5 $M_{\odot}$, variations lower than $\sim$ 0.2 dex
are produced.  
Note that the neutron poison effect of the very abundant primary $^{12}$C is almost cancelled by 
$^{13}$C($\alpha$, n)$^{16}$O recycling.
\\
 For AGB models with $M$ = 1.4 $M_{\odot}$ (10$^{\rm th}$ TDU) we obtain results similar to $M$ = 1.5 
$M_{\odot}$ 
models.

\onecolumn

\twocolumn


\section{Summary and conclusions}\label{conclusions}

We have presented a detailed discussion of 94 CEMP-$s$ and CEMP-$s/r$ stars
collected from the literature. 
This paper is strictly related to Paper II, in which we provided 
a general description of the sample and the main results obtained.
\\
The theoretical interpretation of CEMP-$s/r$ stars is still largely debated,  
because the $s$ and $r$-processes are ascribed to different physical environments. 
We remind the hypothesis adopted here and described in Paper II:  
we assumed an initial $r$-process enhancement in the molecular cloud from
which the binary system formed, followed by $s$-process
nucleosynthesis during the TP-AGB phase.
\\
In the analysis we followed the star classification provided in Paper II.
We considered all single species, the number of lines detected and the error bars 
determined by the authors, 
with particular attention to the three $s$-process peaks, ls, 
hs and Pb.
For each star, AGB models that better interpret the observations have 
exhaustively discussed (a summary of the results has already been presented
in Paper II, Tables~10 and~11).
We considered separately those stars for which a limited number of observations among 
the $s$-process elements are available (Appendix~A). 
Five stars with [Fe/H] $>$ $-$1.5 were discussed in a separate Section: 
HD 206983, HD 26 and HE 1152--0355,
as well as CS 29503--010 and HE 0507--1653, which have a limited number of $s$-element
observations.
Note that some CEMP-$s$ stars analysed here were previously classified
as CH stars (e.g., \citealt{vaneck03}). As firstly suggested by \citet{McClure89} and
\citet{LB91}, these objects, showing strong features of CN and CH bands as well as high $s$-process
abundances, may be considered a link between barium stars (with disk metallicity) 
and CEMP-$s$ stars (with halo metallicity).
Indeed, they all belong to binary systems with mass transfer from the most evolved
companion having already undergone the TP-AGB phase. 

As discussed in Paper II, interpretations of spectroscopic data 
are obtained for AGB initial masses in the range $M$ $\sim$ 1.3 to 2 $M_{\odot}$,
provided that different dilutions and $^{13}$C-pocket strengths are chosen.
The AGB initial mass is mainly defined by the occurrence of the FDU 
(which implies a dilution of at least 1 dex) and by the observed [Na/Fe] (and 
[Mg/Fe]). Major information on the efficiency of the $^{13}$C-pocket is provided by both 
[hs/ls] and [Pb/hs]. 
For a given metallicity, a range of $^{13}$C-pocket strengths is required in order 
to interpret observations of CEMP-$s$ and CEMP-$s/r$ stars:
for models with $M^{\rm AGB}_{\rm ini}$ $\sim$ 1.3 $M_{\odot}$, $^{13}$C-pockets 
close to case ST/12 (with a range from cases ST/6 to ST/15) are needed, 
while for $M^{\rm AGB}_{\rm ini}$ $\sim$ 1.5 $M_{\odot}$, $^{13}$C-pockets close to 
case ST/3 (with a range from cases ST/2 to ST/12)\footnote{The
three giants HD 5223, HD 209621 and HE 1305+0007 may be interpreted with 
$M^{\rm AGB}_{\rm ini}$ $\sim$ 2 $M_{\odot}$ and case ST/15, which is slightly lower
than the given range, but still within the uncertainties considering
that the spectra of these stars are highly blended.} are required (see  
Section~\ref{poisons}). 
However, as highlighted in this study, such a classification is only indicative.
Indeed, the analysis of individual stars provided in this paper is necessary 
in order to point out the peculiar characteristics of each star. 
For instance, several stars do not have Na or Pb detected, which are useful 
to constrain the AGB initial mass or the $^{13}$C-pocket strength.
Also, for many stars we do not have information about Eu, and we can not argue
about any possible initial $r$-process enhancement. 
Finally, 40\% of the star sample has a limited number of spectroscopic observations. 
 For all these reasons, an averaged analysis could not be realistic.
Actually,  
six stars need $^{13}$C-pockets $\leq$ case ST/24 (Table~\ref{6stars}), below the
ranges indicated above: CS 22942--019, CS 31062--012, HE 0336+0113
as well as V Ari, HE 1135+0139 and HD 189711, having more uncertain observations 
because their spectra are veiled by strong molecular lines \citep{vaneck03} 
or have lower resolution (HE 1135+0139 with $R$ = 20\,000, \citealt{barklem05}).\footnote{Among 
the star with a limited number of data, 
three CEMP-$s$ need low $s$-process efficiencies, CS 22956--28, CS 22891--171 and 
HE 1001--0243. Further spectroscopic investigations are desirable for these objects
(e.g., no ls elements are available for CS 22891--171 and HE 1001--0243).} 
For CS 22942--019, CS 31062--012 and HE 0336+0113, we do not exclude different
interpretations.
For instance, in AGBs of low mass and metallicity ([Fe/H] $\la$ $-$2.5),
\citep{hollowell90,iwamoto04,campbell08,cristallo09pasa,campbell10}, 
a proton ingestion episode may modify the $s$-process pattern. 
However, the AGB initial mass and the metallicity at which this proton ingestion starts
to occur would be decreased if an initial oxygen enhancement is adopted \citep{straniero10}. 
Moreover, a possible contribution from AGB models of initial mass 3 $\la$ 
$M^{\rm AGB}_{\rm ini}/M_{\odot}$ $\la$ 7 (intermediate AGBs, IMS) may be investigated.
In Paper I, we briefly discuss the nucleosynthesis in IMS of disc metallicity.
In IMS the temperatures at the bottom of the convective pulse is higher than in low mass
AGBs (1.2 $\la$ $M^{\rm AGB}_{\rm ini}/M_{\odot}$ $\la$ 2, LMS), and the neutron source
$^{22}$Ne($\alpha$, n)$^{25}$Mg is more efficiently activated.
This may increase the $s$-process contribution to the ls elements
with respect to the other $s$-peaks \citep{travaglio04}, as observed in CS 22942--019
([ls/Fe] $>$ [hs/Fe] $>$ [Pb/Fe]).  
However, the mass of the He-intershell in IMS is smaller than LMS by one
order of magnitude. This reduces the TDU efficiency and the final [El/Fe] abundances 
observed in the envelope.
Moreover, in this AGB mass range, Hot Bottom Burning (HBB) also modifies the final C and 
N abundances (\citealt{sugimoto71}, \citealt{iben73}, \citealt{karakas03}, \citealt{ventura05}),
destroying C and producing a large amount of N (NEMPs, nitrogen-enhanced metal-poor
stars \citealt{johnson07}), contrary to that observed in 
CS 22942--019, CS 31062--012 and HE 0336+0113.
In addition, IMS stars with [Fe/H] $\la$ $-$2.3, undergo an hot TDU, in which 
the envelope deeply penetrates into the intershell burning protons at its base 
\citep{goriely04,herwig04,campbell08,lau09}, possibly 
modifying the structure and the evolution of the star.
Further investigations are desirable.
\\
Note that the spread of the $^{13}$C-pocket strengths obtained here 
is larger than that found by \citet{bonacic07}. However, it 
may reflect the uncertainties affecting the formation of the $^{13}$C-pocket: 
 the H profile (and then the amount of $^{13}$C and $^{14}$N), 
the mass of the pocket, and the physical mechanisms involved 
(e.g., the treatment of the mixing process at the radiative/convective interfaces, 
models including rotation, magnetic fields or gravity waves) are object of study 
\citep{herwig97,langer99,herwig03,denissenkov03,siess04,straniero06,cristallo09}. 
\\
As well as [Na/Fe] (and [Mg/Fe]), in some cases the [ls,hs/Fe] and their ratio [hs/ls] may provides 
indications on the AGB initial mass. In five stars (CS 22183--015, CS 22880--074, CS 22898--027, 
HE 0338--3945, HE 2148--1247) the low observed [ls/Fe] (corresponding to a high [hs/ls]) is better 
interpreted by AGB models with low initial mass ($M$ $\sim$ 1.3 $M_{\odot}$).
On the other side, two stars among the sample show an extremely large $s$-process enhancement ([ls/Fe] 
$\sim$ 2 and [hs/Fe] $\sim$ 3), about 1 dex 
higher than the average of the CEMP-$s$II (or CEMP-$s$II/$r$) stars: CS 29528--028 and 
SDSS 1707+58. Both stars can only be interpreted by AGB models with initial mass $M$ 
= 2 $M_{\odot}$, which undergo a larger number of TDUs.
Further investigations are strongly suggested, especially for SDSS 1707+58, which has
a limited number of observations available.
\\
Among the sample listed in Table~\ref{summary1} (in which stars with a large number of
 observations are reported), 17 stars lie on the main-sequence/turnoff. 
The degree of dilution obtained for these stars may provide information on the effect 
of mixing as thermohaline, gravitational settlings and radiative levitation. 
We find that ten of them (CS 22881-036, CS 22898-027, CS 29497-030, CS 29526-110, 
CS 29528-028, CS 31062-012, HE 0338-3945, HE 2148-1247, SDSS 0216+06, SDSSJ1349-0229) 
have only one theoretical interpretation with negligible dilution, from which we may 
hypothesise that low mixing had occurred. 
Three stars (HE 0143-0441, HE 0430-4404, HE 1152+0027) may have different theoretical 
interpretations with different AGB initial masses (in the range of 1.2 -- 2 $M_{\odot}$). 
For each star, a solution with $dil$ = 0.0 dex may be found: HE 0143-0441 and HE 1152+0027 
with $M$ = 1.3 $M_{\odot}$; HE 0430-4404 with $M$ = 1.2 $M_{\odot}$. Higher dilutions can be found with 
higher initial masses. 
Two stars, CS 22887-048 and BS 16080-175, were analysed by \citet{tsangarides05}, 
PhD Thesis. 
CS 22887-048 has solutions with different AGB models, but with low or negligible dilution:
$M$ = 1.4 $M_{\odot}$ and $dil$ = 0 dex, $M$ = 1.5 or 2 $M_{\odot}$ and $dil$ = 0.3 dex. 
\citet{tsangarides05} estimated a metallicity 1 dex higher than \citet{johnson07}. 
A more detailed analysis is desirable. BS 16080-175 has three possible 
interpretations: $M$ = 1.35 $M_{\odot}$ and $dil$ = 0.6 dex, $M$ = 1.5 $M_{\odot}$ and $dil$ = 1.2
dex, $M$ = 2 $M_{\odot}$ and $dil$ = 1.2 dex.
CS 22964-161 has solutions with $M$ = 1.2 $M_{\odot}$ and $dil$ = 0.4 dex, $M$ = 1.3 $M_{\odot}$ 
and $dil$ = 0.9 dex, suggesting moderate mixing. 
The interpretation of SDSS J0912-0229 is uncertain, and no information on possible mixing 
can be deduced.

The main goal of this paper is to present a detailed
study of spectroscopic observations star by star,
through an analysis of the AGB models presented in Paper I and II.
In general, we found possible agreements between theoretical predictions and spectroscopic 
data. The major discrepancies are summarised here below. 
This aims to provide potential indications for future studies, also of spectroscopic nature,
and suggests important starting points of yet unsolved issues.
\\
One of the main problems concerns C and N.
As highlighted in Paper II (Section~5.3), the observed [C/Fe], [N/Fe] and the carbon isotopic 
ratio $^{12}$C/$^{13}$C can not be interpreted by AGB models. 
Large uncertainties are present in both spectroscopic (NLTE and 3D atmospheric 
models; \citealt{collet07,collet09,GAS07,asplund09,caffau09,frebelnorris11}) 
and AGB models, as extra-mixing processes (CBP), 
thermohaline, or rotation and magnetic fields which
may induce the mixing \citep{stancliffe09,stancliffe10,charbonnel10}.
The hypothesis of the CBP, in order to reconcile 
theoretical predictions and observations in stars (or SiC presolar
grains), has been remarked by different authors 
 \citep{nollett03,dominguez04a,dominguez04b,cristallo07,busso10,palmerini11}.
The effects of the CBP on $^{12}$C and $^{14}$N cannot be exactly quantified by models
and the physical processes involved are not clear yet. 
In several CEMP-$s$ stars the predicted [C/Fe] is much higher
than reported by spectroscopic observations. 
The contemporary measurement of a very low $^{12}$C/$^{13}$C ratio observed,
in the typical range 4 to 10, indicates the impact of 
a strong extra-mixing, which have not been included in our AGB models nor in our
treatment of the envelope of the observed low mass star after mass accretion by 
the more massive AGB companion. 
This will imply a concomitant reduction of the expected [C/Fe] and $^{12}$C/$^{13}$C
ratio, but at the same time a strong increase of the predicted [N/Fe]. 
In several cases one would expect CEMP-$s$ stars to be even more N-rich than 
C-rich. A discussion of this issue is deferred to further work (Bisterzo et al., in 
preparation).
An additional process that may increase $^{13}$C and $^{14}$N is 
the proton ingestion episode occurring in low mass AGBs of low metallicity
(see Section~\ref{CS036}). 
\\
Among the light elements, also fluorine is largely produced by AGB models,
while recent spectroscopic determinations in CEMP-$s$ stars provide [F/Fe] about 
1 dex (or more) lower than theoretical predictions \citep{lucatello11astroph}.
Note that recently, \citet{palmerini11} studied the effect of extra-mixing in AGB stars
on the light-elements, suggesting a possible decrease of [F/Fe] at low metallicities. 
However, further studies both on the theoretical and spectroscopic point of view 
are strongly desirable.
\\
Another discrepancy concerns the elements belonging to the ls and hs peaks.  
AGB models predict (within 0.3 dex) 
 similar abundances for the first $s$-peak (Sr, Y, and Zr) and 
for the second $s$-peak (for Ba, La, Ce, Pr and Nd), (with
 a slightly increasing or decreasing trend 
of [El/Fe] with atomic number depending on the $^{13}$C-pocket, see 
Paper I, Appendix~B). 
This is strongly supported by the reliable neutron cross section measurements and  
solar abundances in the ls and hs regions. 
Instead, some CEMP-$s$ stars show an internal spread among ls and 
hs elements greater than 0.5 dex.
In several stars the observed [Ba/Fe] is higher than the average of [hs/Fe]
(e.g., HD 26, CS 30301--015, HE 0143--0441, HE 0336+0113, LP 625--44);
the most evident example is CS 31062--050 \citep{aoki06}, where [Ba/Fe] is 
about 0.5 dex higher than the other hs elements (Section~\ref{CS050}).
In general, Sr is more uncertain than other ls elements, with a limited number
of lines detected. In our analysis, we
exclude Sr from the ls elements and Ba from the hs peak, which are mainly 
affected by higher spectroscopic uncertainties \citep{busso95} due to NLTE effects 
\citep{andr09,andr11,mashonkina08,short06}, especially by decreasing the metallicity.
However, even excluding Sr and Ba from the analysis, very
 large spreads between Y and Zr and among the hs 
elements have been observed in some stars: [Zr/Y] $\ga$ 1 in the 
cold giants HD 5223, HD 206983, HD 209621 and HE 1305+0007; [La/Ce] $\sim$ 
$-$1 in the main-sequence CEMP-$s$II$/r$ stars SDSS J1349--0229 and SDSS J0912+0216; 
[La/Nd] $\sim$ 1 in the cold giant HE 1152--0355.
Lower [La/Nd] ($\sim$ 0.5) were observed in the giant HE 1031-0020 and
lower [Zr/Y] ($\sim$ 0.5) were detected in CS 29497--030, CS 31062--050 and HE 0143--0441,
 HE 0338--3945, HE 2148--1247, CS 29513--032, HK II 17435--00532, LP 625-44.
Note that the recent estimation of NLTE effects for Zr II lines (+0.3 dex)
in low metallicity stars by \citet{vel10} would further increase the 
discrepancy between Zr and Y observed in some stars.
\\
At the state of the art, differences larger than 0.5 dex among the hs elements 
can not be obtained by AGB models.
 Concerning the ls elements, 
 low metallicity unevolved Galactic stars show a large spread in [Sr,Y,Zr/Fe] for 
[Fe/H] $\leq$ $-$2.5. For instance, [Sr/Fe] has minimum observed values of $-$1 for [Fe/H] 
$\sim$ $-$2.7 down to $-$2 for [Fe/H] $\leq$ $-$3 (see e.g., Fig.~20 of
\citealt{kaeppeler11rmp}; or \citealt{andr11}, who account for NLTE effects).
A similar spread is observed for [Y/Fe] and [Zr/Fe].
This may be due to non homogeneous or incomplete mixing 
of the gas in the Galactic halo or to a multiplicity of primary $r$-process
components (see \citealt{travaglio04}, \citealt{QW08}, \citealt{SCG08}).
Therefore, the hypothesis of an extreme initial deficiency of Sr, Y and Zr
in the molecular cloud seems plausible. In some CEMP-s stars, we assumed an [Sr,Y,Zr/Fe]$^{\rm ini}$ 
(or one of them) = $-$1, in order to interpret the observations (e.g., the CEMP-$s$I star 
CS 30322--023 by \citealt{masseron06} and CS 29513--032 by \citealt{roederer10}). 
However, in general, the $s$-process contribution is large and overcomes initial 
deficient compositions (e.g., CS CS29497--030 by \citealt{ivans05} or CS 31062--012 
\citealt{aoki02d,aoki07,aoki08}).  
Then, the discrepancy within the ls elements remains an open problem. 
From the theoretical point of view, the nucleosynthesis of the ls elements
is highly debated. Specifically, a primary contribution of about $\la$ 20\% to 
solar Sr, Y and Zr (lighter element primary process, LEPP \citealt{travaglio04})
has been hypothesised in order to interpret the observations of [Sr,Y,Zr/Fe]
versus [Fe/H] in of Galactic metal-poor stars\footnote{Galactic Chemical Evolution 
models \citep{travaglio04,serminato09pasa}, 
that account for the main and strong $s$-process in low and intermediate mass
AGB stars of different stellar populations, predict $s$ contributions of 
$\sim$ 64\%, $\sim$ 67\%, and $\sim$ 60\% to solar Sr, Y, and Zr, respectively. 
The weak-$s$ process in massive stars is estimated to contribute to 
$\sim$ 9\% to solar Sr, $\sim$ 10\% to solar Y, and $\sim$ 0\% to solar Zr. 
The $r$-process contribution is $\sim$ 12\%, 8\% and 15\% for solar Sr, Y and
Zr, respectively.}. However, the exact contribution from this primary process to individual 
ls elements is not well established and its origin is still under investigation 
(e.g., \citealt{montes07}; see also the recent \citealt{arcones11} and references therein).
\\
Among CEMP-$s/r$ stars, five stars (CS 22898–027, CS 29497–030, HE 0338–3945, 
HE 1305+0007 and HE 2148–1247) require the highest $r$-enhancement [r/Fe]$^{\rm ini}$
= 2.0, with an observed [La/Eu] $\sim$ 0, together with [La/Fe] $\sim$ 2.
As discussed in Paper II (Section~3), the hypothesis of an initial $r$-process 
enhancement is adopted in the region between Ba and Bi.
Indeed, observations of elements between Mo and Cs are lacking.
 For neutron-capture elements lighter than Ba, different initial $r$-process
enrichment could be introduced under the assumption of a multiplicity of $r$-process 
components \citep{sneden03a}. 
Only \citet{JB04} detected Pd for CS 31062--050. 
About 50\% of solar Pd is produced by the $s$-process (see Paper II, Table~5);
the remaining 50\% is ascribed to the $r$-process. 
The [r/Fe]$^{\rm ini}$ = 1.6, adopted in order to interpret the observed 
$r$-elements from Eu to Ir, would overestimate the [Pd/Fe]$_{\rm obs}$ 
by $\sim$ 1 dex. 
Lower initial light-$r$-enhancements [light-r/Fe]$^{\rm ini}$ 
$\sim$ 0.5 -- 1.0 are assumed (Section~\ref{CS050}), likely confirming a multiplicity of the 
$r$-processes.
This is the only reliable detection among the elements included between Mo 
and Cs in CEMP-$s$ or CEMP-$s/r$ stars. 
The upper limits for Ag in CS 29497--030 \citep{ivans05} and 
HE 0338--3945 \citep{jonsell06} do not provide significant constraints.
When allowed, further investigations on the light-$r$ elements are desirable.
\citet{behara10} detected a very enhanced Ru in the main-sequence star SDSS J0912+0216
([Ru/Fe] = 2.6). Even if no error bar is provided by the authors, the Ru observed
in this star seems to agree with some among the heavy-$r$ process elements, as Gd and Tb.
The abundances observed in SDSS J0912+0216 are peculiar, both for $s$- and $r$-process elements:
indeed, in addition to the already mentioned large spread among the hs peak, this star
shows a large spread among the $r$-elements (e.g., [Gd/Eu] $\sim$ 1.5).
A similar behaviour is found in SDSS J1349–0229, studied by
the same authors. The spread is confirmed in both stars, and highlights a crucial problem
from the point of view of the theoretical interpretation.
Other three stars show discrepancies of about 0.5 dex between observed and predicted
$r$-elements: CS 31062--050 (Er, Yb, Lu), LP 625--44 (Gd, Er) and HE 0338--3945 
(Dy, Tm). These stars may be considered important starting point for future studies.
In general, further investigations on the $r$-elements in CEMP-$s$ and CEMP-$s/r$ stars
would be strongly useful.

We recall that the theoretical interpretations presented here are thought as test 
for AGB models, obtained with post-process nucleosynthesis models based on old FRANEC
 models \citep{gallino98,straniero03}. 
A new generation of FRANEC code is developing.
These new full evolutionary models account for new opacities,
updated reaction rates, a mixing algorithm to obtain the $^{13}$C-pocket, a new
evaluation of the mass loss rate based on the observed correlation with the 
pulsational period (\citealt{straniero06,cristallo09pasa,cristallo09,cristallo11}).
Future investigations are planned in order to update our predictions accounting 
for low metallicity fully FRANEC models, once
 a whole spectrum of masses and metallicities will be completed.

\clearpage
\onecolumn

\begin{table}
\caption{Summary of theoretical interpretations for six CEMP-$s$ and CEMP-$s/r$ stars 
that need low $^{13}$C-pocket efficiencies ($\leq$ case ST/24). 
The AGB initial mass, $^{13}$C-pocket, dilution factor, and 
initial $r$-enhancement are reported. 
Asterisks in column~9 indicate that no Eu has been observed.
 The Figure number associated to the 
theoretical interpretation is given in column~10.}
\label{6stars}
\begin{center}
\resizebox{18cm}{!}{\begin{tabular}{|l|llllllllll|}
\hline
Star            &  Ref.           & [Fe/H] & FDU  & Type & $M^{\rm AGB}_{\rm ini}$    & pocket   & dil  &  [r/Fe]$^{\rm ini}$    & Fig.   \\  
(1)              &  (2)            & (3)    & (4)  & (5)    & (6)  & (7) & (8)      &   (9) &     (10)                      \\ 
\hline 
CS 22942--019 &A02c,d,PS01,Sch08,M10,L11&  -2.64,-2.43 &  yes & sI   & 2    & ST/50  & 0.7  &  0.5     &  \ref{CS22942-019_aoki02+aoki07+schuler08+masseron10_bab10d303540m2z5m5rp0p5alf0p5_dil0p7n26}                \\
CS 31062--012 & I01,A07,A02c,d,A08  &  -2.55 &  no & sII/rII   & 1.3  & ST/30  & 0.0    &  1.5     & 
\ref{mnras_CS31062-012_aoki02+aoki0708+Israelian01_bab10d18m1p5z5m5rp1p5_n345}            \\
HE 0336+0113 &  C06            &  -2.68 &  no & sII & 1.4  & ST/55  & 0.0    &   0.5       & \ref{HE0336+0113_cohen06_bab10d35m1p5d30m2z5m5rp0p5_diffdiln8n26}       \\ 
"            &  "              &  "     &  "   & " & 2  & ST/45  & 0.3    &   "       & "       \\ 
\hline
V Ari   &  VE03           &  -2.40 &  yes & sI/$-$   & 1.5  & ST/30   & 0.9  &  0.5*    &  
\ref{VAri_vaneck03_bab10d16d20d24m1p5z1m4rp0p5_dil0p9n20}                                  \\
HE 1135+0139  &  B05            &  -2.33 &  yes & sI   & 1.3  & ST/24   & 1.2  &  0.0        & 
 \ref{HE1135+0139_Barklem05_bab10d16d4m1p5z1m4nr_diffdiln520}        \\   
              "  &       "          &   "     &  " &"   & 1.5;2  & ST/6  & 1.8  &  "       &
"                       \\
HD 189711  &  VE03           &  -1.80 &  yes  & sI/$-$  & 1.5;2  & ST/24  & 0.9  &  0.5*    & 
\ref{HD189711_vaneck03_bab10d12d16d20m1p5z2m4rp0p5_dil0p9n20}                            \\
\hline
\end{tabular}}
\end{center}
\end{table}                                                                                                                                                                                                                                                       

\clearpage
\twocolumn

\section*{Acknowledgments}

We are deeply indebted to T. C. Beers, J. J. Cowan, I. I. Ivans, 
C. Pereira, G. W. Preston, I. U. Roederer, C. Sneden, I. B. Thompson, S. 
Van Eck, S. Vauclair, for enlightening discussions about 
CEMP-$s$ and CEMP-$s/r$ stars. 
Special acknowledgments are addressed to W. Aoki and P. Bonifacio
for precious comments on four peculiar stars (LP 625--44, HD 209621; 
SDSS J1349--0229, SDSS J0912--0216).
Heartfelt thanks go to Dr Maria Lugaro for helping us to improve
the discussion of the main results of the paper.
This work has been supported by the MIUR and KIT (Karlsruhe Institute of Technology).


\onecolumn

\appendix

\newpage

\clearpage

\newpage

\newpage
\clearpage

\section{Online Material} \label{onlinematerial}

\begin{figure}
\includegraphics[angle=-90,width=8.5cm]{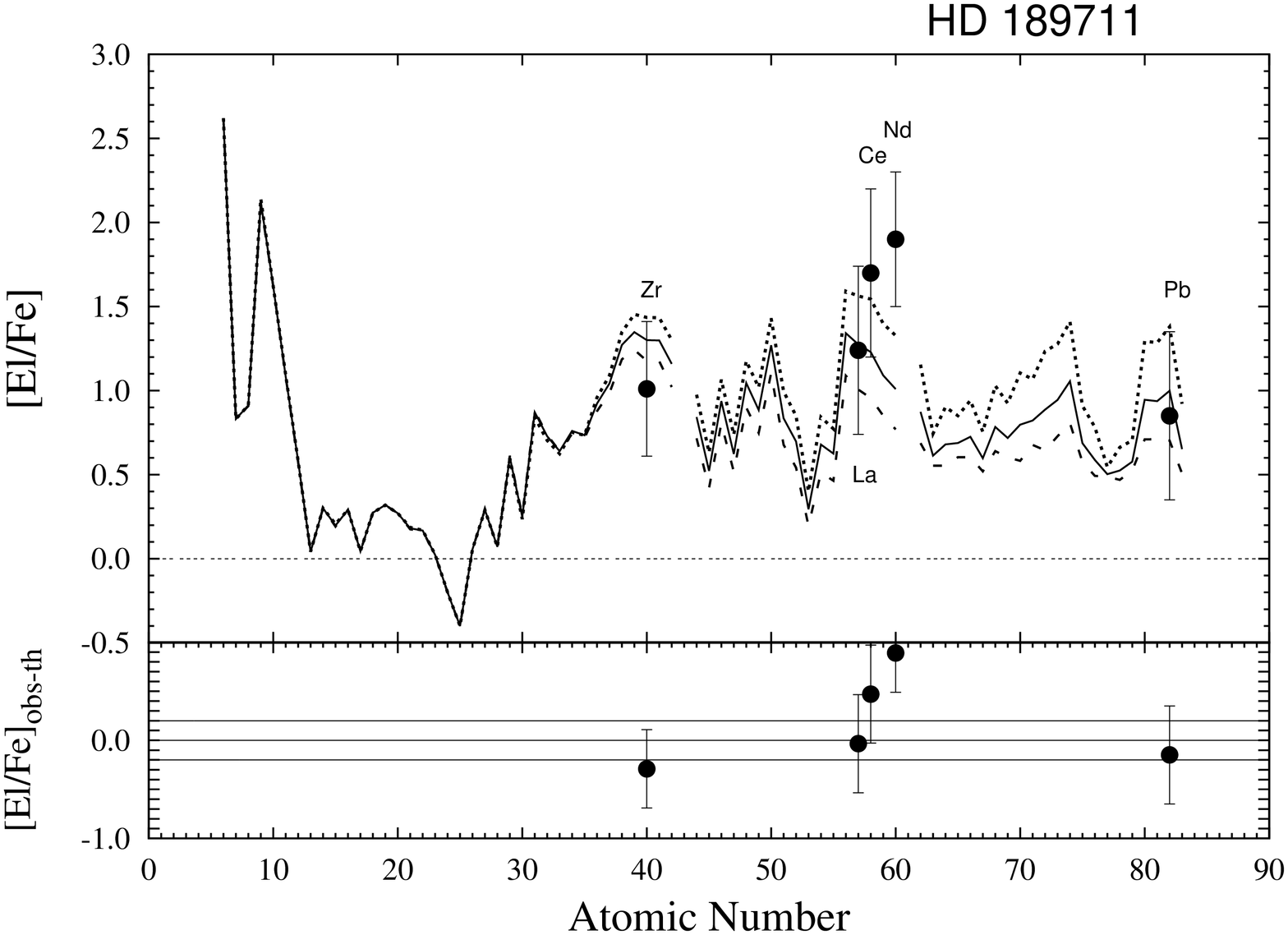}
\vspace{2mm}
\caption{Spectroscopic [El/Fe] abundances of the giant HD 189711 
([Fe/H] = $-$1.8; {\textit T}$_{\rm eff}$ = 3500 K; log $g$ = 0.5)
compared with AGB models of initial mass $M$ = 1.5 $M_{\odot}$, cases ST/18
(dotted line), ST/24 (solid line), ST/30 (dashed line),
and $dil$ = 0.9 dex.
Observations are from \citet{vaneck03}, 
who detected [hs/ls] $\sim$ 0.2 and [Pb/hs] $\sim$ $-$0.4 (see Section~\ref{HD189711}).
An [r/Fe]$^{\rm ini}$ = 0.5 is adopted. }
\label{HD189711_vaneck03_bab10d12d16d20m1p5z2m4rp0p5_dil0p9n20}
\end{figure}

\begin{figure}
\includegraphics[angle=-90,width=8.5cm]{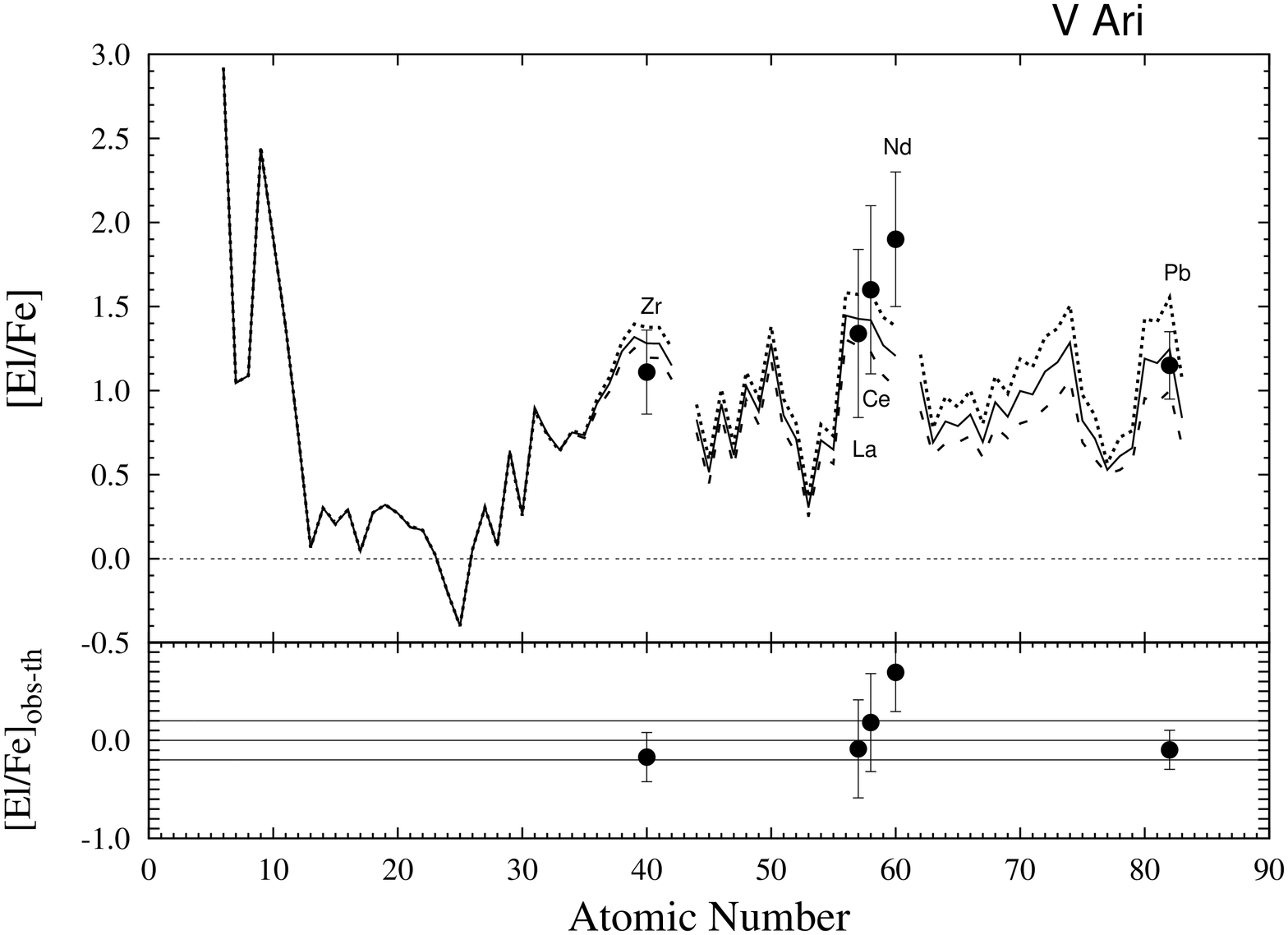}
\vspace{2mm}
\caption{Spectroscopic [El/Fe] abundances of the giant V Ari
 ([Fe/H] = $-$2.4; {\textit T}$_{\rm eff}$ = 3580 K; log $g$ = $-$0.2)
compared with AGB models of initial mass $M$ = 1.5 $M_{\odot}$, 
cases ST/24 (dotted line), ST/30 (solid line), ST/36 (dashed line), 
and $dil$ = 0.9 dex.
Observations are from \citet{vaneck03}, 
who detected [hs/ls] $\sim$ 0.2 and [Pb/hs] $\sim$ $-$0.2.
An [r/Fe]$^{\rm ini}$ = 0.5 is adopted. }
\label{VAri_vaneck03_bab10d16d20d24m1p5z1m4rp0p5_dil0p9n20}
\end{figure}

\begin{figure}                                                                                    
\includegraphics[angle=-90,width=8.5cm]{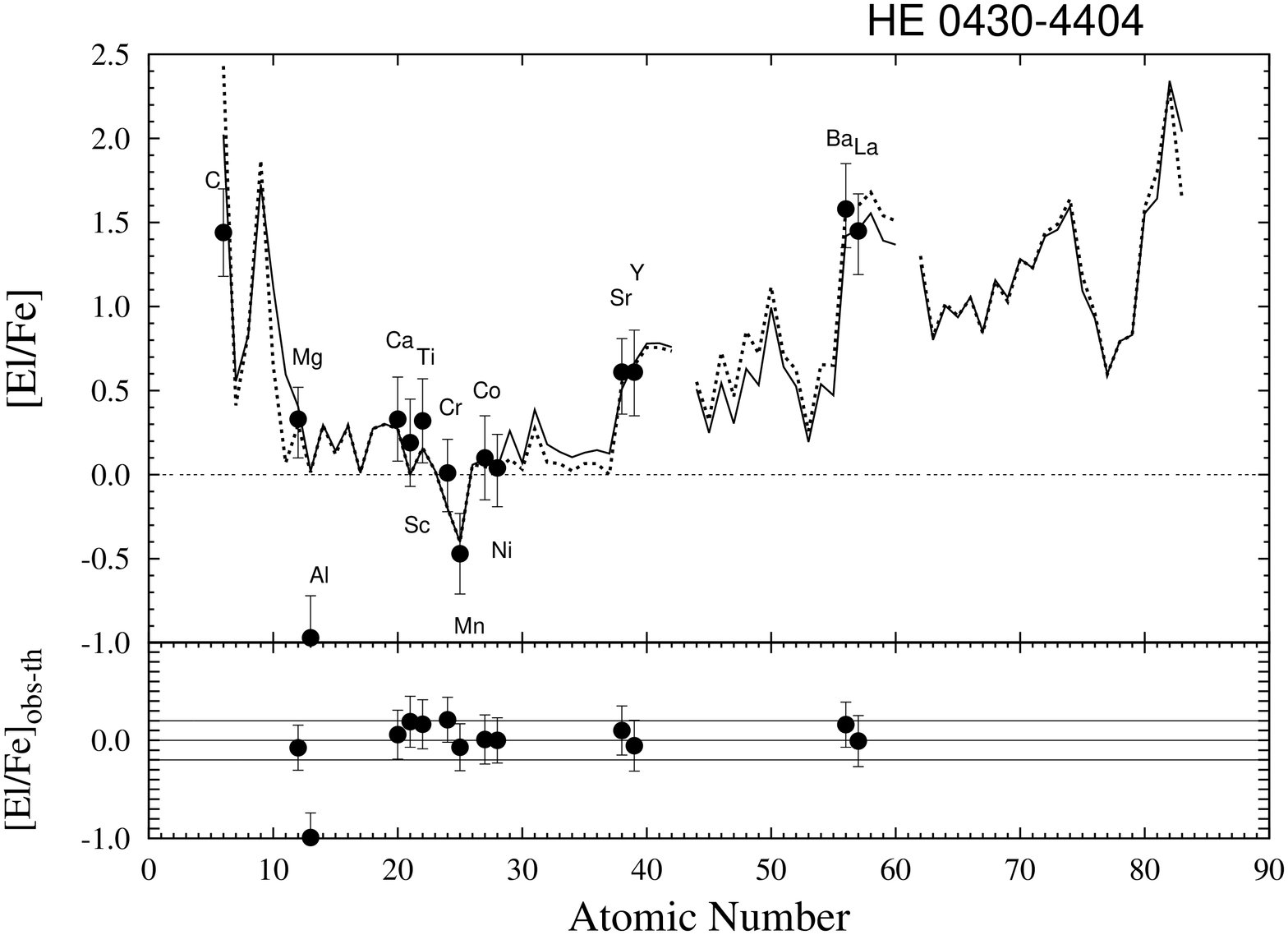} 
\vspace{2mm}
\caption{Spectroscopic [El/Fe] abundances of the main-sequence star HE 0430--4404
([Fe/H] = $-$2.07; {\textit T}$_{\rm eff}$ = 6214 K; log $g$ = 4.27)
compared with two AGB models:                                                   
$M^{\rm AGB}_{\rm ini}$ = 1.2 $M_{\odot}$, case ST/9, $dil$ = 0.0 dex
(dotted line),                                   
or $M^{\rm AGB}_{\rm ini}$ = 1.5 $M_{\odot}$, case ST/3, $dil$ = 1.5 dex
(solid line).
Observations are from \citet{barklem05}, who detected 
[hs/ls] = 0.65. An [r/Fe]$^{\rm ini}$ = 0.5 is adopted. 
The theoretical interpretation provided for this star is similar 
to HE 0231--4016 (see Section~\ref{HE0231}).}                              
\label{HE0430-4404Bark05_bab10d6d2m1p5z2m4rp0p5_diffdiln3n20}                                     
\end{figure}

\begin{figure}                                                                                    
\includegraphics[angle=-90,width=8.5cm]{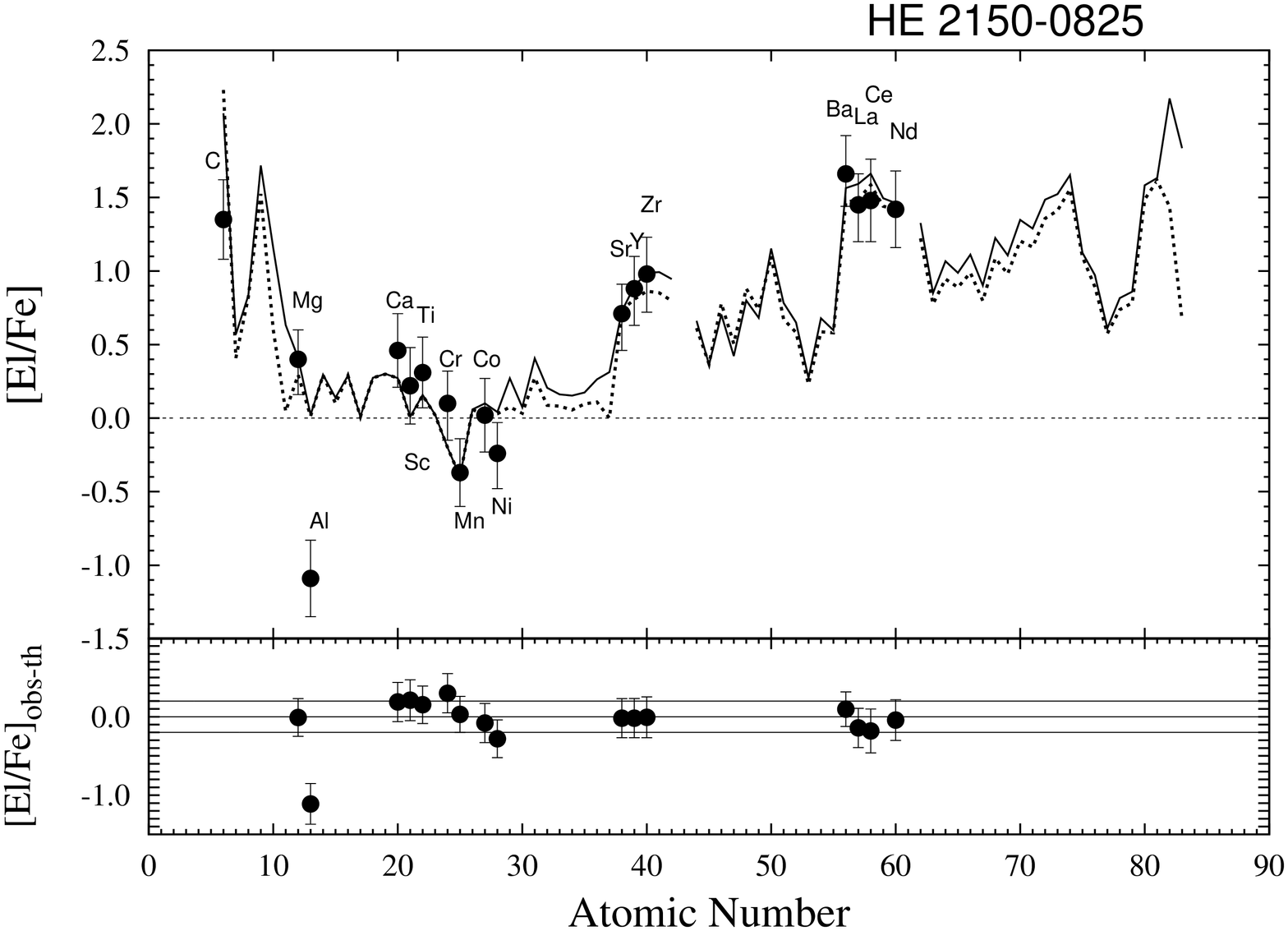}
\vspace{2mm}
\caption{Spectroscopic [El/Fe] abundances of this turnoff/subgiant HE 2150--0825 
([Fe/H] = $-$1.98; {\textit T}$_{\rm eff}$ = 5960 K; log $g$ = 3.67, before the FDU)
compared with two AGB models:                                                
$M^{\rm AGB}_{\rm ini}$ = 1.2 $M_{\odot}$, case ST/15, $dil$ = 0.2 dex
(dotted line),                                  
or $M^{\rm AGB}_{\rm ini}$ = 1.5 $M_{\odot}$, case ST/5, $dil$ = 1.5 dex
(solid line).
Observations are from \citet{barklem05}, who derived
[hs/ls] = 0.5. An [r/Fe]$^{\rm ini}$ = 0.5 is adopted. 
The theoretical interpretation provided for this star is similar 
to HE 0231--4016 (see Section~\ref{HE0231}).}                             
\label{HE2150-0825Bark05_bab10d10d3m1p5z2m4rp0p5_diffdiln3n20}                                    
\end{figure}  

\clearpage
\newpage

\subsection{Interpretations of CEMP-$s$ and CEMP-$s/r$ stars 
with a limited number of spectroscopic observations (see Paper II, Table~3)} 

The number of elements detected for the stars discussed in Sections~\ref{secCEMPs} 
and~\ref{secCEMPs/r} provide important constraints for AGB models.
Unfortunately, many CEMP-$s$ stars have a limited number of $s$-process elements 
available. 
For instance, only Ba is detected 
in nineteen stars
 (\citealt{aoki07}; \citealt{cohen06}); 
for other stars, Sr among the ls and Ba among the hs elements are measured. 
Possible solutions have been provided in Paper II, Table~11.
These theoretical interpretations have to be considered as indicative examples,
because many AGB models may interpret the observations as well.

\subsection{CEMP-$s$II} 

Three main-sequence stars with a limited number of data belong to this group:
HE 0024--2523 by \citet{lucatello03}, CS 22967--07
and CS 30323--107 by \citet{lucatello04PhD}, 
for which a very low upper limit is reported for Eu, excluding
 high initial $r$-process enhancements.

\begin{figure}
\includegraphics[angle=-90,width=8.5cm]{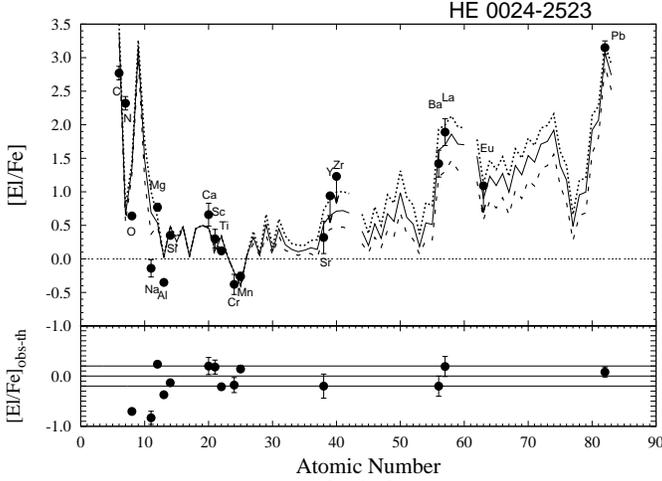}
\vspace{2mm}
\caption{Spectroscopic [El/Fe] abundances of main-sequence star HE 0024--2523 
([Fe/H] = $-$2.72; {\textit T}$_{\rm eff}$ = 6625 K; log $g$ = 4.3)
compared with AGB models of initial mass $M$ = 1.3 $M_{\odot}$, case ST/9 and $dil$ $\sim$ 0.0 dex.
Three thermal pulses with TDU are shown: pulse 4 (dashed line), 5 (solid line), 6 (dotted
line).
Observations are from \citet{lucatello03}, ([hs/ls] = 1.3; [Pb/hs] = 1.7).
An [r/Fe]$^{\rm ini}$ = 0.0 is adopted. }
\label{HE0024-2523_lucatello03_bab10d6m1p5z5m5_nr_alf0p5_dil0p05n456}
\end{figure}

\subsubsection{HE 0024--2523 (Fig.~\ref{HE0024-2523_lucatello03_bab10d6m1p5z5m5_nr_alf0p5_dil0p05n456})}
\label{HE0024}

This is a very metal-poor main-sequence star
([Fe/H] = $-$2.7; {\textit T}$_{\rm eff}$ = 6625 K; log $g$ = 4.3), 
analysed by \citet{cohen02}, \citet{carretta02} 
(who detected Sr and Ba among the $s$-elements),
and \citet{lucatello03}.
\citet{lucatello03} provide the most complete study of this star,
with a deep discussion about the orbital parameters.
HE 0024--2523 is a short-period spectroscopic binary,
with $P$ = 3.4 days.
Although the exact values of the individual masses are unknown, 
limits have been derived from the available photometric and 
spectroscopic informations: the observed star 
has a mass of about 0.9 $M_{\odot}$; for the AGB companion (now 
a white dwarf) \citet{lucatello03} estimated a mass range 0.6 $\leq$ $M/M_{\odot}$ $\leq$ 1.4.
The values of [O/Fe], [Na/Fe] and [Al/Fe] take into account the
NLTE effects according to the prescription by \citet{gratton99}. 
The lines of C, N, Eu, La, and Pb were derived from spectral synthesis,
because they are very weak or somewhat blended with nearby 
lines of other species \citep{lucatello03}.
Only Sr with two lines is detected among the ls elements,
while for Y and Zr upper limits are provided.
The authors estimated an uncertainty of 0.1 dex for Pb
because hyperfine structure and isotopic splitting were not included
 in the analysis. 
Fig.~\ref{HE0024-2523_lucatello03_bab10d6m1p5z5m5_nr_alf0p5_dil0p05n456}
shows possible theoretical interpretations with AGB models 
of initial mass $M^{\rm AGB}_{\rm ini}$ = 1.3 $M_{\odot}$, 
case ST/9 and no dilution. 
No initial $r$-process enrichment is necessary 
to interpret the
observed [La/Eu] $\geq$ 0.7.
For higher initial masses 
 ($M^{\rm AGB}_{\rm ini}$ = 1.5 -- 2 $M_{\odot}$) 
dilutions higher than 1 dex are necessary. With these models,
Mg and the neutron capture elements Ba, La and Pb would be equally
fitted, while the observed [Na/Fe] approaches to 
the solution with $M^{\rm AGB}_{\rm ini}$ = 1.3 $M_{\odot}$.
The low carbon isotopic ratio
$^{12}$C/$^{13}$C = 6 $\pm$ 1 sustains the hypothesis of efficient mixing.
 
%

\subsubsection{CS 22967--07} 
\label{CS07}

This is a main-sequence star (Fe/H] = $-$1.81; {\textit T}$_{\rm eff}$ = 6479 K; 
log $g$ = 4.2), \citep{lucatello04PhD}.
The solar [Na/Fe] agrees with AGB models of initial mass 1.3 $M_\odot$ 
(case ST/9, dil = 0.0 dex). 
No initial $r$-process enhancement is needed in order to interpret the observed 
[La/Eu] = 0.7 dex.

\subsubsection{CS 30323--107} 
\label{CS107}

CS 30323--107 is a main-sequence star ({[Fe/H] = $-$1.75; {\textit T}$_{\rm eff}$ = 6126 K; 
log $g$ = 4.4) studied by \citet{lucatello04PhD}.
A difference of 0.6 dex is observed between [Ba/Fe] and [La,Ce/Fe]. 
The authors detected [Na/Fe] = $-$0.7, which can not be explained by 
an initial solar Na scaled with the metallicity.
We suggest a solution with $M^{\rm AGB}_{\rm ini}$ = 1.3 $M_{\odot}$
(case ST/3, $dil$ = 0.3 dex), which approaches [Na/Fe] $\sim$ 0. 
A low upper 
 limit is measured for Eu, excluding a high initial $r$-process 
enhancement ([La/Eu] $\geq$ 0.5).

\subsection{CEMP-$s$I} 

In this Section we describe three giants, CS 29495--42 by \citet{lucatello04PhD},
HE 1001--0243 and HE 1419--1324 by \citet{masseron10}, without 
observations of ls elements, and one star with uncertain occurrence of 
the FDU, CS 30315--91 \citep{lucatello04PhD}.

\subsubsection{CS 29495--42} 
\label{CS42}

CS 29495--42 is classified as a CEMP-$s$I star, with [Sr/Fe] = 0.2, [La/Fe] = 1.3 
and [Pb/Fe] = 1.3. As observed in other stars, Ba (three lines) is about 0.5 dex higher 
than La and Ce (three and two lines, respectively).
This star has $T_{\rm eff}$ = 5544 K and log $g$ = 3.4 \citep{lucatello04PhD},
([Fe/H] = $-$1.88). The occurrence of the FDU 
is uncertain in this star. Recently, \citet{johnson07} obtained $T_{\rm eff}$ = 5400 K and log $g$ = 3.3
with low resolution spectra, surely after the occurrence of the FDU. 
Solutions are found with AGB models of initial mass 1.3 $M_{\odot}$ and $dil$ = 0.9 dex.
A low $^{13}$C-pocket is needed 
to interpret the negative [Pb/hs]
(ST/18), but the observed [Sr/Fe] (one detected line) is overestimated by the models. 
Higher $^{13}$C-pockets agree with the observed [hs/Sr], but the predicted
[Pb/Fe] would be higher than observed.
The [Na/Fe] prediction may be considered in agreement with the observed
value within an uncertainty of 0.2 dex.
AGB models with higher initial mass would overestimate the observed [Na/Fe]
and [Sr/Fe].

\subsubsection{HE 1001--0243 and HE 1419--1324 
(Fig.s~\ref{HE1001-0243_masseron10_bab10d20d50m1p5z2m5_nr_diffdiln5},~\ref{HE1419-1324_masseron10_bab10d9d1p3m1p5z2m5_rp0p5_diffdiln5n20})}
\label{HE1001}
\label{HE1419}

These two newly discovered giants have been recently studied by 
\citet{masseron10}. Unfortunately no ls elements have been published so far.
Observations by \citet{masseron10} will be discussed by the authors in a forthcoming paper.
Both stars are giants with very low metallicity: HE 1001--0243 has 
[Fe/H] = $-$2.88, {\textit T}$_{\rm eff}$ = 5000 K and log $g$ = 2.0, 
while HE 1419--1324 shows [Fe/H] = $-$3.05, {\textit T}$_{\rm eff}$ = 4900 K
and log $g$ = 1.8.
For HE 1001--0243 the hs elements are very low ([hs/Fe] = 0.6)
and only an upper limit for Pb is measured, but HE 1419--1324 shows high lead 
([Pb/Fe] = 2.15; [Pb/hs] = 1.31) in accordance with an efficient $s$-process.
We present here possible theoretical interpretations in
Figs.~\ref{HE1001-0243_masseron10_bab10d20d50m1p5z2m5_nr_diffdiln5}
and~\ref{HE1419-1324_masseron10_bab10d9d1p3m1p5z2m5_rp0p5_diffdiln5n20}.

\begin{figure}
\includegraphics[angle=-90,width=8.5cm]{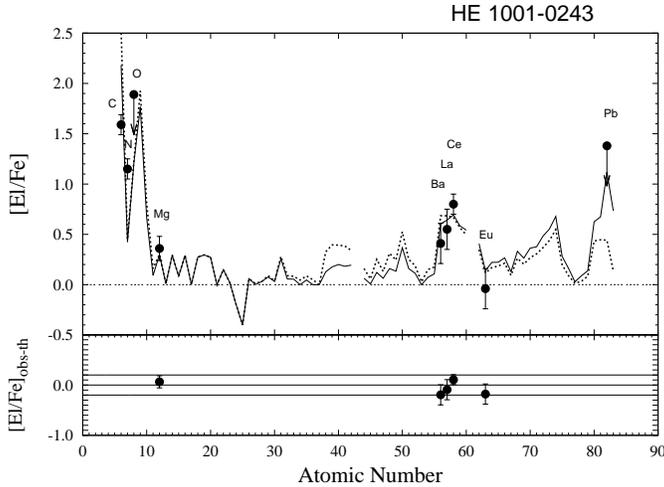}
\vspace{2mm}
\caption{Spectroscopic [El/Fe] abundances of the giant HE 1001--0243  
([Fe/H] = $-$2.88; {\textit T}$_{\rm eff}$ = 5000 K; log $g$ = 2.0)
compared with AGB models of initial mass $M$ = 1.3 $M_{\odot}$, cases ST/30 and ST/75, 
$dil$ $\sim$ 1.7 -- 1.3 dex, solid and dotted lines, respectively. 
Similar solutions can be obtained by AGB models 
of higher initial mass ($M$ = 1.5 and 2 $M_{\odot}$) and $dil$ $\sim$ 2 dex.
Observations are from \citet{masseron10}, who detected
[Pb/hs] $\la$ 0.9.
An [r/Fe]$^{\rm ini}$ = 0.0 is adopted. }
\label{HE1001-0243_masseron10_bab10d20d50m1p5z2m5_nr_diffdiln5}
\end{figure}

\begin{figure}
\includegraphics[angle=-90,width=8.5cm]{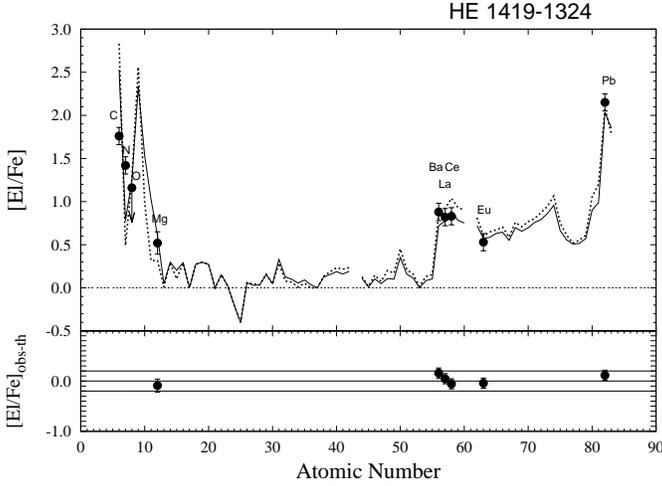}
\vspace{2mm}
\caption{Spectroscopic [El/Fe] abundances of the giant HE 1419-1324  
([Fe/H] = $-$3.05; {\textit T}$_{\rm eff}$ = 4900 K; log $g$ = 1.8) 
compared with AGB models of initial mass $M$ = 1.3 and 1.5 $M_{\odot}$, 
cases ST/14 and ST/2, dil $\sim$ 1.0 - 2.0 dex, dotted and solid lines,
respectively.
Observations are from \citet{masseron10} ([Pb/hs] = 1.31).
An [r/Fe]$^{\rm ini}$ = 0.5 is adopted. }
\label{HE1419-1324_masseron10_bab10d9d1p3m1p5z2m5_rp0p5_diffdiln5n20}
\end{figure}

\subsubsection{CS 30315--91} 
\label{CS91}

As for CS 29495--42, for the subgiant CS 30315--91 the
occurrence of the FDU remains uncertain 
([Fe/H] = $-$1.68; {\textit T}$_{\rm eff}$ = 5536 K, log $g$ = 3.4,
\citealt{lucatello04PhD}).
Due to the mild $s$-process enhancement (CEMP-$s$I),
a large dilution has to be applied even with AGB models of low initial mass.
Solutions with negligible dilution may be found only with 
 $M^{\rm AGB}_{\rm ini}$ = 1.2 $M_\odot$ models. 
AGB models with higher initial mass ($M^{\rm AGB}_{\rm ini}$ = 1.5
and 2 $M_\odot$) agree with the observed [Na/Fe] $\sim$ 0.
 No initial $r$-process enrichment is needed in order to interpret the observed 
[La/Eu] $<$ 0.89.

\subsection{CEMP-$s$II$/r$II with [r/Fe]$^{\rm ini}$ $\sim$ 1.5}

The turnoff star HE 0131-3953 observed by \citet{barklem05}, with a limited 
number of spectroscopic data among the ls elements (only Sr with one detected line), 
may belong to this group due to the high $s$- and $r$-process enhancement.
For the giant CS 22891--171 by \citet{masseron10} no observations
are available among the ls elements, but the authors detected Ba, La, Ce
and Eu, adding this star to the CEMP-$s$II$/r$II class.
The giant CS 30338--089 has been detected by \citet{lucatello04PhD}
and by \citet{aoki07}, with discrepant metallicities ($\Delta$[Fe/H] 
$\sim$ 0.7 dex). \citet{lucatello04PhD} found a high Eu enhancement, 
not confirmed by \citet{aoki07}, who detected only Ba and Na.
At the state of the art this star belongs to CEMP-$s$II$/r$II, but further measurements are needed.

\subsubsection{HE 0131--3953 (Fig.~\ref{HE0131-3953Bark05_bab10d7d8d10m1p5z5m5rp1p5_dil0p1n5})}
\label{HE0131}

\begin{figure}
\includegraphics[angle=-90,width=8.5cm]{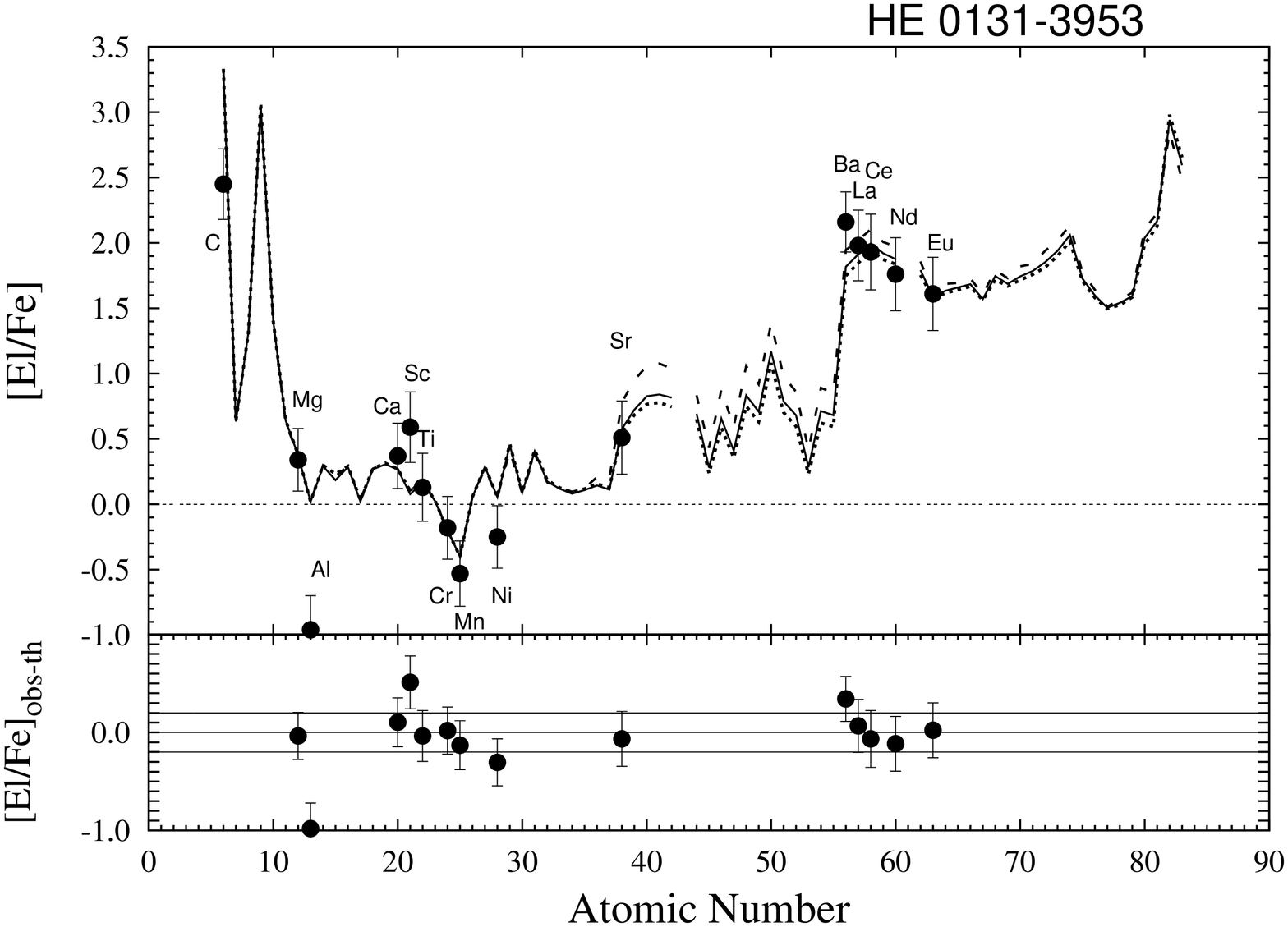}
\vspace{2mm}
\caption{Spectroscopic [El/Fe] abundances of the turnoff/subgiant HE 0131--3953 
([Fe/H] = $-$2.71; {\textit T}$_{\rm eff}$ = 5928 K; log $g$ = 3.83, before the FDU)
compared with AGB models of $M^{\rm AGB}_{\rm ini}$ 
= 1.3 $M_{\odot}$, cases ST/10 (dotted line), ST/12 (solid line), ST/15 
(dashed line), and no dilution.
Observations are from \citet{barklem05}, who detected
[hs/ls] = 1.5.
We predict [Pb/Fe]$_{\rm th}$ $\sim$ 3.
An initial $r$-process enrichment [r/Fe]$^{\rm ini}$ = 1.5
is assumed.}
\label{HE0131-3953Bark05_bab10d7d8d10m1p5z5m5rp1p5_dil0p1n5}
\end{figure}

This star lies close to the turnoff (\citealt{barklem05},
[Fe/H] = $-$2.71; {\textit T}$_{\rm eff}$ = 5928 K; log $g$ = 3.83). 
The low [Mg/Fe] = 0.3 (three lines) is 
interpreted with AGB models of initial mass
$M^{\rm AGB}_{\rm ini}$ = 1.3 $M_{\odot}$, 
low $s$-process efficiencies (ST/10, ST/12 and ST/15) and 
no dilution (Fig.~\ref{HE0131-3953Bark05_bab10d7d8d10m1p5z5m5rp1p5_dil0p1n5}). 
A [Na/Fe] lower than 0.6 dex would confirm this interpretation.
Solutions with higher initial mass predict a large [Mg/Fe]$_{\rm th}$,
about 0.2 dex out of the error bars, and [Na/Fe]$_{\rm th}$
$\geq$ 1 dex.
The lead prediction is very uncertain because it depends strictly on the 
ls elements, for which only Sr is observed. At this state, we predict
[Pb/Fe]$_{\rm th}$ $\sim$ 3.
An initial $r$-process enrichment of [r/Fe]$^{\rm ini}$ = 1.5 
is needed in order to predict [La/Eu] = 0.3.

\subsubsection{CS 22891--171 (Fig.~\ref{mnras_CS22891-171_masseron10_bab10d30m2z5m5_rp1p8_dil0p3n26})}
\label{CS171}

The giant CS 22891--171 ([Fe/H] = $-$2.25; {\textit T}$_{\rm eff}$ = 5100 K; log $g$ = 1.6)
has been analysed by \citet{masseron10}.
No ls elements have been detected, while the observed [Pb/hs] $\sim$ 0 suggests
a low $s$-process efficiency.
In Fig.~\ref{mnras_CS22891-171_masseron10_bab10d30m2z5m5_rp1p8_dil0p3n26}
 we provide a possible theoretical interpretation with
an AGB model of initial mass $M$ = 2 $M_{\odot}$ and a very
low $^{13}$C-pocket efficiency (case ST/45).
High [hs/Fe] ($\ga$ 2 dex), together with a low [Pb/hs] ratio ($\la$ 0), 
can not be obtained by AGB models with low initial mass ($M^{\rm AGB}_{\rm ini}$ 
= 1.3 -- 1.4 $M_{\odot}$), which undergo a lower number of thermal pulses with TDU.
Even for the case shown in Fig.~\ref{mnras_CS22891-171_masseron10_bab10d30m2z5m5_rp1p8_dil0p3n26}
a low dilution is applied, $dil$ = 0.3 dex, hardly compatible with a giant having
suffered the FDU.
The low upper limit for oxygen is about 1 dex
lower than the predictions.
No satisfactory theoretical interpretations may be
found for the observed C, N and O. However, we highlight 
that in CEMP stars these light elements are affected by large uncertainties. 
A high initial $r$-process enhancement [r/Fe]$^{\rm ini}$ = 1.8
is required to explain the the observed [La/Eu] ratio ($\sim$ 0.4). 
For a further discussion of this star, \citet{masseron10} refer to a paper in preparation.

\begin{figure}
\includegraphics[angle=-90,width=8.5cm]{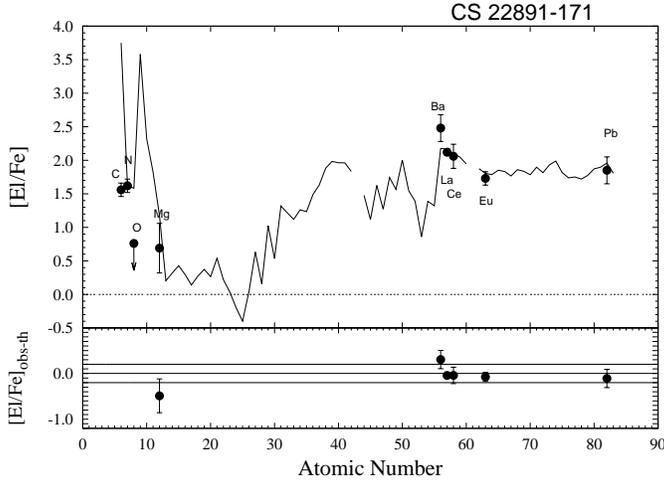} 
\vspace{2mm}
\caption{Spectroscopic [El/Fe] abundances of the giant CS 22891--171 
([Fe/H] = $-$2.25; {\textit T}$_{\rm eff}$ = 5100 K; log $g$ = 1.6) 
compared with AGB models of initial mass $M$ = 2 $M_{\odot}$, 
case ST/45 and dil $\sim$ 0.3 dex.
Observations are from \citet{masseron10}, who detected [Pb/hs] = $-$0.4.
An initial $r$-process enrichment of [r/Fe]$^{\rm ini}$ = 1.8 is adopted. }
\label{mnras_CS22891-171_masseron10_bab10d30m2z5m5_rp1p8_dil0p3n26}
\end{figure}

\subsubsection{CS 30338--089} 
\label{CS089}

For this giant, discrepant metallicities have been detected by 
\citet{aoki07} ([Fe/H] = $-$2.45) and \citet{lucatello04PhD} ([Fe/H] = $-$1.75).
As shown in Paper II, Table~A.1, \citet{lucatello04PhD} 
found lower [Fe/H] ratios also for other stars compared to other authors. 
This may be due to a systematic effect.
The high [Eu/Fe] detected by \citet{lucatello04PhD} classifies this star as a
possible CEMP-$s$II$/r$, but further investigations are desirable.
We considered here the spectroscopic data by \citet{aoki07}.
The occurrence of the FDU ({\textit T}$_{\rm eff}$ = 5000 K and log $g$ = 2.1) 
together with a high $s$-process enhancement, would exclude solutions with AGB models 
of $M^{\rm AGB}_{\rm ini}$ = 1.3 $M_{\odot}$. Indeed, the large mixing occurring 
during the FDU are simulated by a large dilution, which can not be applied 
if the AGB undergoes a limited number of TDUs.
Possible solutions are listed in Paper II, Table~11, for $M^{\rm AGB}_{\rm ini}$ = 1.5
and 2 $M_{\odot}$ with $dil$ = 0.5 dex (case ST/2).
However, these models disagree with the low observed [Na/Fe] ($\sim$ 0.46).

\subsection{CEMP-$s$II without Eu detection} 

Stars with a limited number of data may be classified in this group:
the main-sequence/turnoff stars SDSS 0924+40, SDSS 1707+58, SDSS 2047+00 
by \citet{aoki08};
the giants HE 0206--1916, HE 0400--2030,  HE 1157--0518,
HE 1319--1935, HE 1429--0551, HE 1447+0102, HE 1523--1155, HE 1528--0409,
HE 2221--0453, HE 2228--0706 by \citet{aoki07}.
The main-sequence/turnoff star CS 29503--010 and the giant HE 0507--1653 by 
\citet{aoki07}, with [Fe/H] $\sim$ $-$1.2, have been discussed in
Section~\ref{CH}.

\subsubsection{SDSS 0924+40, SDSS 1707+58, and SDSS 2047+00
(Figs.~\ref{SDSS0924+40_aoki07B_bab10d6d3m1p5m2z5m5_rp0p5_diffdiln7n26},~\ref{SDSS1707+58_aoki07B_bab10d8d12d16m2z5m5_rp0p5_alf0p5_n26},~\ref{SDSS1707+58_aoki07B_bab10d8d12d16m2z5m5_rp0p5_alf0p5_n26})}
\label{SDSS+40}
\label{SDSS+58}
\label{SDSS+00}

The three main-sequence/turnoff stars analysed by \citet{aoki08} are newly 
discovered CEMP-$s$ stars: 
SDSS 0924+40 ([Fe/H] = $-$2.51; {\textit T}$_{\rm eff}$ = 6200 K; log $g$ = 4.0), 
SDSS 1707+58 ([Fe/H] = $-$2.52; {\textit T}$_{\rm eff}$ = 6700 K; log $g$ = 4.2), 
and SDSS 2047+00 ([Fe/H] = $-$2.05; {\textit T}$_{\rm eff}$ = 6600 K; log $g$ = 4.5).
For SDSS 0924+40 and SDSS 1707+58 only Sr (two lines) among the ls elements and Ba 
(five lines) among the hs elements have been detected. 
As well as Ba, SDSS 2047+00 has Y and Zr measurements (one line).
Lead abundance is provided for SDSS 0924+40, while only an upper limit is available
for SDSS 1707+58.
The spectroscopic abundances of \citet{aoki08} 
were not corrected by a 3D analysis based on NLTE calculations.

SDSS 0924+40 needs AGB models with higher initial mass than 
1.3 $M_{\odot}$ due to the enhanced [Na/Fe] $\sim$ 1.2 observed.
Interpretations with AGB stellar
models are shown in Fig.~\ref{SDSS0924+40_aoki07B_bab10d6d3m1p5m2z5m5_rp0p5_diffdiln7n26}.

\begin{figure}
\includegraphics[angle=-90,width=8cm]{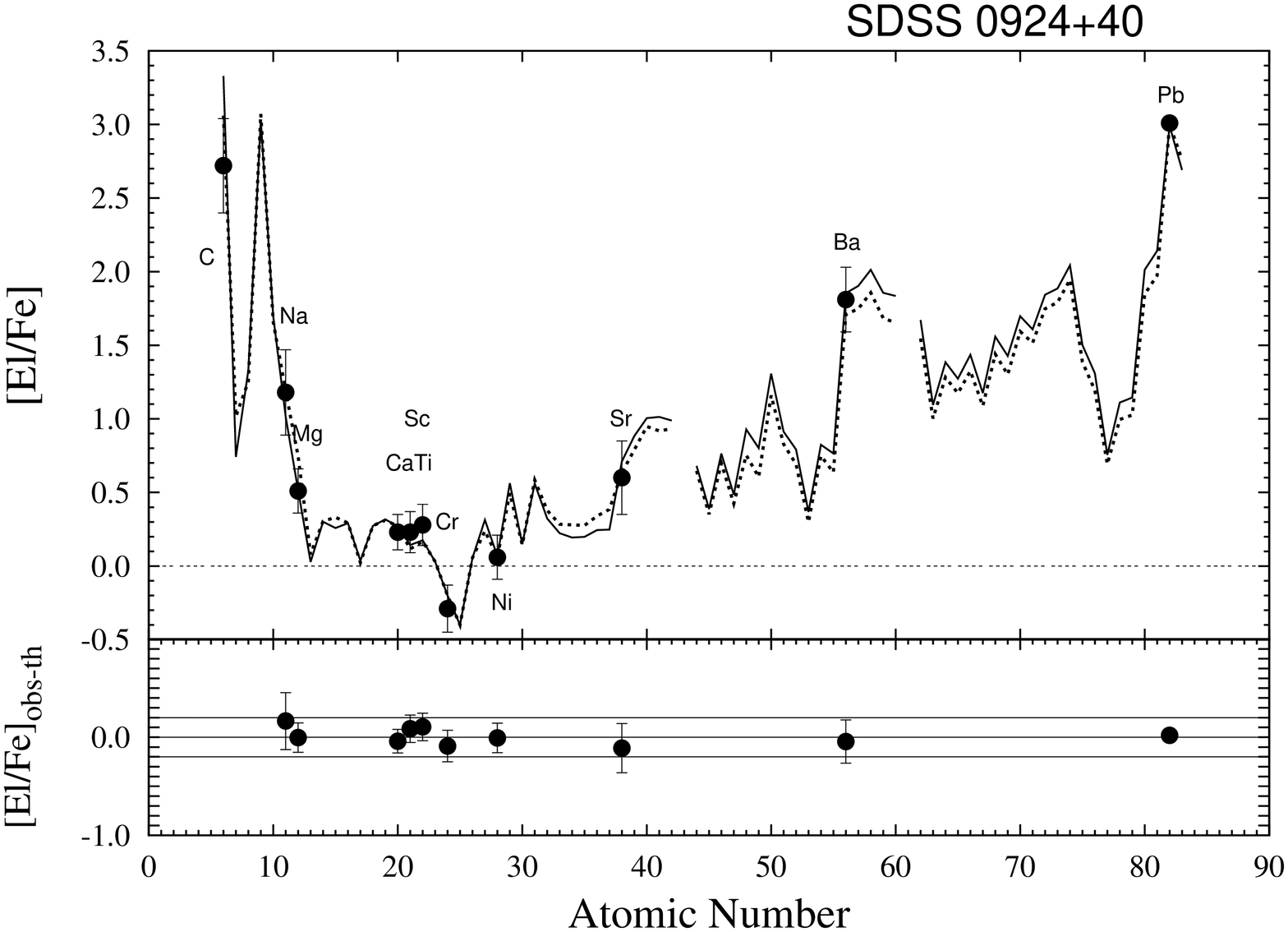}
\vspace{2mm}
\caption{Spectroscopic [El/Fe] abundances of the main-sequence star SDSS 0924+40
([Fe/H] = $-$2.51; {\textit T}$_{\rm eff}$ = 6200 K; log $g$ = 4.0)
compared with AGB models of initial mass $M$ = 1.35 and 2 $M_{\odot}$,
 cases ST/9 and ST/5, dil $\sim$ 0.4 -- 1.0 dex, solid and dotted lines, respectively.
Observations are from \citet{aoki08}, who detected 
[Ba/Sr] $\sim$ [Pb/Ba] $\sim$ 1.2.
The observed [Mg/Fe] agrees better with the $M^{\rm AGB}_{\rm ini}$ = 1.35 $M_{\odot}$
model as shown in the lower panel.
 An initial $r$-process enrichment of [r/Fe]$^{\rm ini}$ = 0.5 is adopted.}
\label{SDSS0924+40_aoki07B_bab10d6d3m1p5m2z5m5_rp0p5_diffdiln7n26}
\end{figure}

SDSS 1707+58 shows rapid radial velocity variations:
according to \citet{aoki08} it belongs probably to a close binary system 
(similarly to HE 0024--2523 studied by \citealt{lucatello03})  
or to a triple system \citep{preston09pasa}.
SDSS 1707+58 is a main-sequence star, with {\textit T}$_{\rm eff}$ = 6700 K
and log $g$ = 4.2 (Fe/H] = $-$2.52).
This star exhibits a very high $s$-process enhancement ([Sr/Fe] = 2.25 and 
[Ba/Fe] = 3.4; \citealt{aoki08}). At present, only an upper limit is measured for Pb,
[Pb/Fe] $\leq$ 3.7. 
A comparable $s$ enhancement was found for CS 29528--028 by \citet{aoki07}.
In Fig.~\ref{SDSS1707+58_aoki07B_bab10d8d12d16m2z5m5_rp0p5_alf0p5_n26},
 we present possible solutions with AGB models of initial
mass $M$ = 2 $M_{\odot}$ and no dilution.
In order to interpret the observed Pb upper limit, $^{13}$C$-$pockets lower than
 case ST/12 are needed,
but the observed [Ba/Fe] is underestimated by about 0.4 dex.
The observed [Na/Fe] ratio is not corrected for a 3D analysis based on NLTE 
calculation, which may reduce this value \citep{aoki07}.
[C/Fe]$_{\rm obs}$ is about 2 dex higher than theoretical predictions. 
However, \citet{aoki08} specified that the C abundance is very uncertain
in this star, because of the relatively low $S/N$ of the spectrum.

\begin{figure}
\includegraphics[angle=-90,width=8cm]{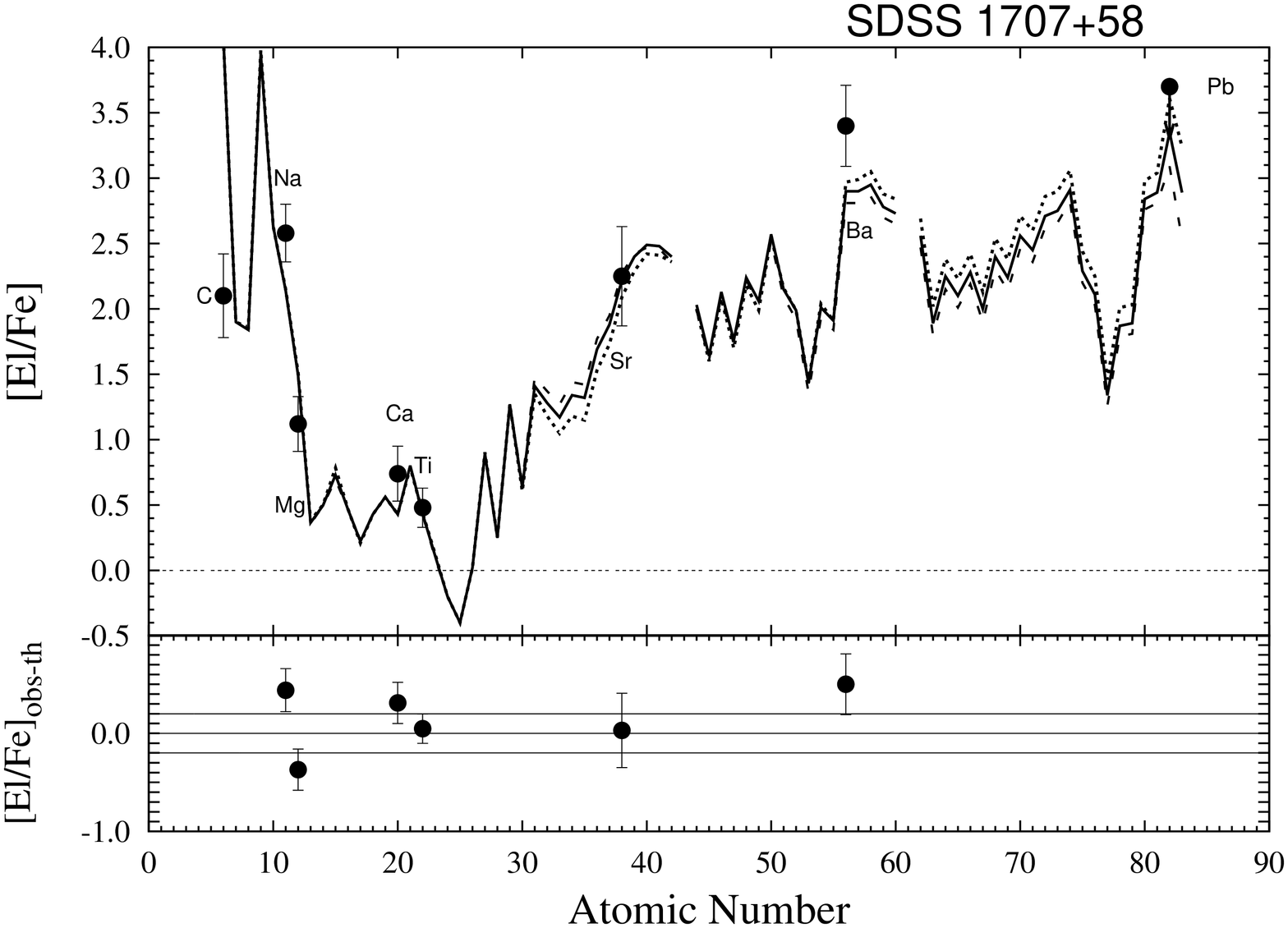} 
\vspace{2mm}
\caption{Spectroscopic [El/Fe] abundances of the main-sequence star SDSS 1707+58 
([Fe/H] = $-$2.52; {\textit T}$_{\rm eff}$ = 6700 K; log $g$ = 4.2)
compared with AGB models of initial mass $M$ = 2 $M_{\odot}$, cases ST/12
(dotted line), ST/18 (solid line), ST/24 (dashed line), and no dilution.
Observations are from \citet{aoki08}, who detected 
[Ba/Sr] = 1.2 and [Pb/Ba] $\leq$ 0.3.
Note that [Na/Fe] has not been corrected with a 3D analysis based on NLTE calculations.
An initial $r$-process enrichment of [r/Fe]$^{\rm ini}$ = 0.5 is adopted. }
\label{SDSS1707+58_aoki07B_bab10d8d12d16m2z5m5_rp0p5_alf0p5_n26}
\end{figure}

For SDSS 2047+00, AGB models with initial masses in the range $M$ = 1.2 -- 2 
$M_\odot$ may equally interpret the $s$-process elements.
Possible solutions are shown in
Fig.~\ref{SDSS2047+00_aoki07B_bab10d8d3m1p5z2m4_rp0p5_diffdiln3n20}.
The observed [Na/Fe] ratio is overestimated by a $M^{\rm AGB}_{\rm ini}$ = 1.5 
$M_{\odot}$ model. We predict [Pb/Fe]$_{\rm th}$ around 2 dex.

\begin{figure}
\includegraphics[angle=-90,width=8.5cm]{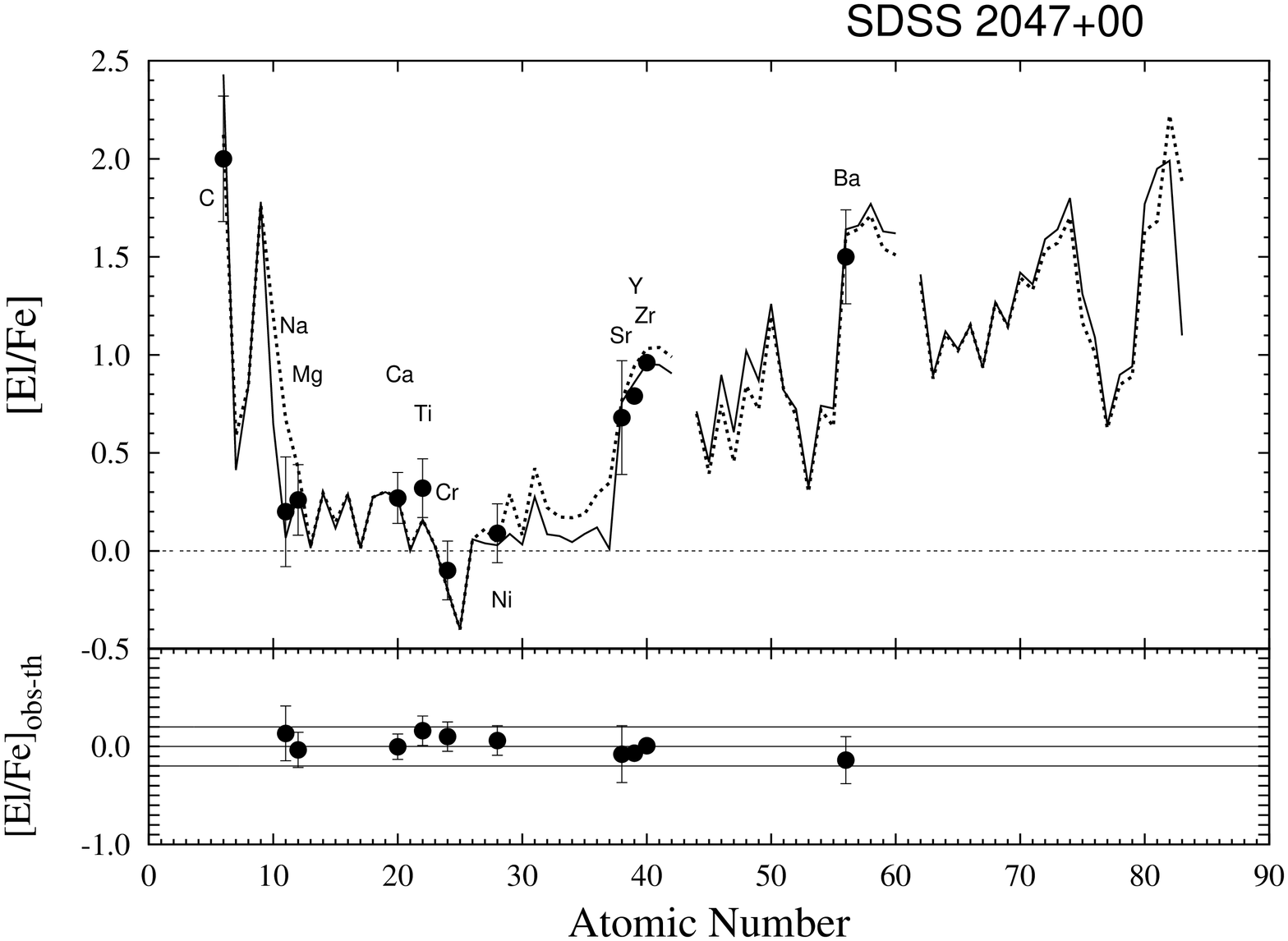}
\vspace{2mm}
\caption{Spectroscopic [El/Fe] abundances of the main-sequence star SDSS 2047+00
 ([Fe/H] = $-$2.05; {\textit T}$_{\rm eff}$ = 6600 K; log $g$ = 4.5)
compared with AGB models of initial mass $M$ = 1.2 and 1.5 $M_{\odot}$, 
cases ST/12 and ST/5, dil = 0.0 - 1.4 dex, solid and dotted lines, 
respectively. Observations are from \citet{aoki08}, who detected 
[hs/ls] = 0.7.
The differences plotted in the lower panel refer to a
$M^{\rm AGB}_{\rm ini}$ = 1.2 $M_{\odot}$ model.
We predict [Pb/Fe]$_{\rm th}$ $\sim$ 2 dex.
An initial $r$-process enrichment of [r/Fe]$^{\rm ini}$ = 0.5 is adopted. }
\label{SDSS2047+00_aoki07B_bab10d8d3m1p5z2m4_rp0p5_diffdiln3n20}
\end{figure}

\subsubsection{Ten CEMP-$s$II giants}
\label{HE0206}
\label{HE0400}
\label{HE1157}
\label{HE1319}
\label{HE1429}
\label{HE1447}
\label{HE1523}
\label{HE1528}
\label{HE2221}
\label{HE2228}

Ten giants studied by \citet{aoki07} (HE 0206--1916, HE 0400--2030, 
HE 1157--0518, HE 1319--1935, HE 1429--0551, HE 1447+0102, HE 1523--1155, 
HE 1528--0409, HE 2221--0453, HE 2228--0706) are briefly discussed here.
All stars have [Fe/H] $\leq$ $-$2.0, with the exception of HE 0400--2030 ([Fe/H] = $-$1.73). 
 Among the $s$-process elements only barium has been measured.  
Due to the barium uncertainty, a range higher than 1 dex may be predicted for 
the ls peak and for lead.
The only indication about the AGB initial mass may come from Na.
However, we underline the spectroscopic uncertainties which affects
Na via NLTE and 3D corrections.
As for similar cases, further studies are required because presently several AGB 
masses may equally fit the observations by changing the $^{13}$C-pocket and the dilution.
A constraint derives from the occurrence of the FDU, for which
a dilution of about 1 dex is needed.
All these stars are giants, with the exception of HE 0400--2030,
for which the FDU is uncertain.
Possible theoretical interpretations have been provided in Paper II, Table~11.
\subsection{CEMP-$s$I without Eu detection} 

Stars with a limited number of data may be classified in this group:
six main-sequence/turnoff stars,
HE 0012--1441 and HE 1410--0004 by \citet{cohen06}, HE 2240--0412 by \citet{barklem05},
CS 22956--28 by \citet{masseron10,sneden03b},
CS 29509--027 by \citet{sneden03b},
G 18--24 by \citet{ish10},
SDSS 0817+26 by \citet{aoki08},
seven giants HE 1305+0132 by \citet{schuler07,schuler08nicX},
HE 1443+0113 by \citet{cohen06},
HE 2227--4044 by \citet{barklem05},
CS 22960--053, HE 0441--0652, HE 1005--1439, HE 2330--0555 by \citet{aoki07}.

\subsubsection{HE 0012--1441, HE 1410--0004 and HE 1443+0113}
\label{HE1410}
\label{HE0012}
\label{HE1443}

\citet{cohen06} reported 9 elements for HE 1410+0213 and 5 for HE 1443+0113. 
Oxygen is detected only for HE 1410$-$0004 ([O/Fe] = 1.18).

HE 1410--0004 is a mild $s$-process star with only
Sr and Ba observed among the $s$-process elements. A very high
upper limit is given for Pb ([Pb/Fe] $\leq$ 3) and Eu.
The only constraint comes from the low [Na/Fe] observed that 
suggests solutions with AGB models of low initial mass. 
Further informations about the three $s$-peaks are needed for
a deeper discussion.

Because of the limited number of elements analysed, all initial
masses included in the range 1.3 $\leq$ $M/M_{\odot}$ $\leq$ 2
can equally interpret HE 1443+0113. 
The high Mg ([Mg/Fe] = 0.9, with 4 lines) observed in HE 0012--1441, 
a double lined spectroscopic binary, rules out solutions
with models of $M^{\rm AGB}_{\rm ini}$ $\leq$ 1.3 $M_{\odot}$. Also for this star
the spectroscopic data are too scarce for speculations about 
possible AGB solutions.

\subsubsection{CS 22956--028 (Fig.~\ref{CS22956-28_sneden03+masseron10_bab10d505560m1p5z1m4_rp0p5_n6})}
\label{CS28}

This main-sequence star has been analysed by \citet{sneden03b}, \citet{lucatello04PhD}
and recently by \citet{masseron10} ([Fe/H] = $-$2.33; {\textit T}$_{\rm eff}$ = 6700 K; 
log $g$ = 3.5). 
It shows the lowest [hs/ls] ratio of the sample, with $\sim$ $-$0.6.
An upper limit is detected for Pb by \citet{masseron10}.
Theoretical AGB models with initial masses in the range 1.3 $\leq$ $M/M_\odot$
$\leq$ 2 predict negative [hs/ls] values at [Fe/H] $\sim$ $-$2.3 only if very
low $^{13}$C-pocket efficiencies are assumed.
These models predict [Pb/Fe]$_{\rm th}$ $\sim$ 0.2 -- 0.5 dex.
An example is shown in 
Fig.~\ref{CS22956-28_sneden03+masseron10_bab10d505560m1p5z1m4_rp0p5_n6}
with AGB models of $M^{\rm AGB}_{\rm ini}$ = 1.3 $M_{\odot}$, cases ST/75, 
ST/80, ST/90, and dil = 0.0 dex.

\begin{figure}                                                                                    
\includegraphics[angle=-90,width=8.5cm]{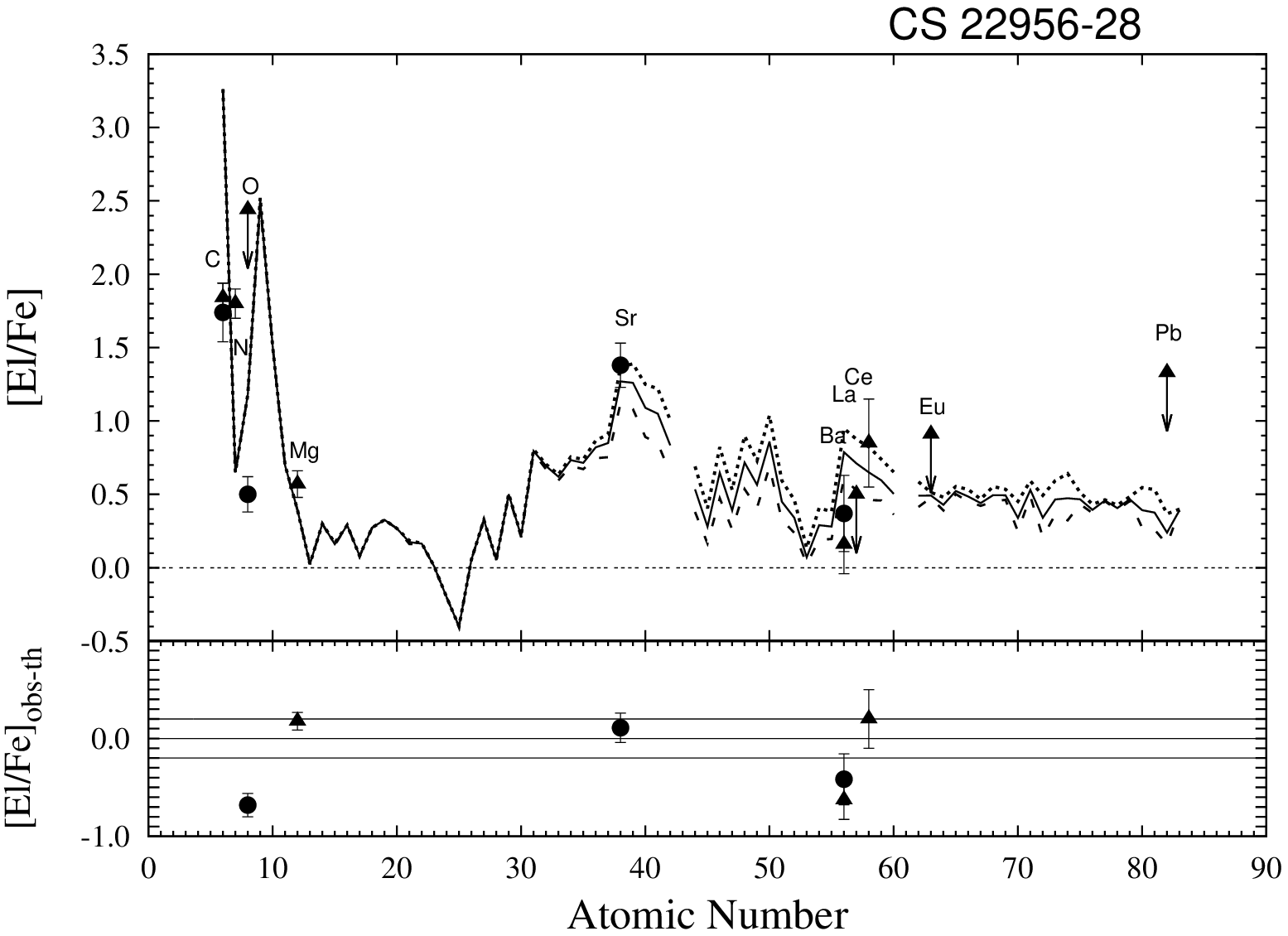}
\vspace{2mm}
\caption{Spectroscopic [El/Fe] abundances of the main-sequence/turnoff star CS 22956--28
([Fe/H] = $-$2.33; {\textit T}$_{\rm eff}$ = 6700 K; log $g$ = 3.5,
 \citealt{masseron10})       
compared with AGB models of $M^{\rm AGB}_{\rm ini}$ = 1.3 $M_{\odot}$, 
cases ST/75 (dotted line), ST/80 (solid line), ST/90 (dashed line), 
and no dilution. Observations are from
\citet{sneden03b} (filled circles) and \citet{masseron10} (filled
triangles), who reported [hs/ls] $\sim$ $-$1.0 and [Pb/hs] $\leq$ 0.8.
An initial $r$-process enrichment of [r/Fe]$^{\rm ini}$ = 0.5 is adopted. } 
\label{CS22956-28_sneden03+masseron10_bab10d505560m1p5z1m4_rp0p5_n6}               
\end{figure}

\subsubsection{CS 29509-027}
\label{CS027S03}

The few data measured for this main-sequence star 
by \citet{sneden03b} (C, O, Sr and Ba) agree with all solutions
with $M^{\rm AGB}_{\rm ini}$ = 1.2 -- 2 $M_\odot$.
No constraints are provided by the limited number of spectroscopic data, 
in particular we can not provide an accurate [Pb/Fe] prediction.

\subsubsection{G 18--24 (= BD 42$^{\circ}$2667)}
\label{G18}

The first spectroscopic observations for G 18--24 at high resolution
by \citet{stephens02} provided [Y/Fe] = 0.18 $\pm$ 0.16; 
[Ba/Fe] = 0.36 $\pm$ 0.46 ([Fe/H] = $-$1.4).
Recently, also \citet{ish10} analysed this giant
([Fe/H] = $-$1.62; {\textit T}$_{\rm eff}$ = 6700 K; log $g$ = 3.5): 
they found a mild enhancement in Y and Ba ([Y/Fe] = 0.6 $\pm$ 0.13; 
[Ba/Fe] = 1.2 $\pm$ 0.17).
No informations about $r$-process elements are available.

We suggest to interpret this star with caution, because C and N 
have not been detected and no conclusive evidence of its binary nature has been found.
Further investigations on this star would be desirable and 
any possible AGB contribution (e.g., Paper II, Table~11) 
needs to be confirmed by carbon and europium detections.

\subsubsection{Four CEMP-$s$I stars}
\label{CS053} %
\label{HE0441}%
\label{HE1005}%
\label{HE2330}%

Among the s-process elements only Ba is measured \citep{aoki07}
 and no accurate theoretical predictions may be provided for these stars.
In similar cases further studies are required, and at present all the 
AGB initial mass may equally fit the observations.
Possible solutions for these star are listed in Paper II, Table~11,
 (CS 22960--053, HE 0441--0652, HE 1005--1439, HE 2330--0555).

\subsubsection{HE 2227--4044 and HE 2240--0412}
\label{HE2227}
\label{HE2240}

For the subgiant HE 2227--4044
only the s-process elements Sr, Ba, and La were detected by 
\citet{barklem05}. Similar spectroscopic data and a similar metallicity
were reported for the subgiant HE 2240-0412. All AGB models in the range between 1.3 $\leq$ 
$M^{\rm AGB}_{\rm ini}$/$M_{\odot}$ $\leq$ 2 may equally interpret the observations,
 with dilutions of $\sim$ 0.8 - 1.7 dex.
We can give a lead estimation of [Pb/Fe]$_{\rm th}$ $\sim$ 2, but
this prediction is very uncertain.
Similar spectroscopic data and metallicity are measured for another
subgiant, HE 2240--0412. 
The same solutions can be adopted for both stars.

\subsubsection{HE 1305+0132}
\label{HEsch}

A preliminary discussion about this star has been provided
by \citet{gallino10}.
This giant ({\textit T}$_{\rm eff}$ = 4462 $\pm$ 100 K; 
log $g$ = 0.80 $\pm$ 0.30)
 was studied by \citet{schuler07,schuler08}.
Discrepant metallicities ([Fe/H] = -2.5 $\pm$ 0.5;
[Fe/H] = -1.9 with higher resolution in 2008)
were found by the authors.
Probably the [F/Fe] = 2.9 detected by \citet{schuler07} is overestimated.
No further investigations were provided ever since. As well
as an overabundant [Ba/Fe] = 0.9 dex, no informations about 
the other $s$-process elements are available for this star.

\subsubsection{SDSS 0817+26}
\label{SDSS+26}

This main-sequence star has been analysed by \citet{aoki08}.
Unfortunately, a low signal-to-noise (S/N) spectra are detected 
with few iron lines available. Large errors have been estimated
for the radial velocity and the difference between the values
obtained from the SDSS spectrum and the HDS spectrum did not
provide conclusive informations about the binarity of this object. 
The limited data do
not permit to speculate about possible contributions from 
an AGB companion: an upper limit for carbon is detected, and only 
a solar Sr abundance ([Sr/Fe] = 0.14 $\pm$ 0.4) and a sligthly enhanced
Ba have been measured ([Ba/Fe] = 0.77 $\pm$ 0.35).
No europium or lead lines are available.
In SDSS 0817+26 no clear excess of carbon and neutron capture 
elements has been found and it is, therefore, excluded from the CEMP-$s$ sample.

\label{onlinematerial}

%
%

%
%
%
\twocolumn

\bsp

\label{lastpage}

\end{document}